\documentclass[12pt,preprint]{aastex}
 \usepackage{emulateapj5,epsfig}
\usepackage{epsfig}
\newcommand{\pf}[1]{\begin{minipage}[t]{0.30\textwidth}
   \psfig{figure={#1},height=0.30\textheight,width=\columnwidth}\end{minipage}
   \quad
}
\def\etal{{\it et al.}\thinspace}

\shorttitle{Multi-frequency Pulsar Profiles}
\shortauthors{T.~H.~Hankins and J.~M.~Rankin}

\begin{document}
\pagestyle{myheadings}
\markboth{ 
{\protect \footnotesize {\protect \bf Time-Aligned Profiles} \hfill V2.5 \hfill \today \hfill }}
{\protect \footnotesize {\protect \bf Time-Aligned Profiles} \hfill V2.5 \hfill \today \hfill Page }
\title{\bf Arecibo Multi-frequency Time-Aligned Pulsar Average-Profile and Polarization Database}
 
\author{Timothy H. Hankins}
\affil{Physics Department, New Mexico Tech}
\affil{Socorro, NM 87801}
\email{thankins@nrao.edu}
\and
\author{Joanna M. Rankin}
\affil{Physics Department, University of Vermont}
\affil{Burlington, VT  05401}
\email{Joanna.Rankin@uvm.edu}

\begin{abstract}

We present Arecibo time-aligned, total intensity profiles for 46 pulsars over an
unusually wide range of radio frequencies and multi-frequency, polarization-angle 
density diagrams and/or polarization profiles for 58 pulsars pulsars at some or all 
of the frequencies 50, 111/130, 430 and 1400 MHz.  The frequency-dependent 
dispersion delay has been removed In order to align the profiles for study of their 
spectral evolution and wherever possible the profiles of each pulsar are displayed 
on the same longitude scale.  Most of the pulsars within Arecibo's declination range 
that are sufficiently bright for such spectral or single pulse analysis are included in 
this survey---and the calibrated pulse sequences are available by web download 
for further study.  

\end{abstract}

\section{Introduction}

As part of a program to study the geometrical constraints on the
emission mechanism of pulsars, we have recorded a set of total
intensity average profiles at a number of frequencies from 25 to
4800\,MHz at the Arecibo Observatory. The data were time-tagged 
so that through use of a suitable timing model for the pulsar, we 
were able to shift the arrival times of the pulses, modulo one stellar 
rotation period, to the solar system barycenter. Then by assuming 
a dispersion measure {\it DM}, we shifted all the profiles to infinite 
frequency arrival time. In many cases the dispersion measure was 
the only free parameter for alignment.  The techniques used were 
similar to those of Hankins \& Rickett (1986)\nocite{hr86}. In addition 
we have observed a large overlapping group of pulsars at selected 
frequencies with full Stokes' parameters. We present these two sets 
of observations in a coordinated fashion, so that they can be used 
for studies of the pulsar magnetospheric structure.

\section{Observations}
The two main programs of time-aligned total-power (P868) and 
polarimetric (P1260) observations presented here were made at 
the Arecibo Observatory during a number of sessions between 
1988 and 1992.  A few observations from previous polarimetric 
surveys (P110 at 430 MHz and P467 at 21 cm), however, are also 
presented in order to extend the completeness of our coverage. 

\subsection{Multiple frequency total intensity profiles}

For the total intensity observation program, circularly polarized feeds 
were used at 430, 1400, 2380, and 4800\,MHz and crossed-dipole 
linearly polarized feeds (with circular hybrids) were used at 130/111, 
49, and 26\,MHz. All profiles were obtained using the then existing 
Arecibo 40-MHz Correlator (Hagen 1987); the autocorrelation 
functions (ACFs) were accumulated into typically 1024 
pulsar-synchronous phase bins spanning one pulsar period, and 
then dumped to tape every two minutes.
 
These ACFs were then Fourier transformed off-line to obtain power
spectra, and the appropriate dispersion delays between adjacent
spectral channels were removed. The profiles from all spectral 
channels were then co-added to produce a dedispersed average 
profile. The two-minute averages were edited to remove interference 
and other corruption, then synchronously added together to form (in 
many cases) a high-signal-to-noise ratio (hereafter S/N) average 
profile. The time-tag of the first sample was used to relate the times 
of arrival to the solar system barycenter in terms of the pulsar phase 
at infinite frequency.  Thus if the dispersion delays strictly obey the 
cold plasma dispersion law, then the pulses align in pulse phase 
exactly as they would if the observer were in the immediate pulsar 
neighborhood.  The dispersion measure {\it DM} is, however, a free 
parameter in the phase alignment, and we found that adjustments 
from the tabulated values (Taylor, Manchester \& Lyne 1993, 
Manchester 2002) \nocite{tml93,man2002} were required for optimal 
alignment for many of the pulsars. The dispersion measures and 
timing models used for alignment are listed in Table \ref{table1}.

To test and monitor the timing stability of the system during each
observing session, we recorded a reference profile from a strong 
pulsar using the same sampling parameters each time.  From these 
profiles systematic timing offsets could be determined and applied 
to data from other pulsars. Since it was not possible to observe 
throughout the whole range of frequencies in a single observing 
session, we recorded reference profiles for each pulsar, usually at 
430 MHz, to confirm its tabulated timing model.  In a number of cases 
we had to make minor adjustments in the timing model to achieve 
alignments among profiles recorded at the same frequency. After 
any timing issues were resolved, the dispersion measure was 
adjusted to obtain pulse-phase alignment among profiles at 
different frequencies. Since the change in pulse phase $\phi$ as 
a function of radio frequency $\nu$, $d\phi/d\nu \propto \nu^{-3}$, 
we used the lowest frequency pairs with good S/N to determine 
any adjustment to {\it DM}. 

The fiducial point of a pulse profile to be used for inter-frequency
alignment depends upon the pulse shape. Usually aligning core
components works well for pulsars with odd numbers of components. 
We aligned either the peak of the core component or the (symmetrical)
half-power points for these pulsars. For double profiles, where the
separation between the main components is clearly frequency 
dependent, we used the centroid between the two components.

\subsection{Polarimetry}

Most of the primary polarimetry observations were carried out during 
a set of sessions in October 1992; a few, however, come from earlier 
observations in March 1992 and January 1990.  All used the then 
existing Arecibo 40-MHz Correlator, such that the average polarimetry
was carried out in a continuous mode, whereas the pulse-sequence
observations entailed use of special programs which gated the
correlator synchronously with the pulsar.  In both cases the basic data
recorded at the telescope were the auto- and cross-correlation functions
of the left- and right-hand channel voltages.  The 21-cm observations 
used a 20-MHz bandwidth and the lower a maximum of 10-MHz at 430 
MHz, 2.5 MHz at 111/130 MHz, and 625 kHz at 49 MHz. A minimum of 
32 correlation lags were retained in order to reduce the dispersion 
delay across the bandpass to usually negligible levels.  The resolution 
was then often essentially the correlator dump time for the 
pulse-sequence observations and the averaging-bin size for the 
profile-polarization measurements.  These parameters are tabulated in 
Table \ref{table_polarization}.  The Arecibo 40-MHz Correlator is 
described by Hagen (1987)\nocite{hag1987} and the continuous and 
gated observing software by Perillat (1988, 1992)\nocite{per1988,per1992}.  
The measured correlation functions were scaled, 3-level-sampling 
corrected, and Fourier transformed to produce raw Stokes parameters, 
which were in turn corrected (channel by channel) for dispersion, 
Faraday rotation, instrumental delays, and all of the known feed 
imperfections as determined by full-sky tracks of pulsar B1929+10 and 
other sources.  These methods are more completely described in the 
Appendix.  During the course of our analyses, we discovered that the 
instrumental polarization is highly frequency dependent, particularly at 
430 MHz; therefore, the most recent observations represent some of 
the best calibrated ever made at the Arecibo Observatory.  For some 
few, however, not all of the calibration information was available for 
one reason or another.  We found, for instance, that interference rather 
easily corrupted our continuum-source observations, which are needed 
to determine the relative left- and right-hand channel gains, so as to 
calibrate the circular polarization $V$.  In any case, Table \ref{table_polarization}
indicates this gain calibration with a ``c'' when a continuum source 
was used and and ``n'' when computed from the off-pulse noise 
level.  The correction of the Stokes parameters for cross-coupling in 
the feed is denoted by a ``p''.
  
Profile and single-pulse polarimetry were also carried out during these
three sessions at 111.5/130 MHz and near 50 MHz.  The techniques were
virtually identical to those described above except that additional
corrections were required for the changing ionospheric Faraday rotation
and many fewer of our efforts to obtain reliable feed cross-coupling data 
were successful.
  
The older 430-MHz observations were carried out in the early 1970s with
a single-channel polarimeter using various bandwidths and integration
times as indicated by the particular star being observed; these are
tabulated in Table \ref{table_polarization}.  The polarimetry scheme is
described in Rankin \etal\ (1975)\nocite{rcs1975}; and while the nominal 
0.25\% voltage amplitude of the feed crosscoupling---which can produce 
spurious circular polarization at typical levels of about 10\% of the linear 
polarization---was known from radar observations, correction of the Stokes 
parameters was then impossible because the cross-coupling phase was 
unknown.
  
The older 1400-MHz observations were carried out in October 1981, 
again with a single-channel, adding polarimeter.  Here, the bandwidths 
ranged up to 40 MHz, with these and the integration times chosen to 
provide a resolution of about a milliperiod. A serious effort was made for 
the first time to correct the measured Stokes parameters for instrumental
cross-coupling distortion using the ``orthogonal'' approximation described 
by Stinebring \etal\ (1984).

In all cases we make no effort to give absolute polarization position angles.  
Therefore, the position angles in our polarization displays are arbitrary 
within a constant value.  
  
\section{Results}

Our results are summarized in the tables and figures. Table \ref{table1} 
gives the period, period derivative, timing epoch and dispersion-measure 
value used for the multi-frequency alignments, and Tables~\ref{table_polarization} 
\& \ref{table_polarization_a} give the observations date, configuration, 
resolution and rotation-measure value used in the polarimetric analyses. 
Comments on Tables \ref{table1} \& \ref{table_polarization} are given in 
Table~\ref{table2}.  

Rotation-measure values were generally taken from Manchester /etal/ (2002); 
however, those given in boldface with errors in Tables~\ref{table_polarization} 
\& \ref{table_polarization_a} were determined in the course of our polarimetric 
analyses by fitting the polarization position-angle swing across the available 
bandwidth.  The errors in the rotation-measure values were computed from 
the position-angle errors which in turn were computed from the off-pulse noise 
level and represent two standard deviations.  

\begin{table*}[htb]
\caption{Alignment Periods, Period Derivatives, Reference Epoch, and Dispersion Measures\label{table1}}
\begin{center}
\begin{tabular}{clr@{.}llr@{.}lllr}
\tableline
Pulsar     & \multicolumn{1}{c}{$P$ }
                             & \multicolumn{2}{c}{$\dot{P}\times 10^{-15}$ } 
                                         & Epoch        & \multicolumn{2}{c}{{\it DM}}
                                                                 & Reference & Note & Figure\\
           & \multicolumn{1}{c}{(s) }  
                             & \multicolumn{2}{c}{ (s/s)}                     
                                         & (MJD)        & \multicolumn{2}{c}{(pc\,cm$^{-3}$)} &&&\\
\tableline
\tableline
B0301$+$19 & 1.3875836665891 &  1&29613  & 42325.500    & 15&650 & C,C,C,C & {\em a.} & \ref{b1} \\
B0523$+$11 & 0.354437595275  &  0&07362  & 48382.000    & 79&294 & M,M,M,M &          & \ref{b1} \\
B0525$+$21 & 3.74549702041   & 40&0565   & 41993.500    & 51&024 & C,C,C,C & {\em b.} & \ref{b1} \\
B0540$+$23 & 0.2459740892957 & 15&42378  & 48382.000    & 77&698 & M,M,M,M &          & \ref{b2} \\
B0611$+$22 & 0.33492505401   & 59&630    & 42881.000    & 96&86  & M,M,M,H &          & \ref{b2} \\
B0626$+$24 & 0.476622653938  &  1&99705  & 48382.000    & 84&216 & M,M,M,H & {\em c.} & \ref{b2} \\
B0656$+$14 & 0.384885025950  & 55&0134   & 48423.000    & 14&02  & M,M,M,M & {\em d.} & \ref{b3} \\
B0751$+$32 & 1.44234944724   &  1&0802   & 43830.000    & 40&04  & M,M,M,M & {\em e.} & \ref{b3} \\
B0820$+$02 & 0.864872751896  &  0&10366  & 43419.347    & 23&6   & M,M,M,M & {\em f.} & \ref{b3} \\
B0823$+$26 & 0.53066061717618&  1&714776 & 47165.705796 & 19&4750& Mc,Mc,Mc,H & & \ref{b4} \\
B0834$+$06 & 1.273768080098  &  6&7995   & 48362.00     & 12&8579& H,M,M,H & {\em g.} & \ref{b4} \\
B0919$+$06 & 0.43061431085   & 13&7248   & 43890.786    & 27&3091& H,T,T,T & {\em h.} & \ref{b5} \\
B0940$+$16 & 1.08741772517   &  0&1      & 48500.00     & 20&5   & M,M,M,H &          & \ref{b5} \\
B0943$+$10 & 1.097704890211  &  3&4884   & 45781.306    & 15&339 & S,S,S,H & {\em i.} & \ref{b6} \\
B0950$+$08 & 0.253065180710  &  0&229314 & 47165.7973   & 2&9701 & Mc,Mc,Mc,H &       & \ref{b6} \\
B1133$+$16 & 1.187913314434  &  3&7341   & 47177.902    & 4&8472 & Mc,Mc,Mc,H & {\em j.} & \ref{b7} \\
B1237$+$25 & 1.38244915690   &  0&95954  & 47177.91     & 9&277  & Mc,Mc,Mc,H & {\em k.} & \ref{b7} \\
B1530$+$27 & 1.124835519553  &  0&803    & 46387.700    & 14&67  & H,M,M,H &          & \ref{b8} \\
B1541$+$09 & 0.74844817748   &  0&43030  & 42304.00     & 34&99  & T,T,T,T &          & \ref{b8} \\
B1604$-$00 & 0.4218162717899 &  0&30610  & 48419.00     & 10&6831& H,M,M,H &          & \ref{b8} \\
B1612$+$07 & 1.2068002120    &  2&357    & 43891.065    & 21&5   & M,M,M,M &          & \ref{b8}{\it b} \\
B1633$+$24 & 0.490506485351  &  0&1195   & 46075.099    & 24&265 & M,M,M,H &          & \ref{b8}{\it c} \\
B1737$+$13 & 0.803049715913  &  1&454    & 43892.616    & 48&73  & H,M,M,H &          & \ref{b9} \\
B1821$+$05 & 0.752906449119  &  0&225    & 43890.159    & 66&7827& H,M,M,H &          & \ref{b9} \\
B1839$+$09 & 0.38131888145508&  1&0916   & 43890.169    & 49&132 & H,M,M,H &          & \ref{b9} \\
B1842$+$14 & 0.375462510949  &  1&866    & 43986.911    & 41&6   & H,M,M,M &          & \ref{b10} \\
B1900$+$01 & 0.72930163274   &  4&0322   & 42345.50     & 245&   & M,M,M,H &          & \ref{b10} \\
B1907$+$00 & 1.01694545765   &  5&5151   & 42647.00     & 111&   & M,M,M,M &          & \ref{b10}{\it e} \\
B1907$+$03 & 2.330260471     &  4&53     & 43984.933    &  83&5  & M,M,M,H &          & \ref{b10}{\it f} \\
B1915$+$13 & 0.19462634149   &  7&20286  & 42302.00     & 94&494 & T,T,T,M &          & \ref{b11} \\
B1918$+$19 & 0.82103460383   &  0&8952   & 42577.00     & 153&48 & M,M,M,H &          & \ref{b11}{\it g} \\
B1919$+$21 & 1.337301192269  &  1&34809  & 40689.95     & 12&4309& M,M,M,M &          & \ref{b11} \\
B1920$+$21 & 1.07791915514   &  8&1899   & 42547.00     & 217&1  & M,M,M,M &          & \ref{b11}{\it i} \\
B1929$+$10 & 0.226517153473  & 1&15675   & 41704.00     & 3&176  & H,T,T,T &          & \ref{b12} \\
B1933$+$16 & 0.358736247894  & 6&00354   & 42265.00     & 158&53 & H,M,M,M &          & \ref{b12} \\
B1944$+$17 & 0.44061846173   & 0&02404   & 41501.00     & 16&11  & T,T,T,M &          & \ref{b12} \\
B1952$+$29 & 0.426676786488  & 0&00164   & 48415.00     & 7&86   & M,M,M,H &          & \ref{b13} \\
B2016$+$28 & 0.557953408114  & 0&14720   & 40688.50     & 14&1965& C,C,C,H & {\em l.} & \ref{b14} \\
B2020$+$28 & 0.343401720116  & 1&8935    & 47018.182    & 24&623 & H,M,M,H &          & \ref{b14} \\
B2044$+$15 & 1.1382856067    & 0&185     & 43890.254    & 39&71  & M,M,M,H &          & \ref{b14} \\
B2110$+$27 & 1.202851149957  & 2&6225    & 46075.294    & 25&122 & H,M,M,H &          & \ref{b15} \\
B2113$+$14 & 0.440152954660  & 0&290     & 43986.016    & 56&14  & H,M,M,H &          & \ref{b15} \\
B2210$+$29 & 1.00459237450   & 0&4948    & 46074.83     & 74&6   & M,M,M,H &          & \ref{b16} \\
B2303$+$30 & 1.575884744270  & 2&89567   & 42341.00     & 49&575 & M,M,M,H & {\em m.} & \ref{b16} \\
B2315$+$21 & 1.444652673768  & 1&05      & 43987.099    & 20&865 & H,M,M,H & {\em n.} & \ref{b17} \\
\tableline
\end{tabular}
\tablerefs{C: Cordes, 1992, H: this work, M: Manchester, {\em et al.} 2002\nocite{man2002}, Mc: McKinnon, 1990\nocite{mck90}, T: Taylor, {\em et al.} 1993, S: Shibanova (1990)\nocite{sha90}}
\end{center}
\end{table*}

\begin{table*}[htb]
\tablewidth{0pt}
\caption{Polarimetry, Calibration \& Rotation-Measure Information.\label{table_polarization}}
\begin{center}
{\small
\begin{tabular}{cllr@{:}lr@{:}lccr@{.}lcr}
\tableline
Pulsar     & Date A & Date B & \multicolumn{2}{c}{Config A} 
                                             & \multicolumn{2}{c}{Config B} 
                                                        & Resol A & Resol B & \multicolumn{2}{c}{$RM$} & Note & Figure\\
           &          &          & \multicolumn{2}{c}{(MHz/ch)} 
                                             & \multicolumn{2}{c}{(MHz/ch)}
                                                        &($\deg$) &($\deg$) & \multicolumn{2}{c}{(rad-m$^2$)}&&   \\
\tableline
B0301$+$19 & 05/01/74 & 20/10/92 &    2/1&c       & 20/32&cp & 0.85 & 0.31 &$-$8&3& &\ref{b1}\\
B0523$+$11 & 17/10/92 & 20/10/92 &1.25/32&cp & 20/32&cp & 0.41 & 0.41 &78&7   & &\ref{b1}\\
B0525$+$21 & 01/04/74 & 11/10/81 &    2/1&c       &  20/1&cp   & 2.76 & 0.15 &50&96 & &\ref{b1}\\
B0540$+$23 & 05/01/74 & 20/10/92 &   0.1/1&c     & 20/32&cp & 1.19 & 0.38 & 77&58& &\ref{b2}\\
B0611$+$22 & 19/10/92 & 20/10/92 &   5/32&cp    & 20/32&cp & 0.85 & 0.38 & 67&    & &\ref{b2}\\
B0626$+$24 & 16/02/92 & 20/10/92 &  10/64&cp   & 20/32&cp & 1.06 & 0.31 & 82&    & &\ref{b2}\\
B0656$+$14 & 12/02/92 & 17/02/92 &  10/32&cp   & 20/32&cp & 0.43 & 0.37 & 22&    & &\ref{b3}\\
B0751$+$32 & 19/10/92 &                  &  10/32&cp   &           &      & 0.32 &          &$-$7& & &\ref{b3}\\
B0820$+$02 & 19/10/92 & 20/10/92 &  10/32&cp   & 20/32&cp & 0.36 & 0.36 & 13&    & &\ref{b3}\\
B0823$+$26 & 20/10/92 & 15/10/92 &  10/32&cp   & 20/32&cp & 0.43 & 0.34 &  5&9   & &\ref{b4}\\
B0834$+$06 & 01/04/74 & 10/10/81 &    2/1&c        &  20/1&cp  & 0.76 & 0.22 & 3&9   & &\ref{b4}\\
                        &       ?         &        ?        &       ? &          &         &       & 0.70 & 0.70  &   &      & &\ref{b4}\\
B0919$+$06 & 20/10/92 & 11/10/81 &  10/32&cp   &  20/1 &cp & 0.74 & 0.68 &27&25& &\ref{b5}\\
                        & 17/02/92  & 15/02/92 &0.313/256& &0.25/128&     & 2.00 & 1.61 &      &    & &\ref{b5}\\
B0940$+$16 & 17/10/92 &                  &  10/32&cp   &            &     & 0.35 &          & 53&    & &\ref{b5}\\
B0943$+$10 & 17/01/00 & 19/10/92 & 2.5/64&       & 10/32&cp  & 1.18 & 0.33 &  13&3 & &\ref{b6}\\
B0950$+$08 & 06/11/71 & 10/10/81 &   10/1&c      &  20/1&cp   & 0.88 & 0.28 &2&969 & &\ref{b6}\\
                        & 15/02/92 &                   &0.63/128&   &          &       & 1.99 &           &  &        & &\ref{b6}\\
B1133$+$16 & 19/10/92 & 16/10/92 &  10/32&cp  & 20/32&cp  & 0.31 & 0.37 & 3&9    & &\ref{b7}\\
                        & 16/02/92 & 17/01/90 &0.63/128&   &1.25/128&   & 0.50 & 0.70 &    &       & &\ref{b7}\\
B1237$+$25 & 06/01/74 & 11/10/81 &    2/1&c       &  20/1&cp   & 0.50 & 0.10 &9&296 & &\ref{b7}\\
                        &                  & 15/02/92 &         &          & 2.5/64&      & 0.35  &          &   &        & &\ref{b7}{\it d}\\
B1541$+$09 & 14/02/92  & 17/01/00 &2.5/256&     & 10/32&cp & 1.15 & 1.45 & 21&     & &\ref{b8}\\
B1604$-$00 & 14/02/92  & 11/02/73  &2.5/128&     &  10/1&c     & 1.20 & 1.91 & 6&5    &  &\ref{b8}\\
B1737$+$13 & 20/10/92 & 24/10/92 &  10/32&cp  & 20/32&cp  & 0.71 & 0.36  & 73&    & &\ref{b9}\\
B1821$+$05 & 05/01/00 & 19/10/92 &   5/64&cp   & 20/32&cp  & 0.34 & 0.36 &145&   & &\ref{b9}\\
B1839$+$09 &                 & 23/10/92  &           &       & 20/32&cp  &          & 0.38 &53&     & &\ref{b9}\\
B1842$+$14 &                 & 23/10/92  &           &       & 20/32&cp  &          & 0.39 &121&   & &\ref{b10}\\
B1907$+$10 & 21/10/73 & 24/10/92 & 0.05/1&c    & 20/32&cp  & 0.98 & 0.51 &540&   & &\ref{b10}\\
B1915$+$13 & 07/09/73 & 22/10/92 &  0.1/1&c     & 20/32&cp  & 1.82 & 0.33 &233&   & &\ref{b11}\\
B1919$+$21 & 05/01/00 & 22/10/92 & 10/32&cp  & 20/32&cp  & 0.19 & 0.33 &$-$37&& &\ref{b11}\\
B1920$+$21 &                 & 19/10/92 &             &      & 20/32&cp   &         & 0.34 &282&   & &\ref{b11}\\
B1929$+$10 & 24/12/74 & 18/10/92 &      2/1&c    & 20/32&cp   & 1.05 & 0.49 &$-$6&1& &\ref{b12}\\
B1933$+$16 & 24/07/73 &                 &    0.1/1&c   &            &       & 1.66 &           &$-$1&9& &\ref{b12}\\
B1944$+$17 & 27/07/73 & 23/10/92 &      2/1&c   & 20/32&cp    & 2.78 & 0.33 &$-$28&& &\ref{b12}\\
B1946$+$35 & 08/05/74 & 18/10/92 &     ??/1&     & 20/32&??   & 1.69 & 0.35 &           && &\ref{b13}\\
B1952$+$29 & 15/02/00 &                  & 20/32&cp  &           &        & 0.35 &            &$-$18&& &\ref{b13}\\
B2002$+$31 &                 &                   &            &     &            &       & 1.04 & 0.34 &         &  & &\ref{b13}\\
B2016$+$28 & 15/10/92 & 18/10/92 &  10/32&cp & 20/32&cp  & 0.39 & 0.39 &$-$34&6&&\ref{b14}\\
B2020$+$28 & 16/10/92 & 18/10/92 &  10/32&cp & 20/32&cp  & 0.43 & 0.37 &$-$74&8&&\ref{b14}\\
B2044$+$15 & 14/02/92 & 19/10/92 &  10/64&cp & 20/32&cp  & 0.40 & 0.32 & $-$101&&&\ref{b14}\\
B2053$+$21 & 20/10/92 & 18/10/92 &  10/64&cp & 20/32&cp  & 0.35 & 0.35  & {\bf $-$100}&{\bf $\pm$7}& &\ref{b15}\\
B2110$+$27 & 14/02/92 & 22/10/92 &  10/32&cp & 20/32&cp  & 0.35 & ???? &$-$65&    & &\ref{b15}\\
B2113$+$14 & 29/10/92 & 23/10/92 &   5/32&cp & 20/32&cp   & ???  &  0.37  &$-$25&    & &\ref{b15}\\
B2210$+$29 & 16/10/92 & 23/10/92 &  10/32&cp & 20/32&cp & 0.43 &  0.36  &{\bf $-$175}&{\bf $\pm$15}  & &\ref{b16}\\
B2303$+$30 & 15/10/92 & 18/10/92 &  10/32&cp & 20/32&cp & ???  & ???? &$-$84&    & &\ref{b16}\\
                        &                  &                   &            &      &           &      &          & 0.84  &           &     & &\ref{b16}\\
B2315$+$21 & 14/02/92 & 19/10/92 &  10/32&cp & 20/32&cp & 0.35  & 0.35  &$-$37&    & &\ref{b17}\\
\tableline
\end{tabular}
} 
\end{center}
\end{table*}

\begin{table*}[htb]
\tablewidth{0pt}
\caption{Miscellaneous Pulsar Polarimetry Data (Figs.~18 \& 19).\label{table_polarization_a}}
\begin{center}
{\small
\begin{tabular}{ccr@{:}lcr@{.}lc}
\tableline
Pulsar     & Date & \multicolumn{2}{c}{Config} & Resol & \multicolumn{2}{c}{$RM$} & Note \\
                 &           & \multicolumn{2}{c}{(MHz/ch)} 
                                                                                  &($\deg$) & \multicolumn{2}{c}{(rad-m$^2$)}& \\
\tableline
B0045$+$33 & 16/10/92 &  10/32&cp  & 0.40 & \multicolumn{2}{c}{???}    & \\
B0609$+$37 &                  &            &       & 0.53 &       &                                     & \\
B1845$-$01 &                   &            &       & 0.39 &       &                                     & \\
B1859$+$03 &                  &            &       & 0.66 &       &                                     & \\
B1900$+$05 & 24/10/92 &  20/32&cp   & 0.37 & $-$113&                              & \\
B1910$+$20 &                  &            &        & 0.35 &               &                              & \\
B1917$+$00 &                  &            &        & 1.34 &       &                                     & \\
B1919$+$14 & 20/10/92 &    5/64&cp   & 0.44 & {\bf    275}&{\bf $\pm$60}. & \\
B1923$+$04 &                  &            &        & 0.55 &       &                                      & \\
B1951$+$32 & 04/01/00 &            &       & 2.81 & \multicolumn{2}{c}{???}    & \\
B2000$+$32 & 20/10/92 &    5/64&cp   & 1.03 & \multicolumn{2}{c}{$\approx 0$??} & \\
B2028$+$22 & 26/10/92 &    5/64&cp   & 0.40 & {\bf $-$195}&{\bf $\pm$25}& \\
B2034$+$19 & 26/10/92 &  10/32&cp   & 0.35 & {\bf $-$98}&{\bf 5$\pm$8}  & \\
B2053$+$36 & 24/10/92 &  20/32&cp   & ????&  $-$68&                                & \\
B2122$+$13 & 16/10/92 &  10/32&cp   & 0.51 & {\bf  $-$48}&{\bf 3$\pm$3.6}& \\
\tableline
\end{tabular}
} 
\end{center}
\end{table*}

\begin{table*}[htb]
\begin{center}
\caption{Notes to Tables 1 -- 3\label{table2}}
\begin{tabular}{llp{0.8\textwidth}}
Pulsar & Note \\
\tableline
\tableline
B0301$+$19 & {\em a.} & {2380-MHz profile shifted by $-$6.17 ms to correct systematic offset.} \\

B0525$+$21 & {\em b.} & {2380-MHz profile shifted by $-$7.32 ms to correct systematic offset.} \\

B0626$+$24 & {\em c.} & {2380-MHz and 430-MHz profiles shifted by 9.25 and 40.91 ms to correct systematic offset. DM determined from 1408 to 111.5-MHz alignment. The ``bump'' preceding the 430-MHz pulse at longitude $-18^\circ$ is probably spurious.} \\
B0656$+$14 & {\em d.} & {430-MHz profile shifted by $-$11.29 ms to correct systematic offset.} \\
B0751$+$32 & {\em e.} & {Arbitrary alignment by profile centroids.} \\
B0820$+$02 & {\em f.} & {Arbitrary alignment by profile centroids.} \\
B0834$+$06 & {\em g.} & {2380-MHz profile shifted by 8.40 ms to correct systematic offset.} \\
B0919$+$06 & {\em h.} & {4880 and 111.5-MHz profiles shifted by $-$1.42 and $-$9.93 ms to correct systematic offsets.} \\
B0943$+$10 & {\em i.} & {The two 111.5-MHz profiles,  showing the two modes of B0943+10, were recorded 1 year apart. They align well using Shabanova's (1990) timing model\nocite{sha90}. The 430 and 24.8-MHz profiles, recorded 4 days apart, also align well, so the DM determination is based on this frequency pair. Then these two profiles were shifted by 5.92 ms to align with the 111.5-MHz profiles, and the 49.2-MHz profile is arbitrarily shifted by 26.1 ms to align its centroid.}\\
B1133$+$16 & {\em j.} & {49.2-MHz profile shifted by $-$9.48 ms to correct systematic offset.} \\
B1237$+$25 & {\em k.} & {4880-MHz profile shifted by 15.1 ms to correct systematic offset.} \\
B2016$+$28 & {\em l.} & {The dramatic pulse shape change between 111.5 and 430 MHz is consistently observed on different days. The DM used for alignment was adjusted to align the 49.2 and 111.5-MHz profile peaks, though if the profile is actually bifurcating at low frequencies, this alignment may be incorrect.} \\
B2303$+$30 & {\em m.} & {4880 and 1408-MHz profiles shifted by $-$39.40 ms to correct systematic offset.} \\
B2315$+$21 & {\em n.} & {2380-MHz profile shifted by $-$37.12 ms to correct systematic offset.} \\
\tableline
\end{tabular}
\end{center}
\end{table*}

\begin{figure}[htb]
\centerline{ 
\pf{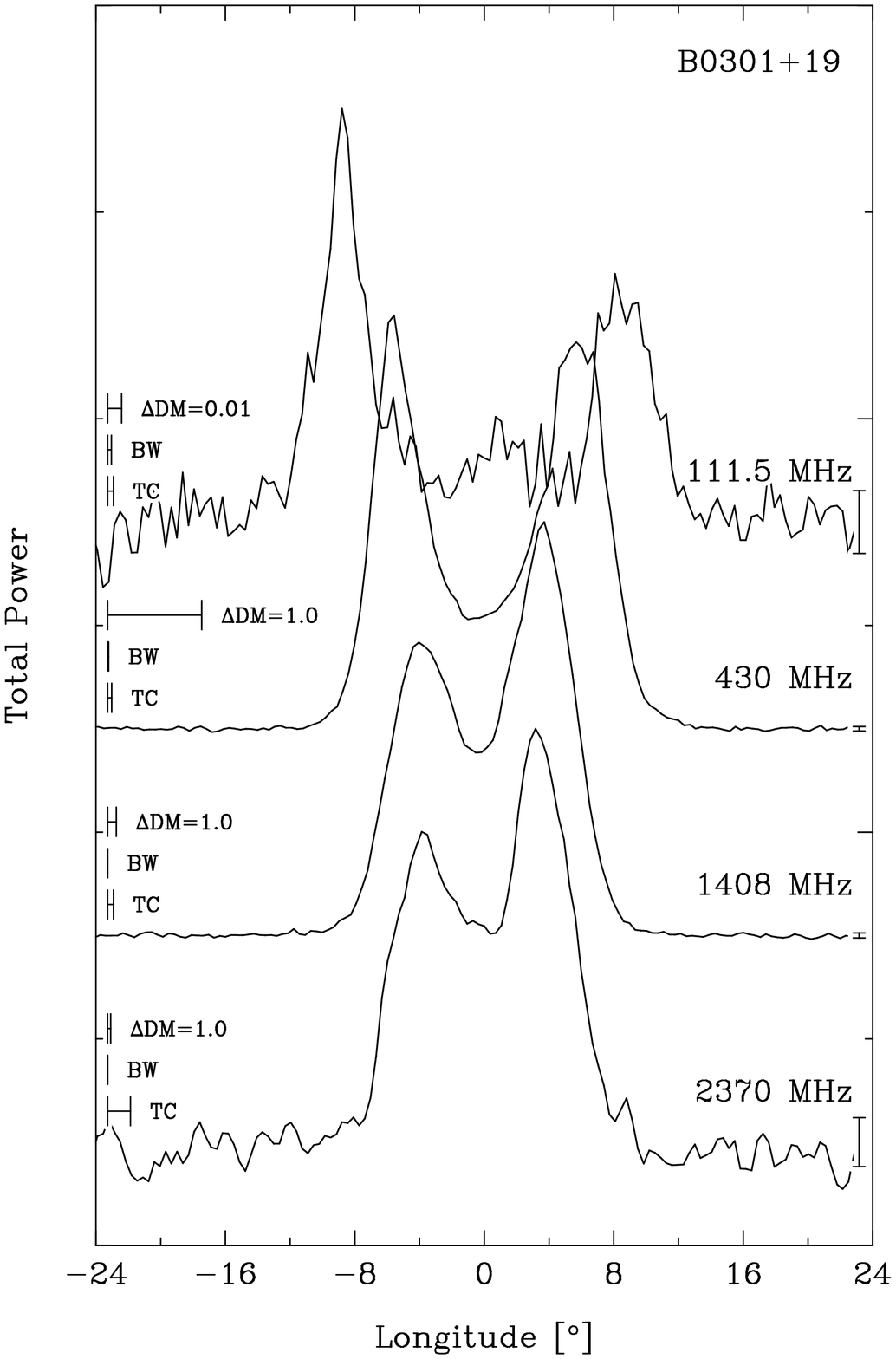}     
\pf{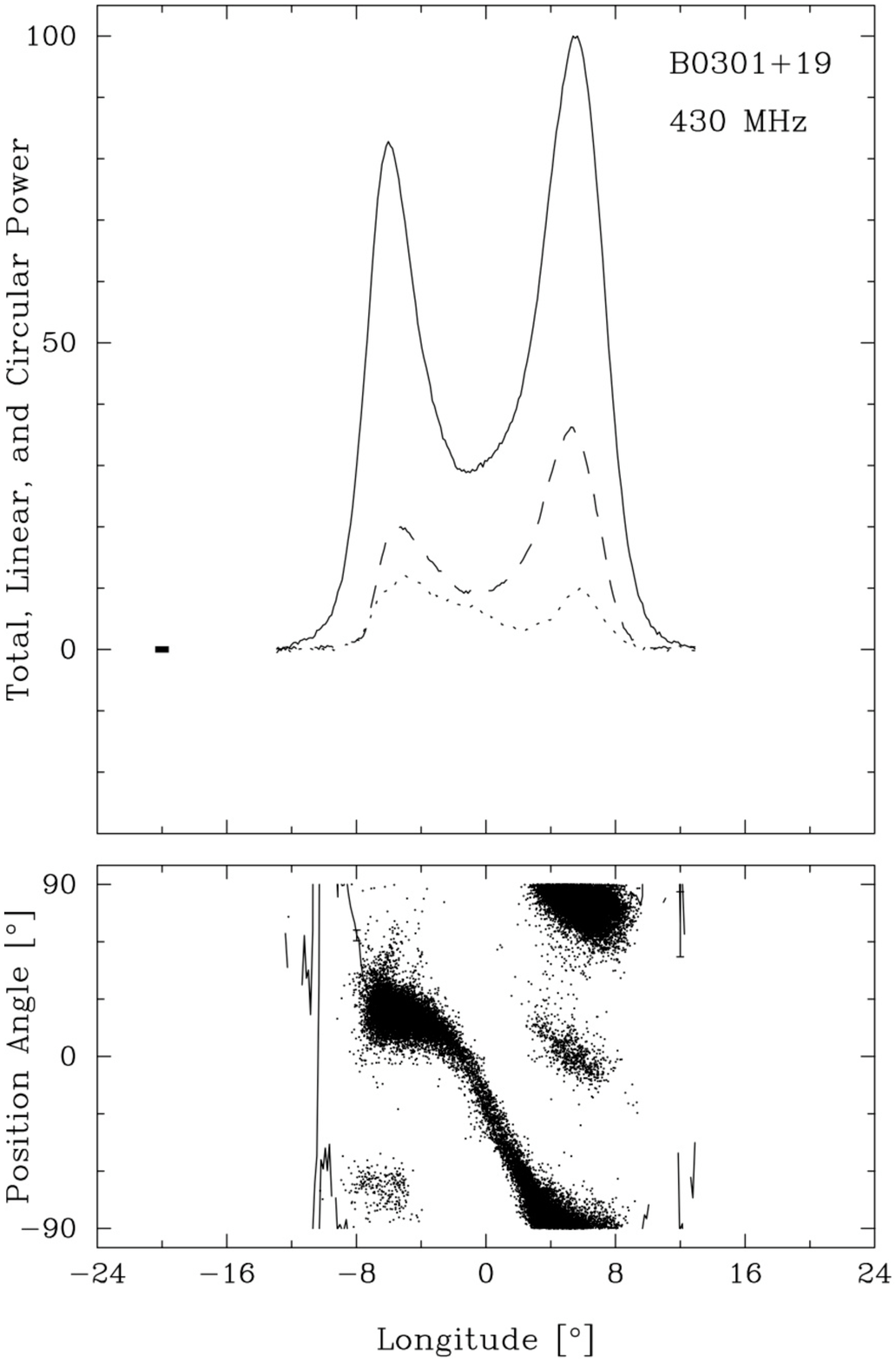} 
\pf{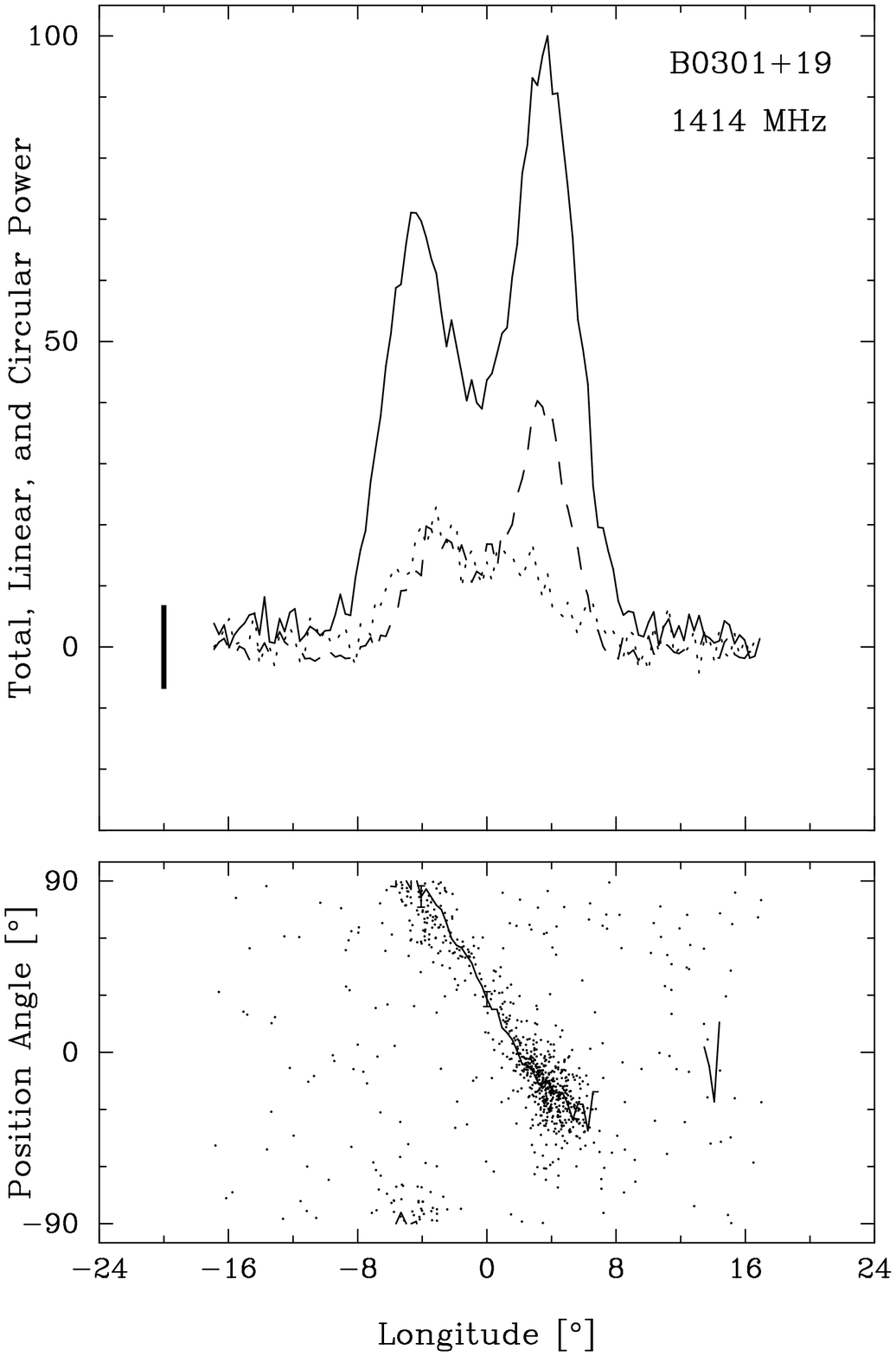} 
}
\quad \\
\centerline{
\pf{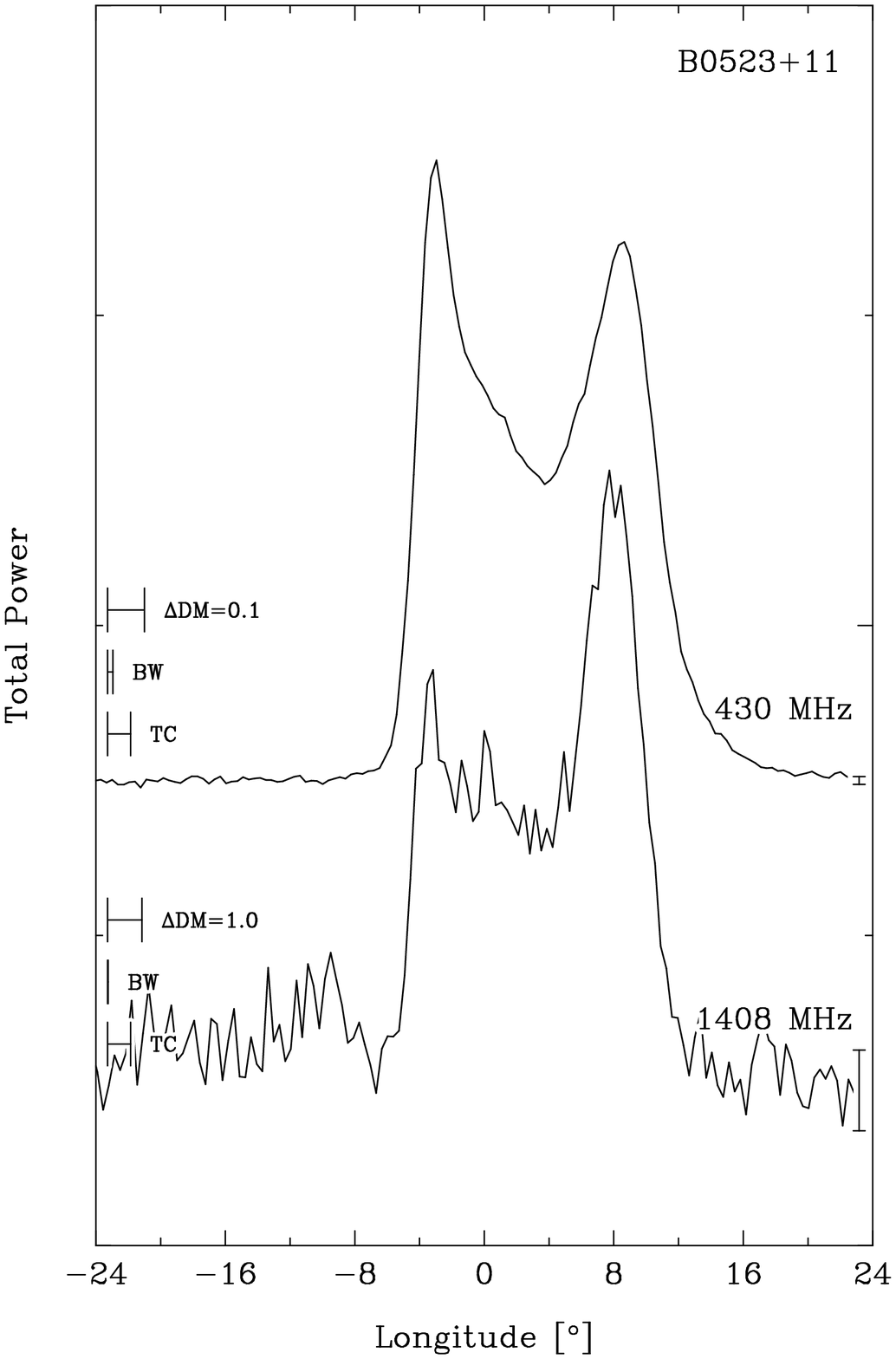}     
\pf{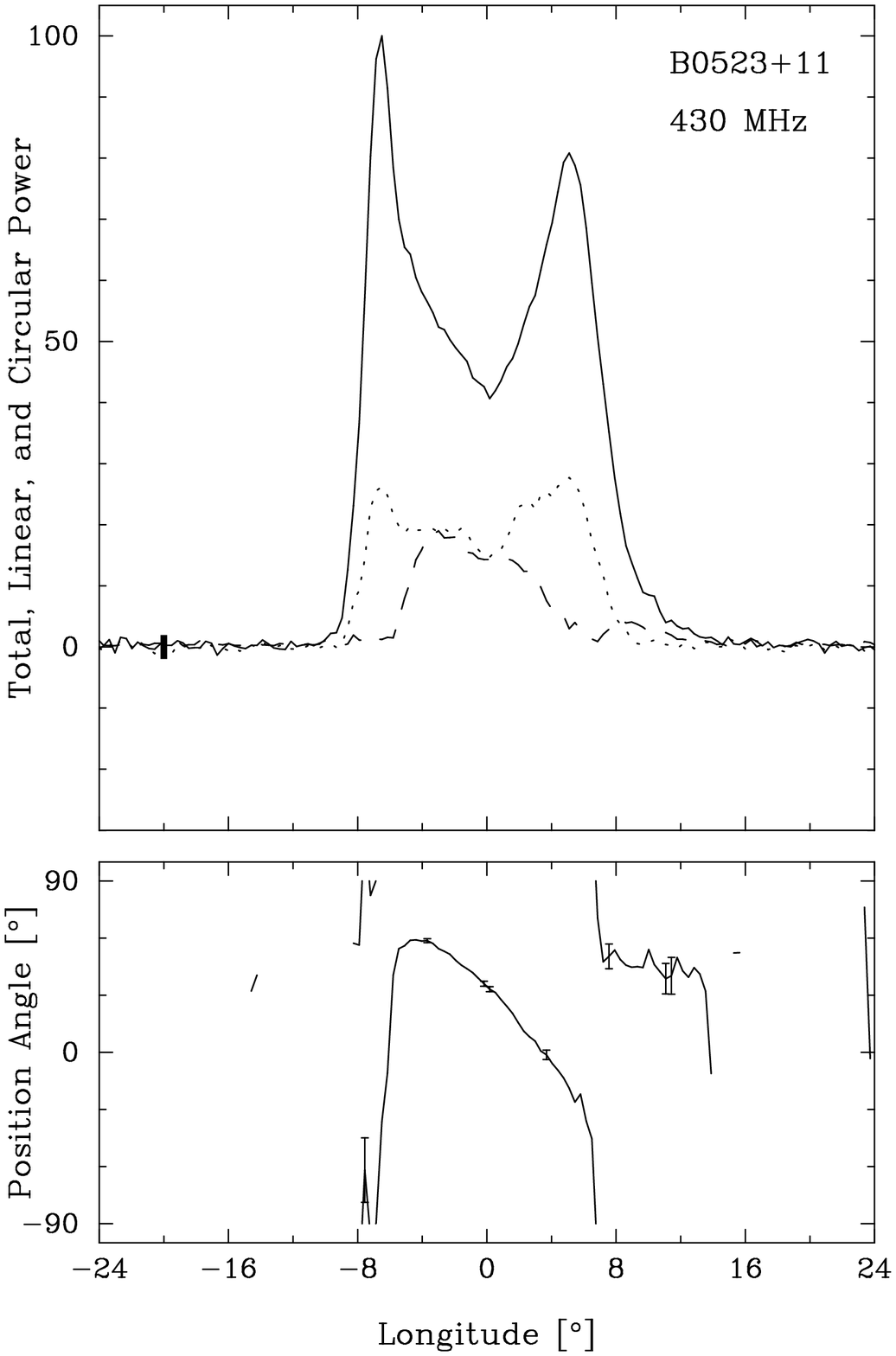} 
\pf{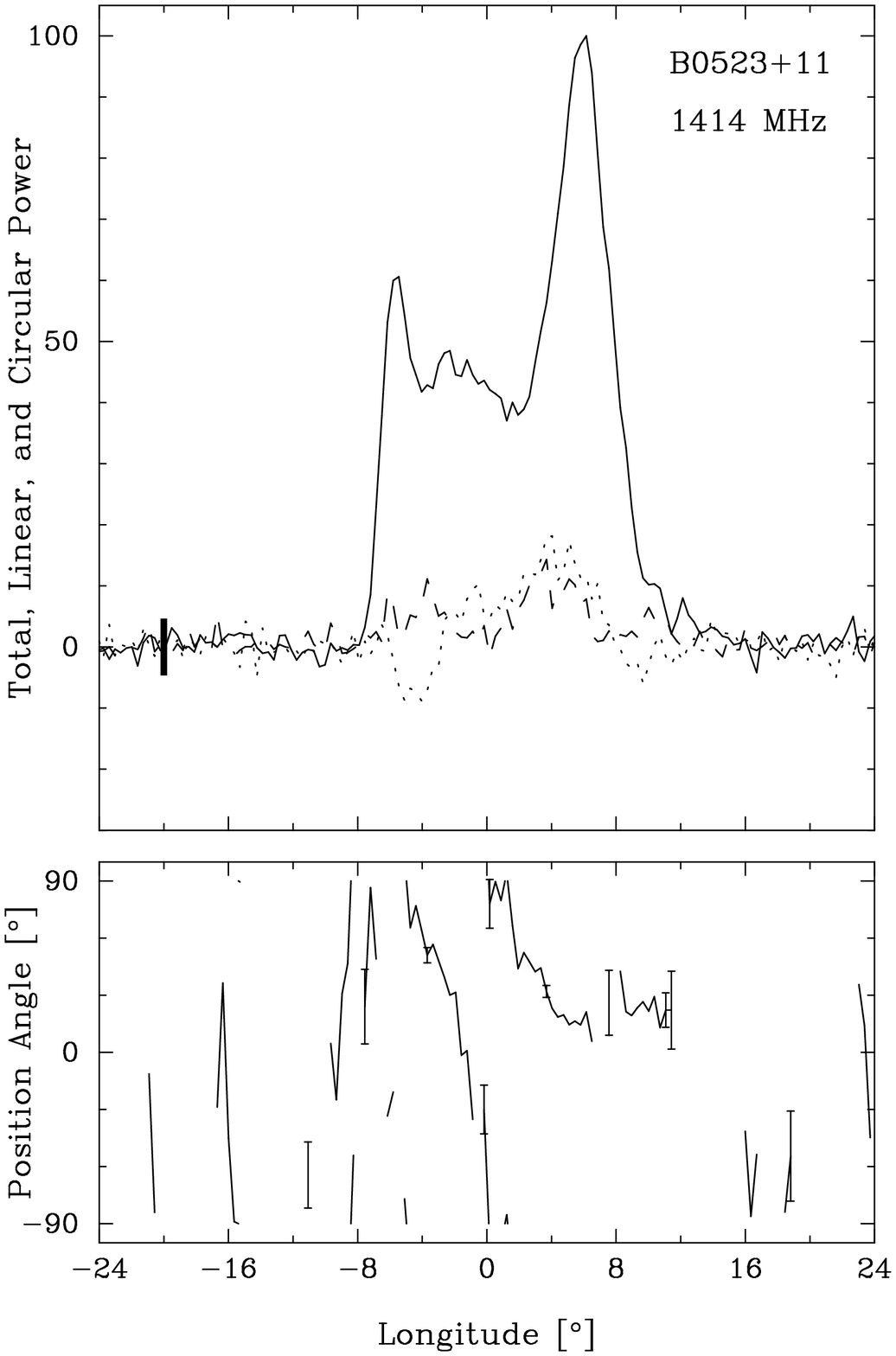} 
}
\quad \\
\centerline{
\pf{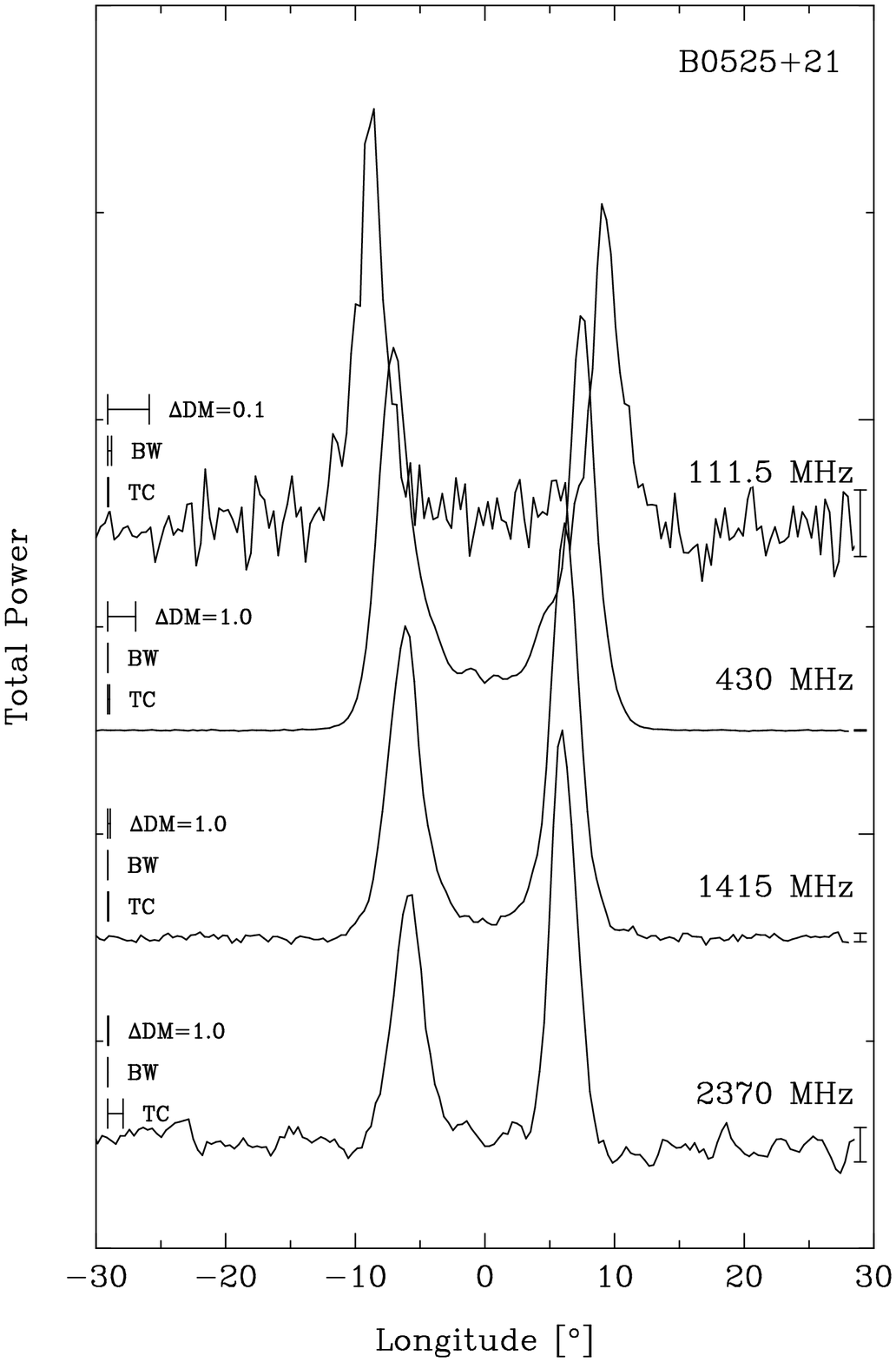}     
\pf{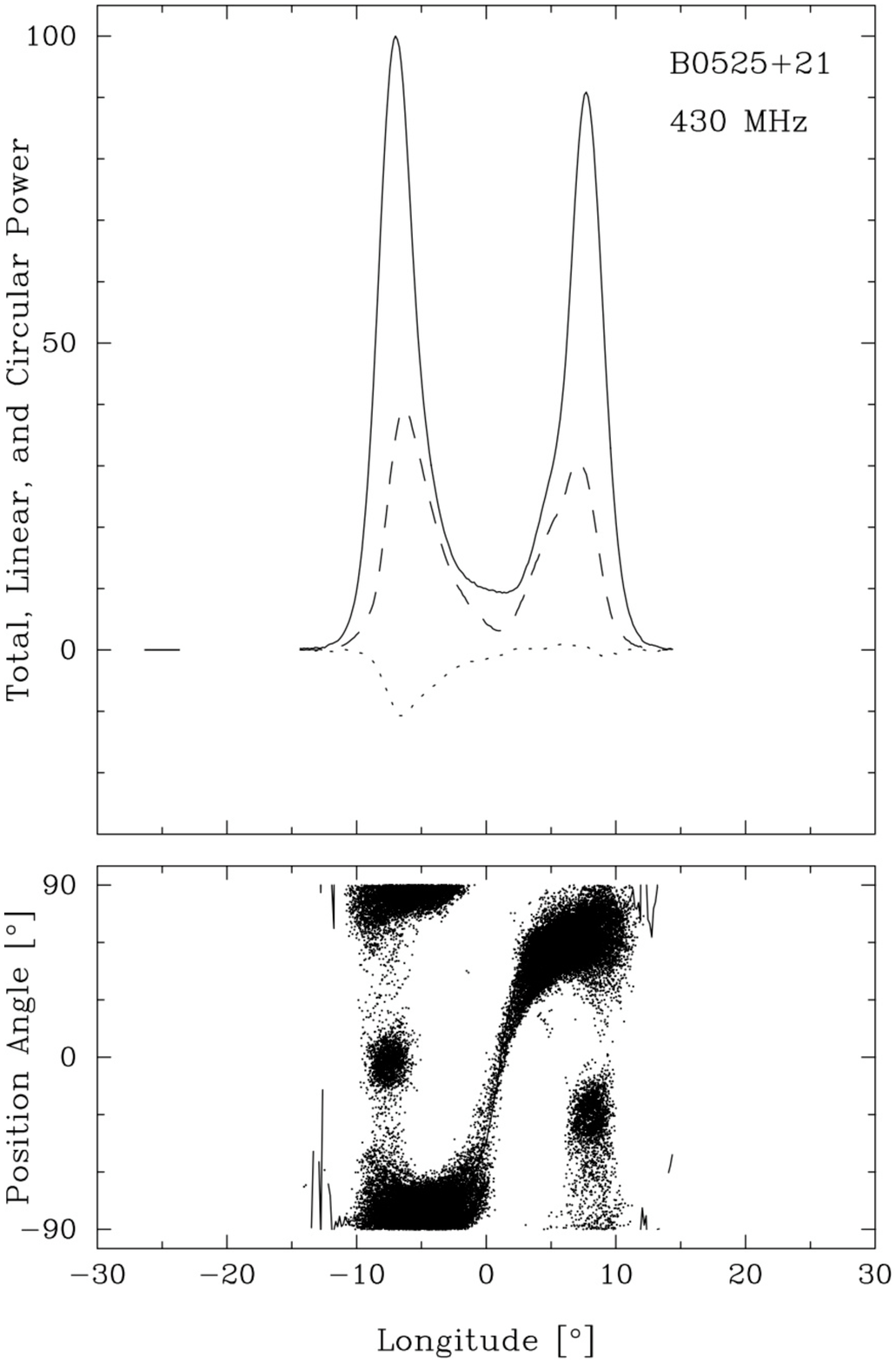} 
\pf{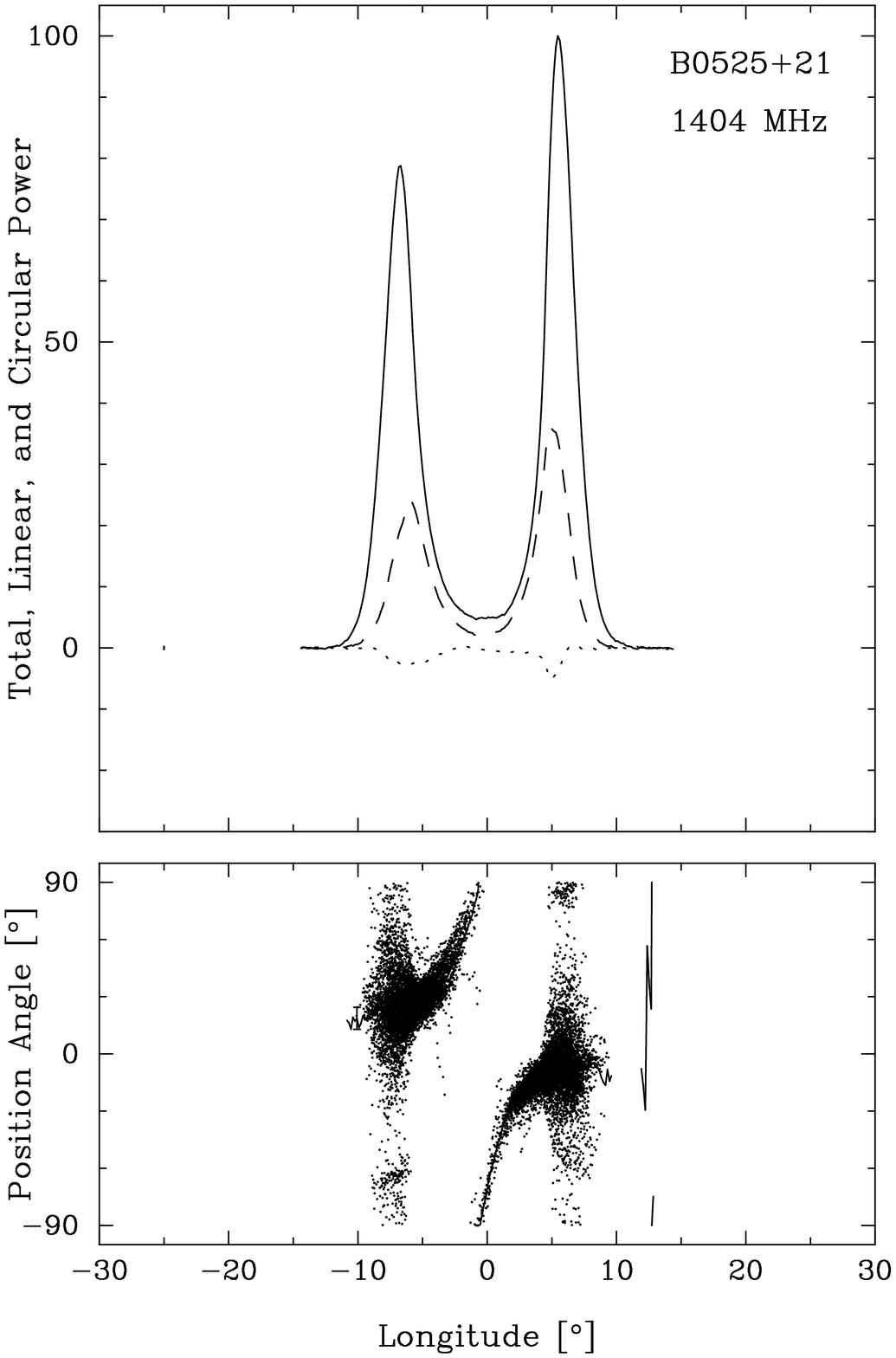} 
}
\caption{Multi-frequency and polarization profiles of B0301+19, B0523+11 and B0525+21. }
\label{b1}
\end{figure}

\begin{figure}[htb]
\centerline{ 
\pf{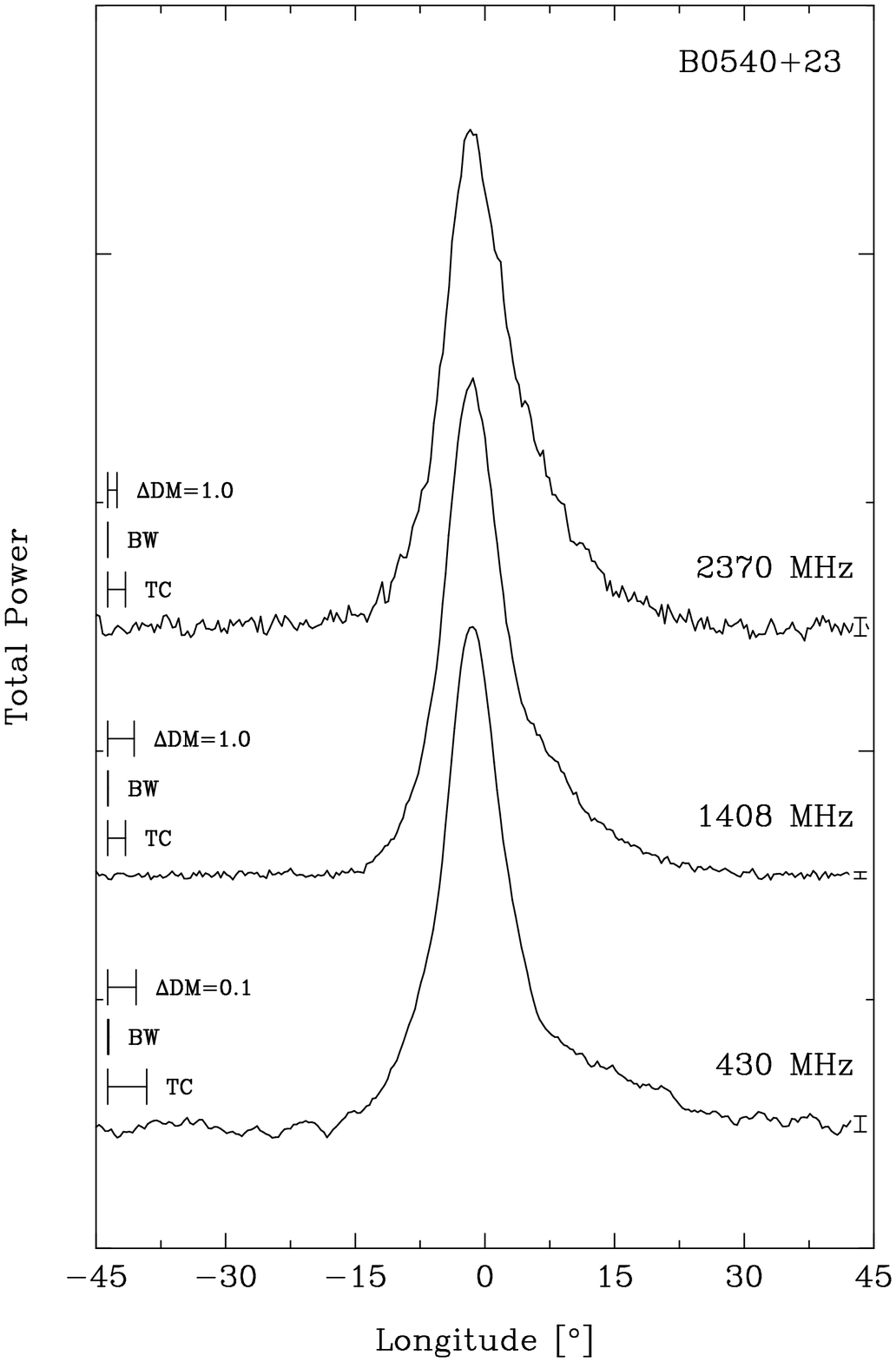}     
\pf{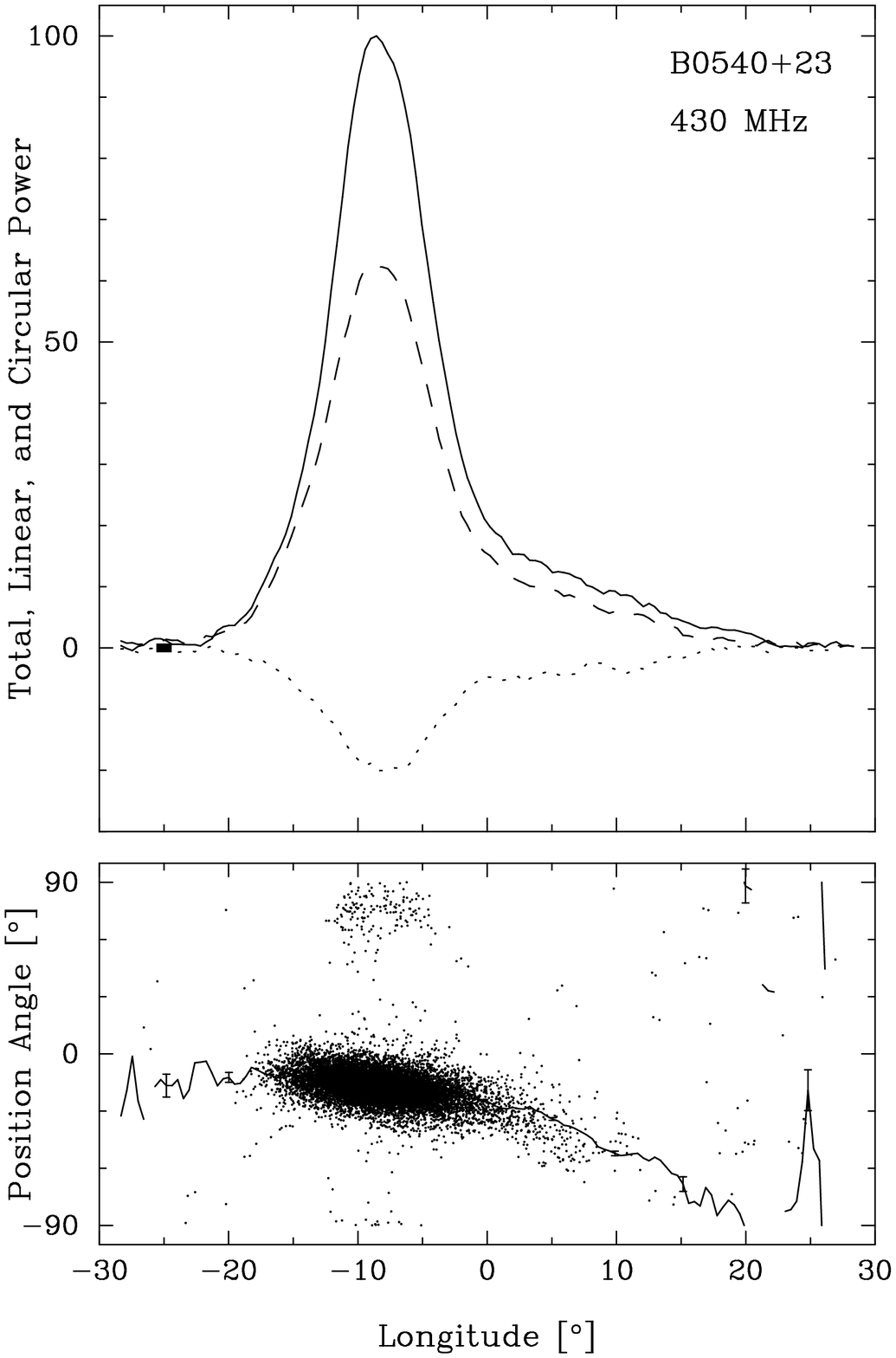} 
\pf{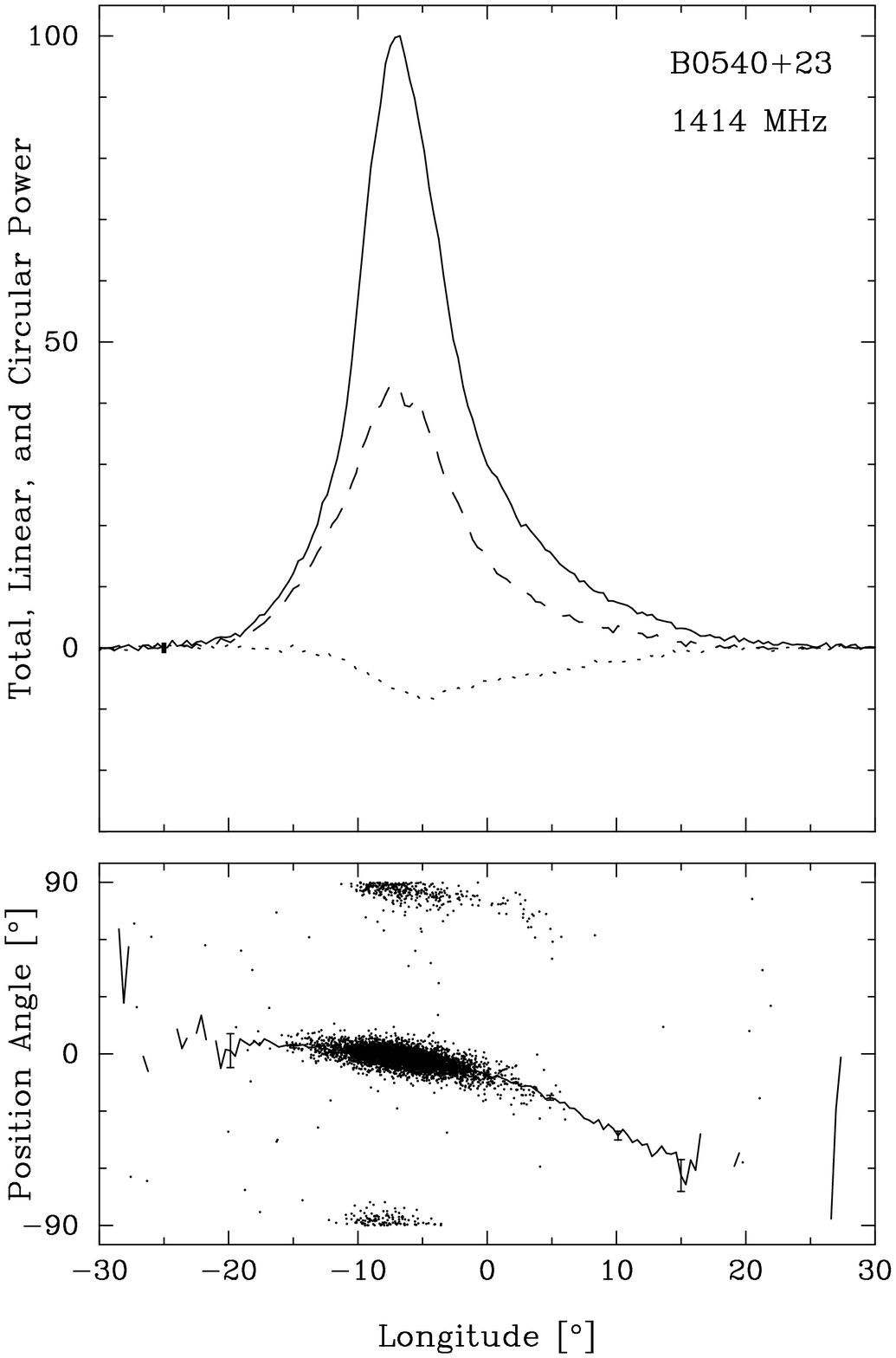} 
}
\quad \\
\centerline{
\pf{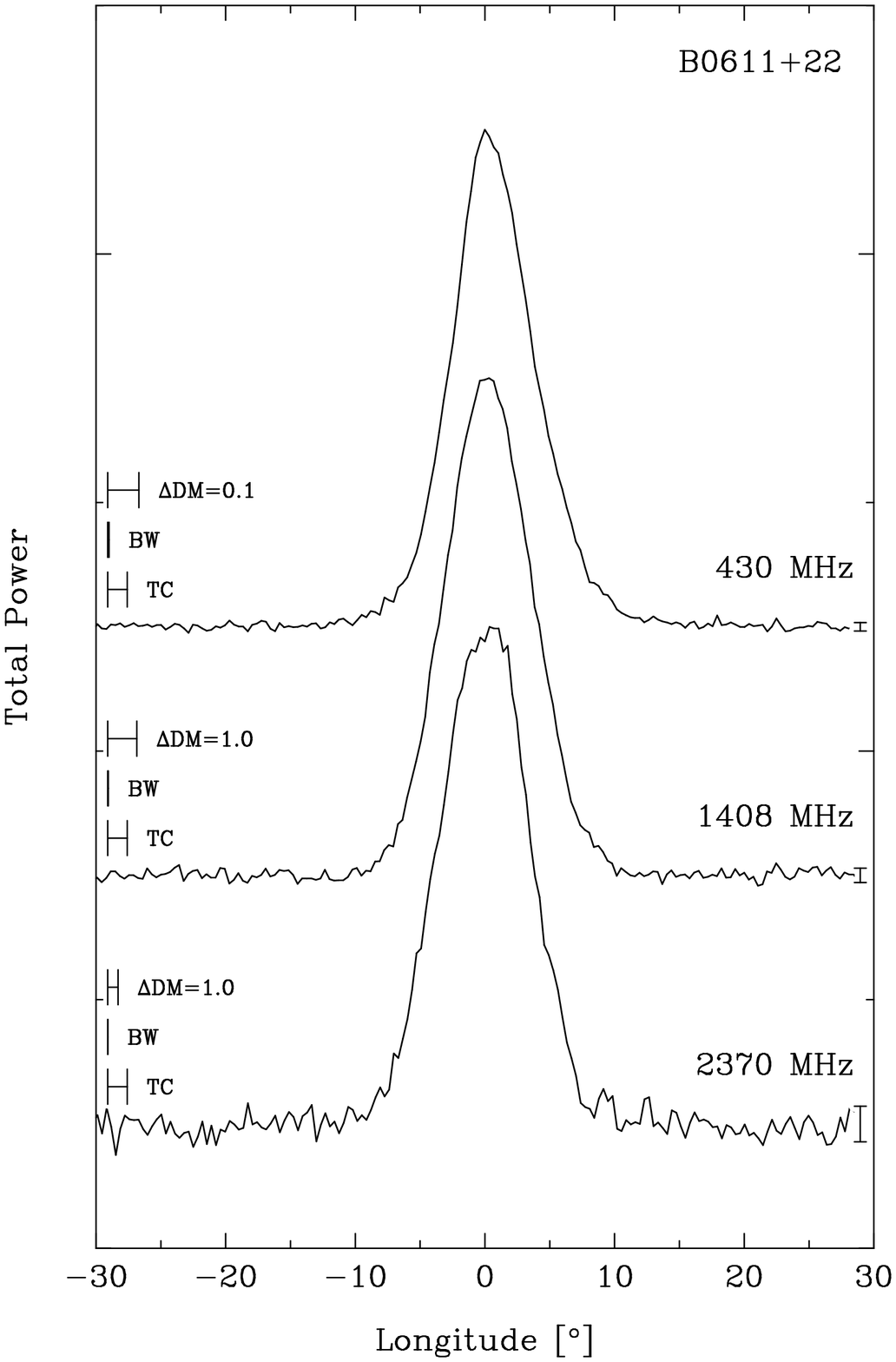}     
\pf{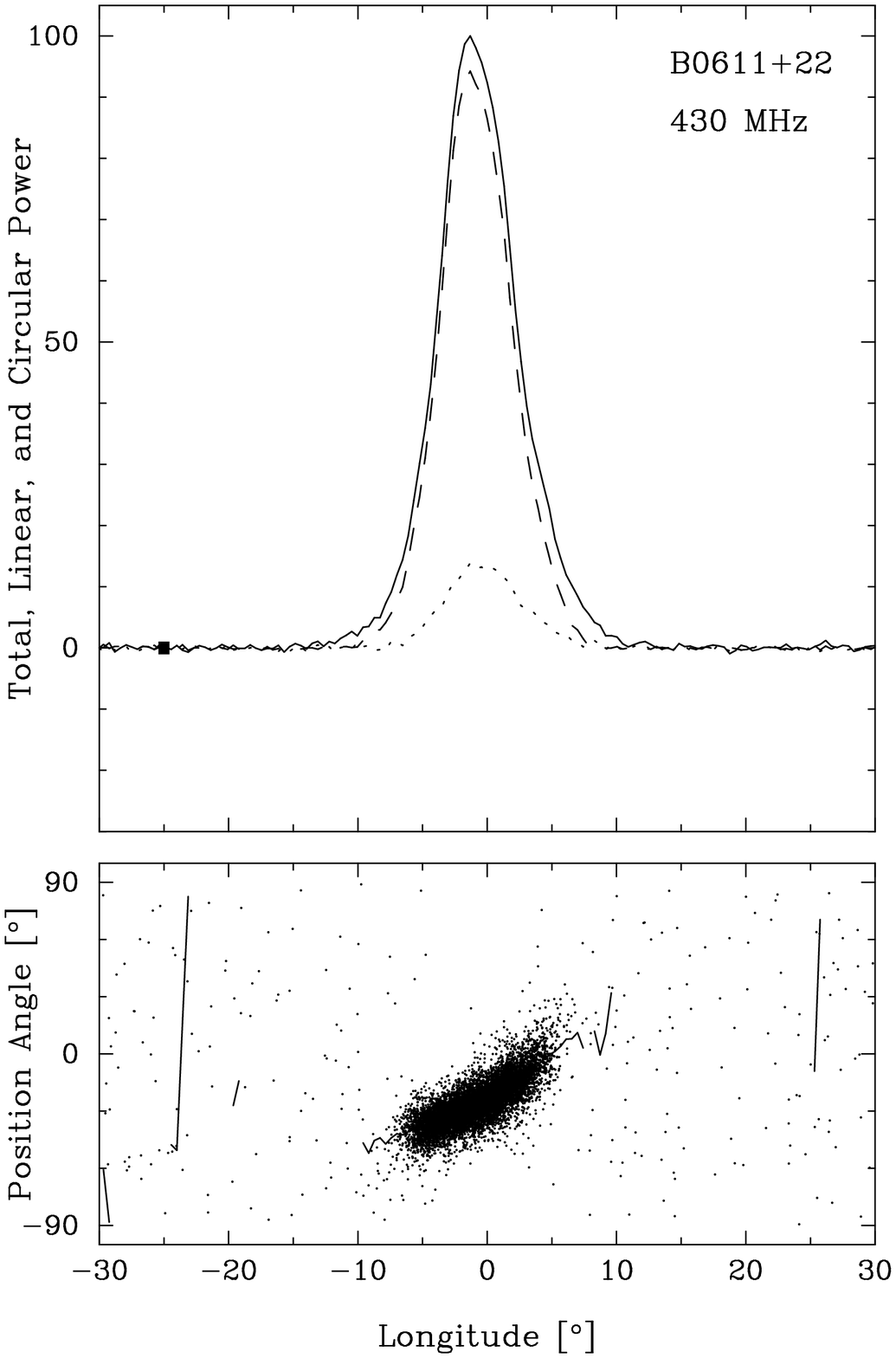} 
\pf{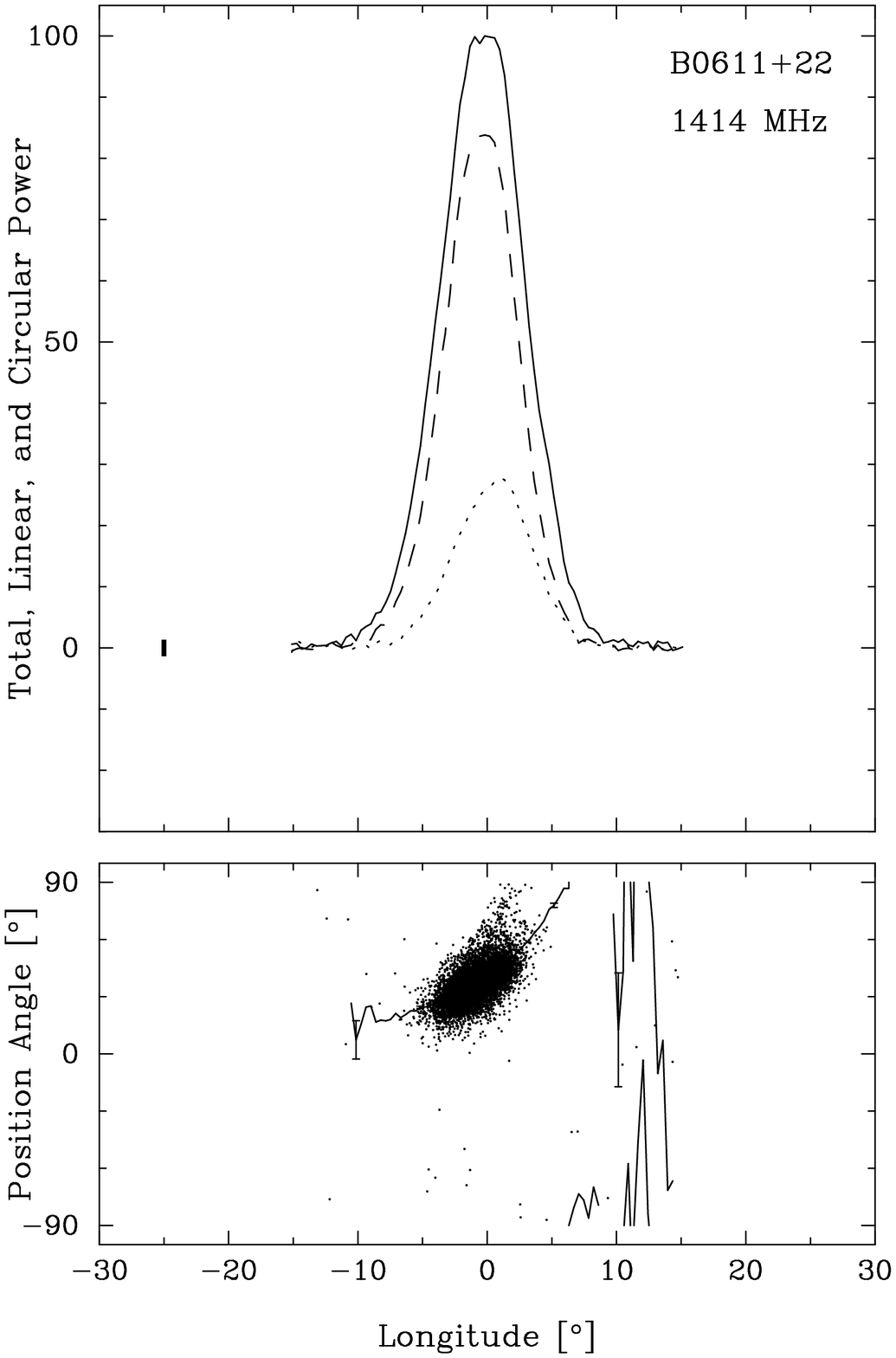} 
}
\quad \\
\centerline{
\pf{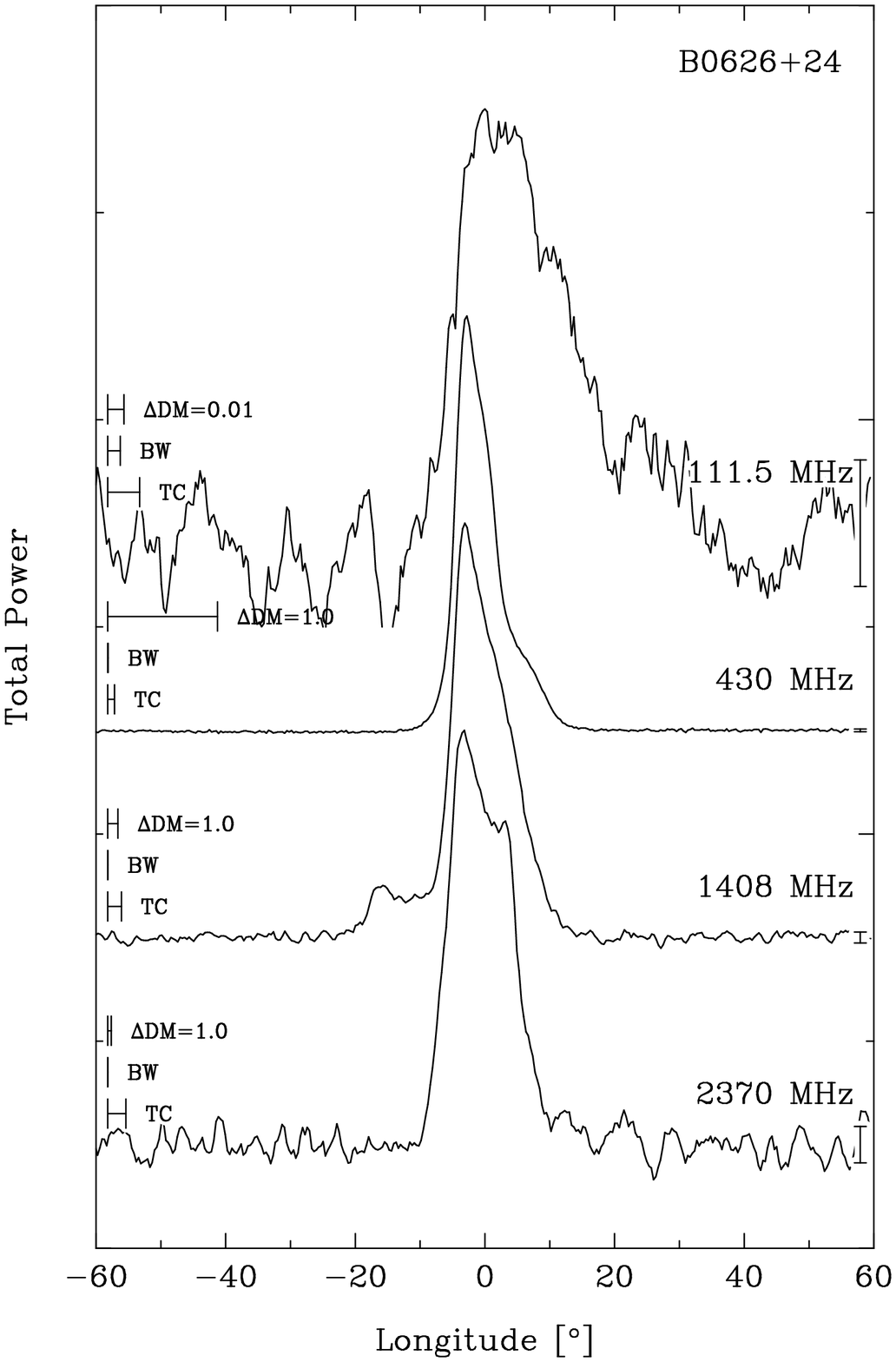}     
\pf{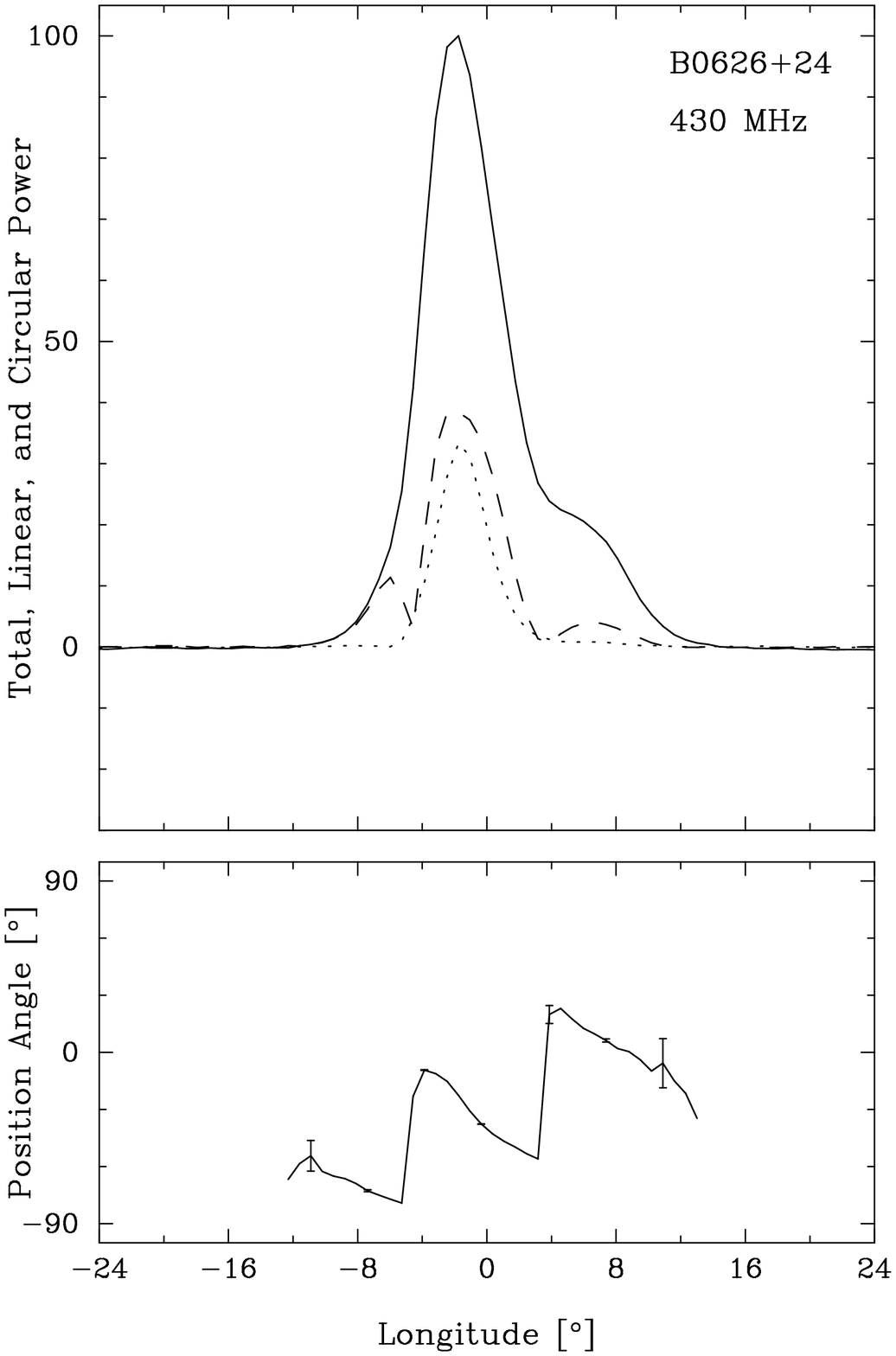} 
\pf{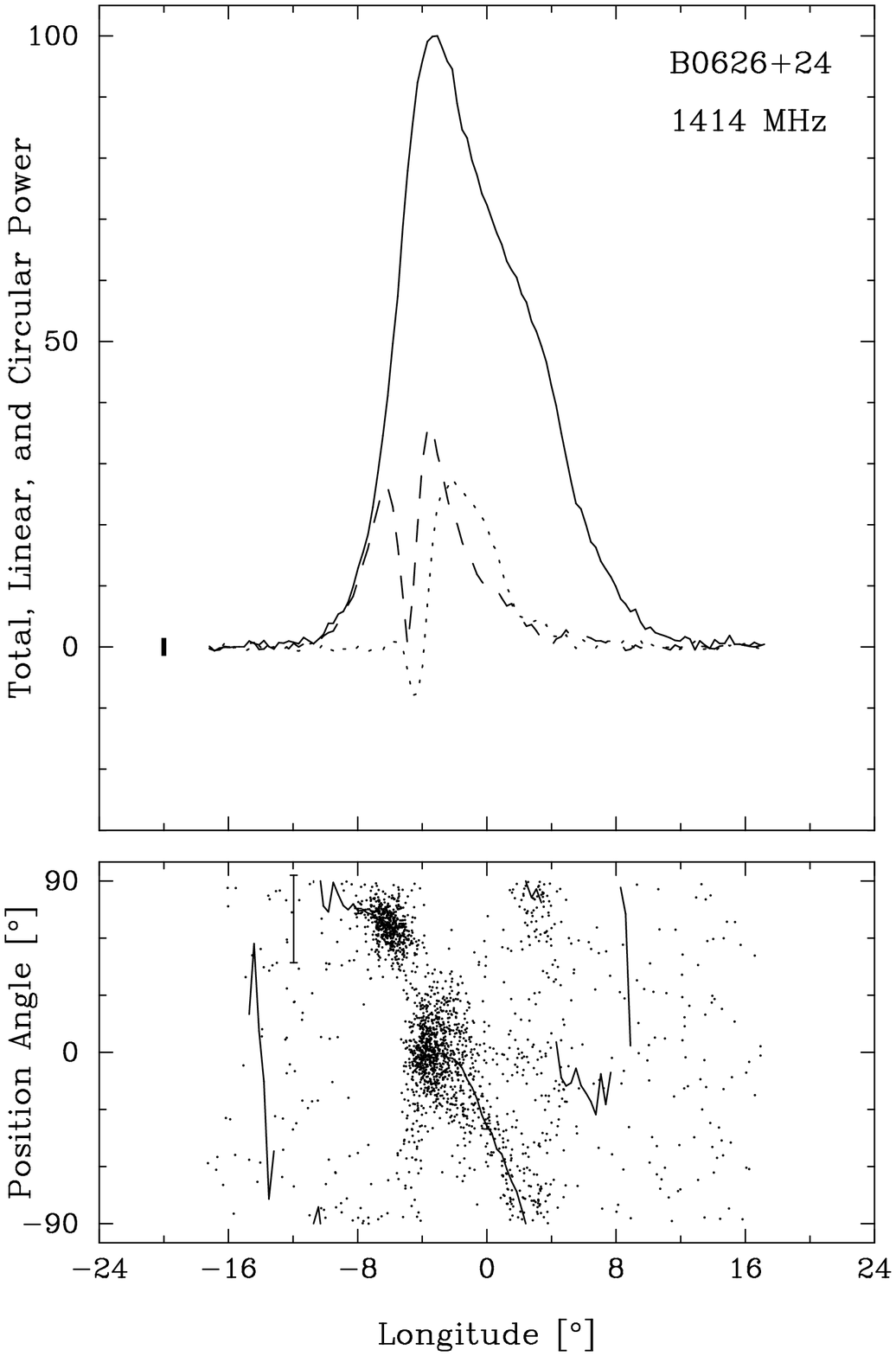} 
}
\caption{Multi-frequency and polarization profiles of B0540+23, B0611+22 and B0626+24.}
\label{b2}
\end{figure}

\begin{figure}[htb]
\centerline{ 
\pf{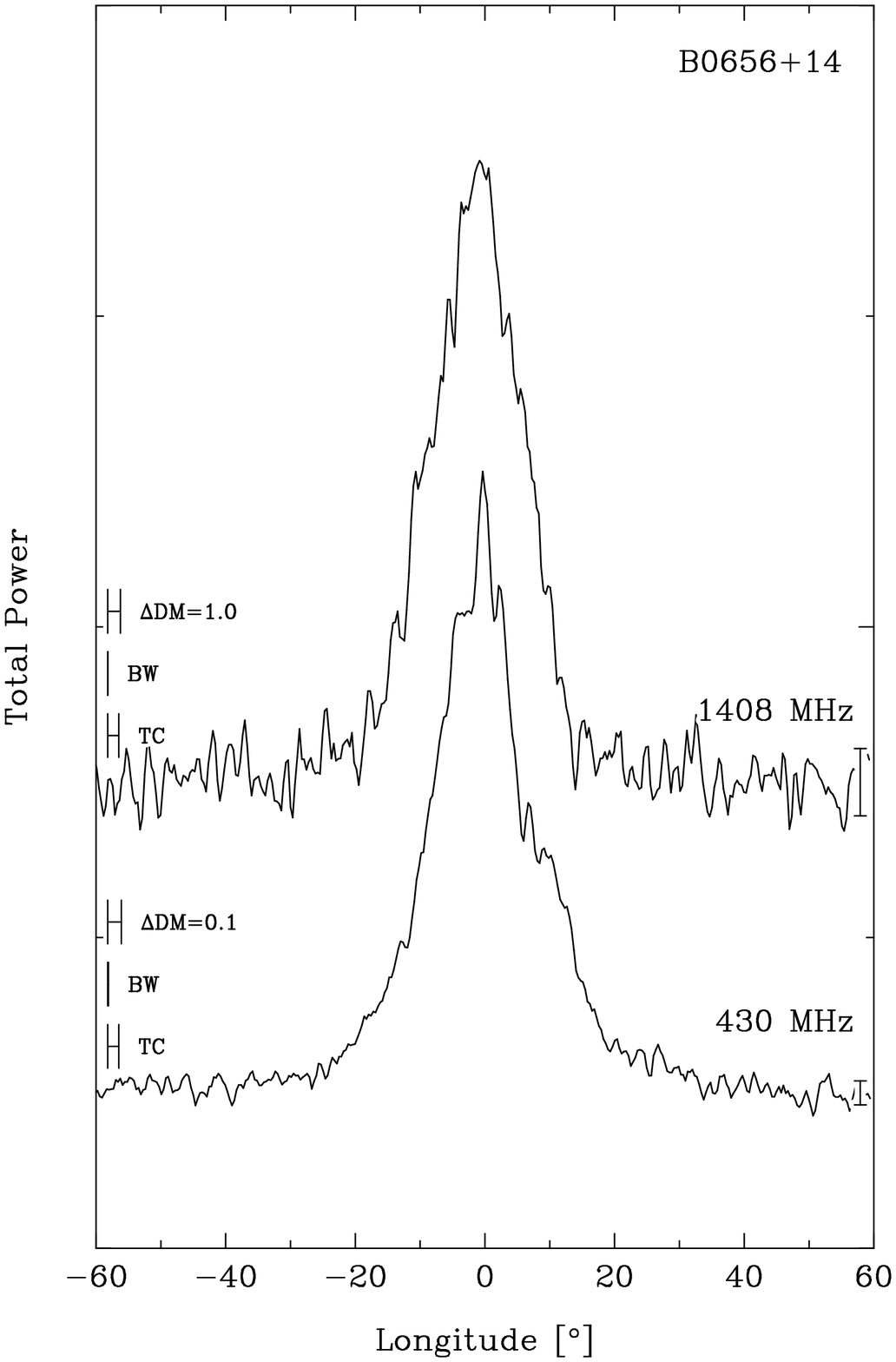}     
\pf{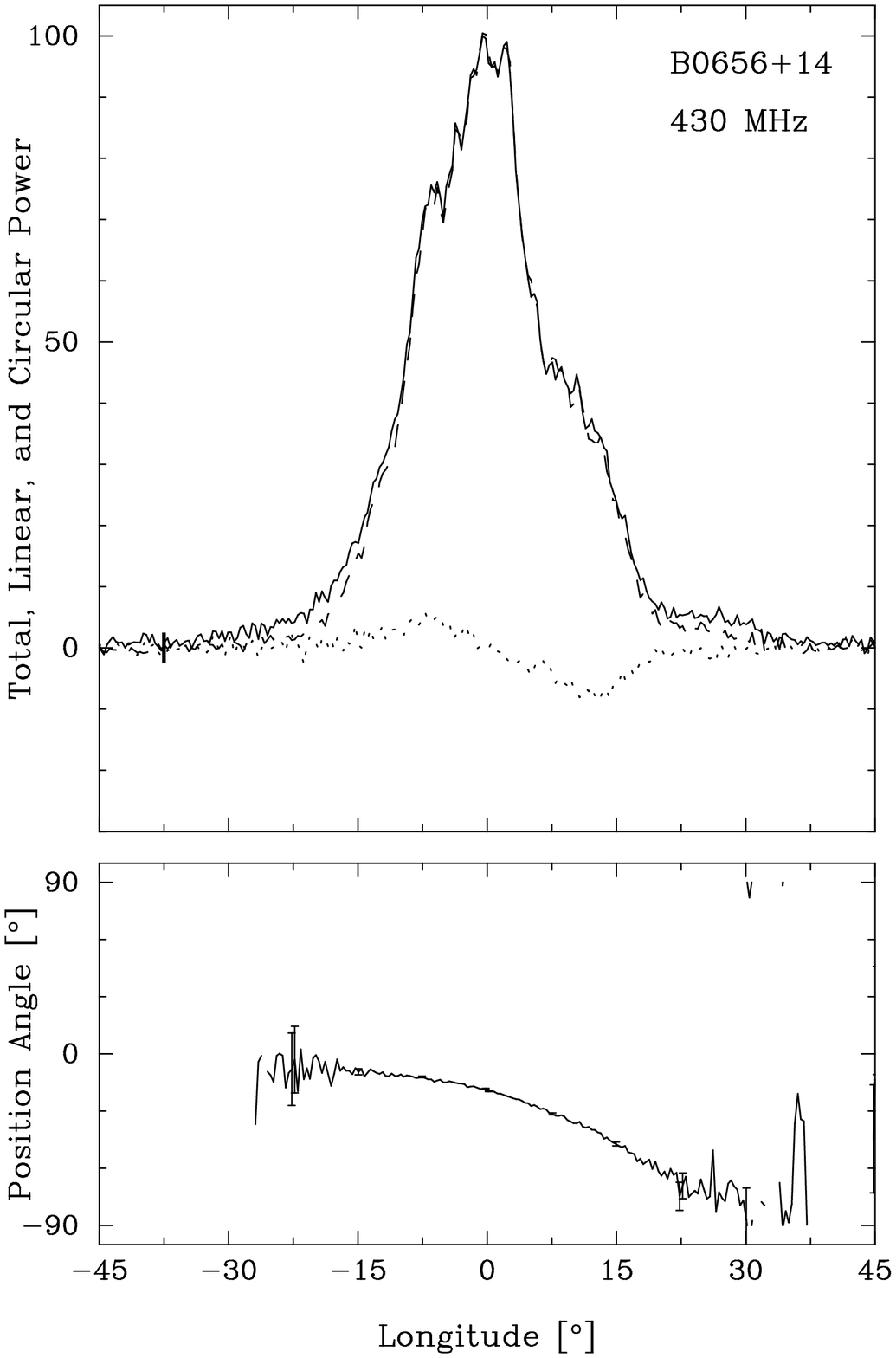} 
\pf{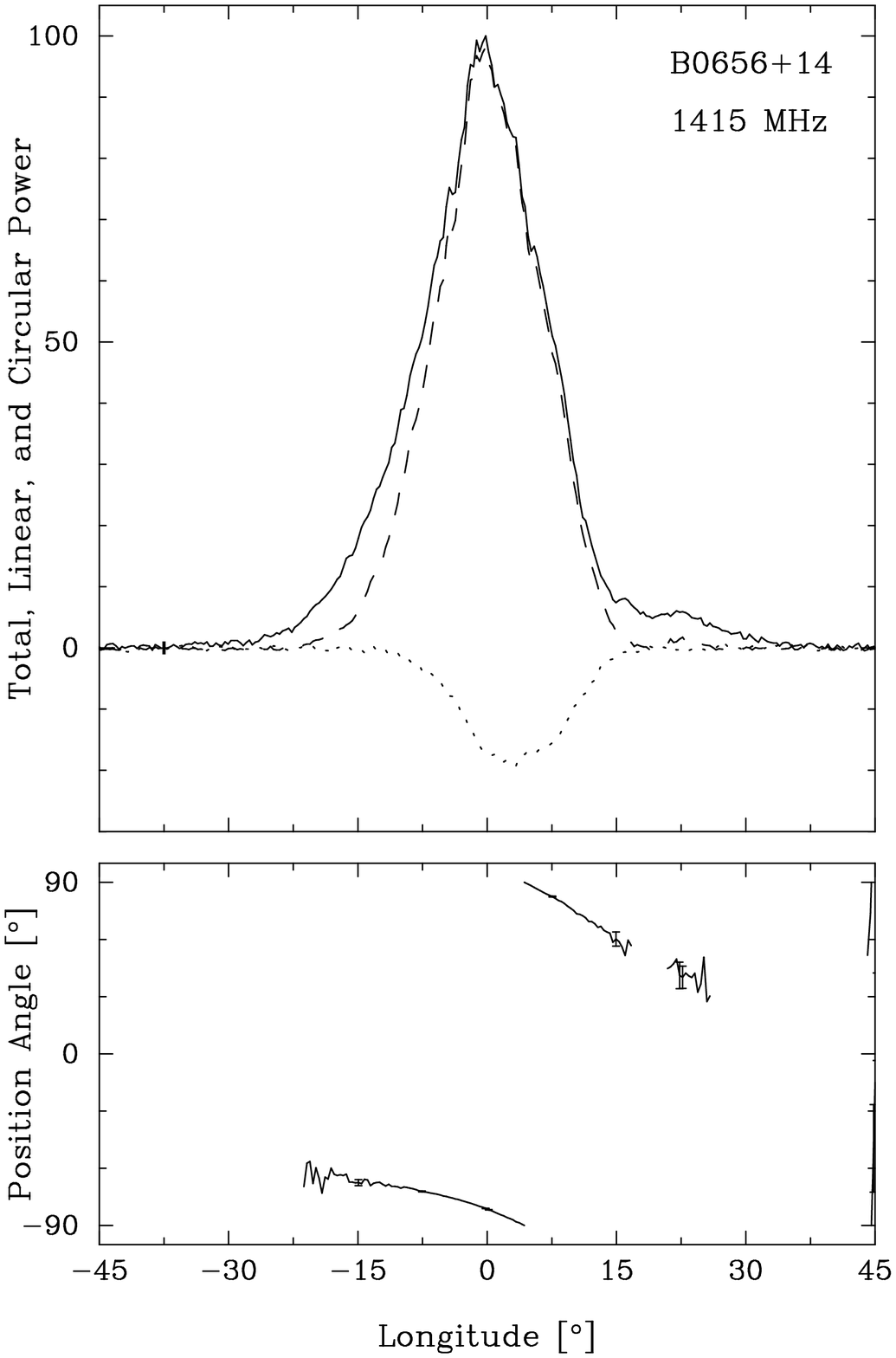} 
}
\quad \\
\centerline{
\pf{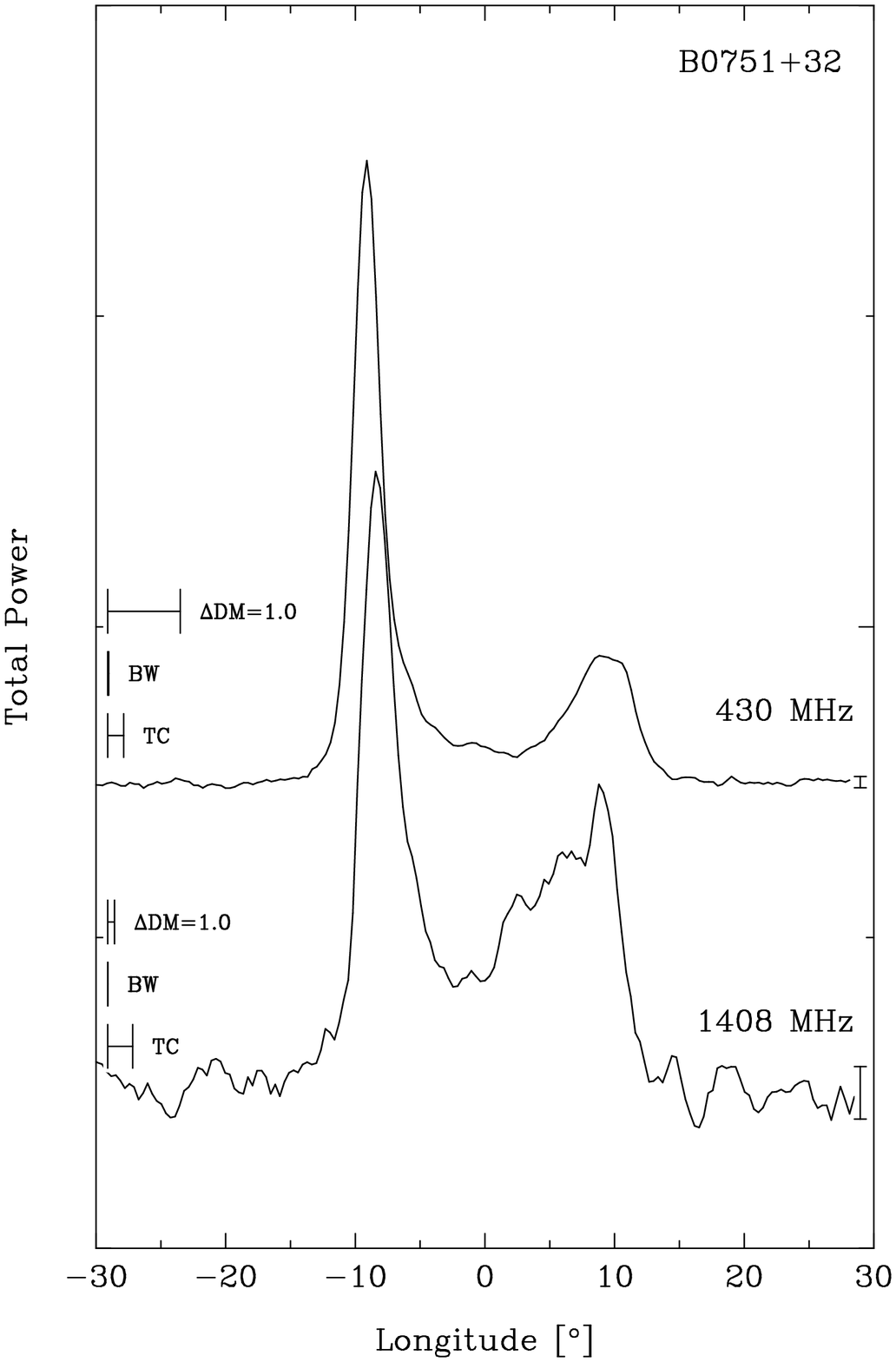}     
\pf{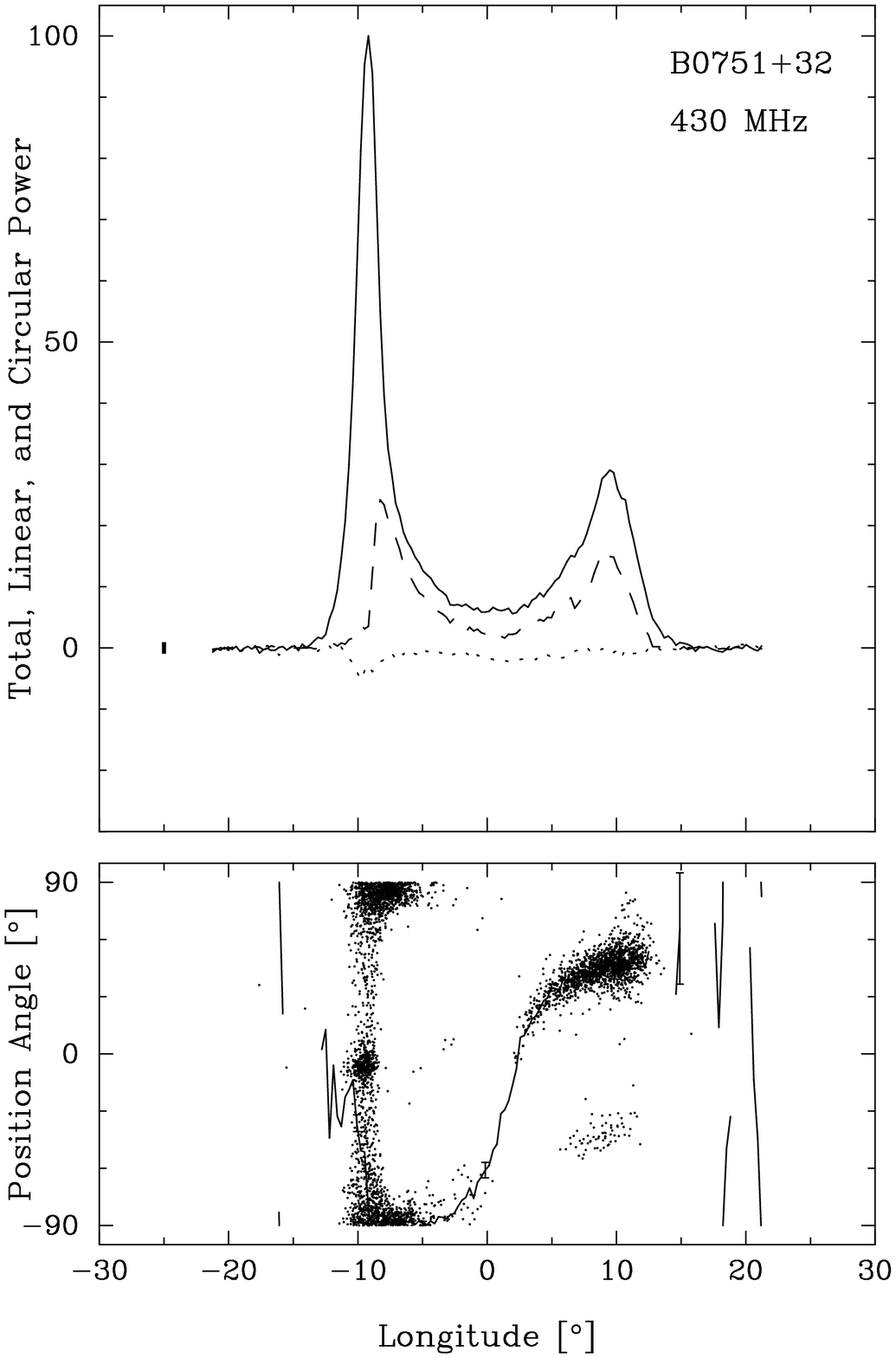} 
\pf{}     
}
\quad \\
\centerline{
\pf{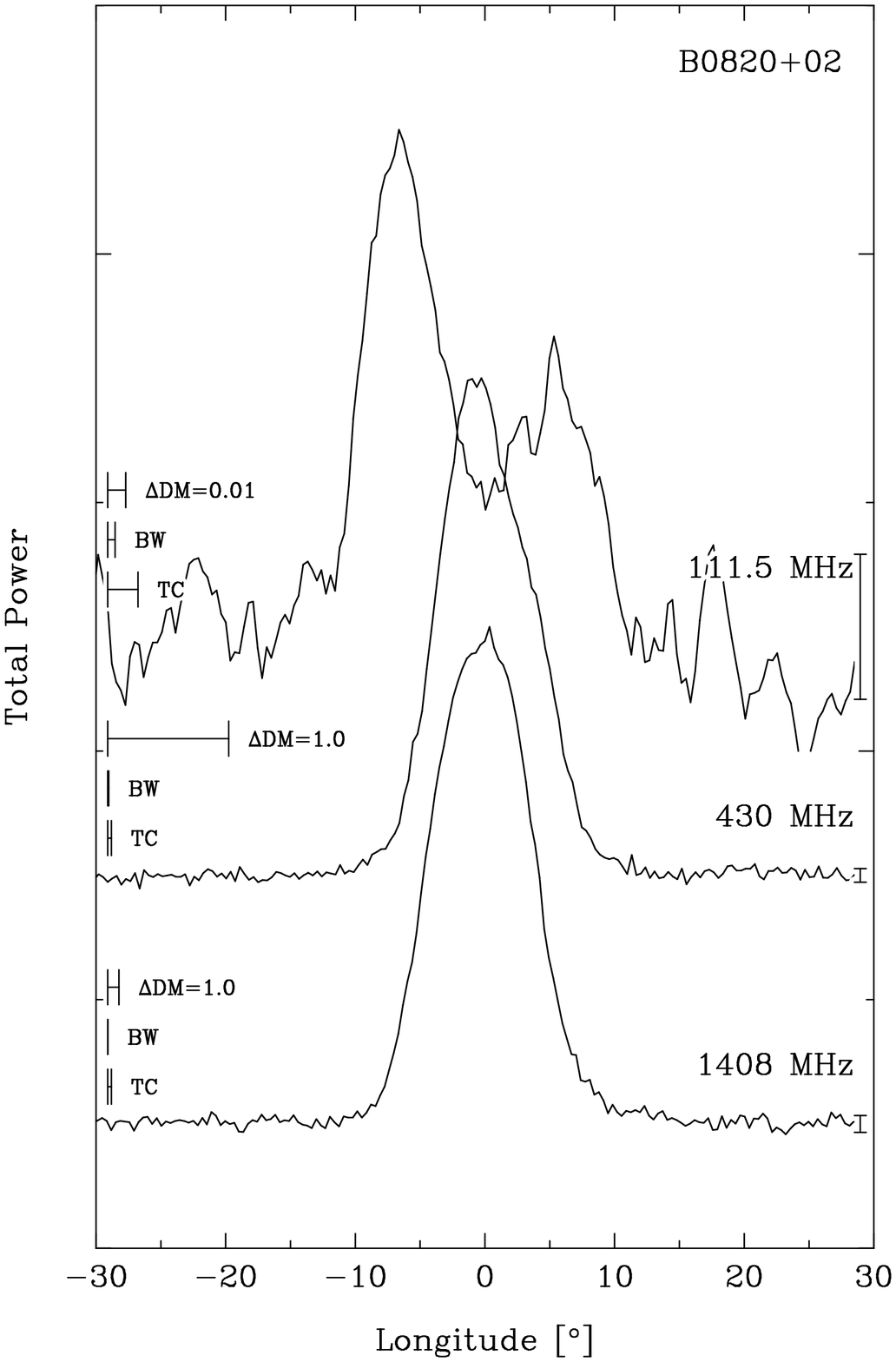}     
\pf{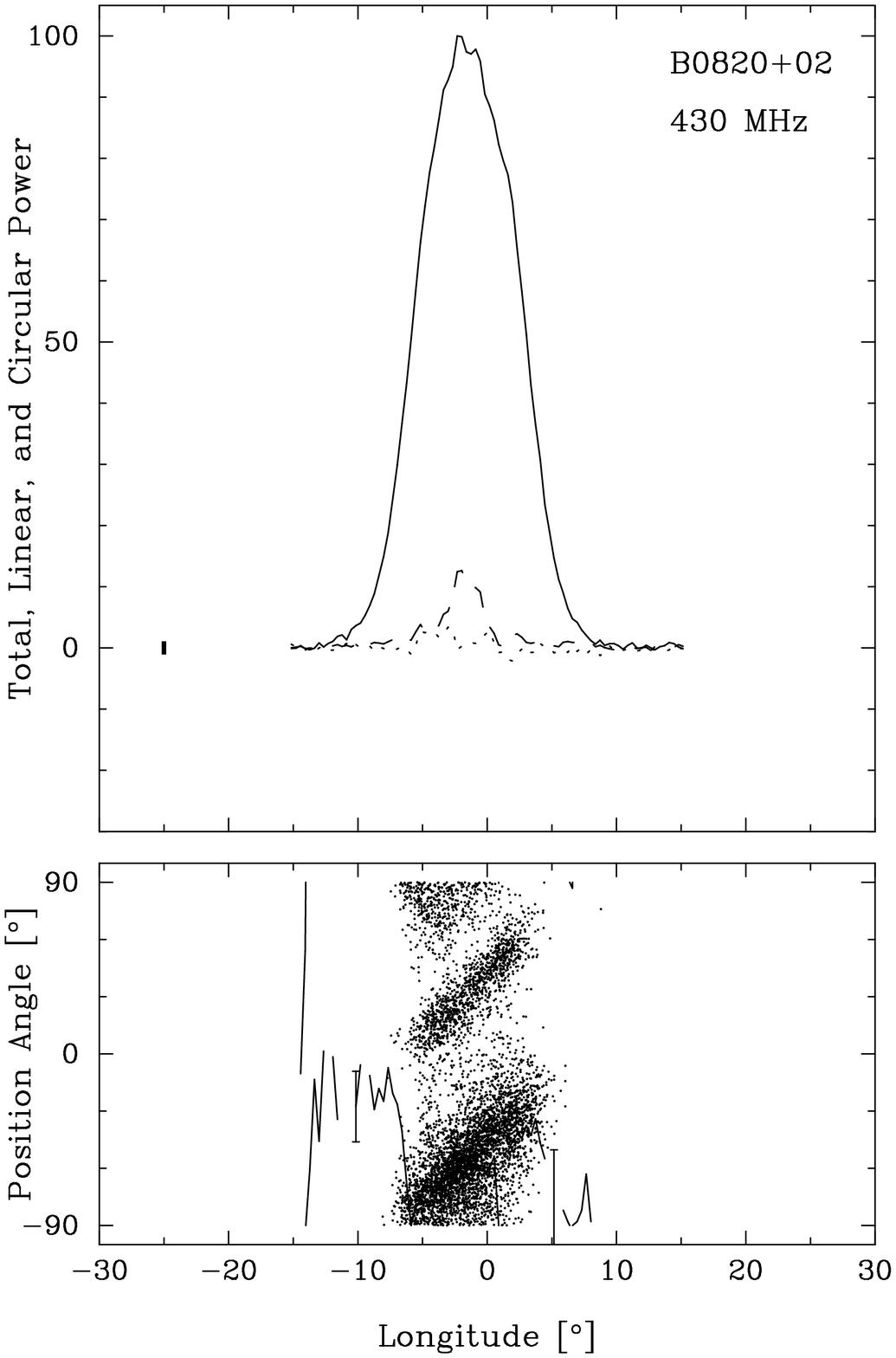} 
\pf{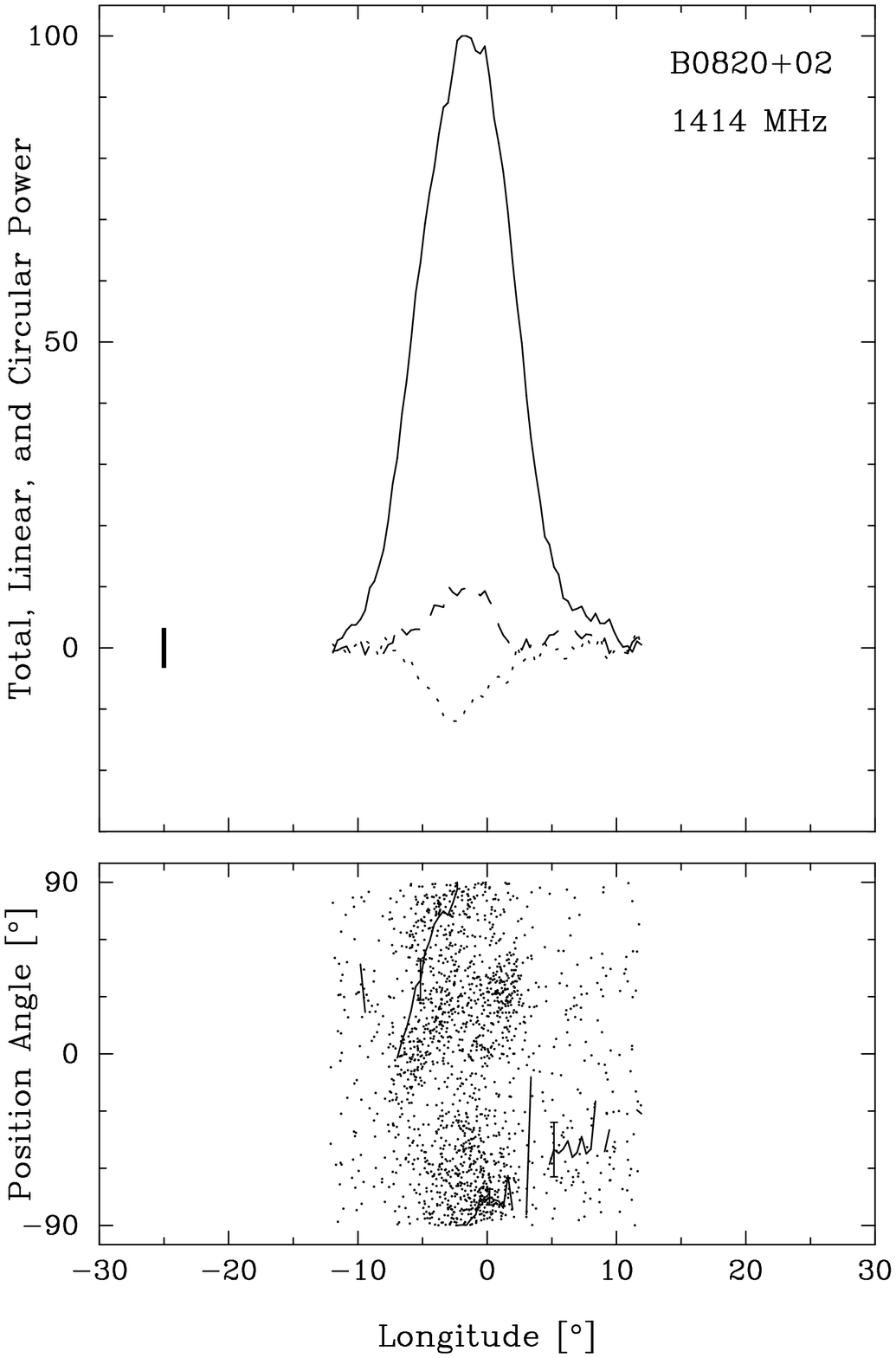} 
}
\caption{Multi-frequency and polarization profiles of B0656+14, B0751+32 and 0820+02.}
\label{b3}
\end{figure}

\begin{figure}[htb]
\centerline{ 
\pf{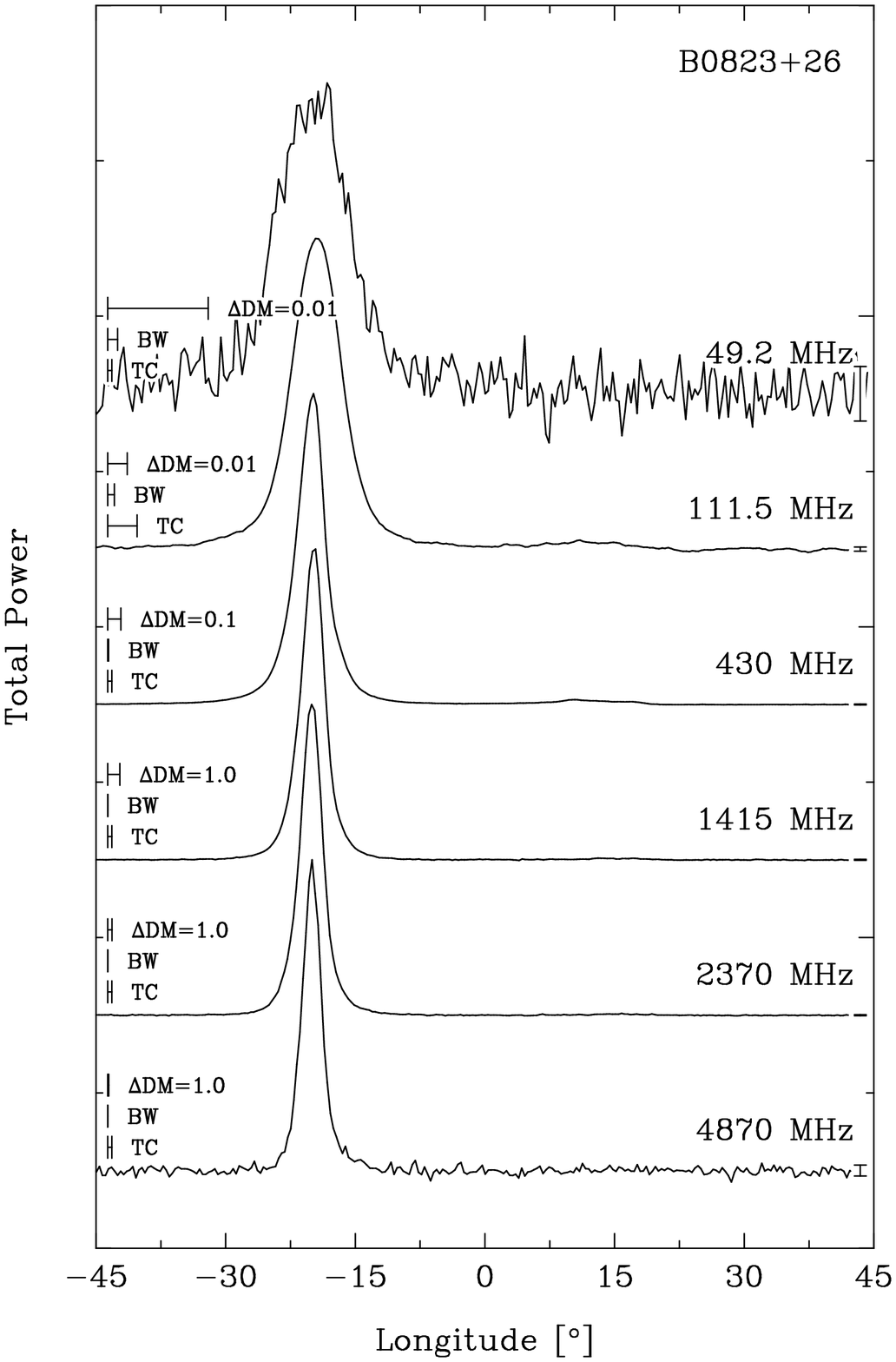}     
\pf{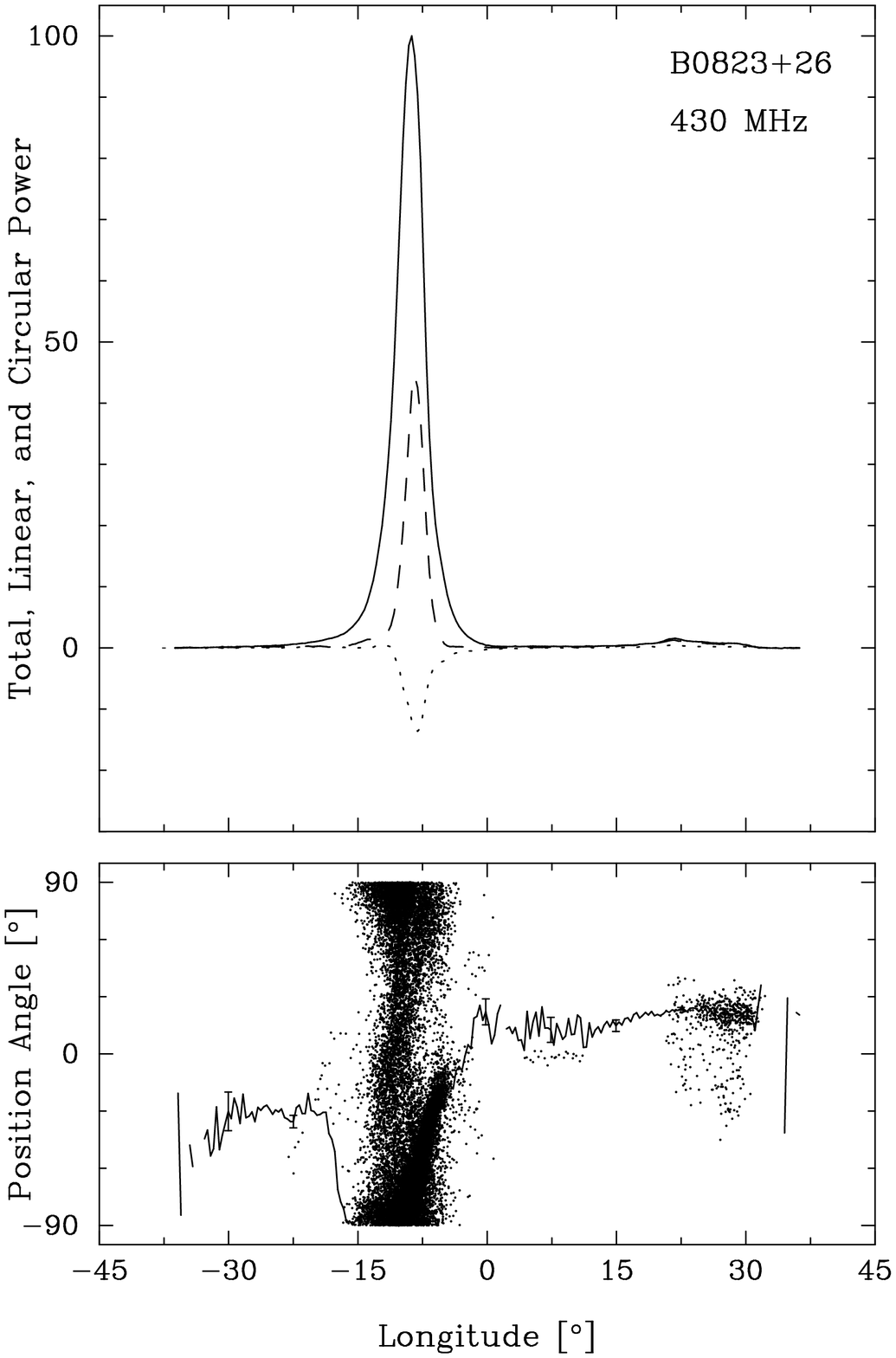} 
\pf{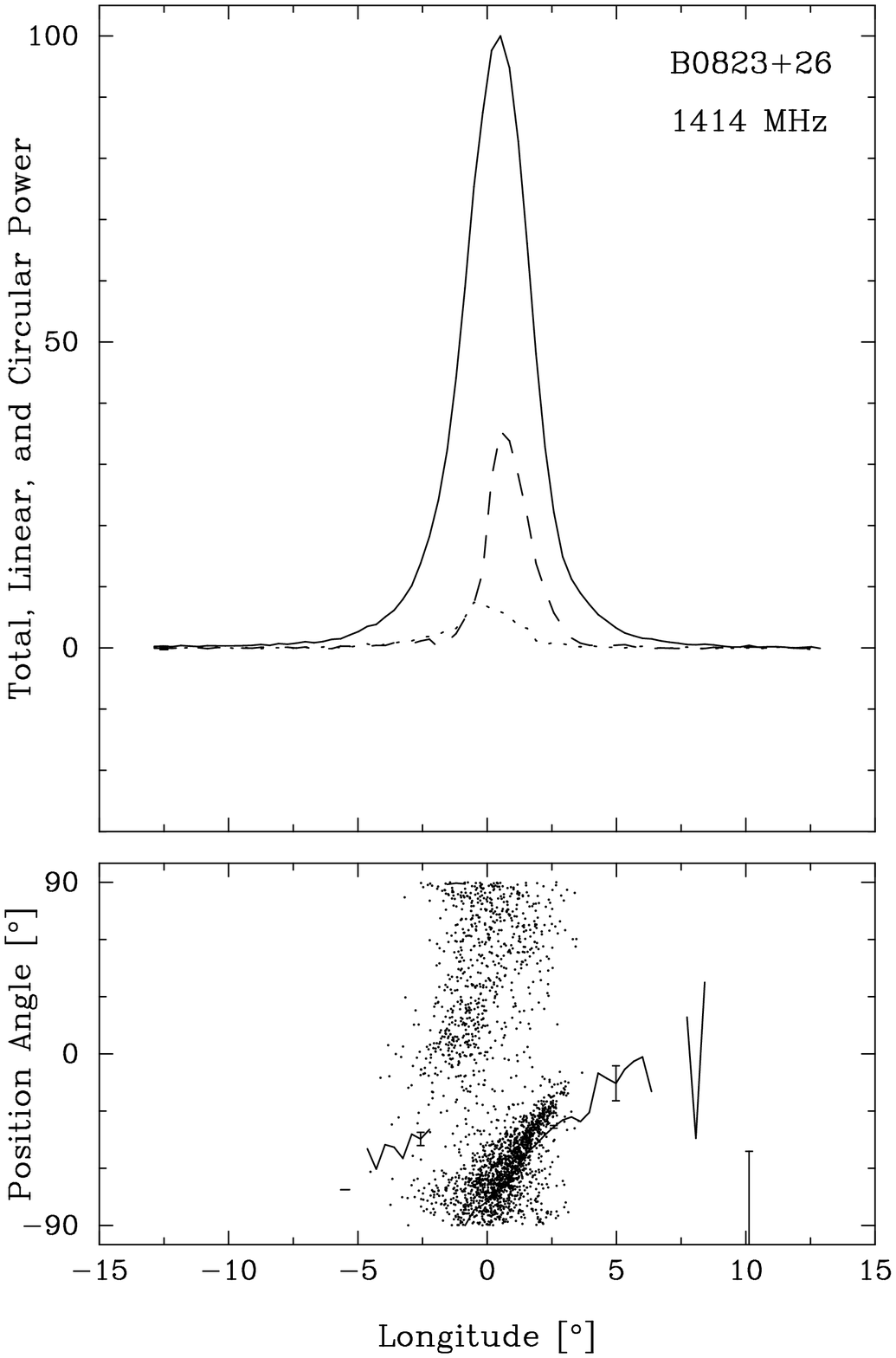}
}
\quad \\
\centerline{
\pf{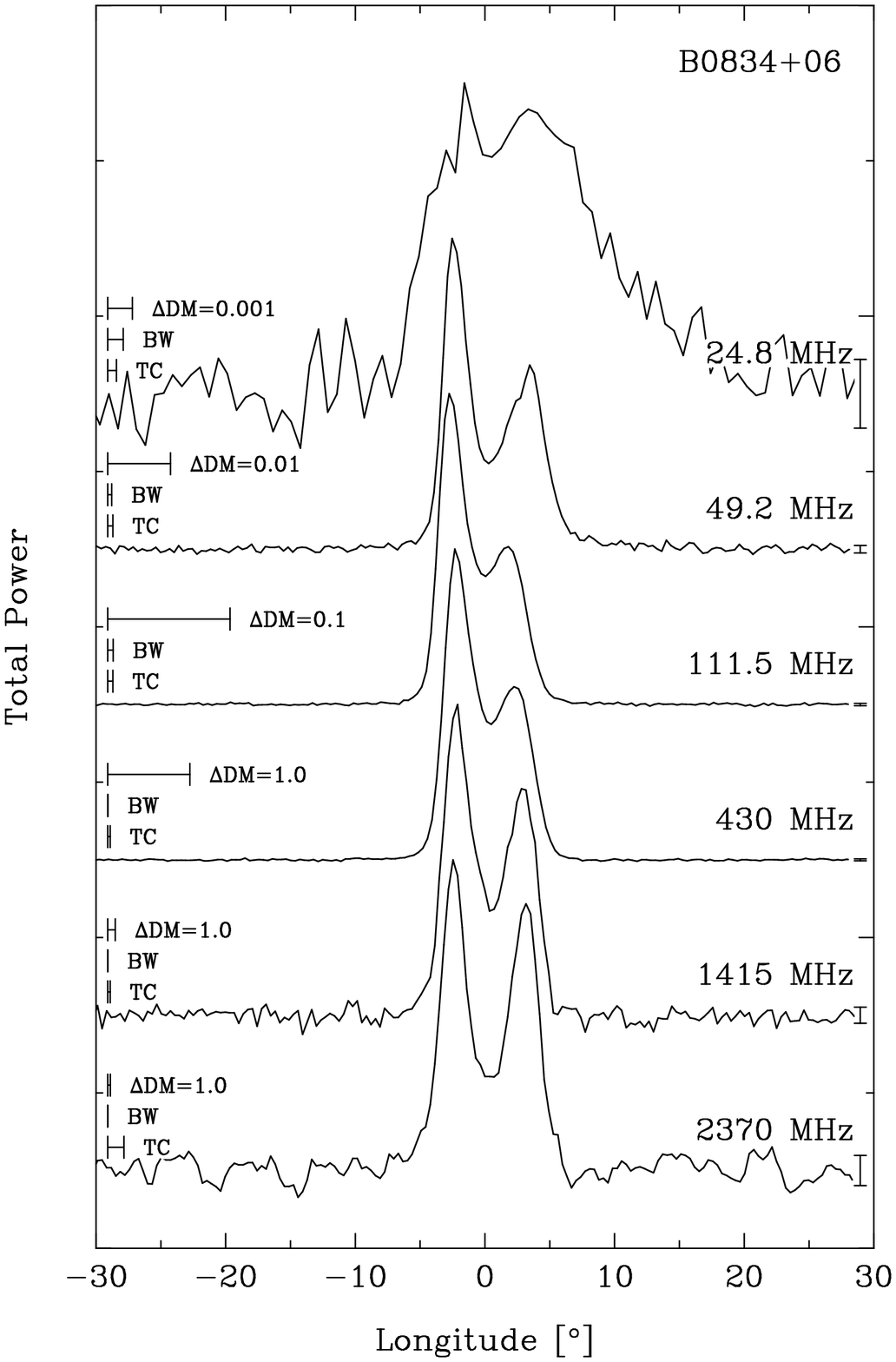}     
\pf{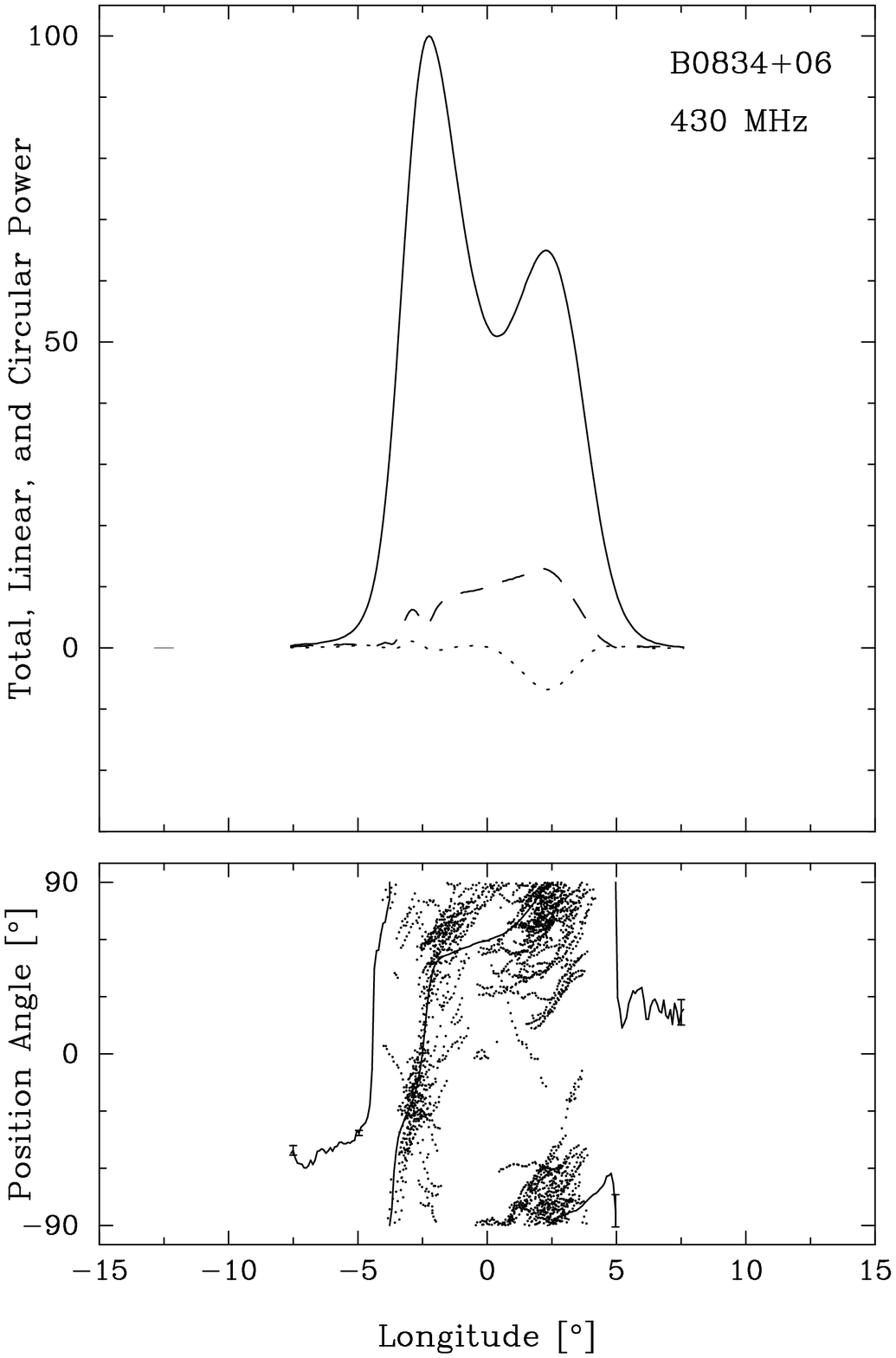} 
\pf{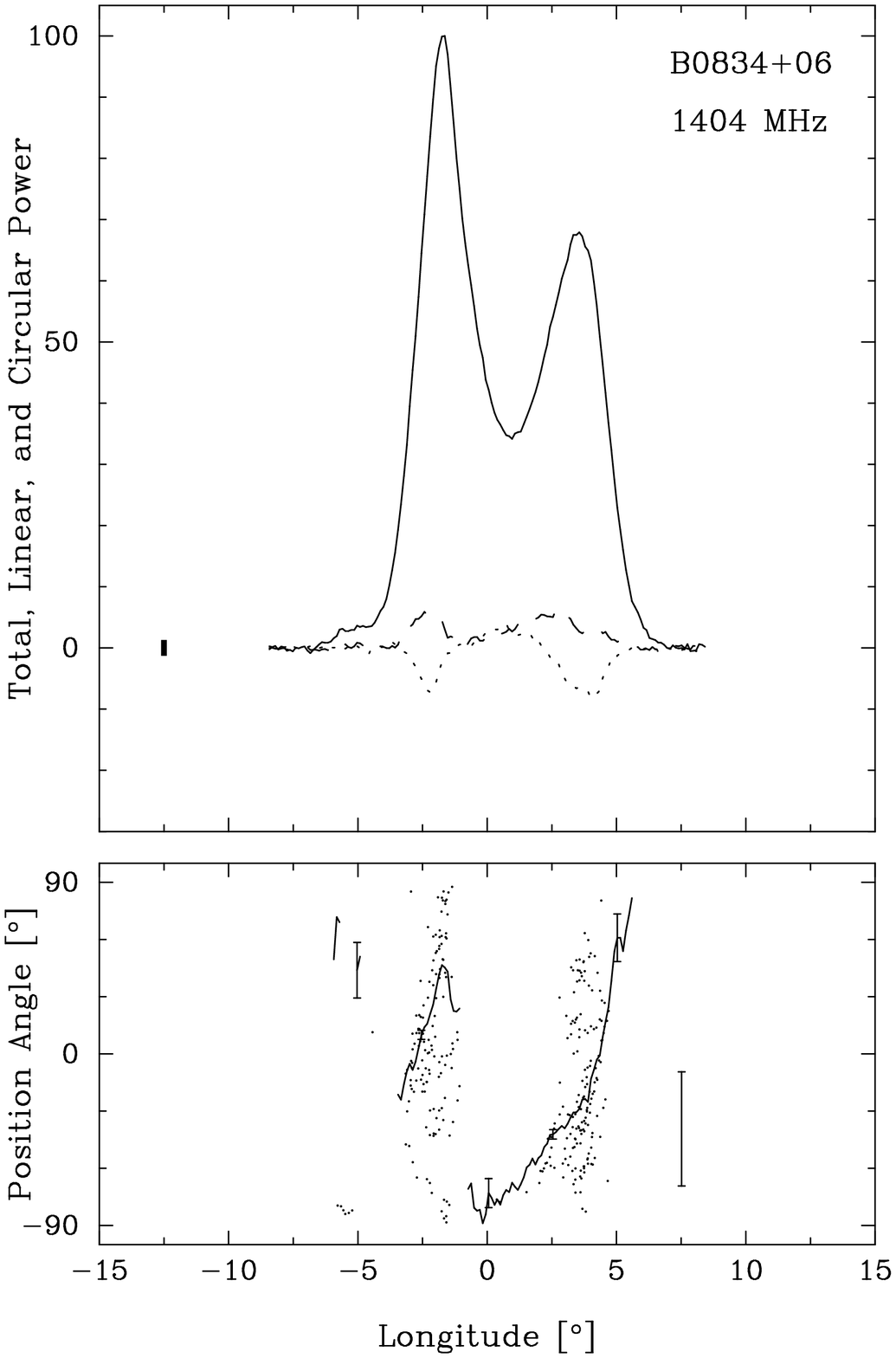} 
}
\quad \\
\centerline{
\pf{figs/dummy_fig.ps}     
\pf{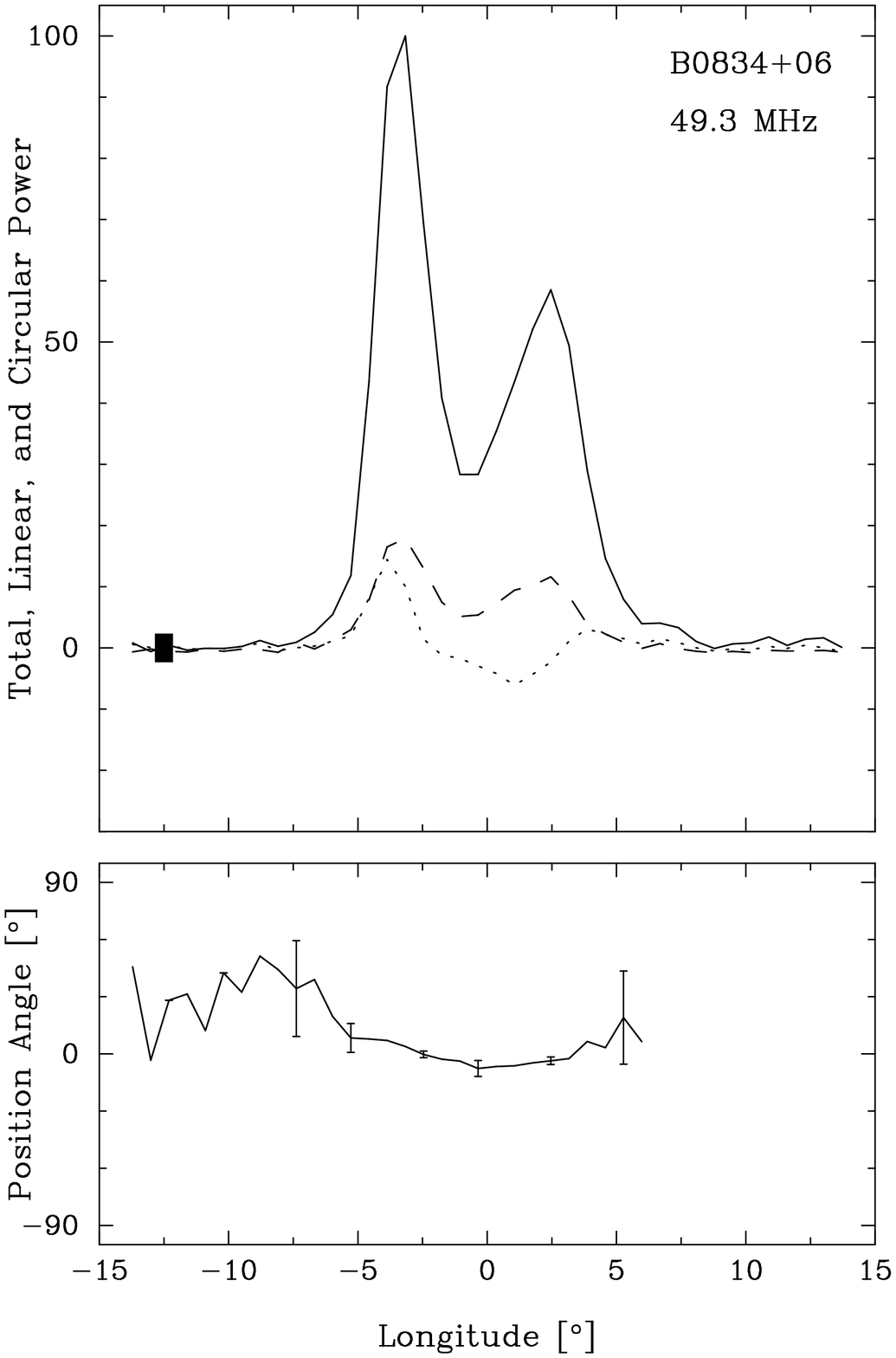} 
\pf{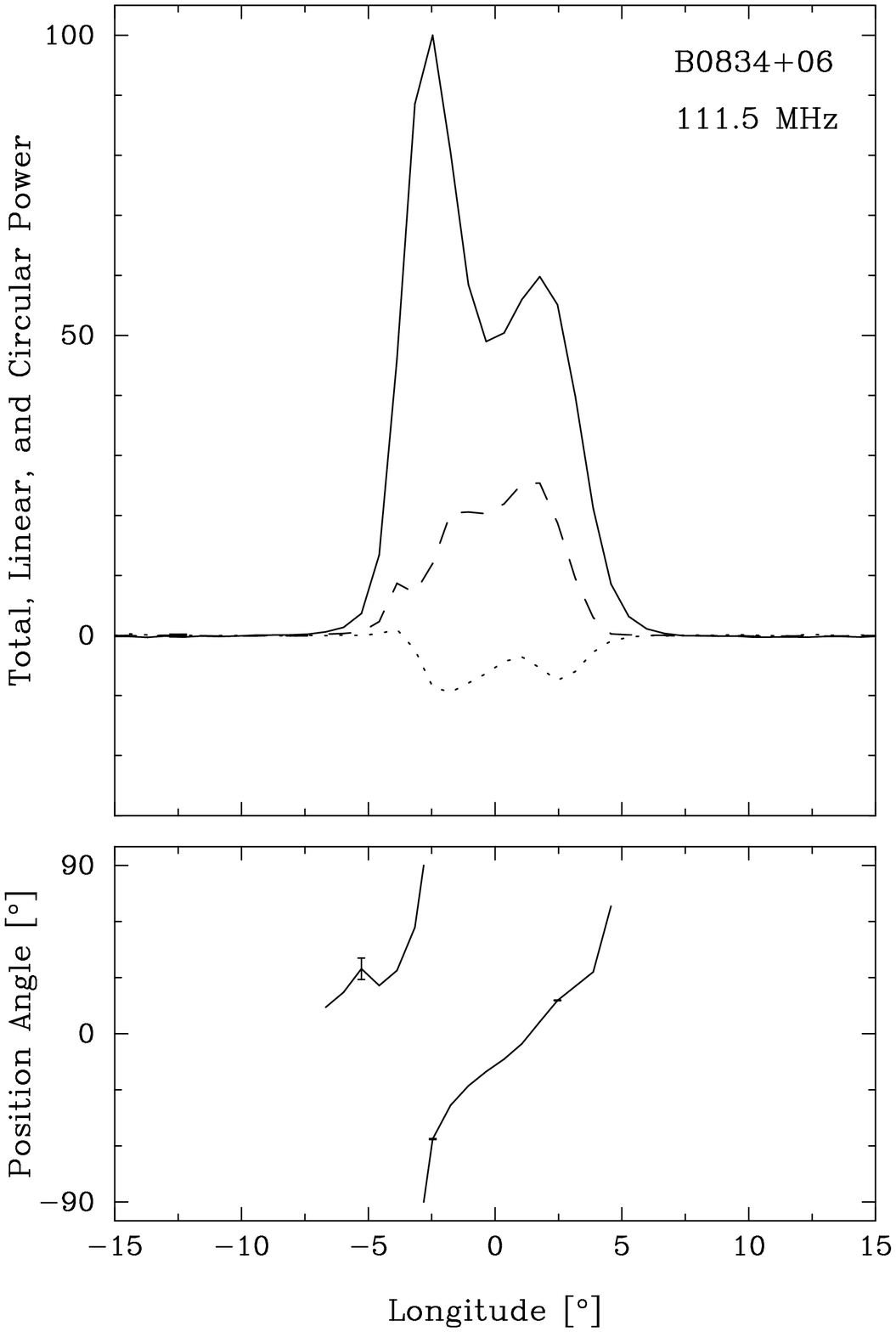} 
}
\caption{Multi-frequency and polarization profiles of B0823+26 and B0834+06.}
\label{b4}
\end{figure}

\begin{figure}[htb]
\centerline{ 
\pf{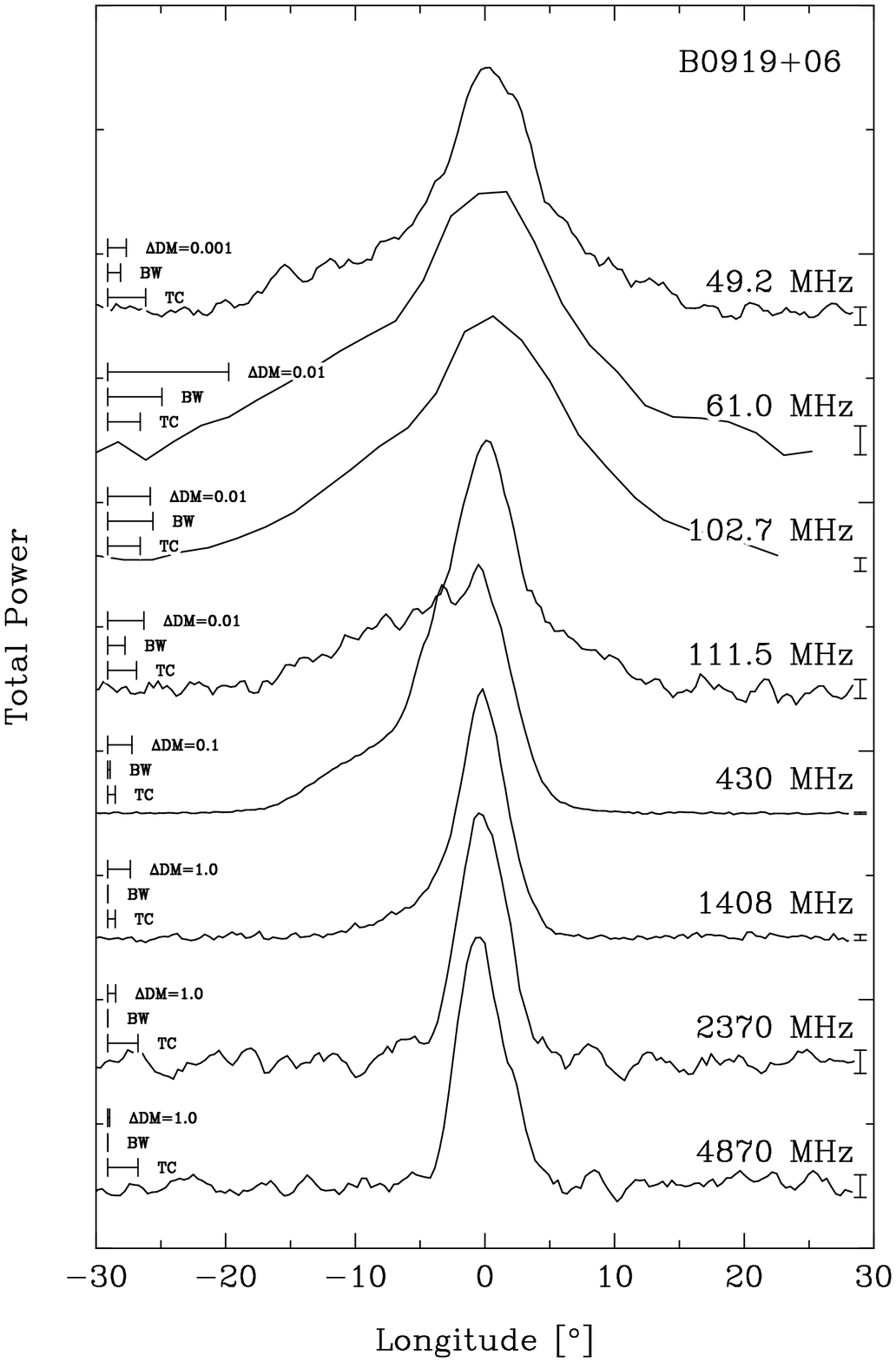}     
\pf{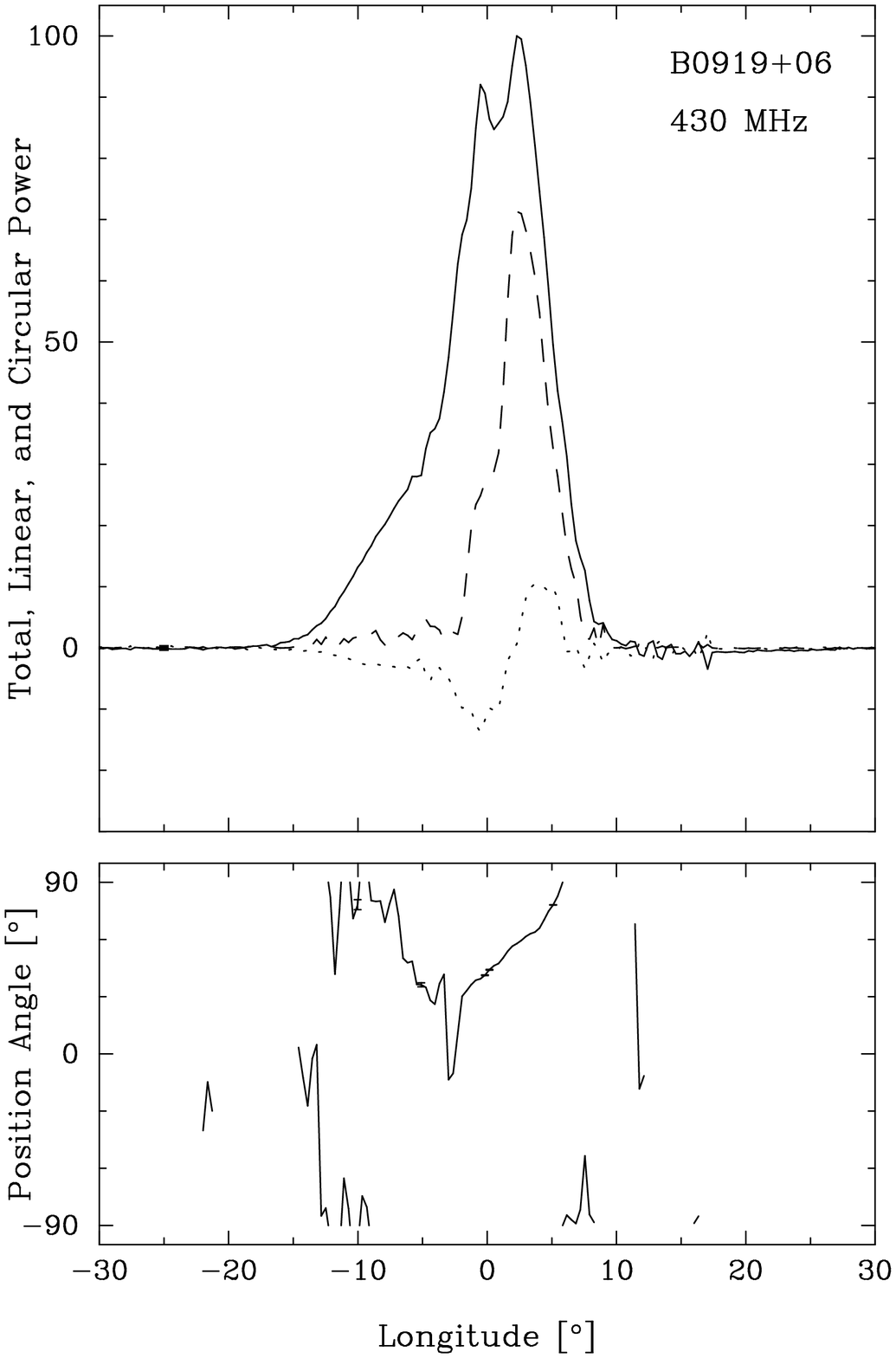} 
\pf{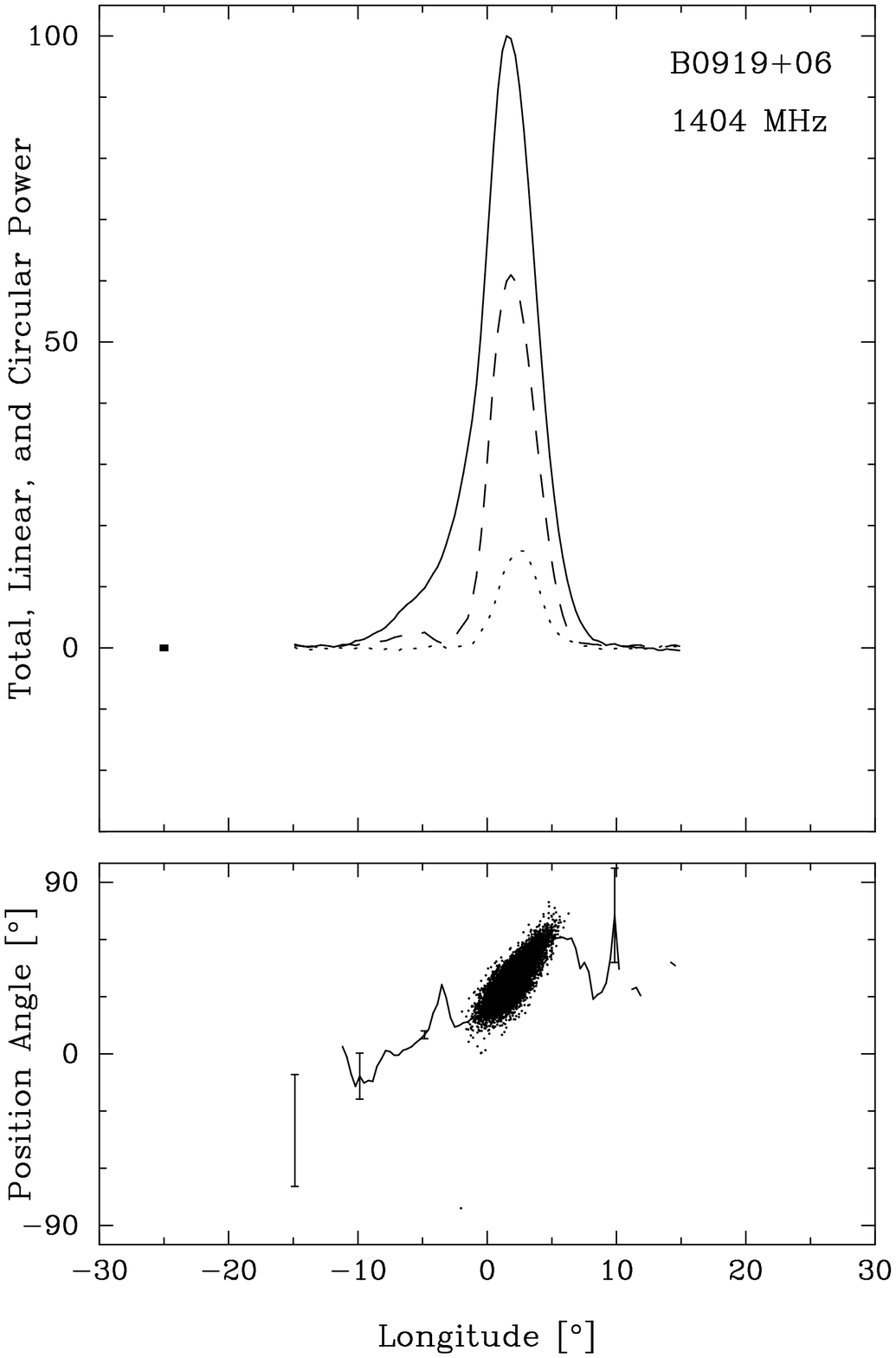} 
}
\quad \\
\centerline{
\pf{figs/dummy_fig.ps}     
\pf{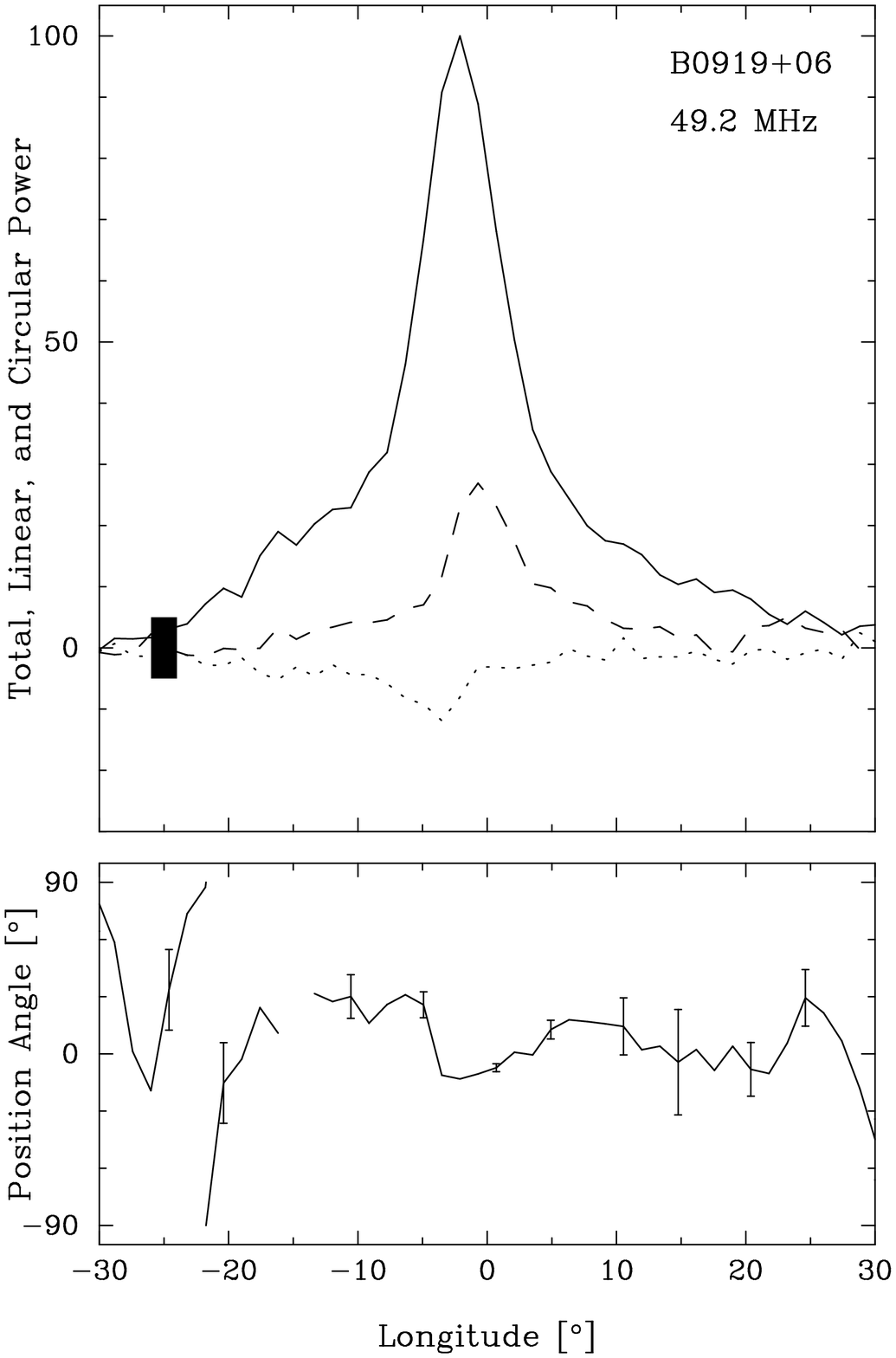} 
\pf{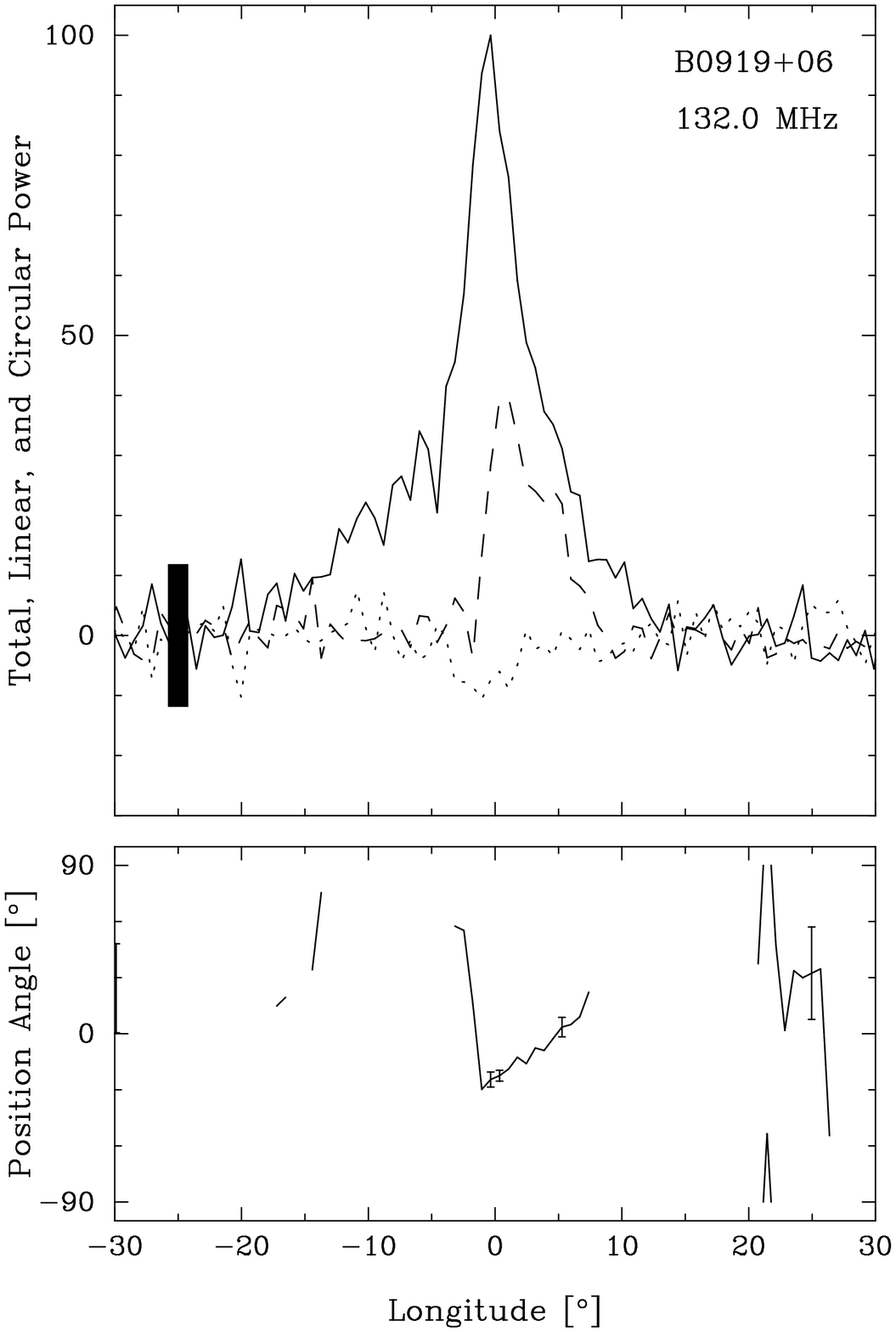} 
}
\quad \\
\centerline{
\pf{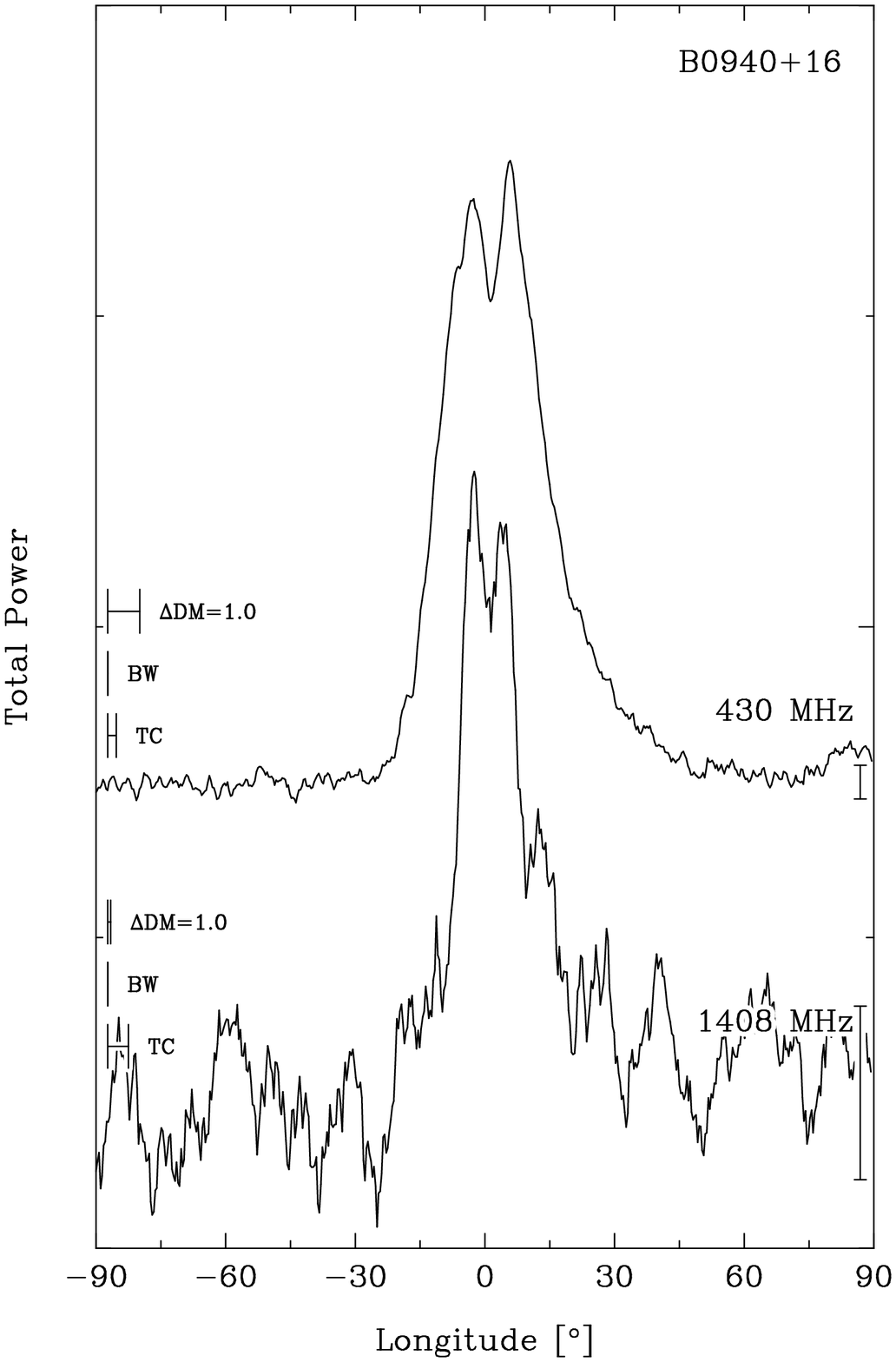}     
\pf{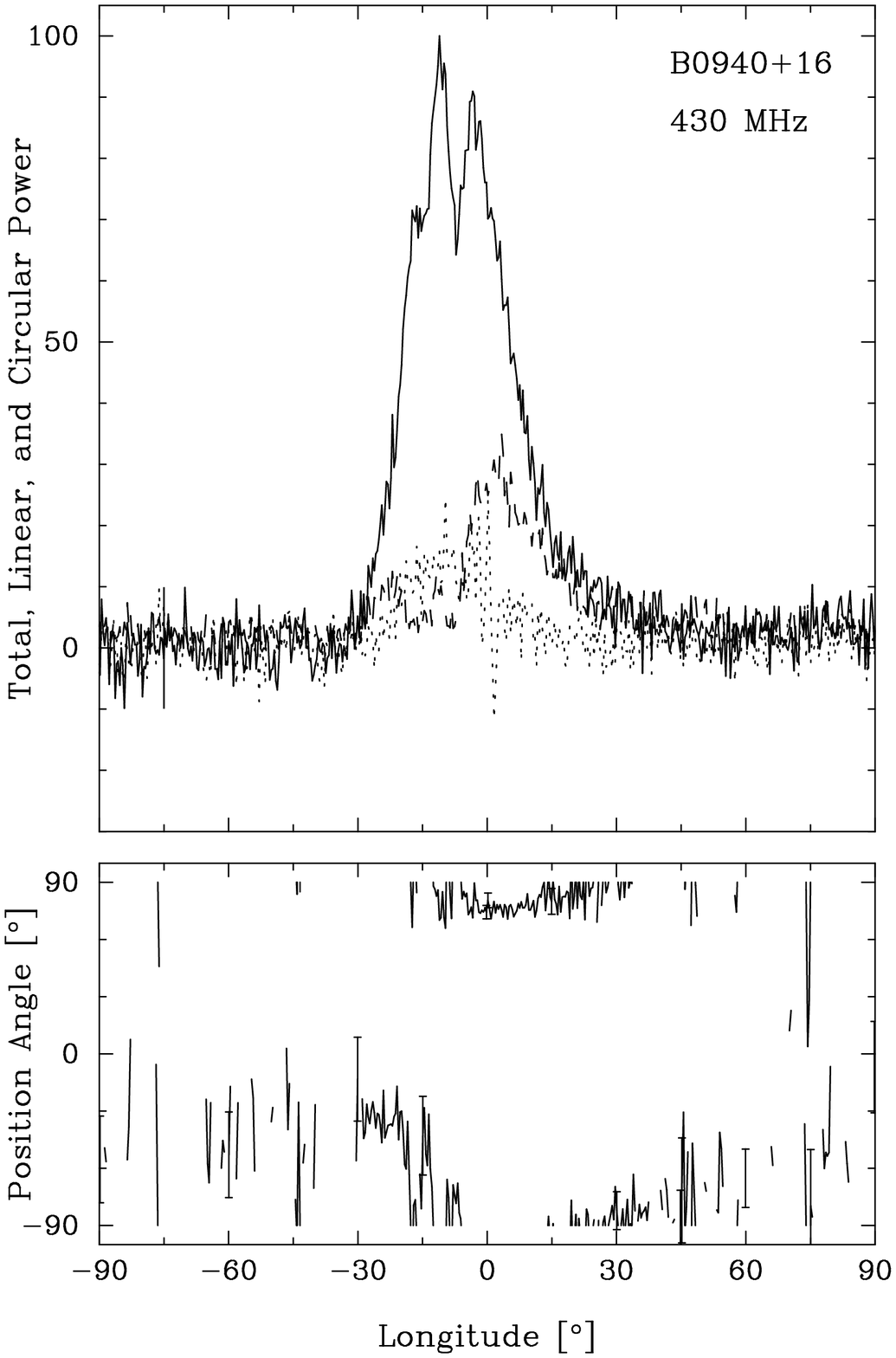} 
\pf{figs/dummy_fig.ps}       
}
\caption{Multi-frequency and polarization profiles of B0919+06 and B0940+16.}
\label{b5}
\end{figure}

\begin{figure}[htb]
\centerline{
\pf{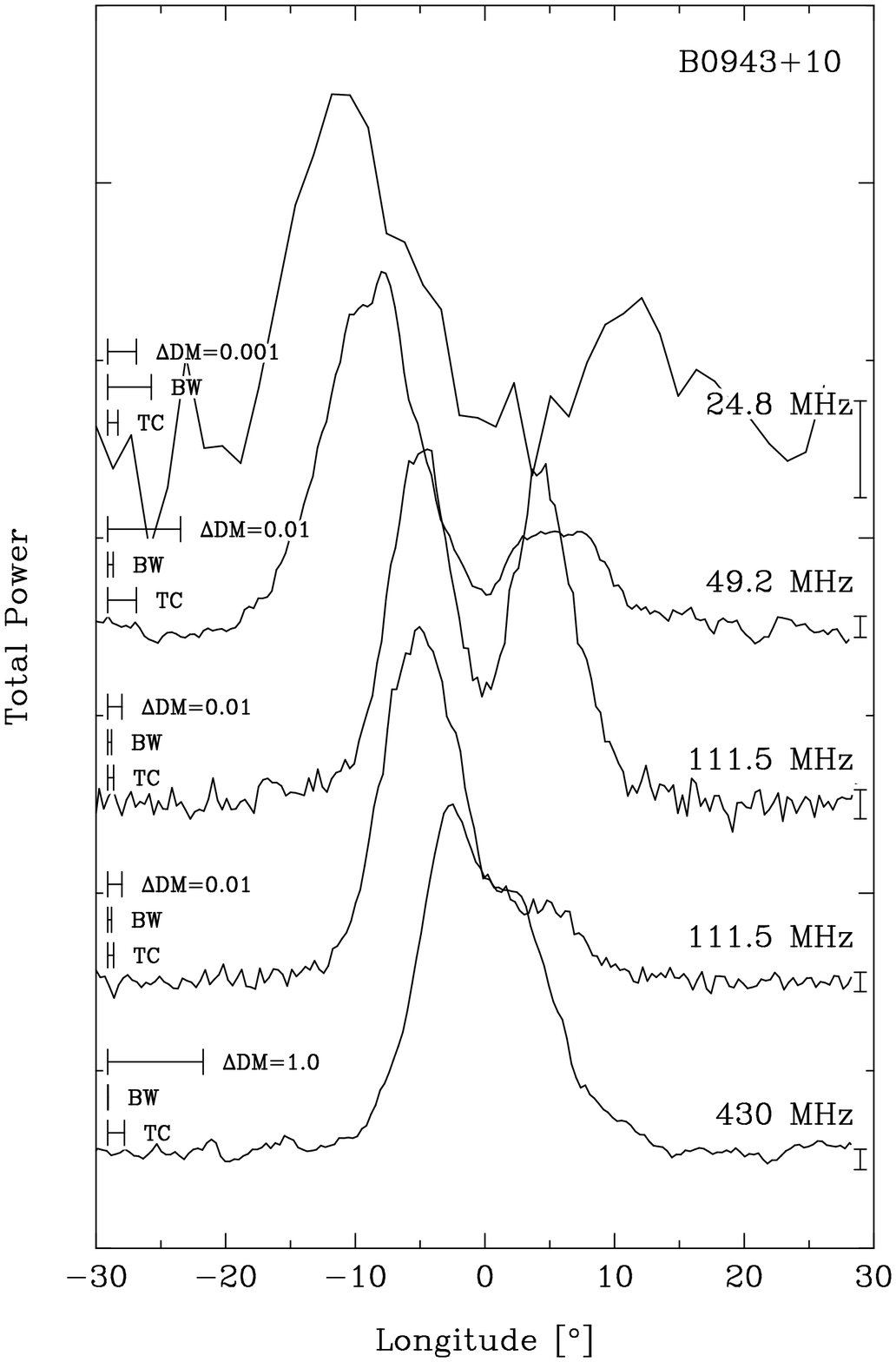}     
\pf{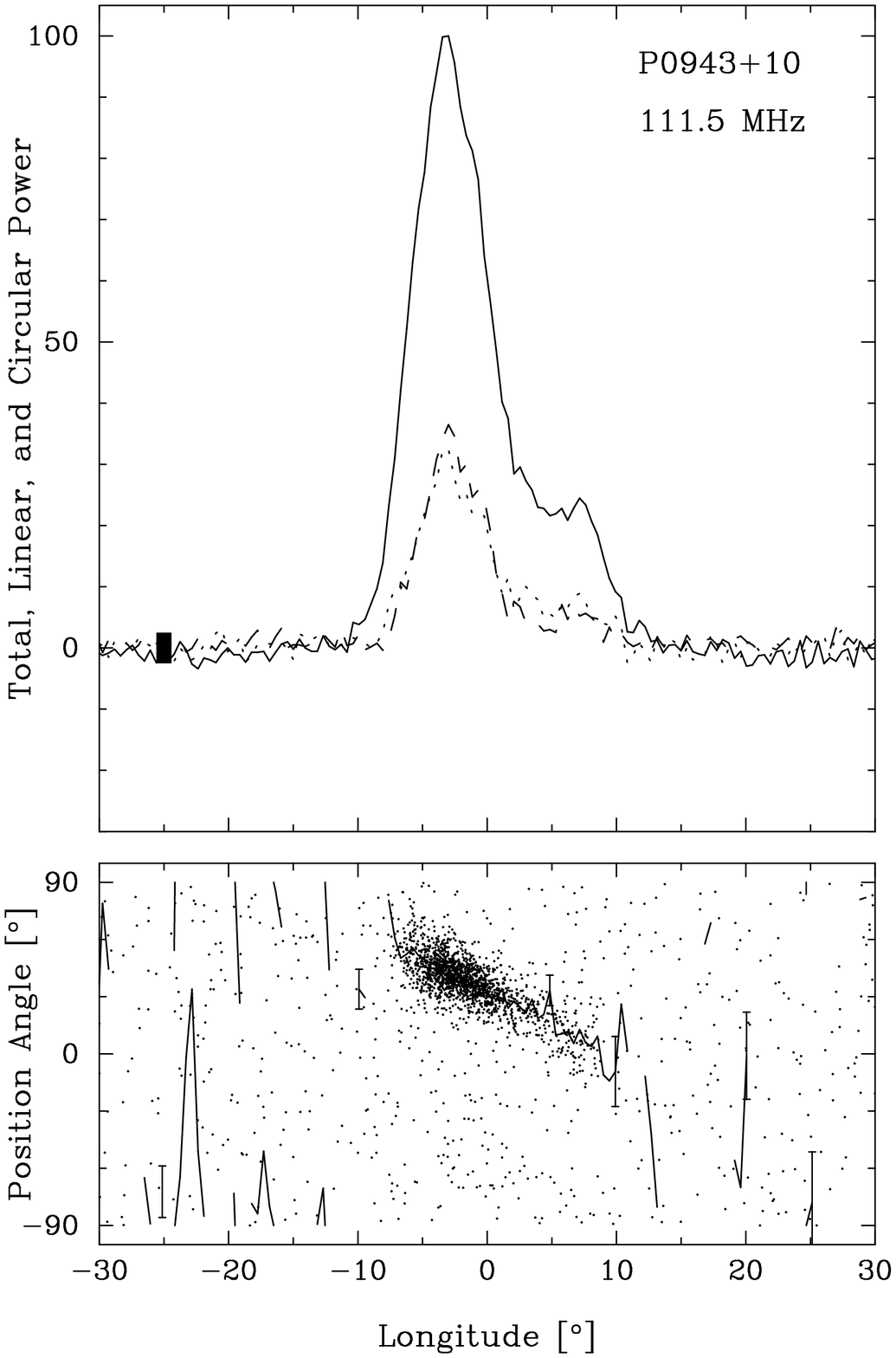} 
\pf{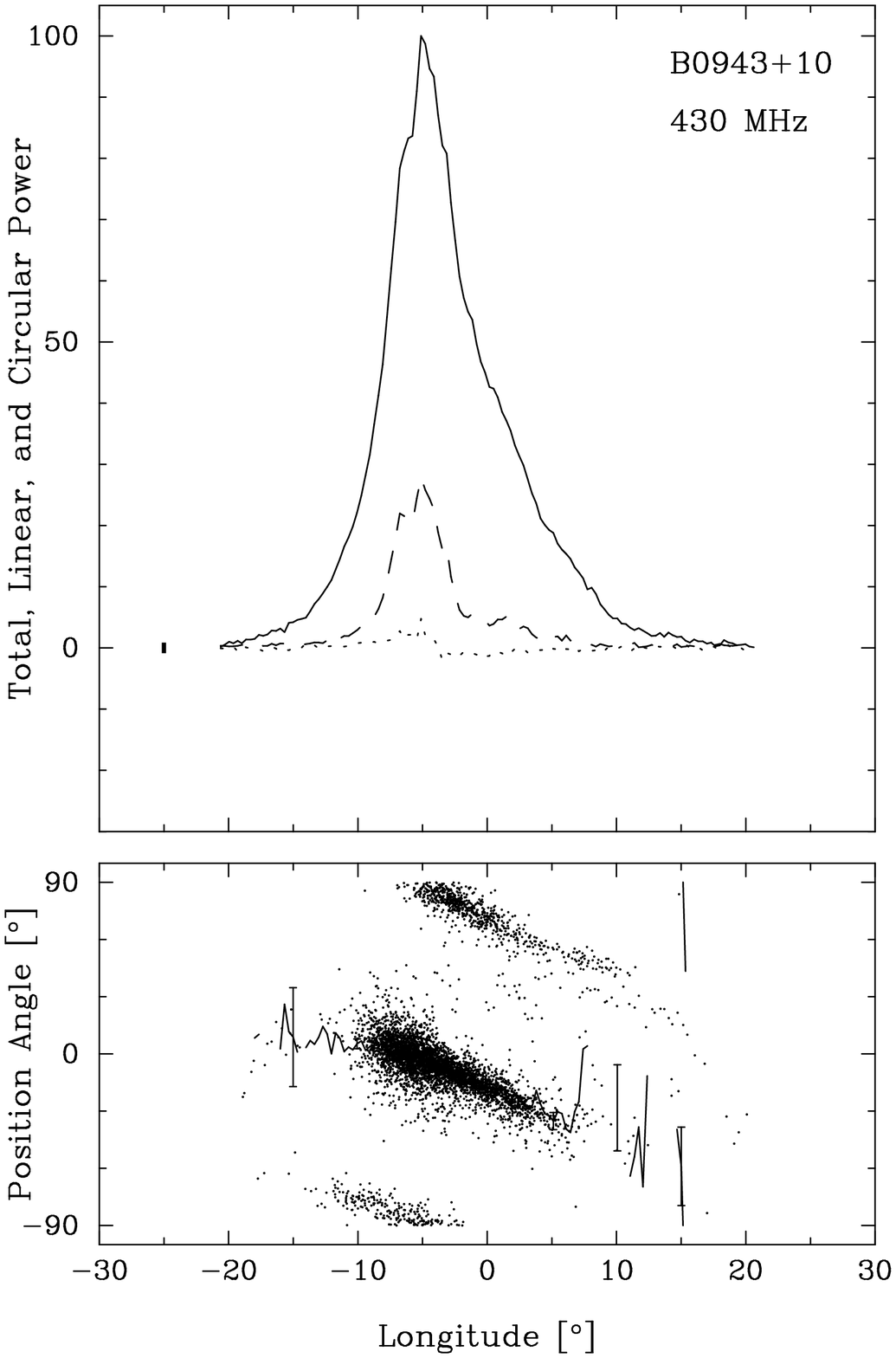} 
}
\quad \\
\centerline{
\pf{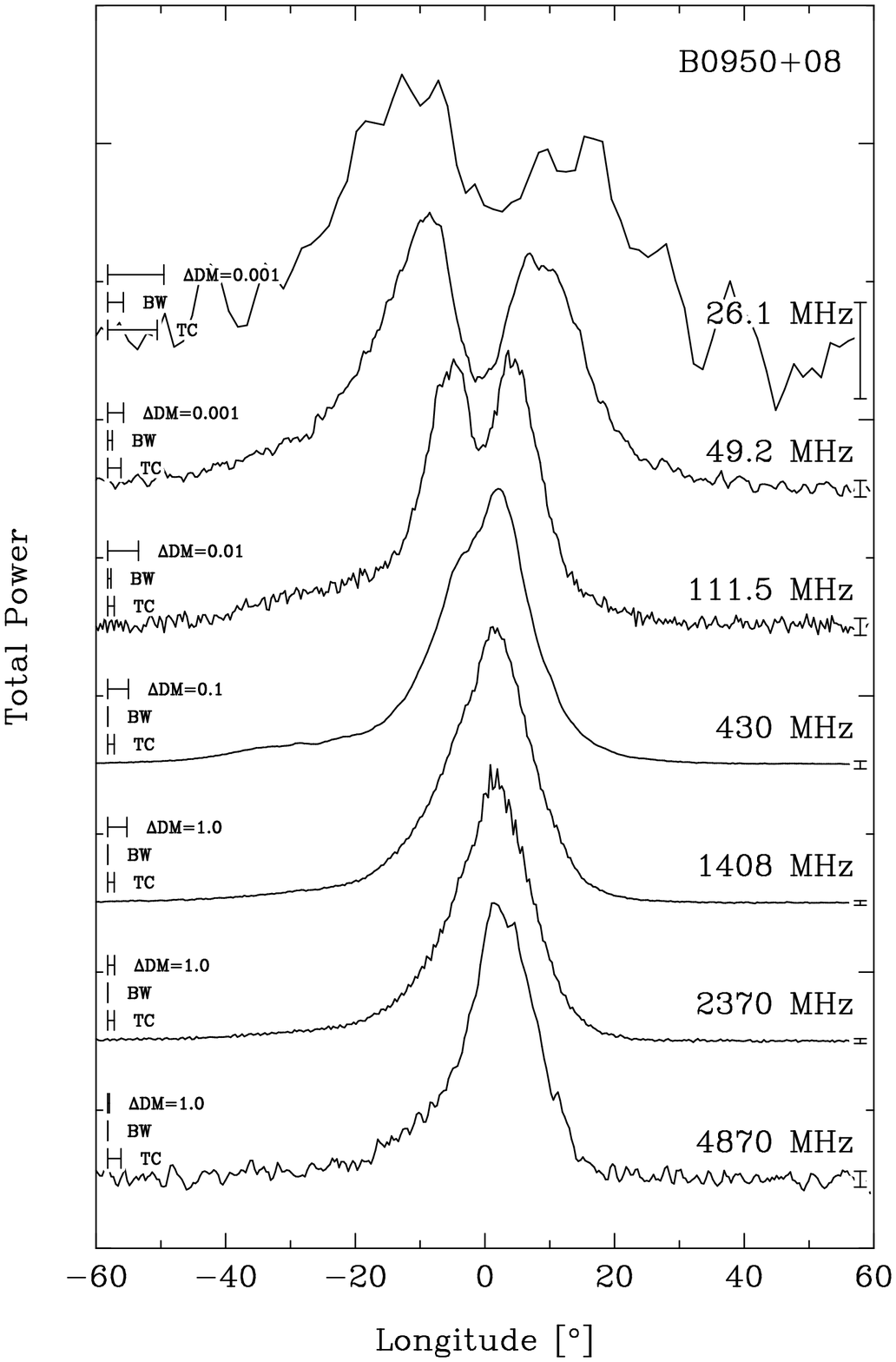}     
\pf{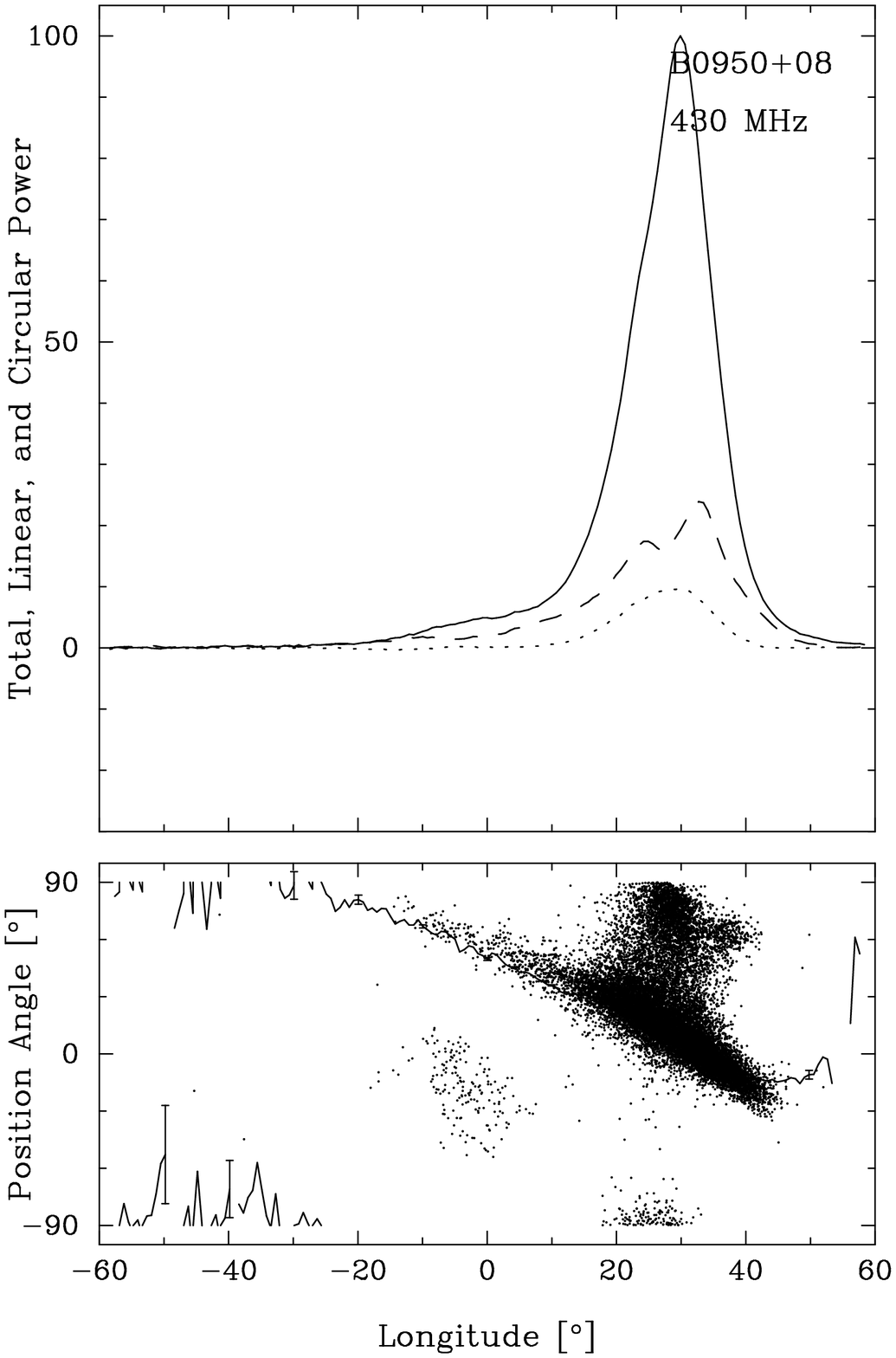} 
\pf{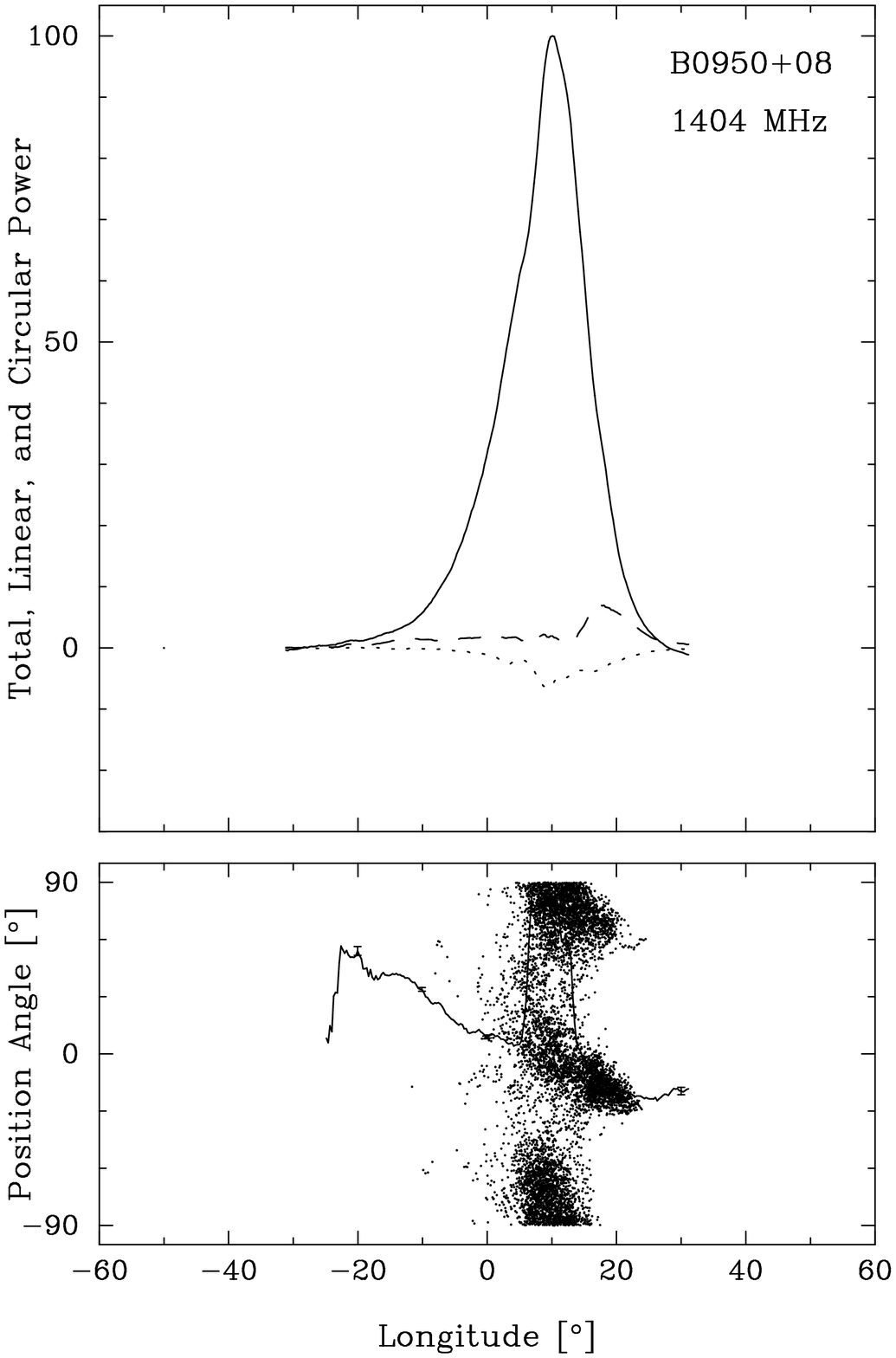} 
}
\quad \\
\centerline{
\pf{figs/dummy_fig.ps}     
\pf{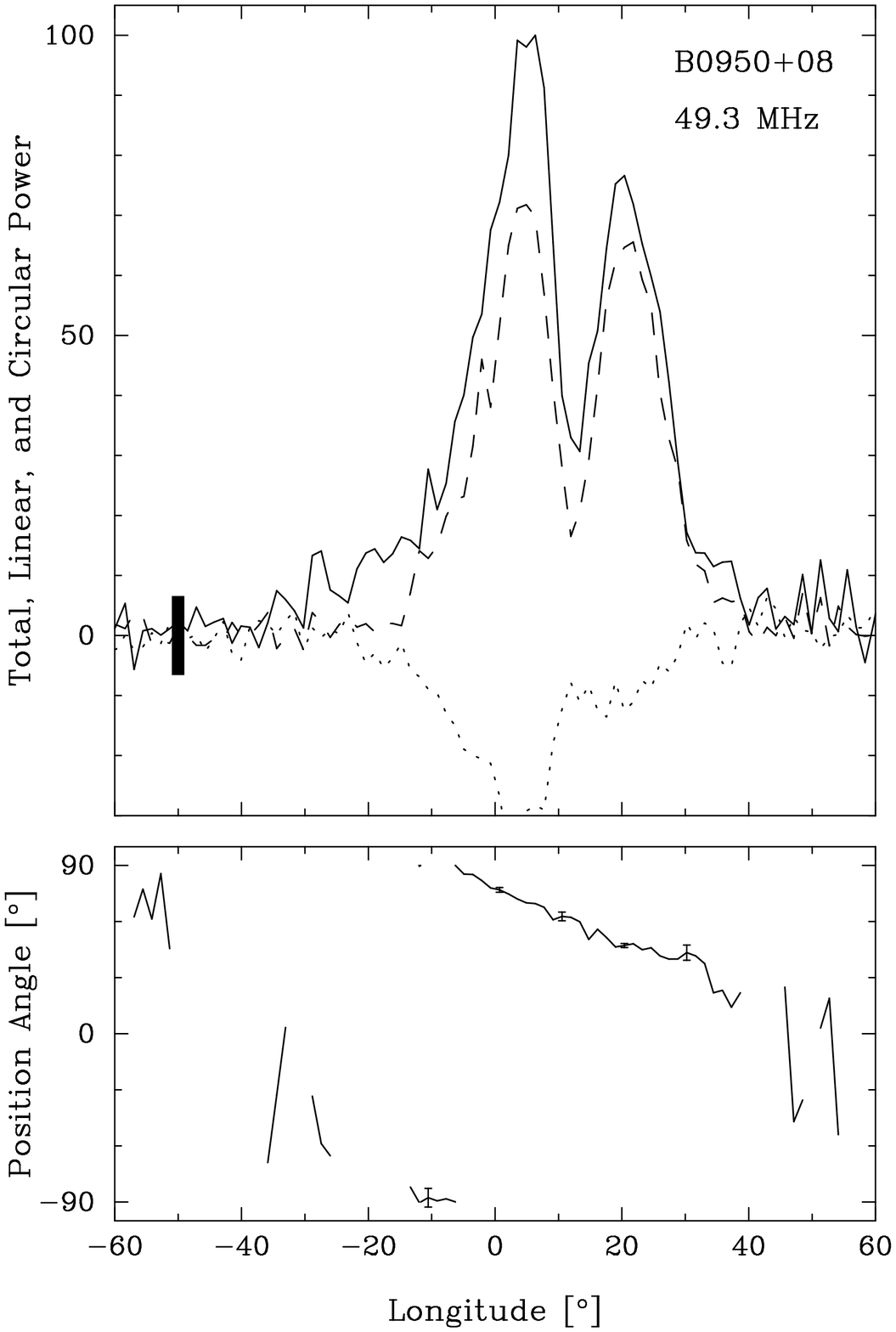} 
\pf{figs/dummy_fig.ps} 
}
\caption{Multi-frequency and polarization profiles of B0943+10 and B0950+08.}
\label{b6}
\end{figure}

\begin{figure}[htb]
\centerline{ 
\pf{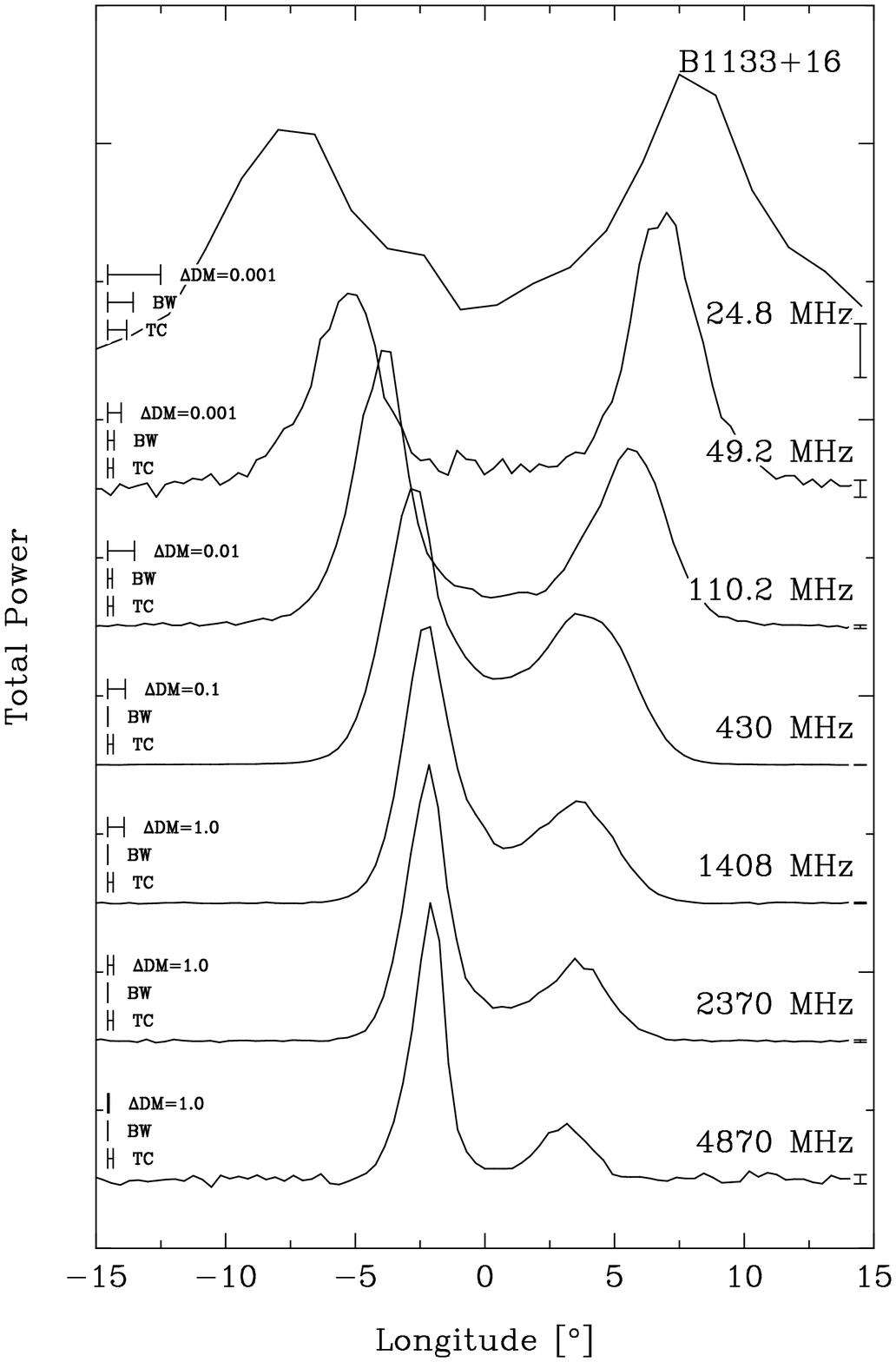}     
\pf{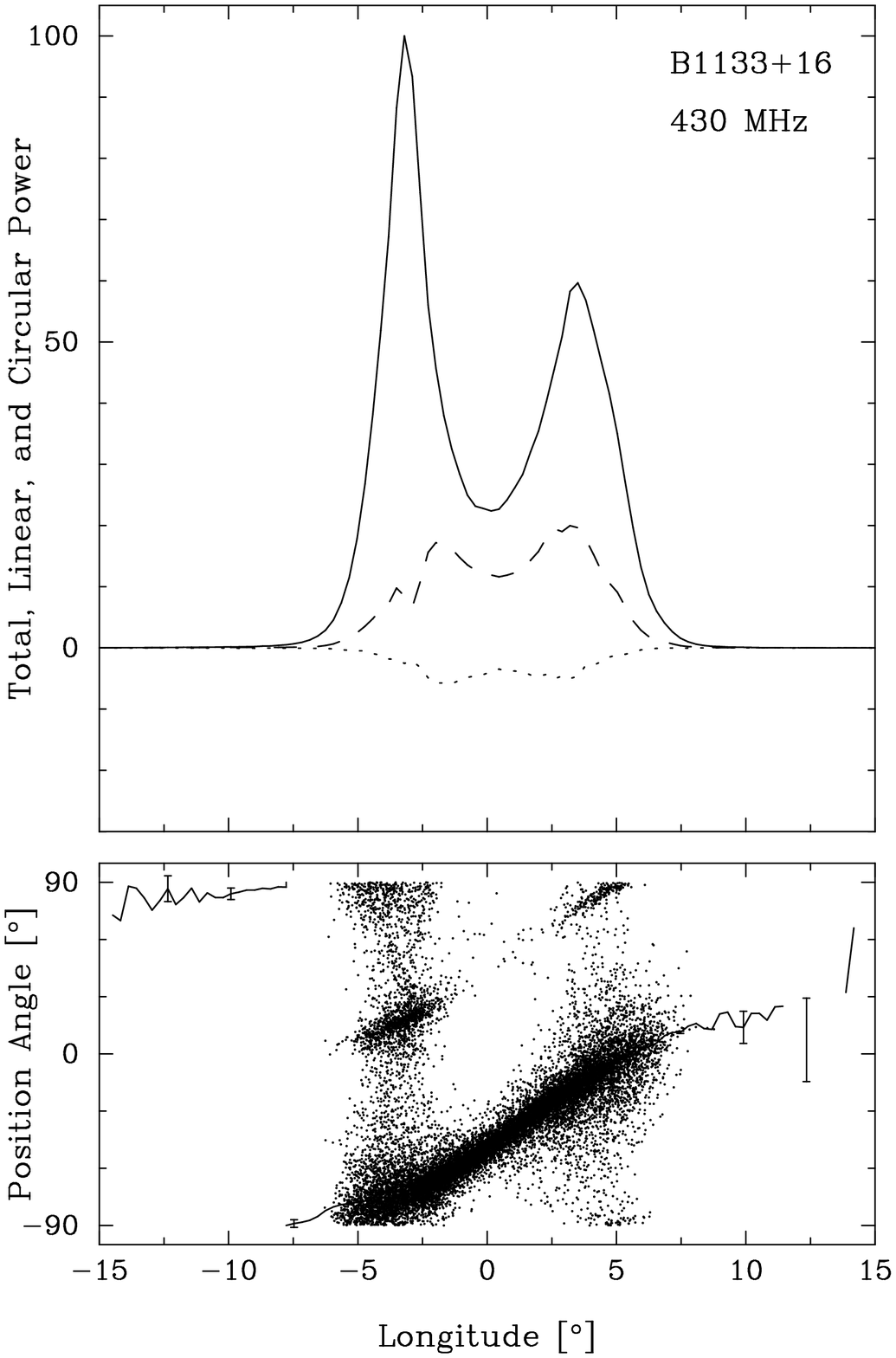} 
\pf{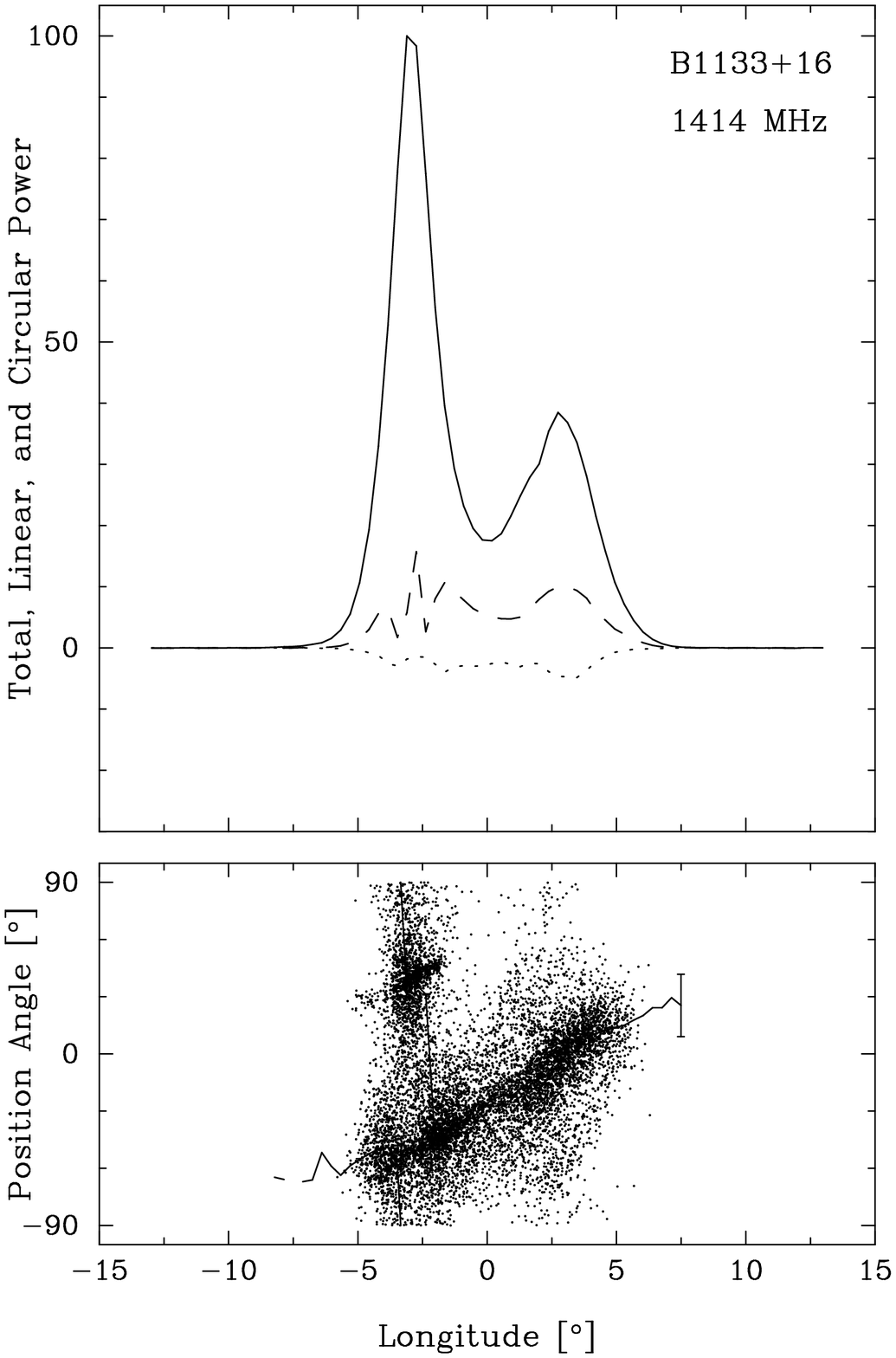} 
}
\quad \\
\centerline{
\pf{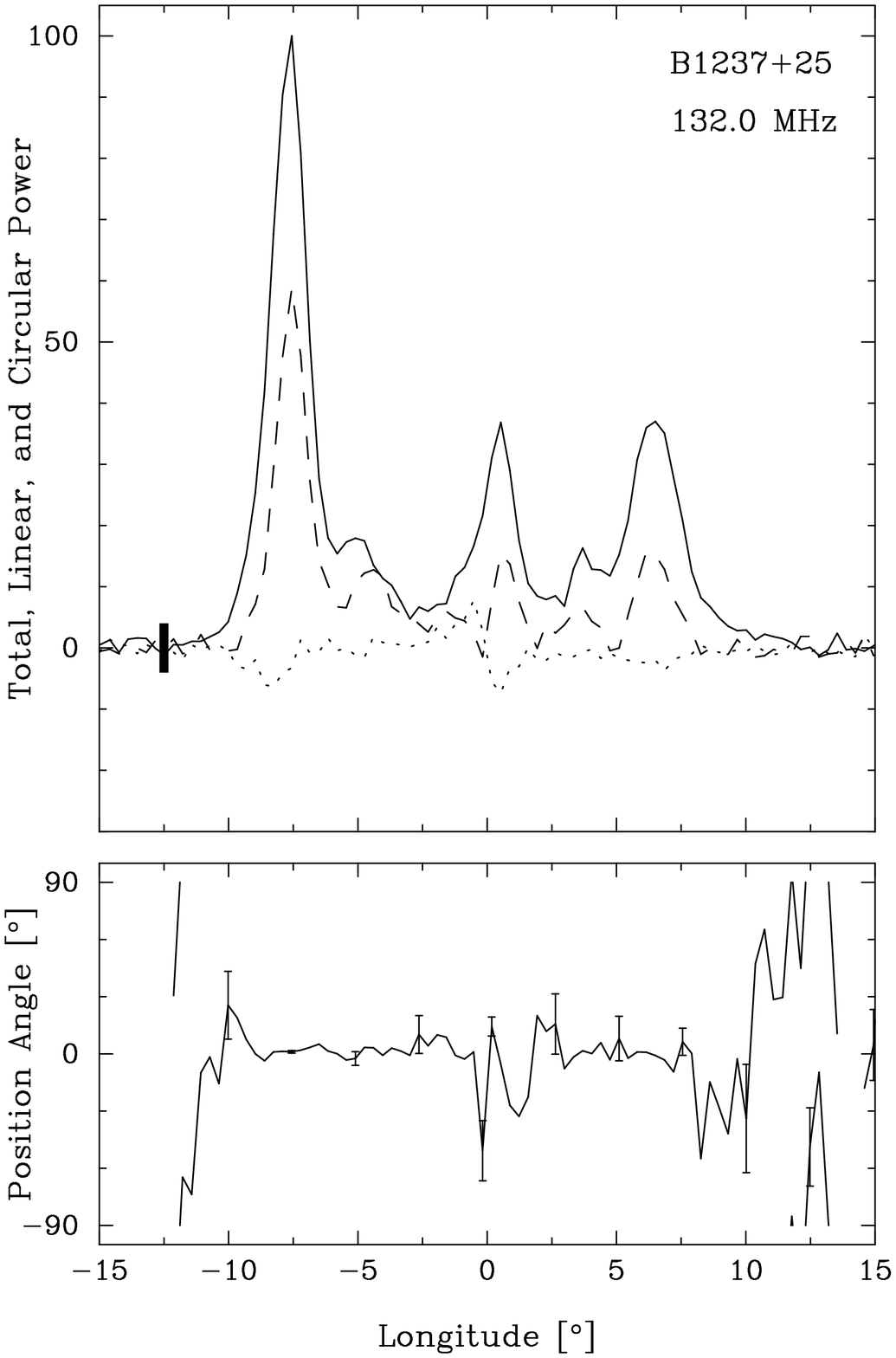}     
\pf{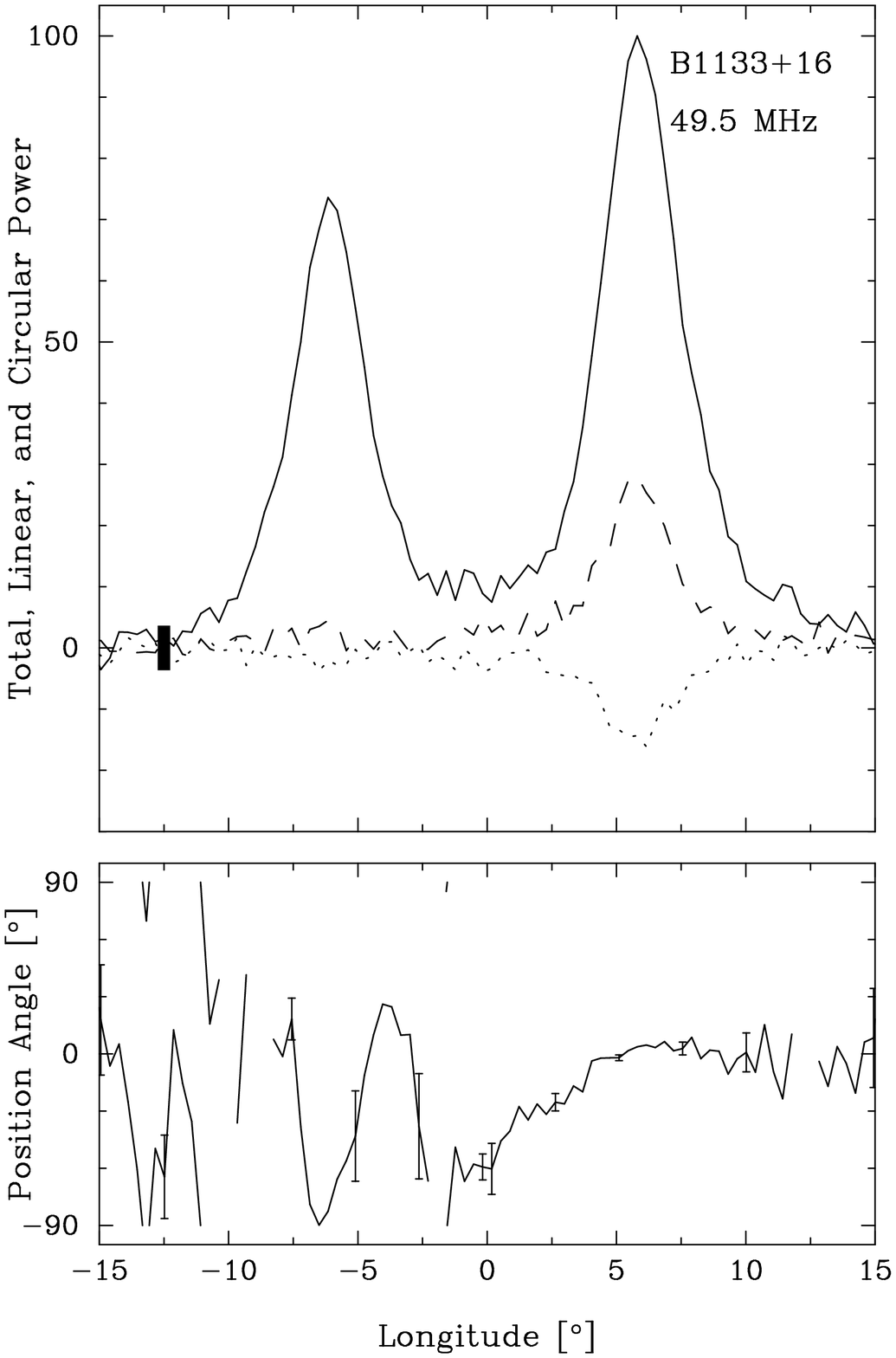} 
\pf{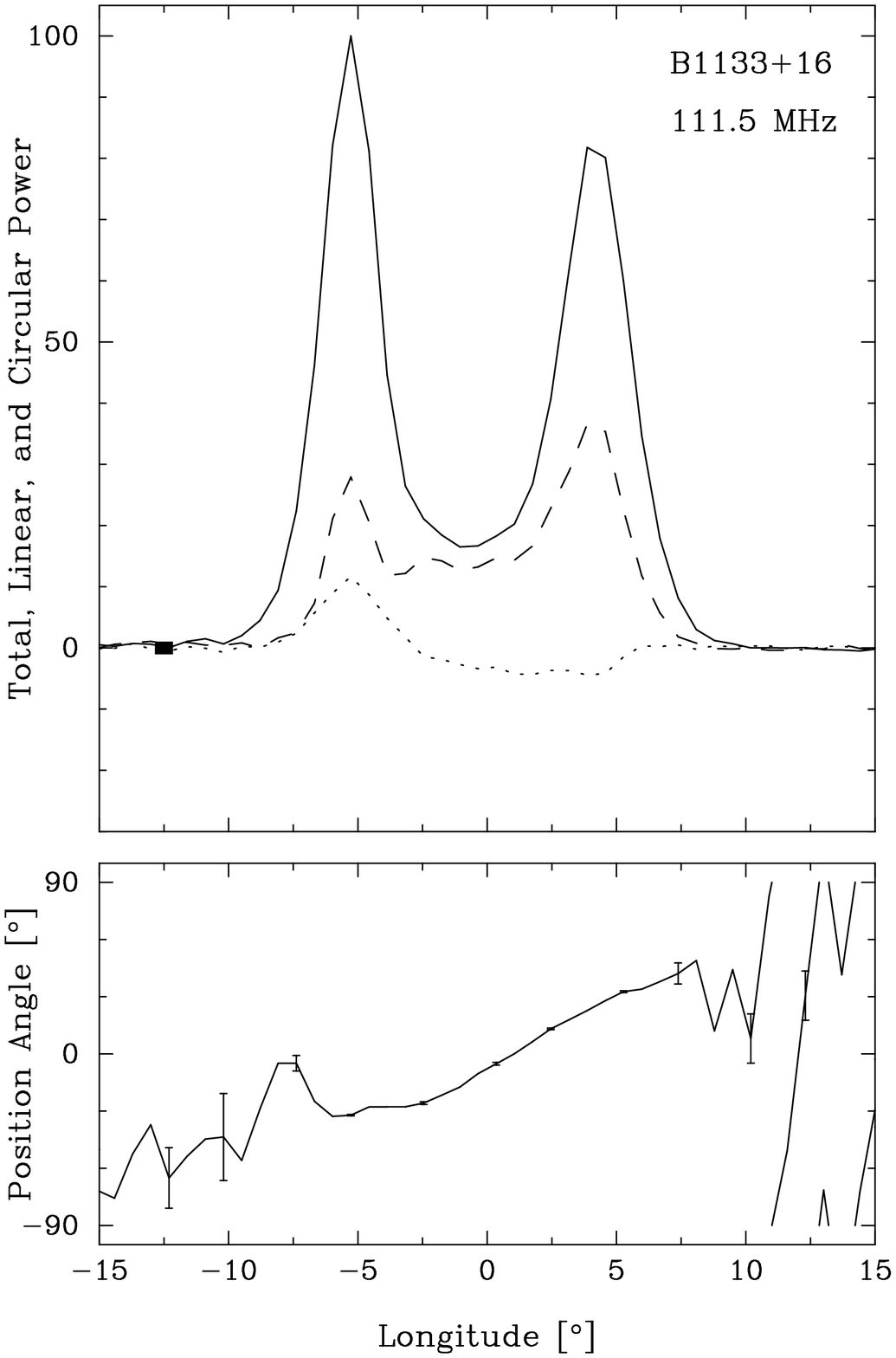} 
}
\quad \\
\centerline{
\pf{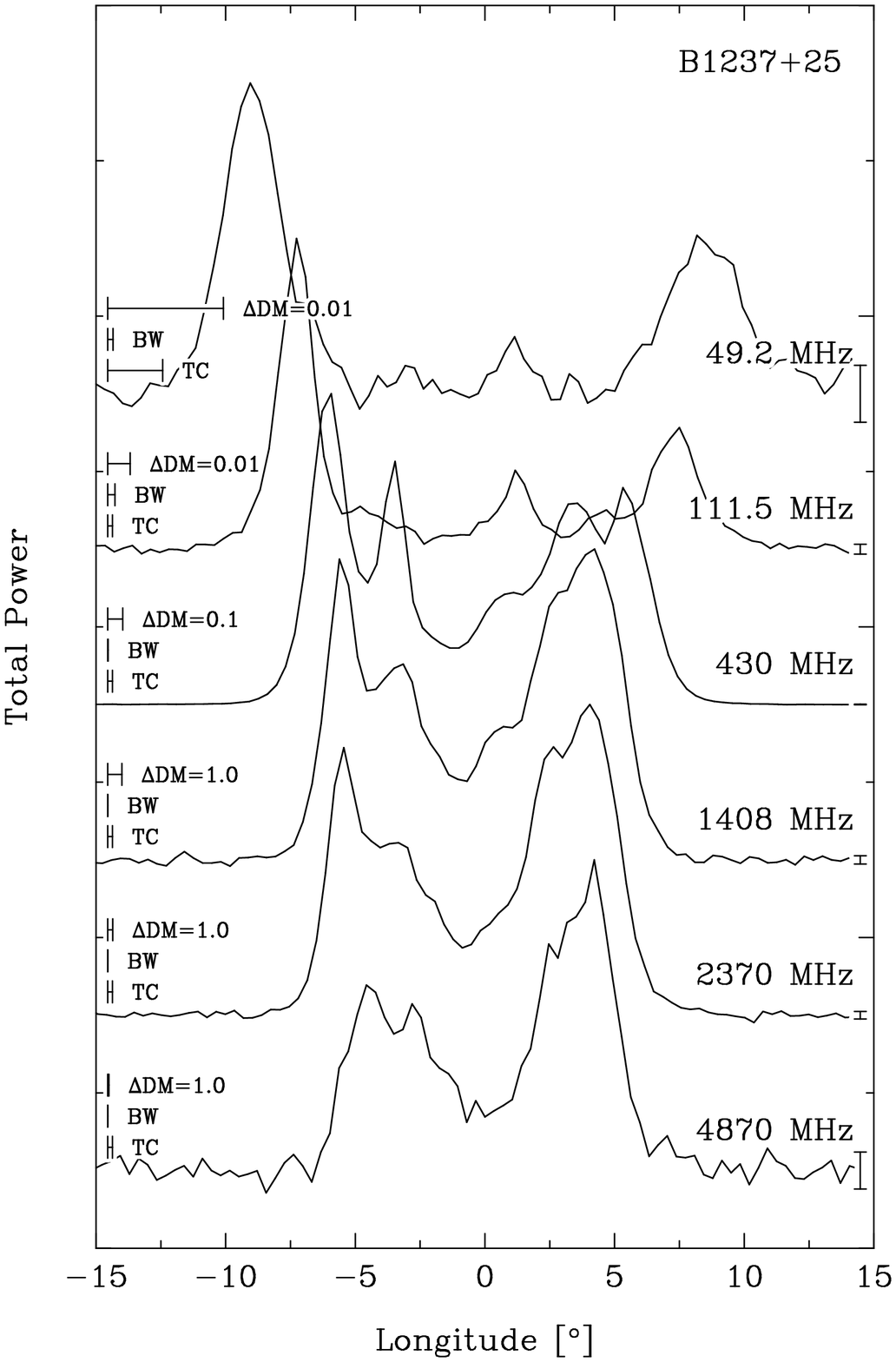}     
\pf{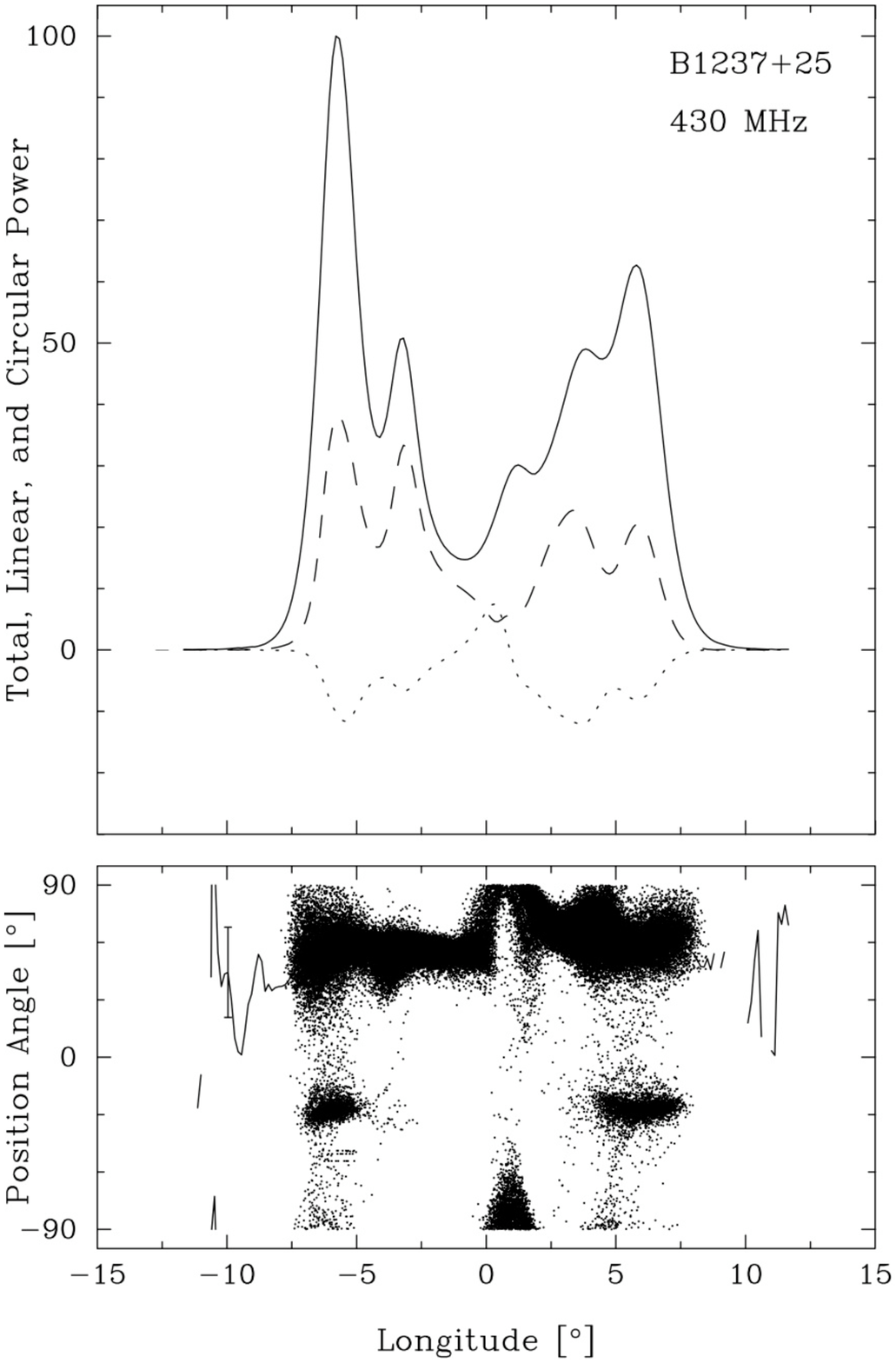} 
\pf{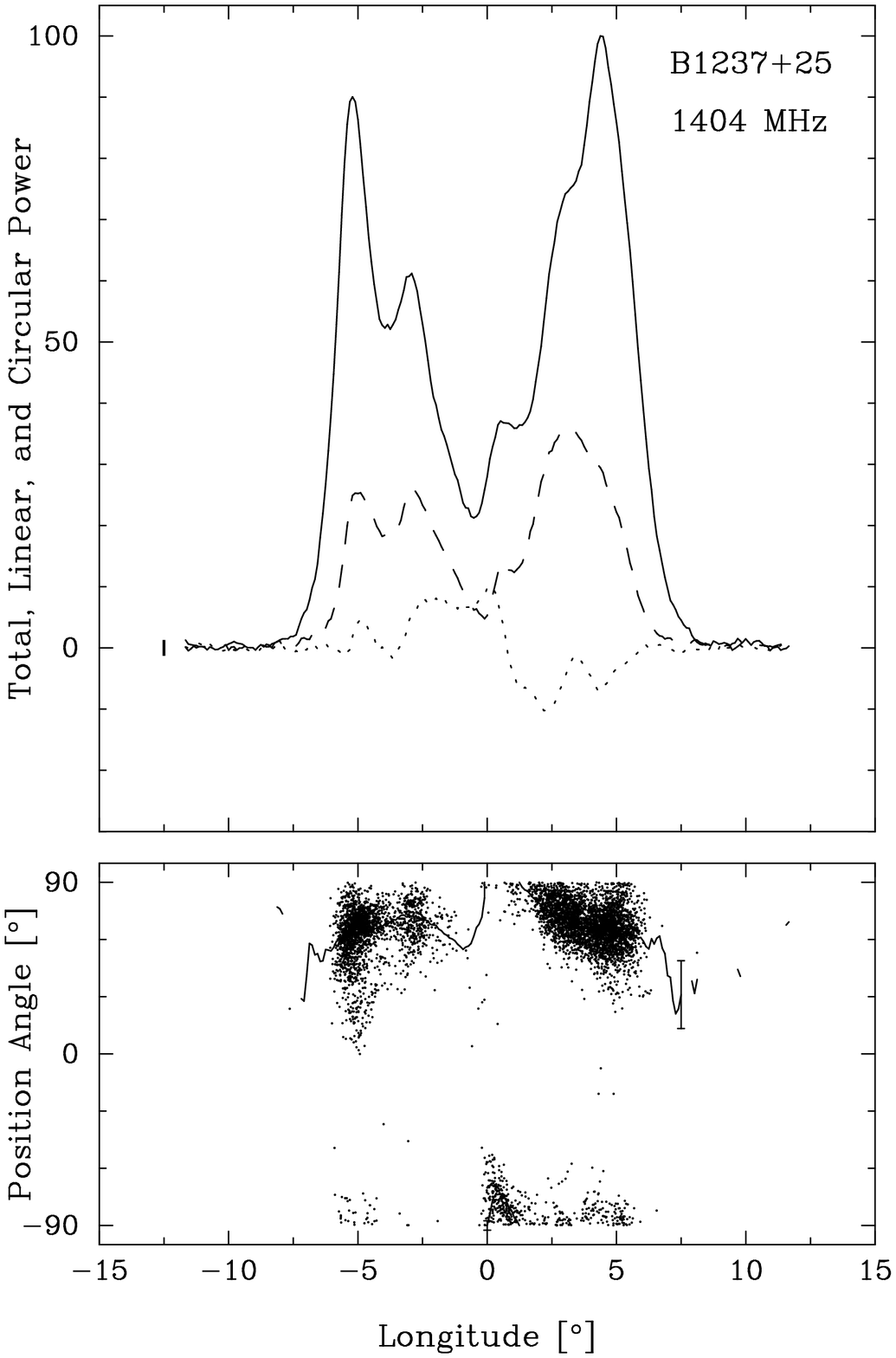} 
}
\caption{Multi-frequency and polarization profiles of B1133+16 and B1237+25.}
\label{b7}
\end{figure}

\begin{figure}[htb] 
\centerline{ 
\pf{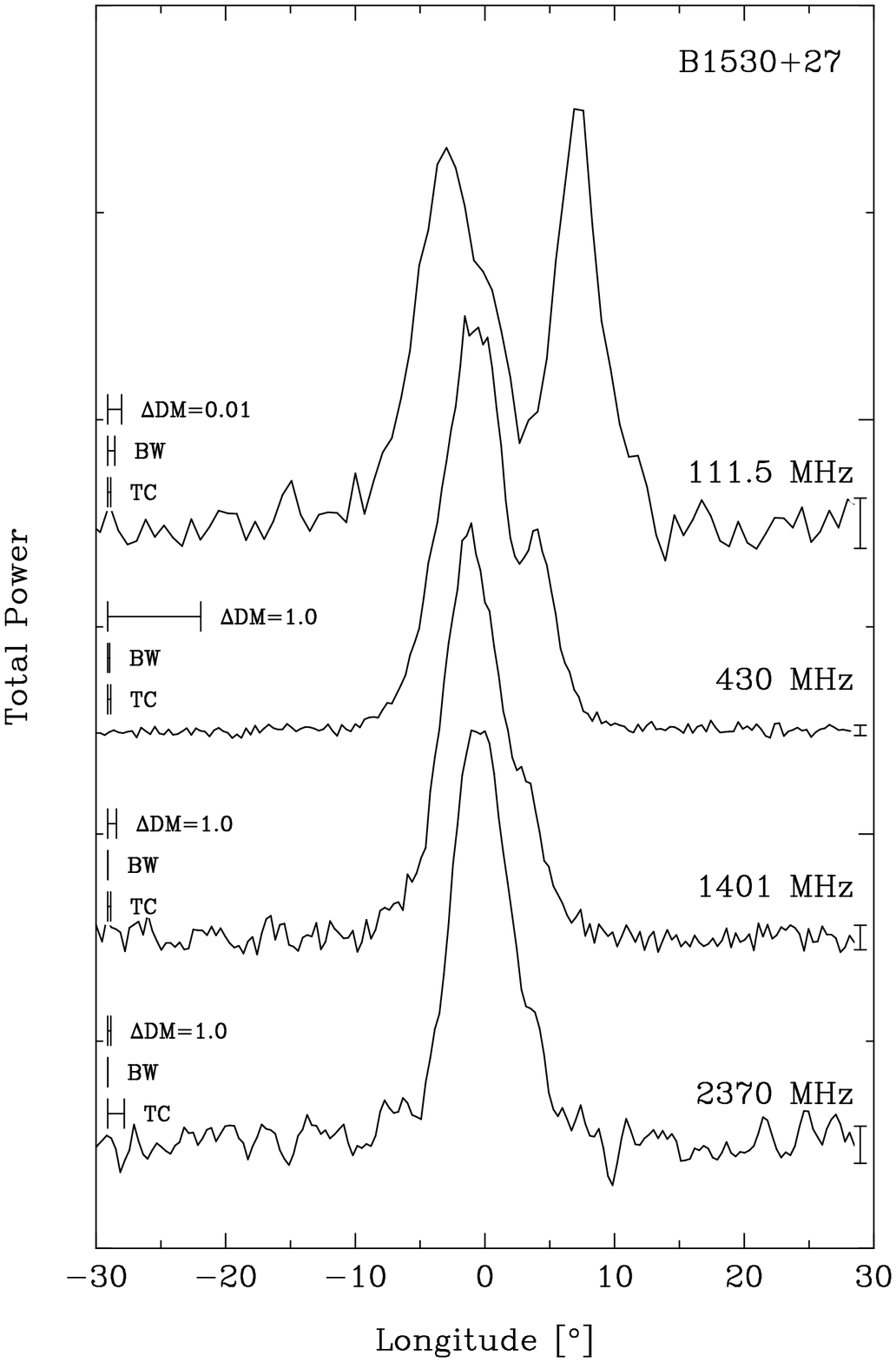}     
\pf{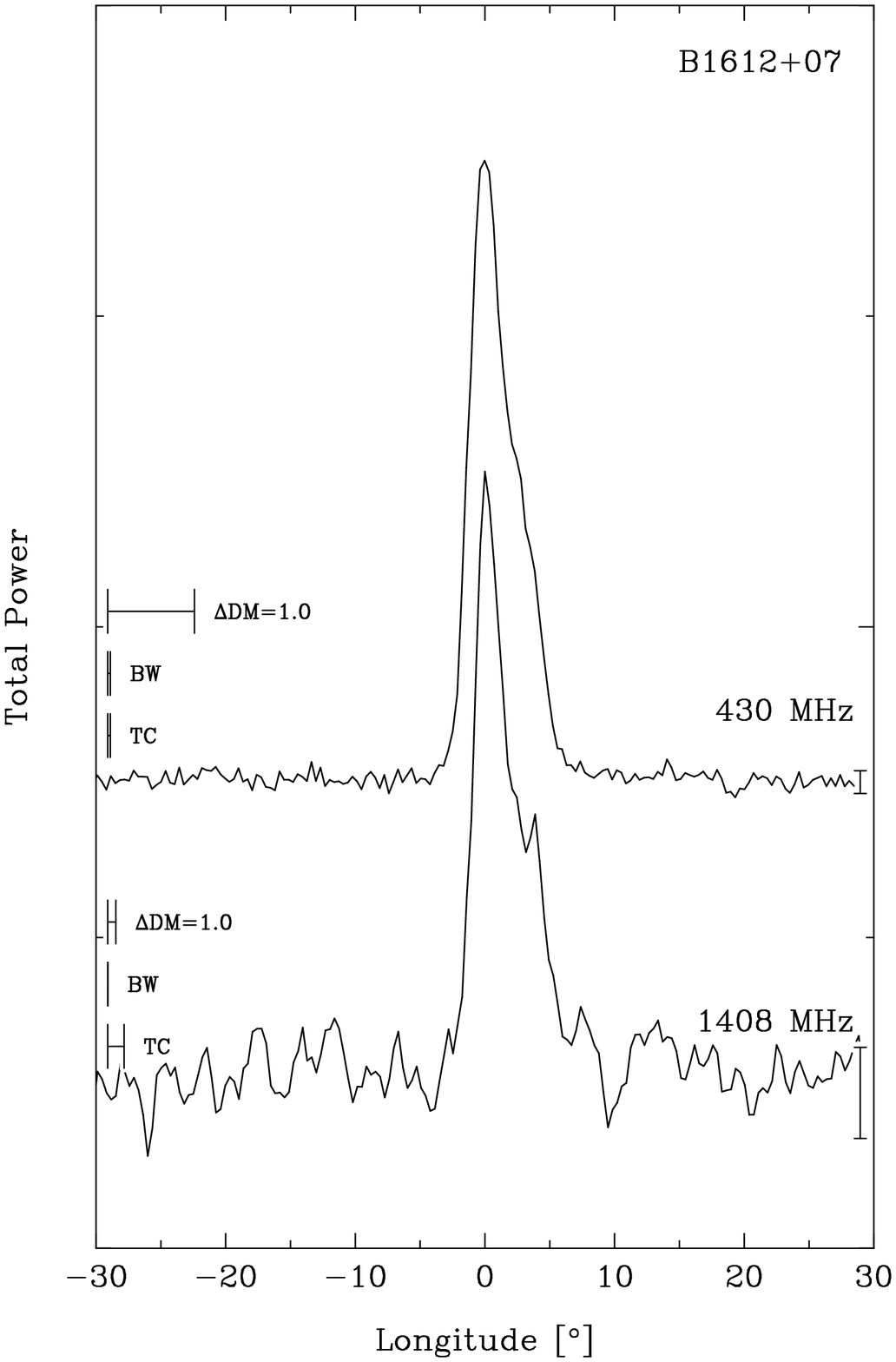}     
\pf{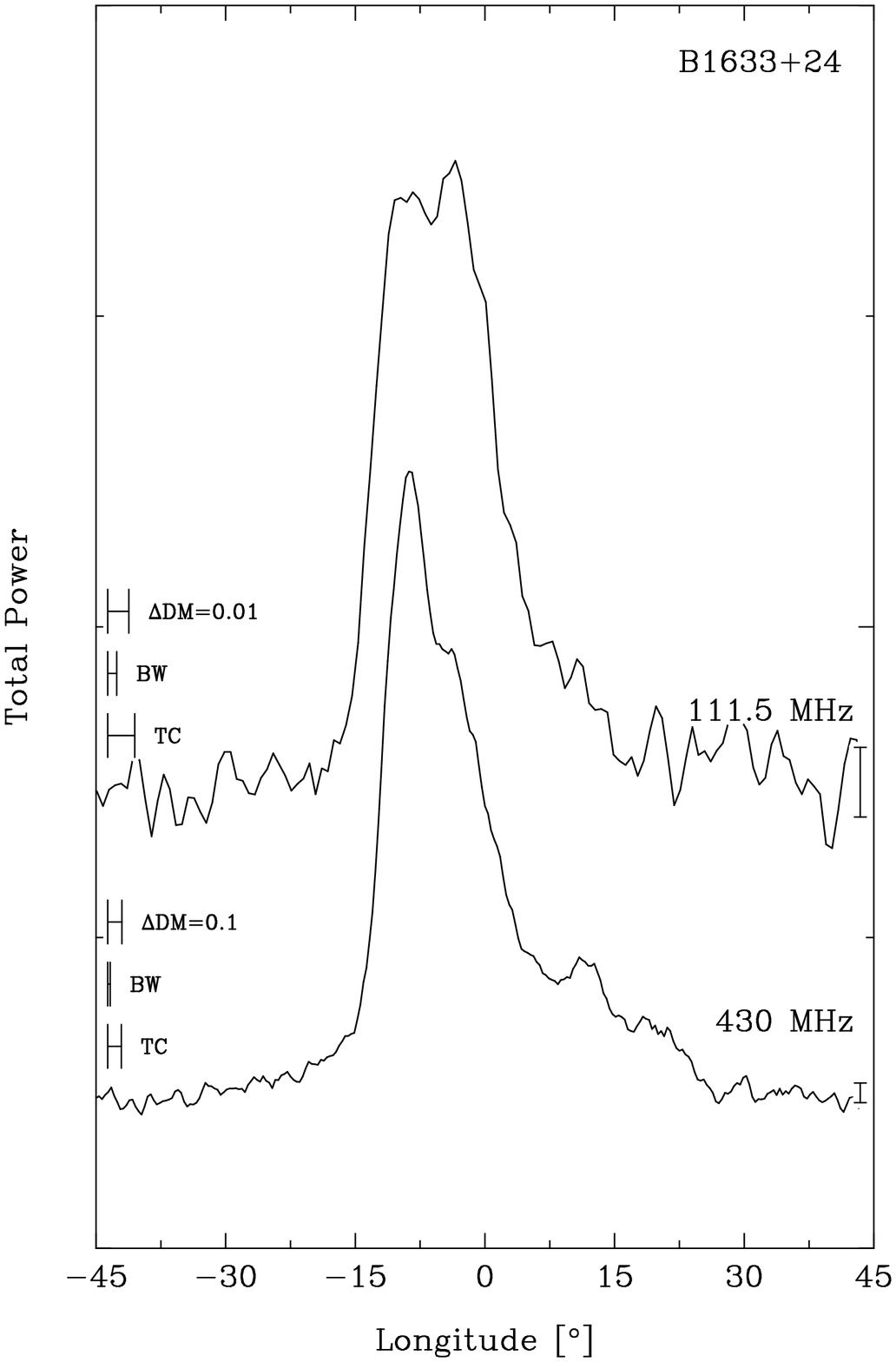}     
}
\quad \\
\centerline{
\pf{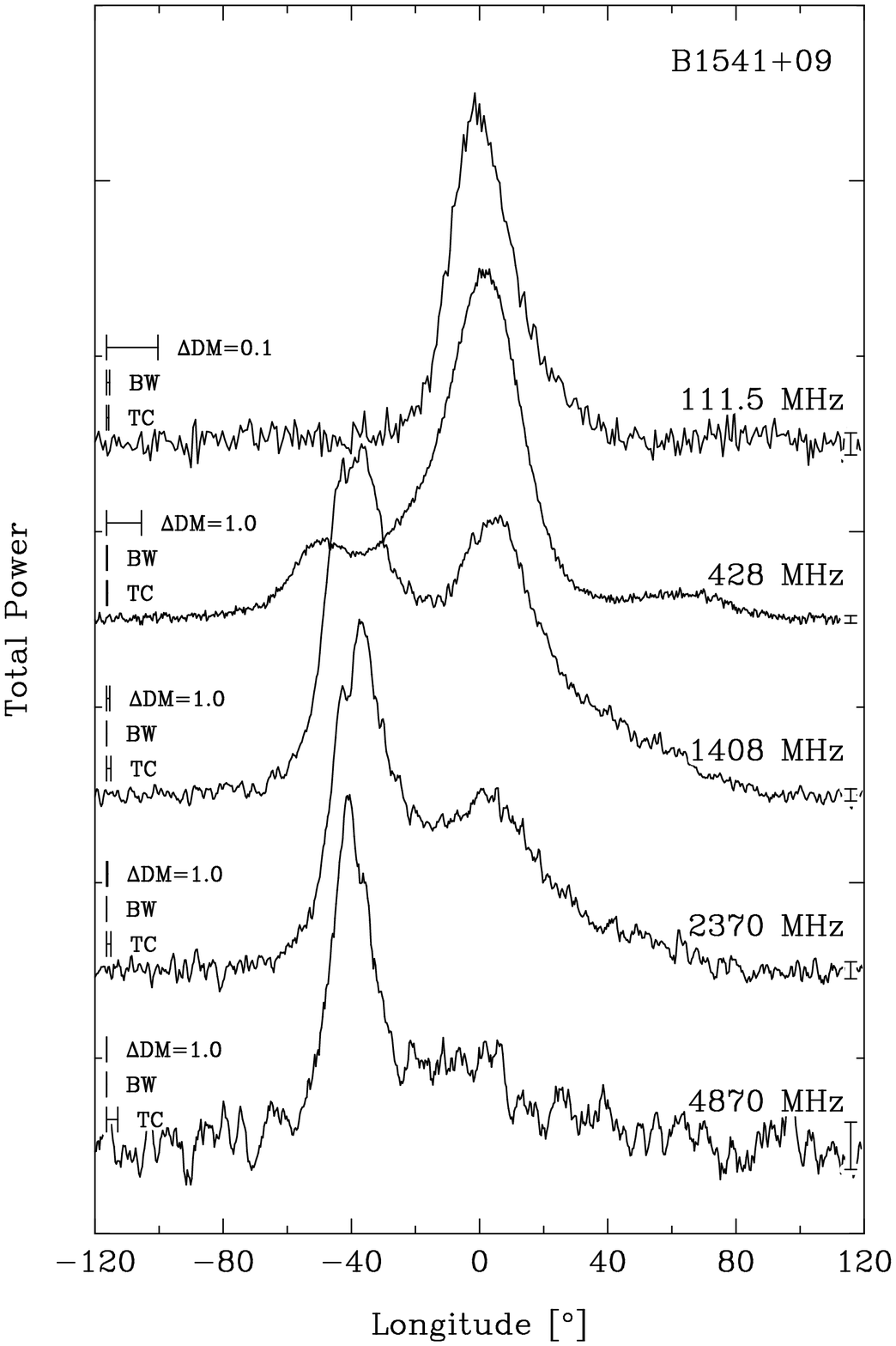}     
\pf{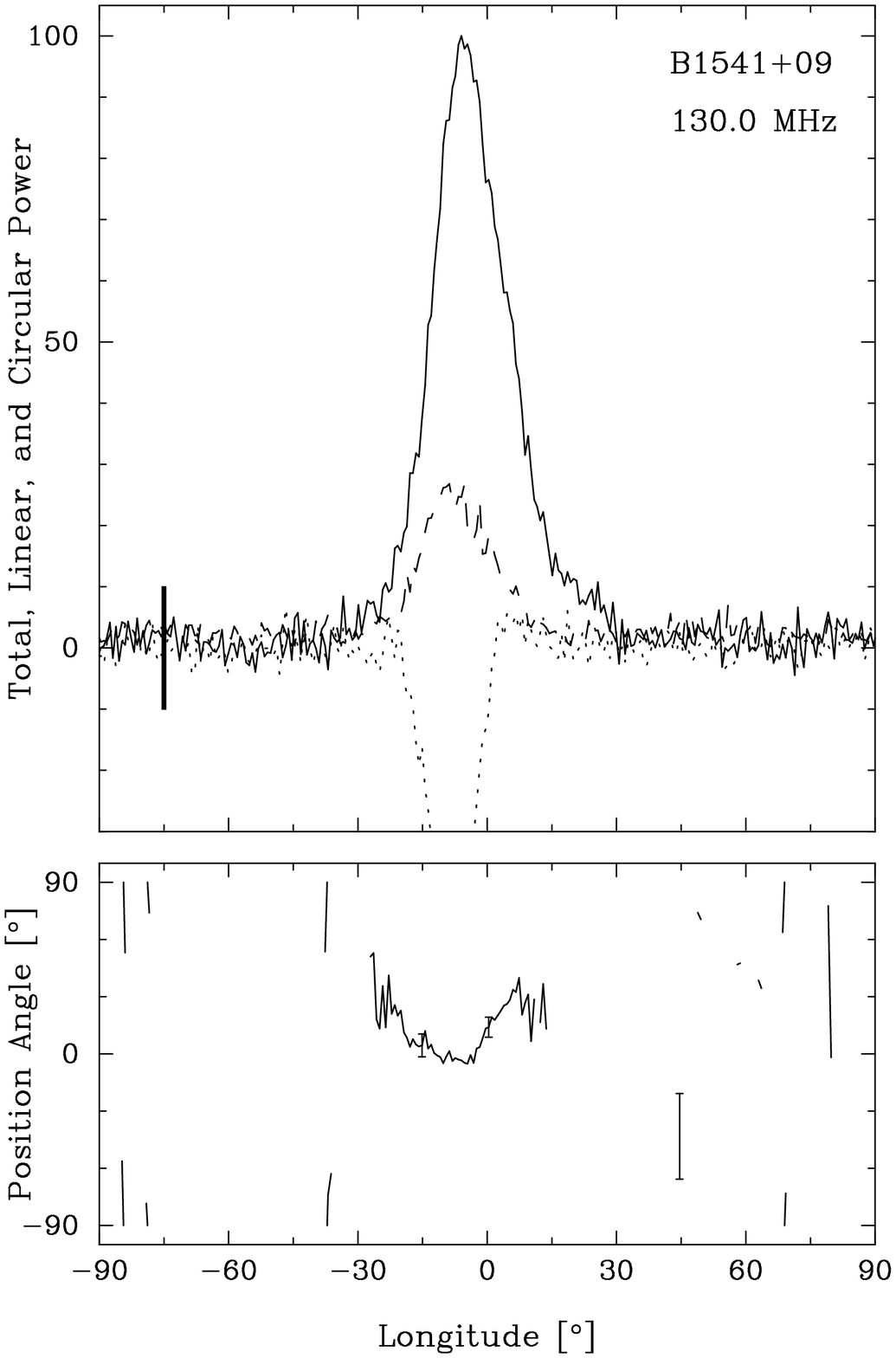} 
\pf{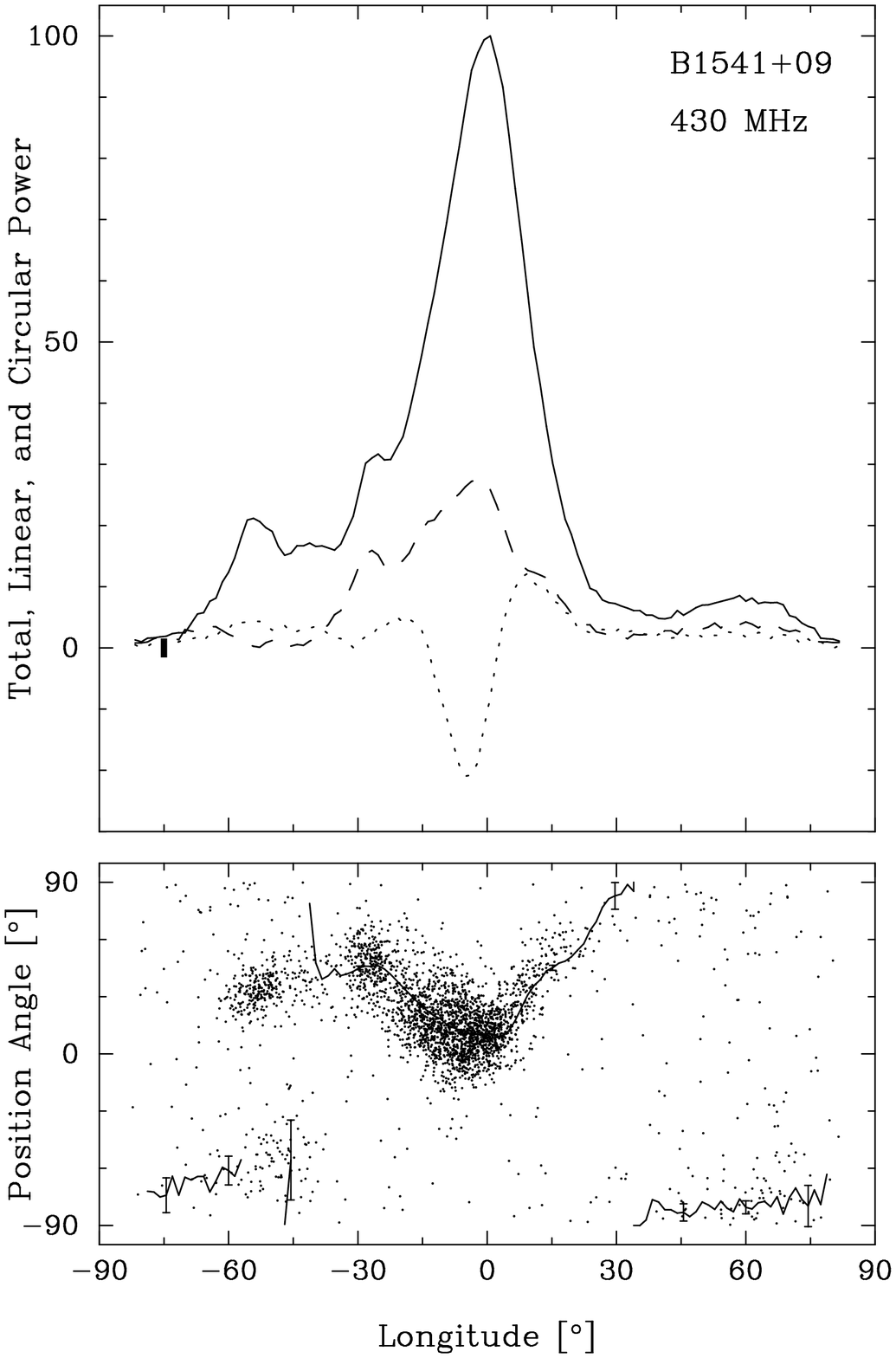} 
}
\quad \\
\centerline{
\pf{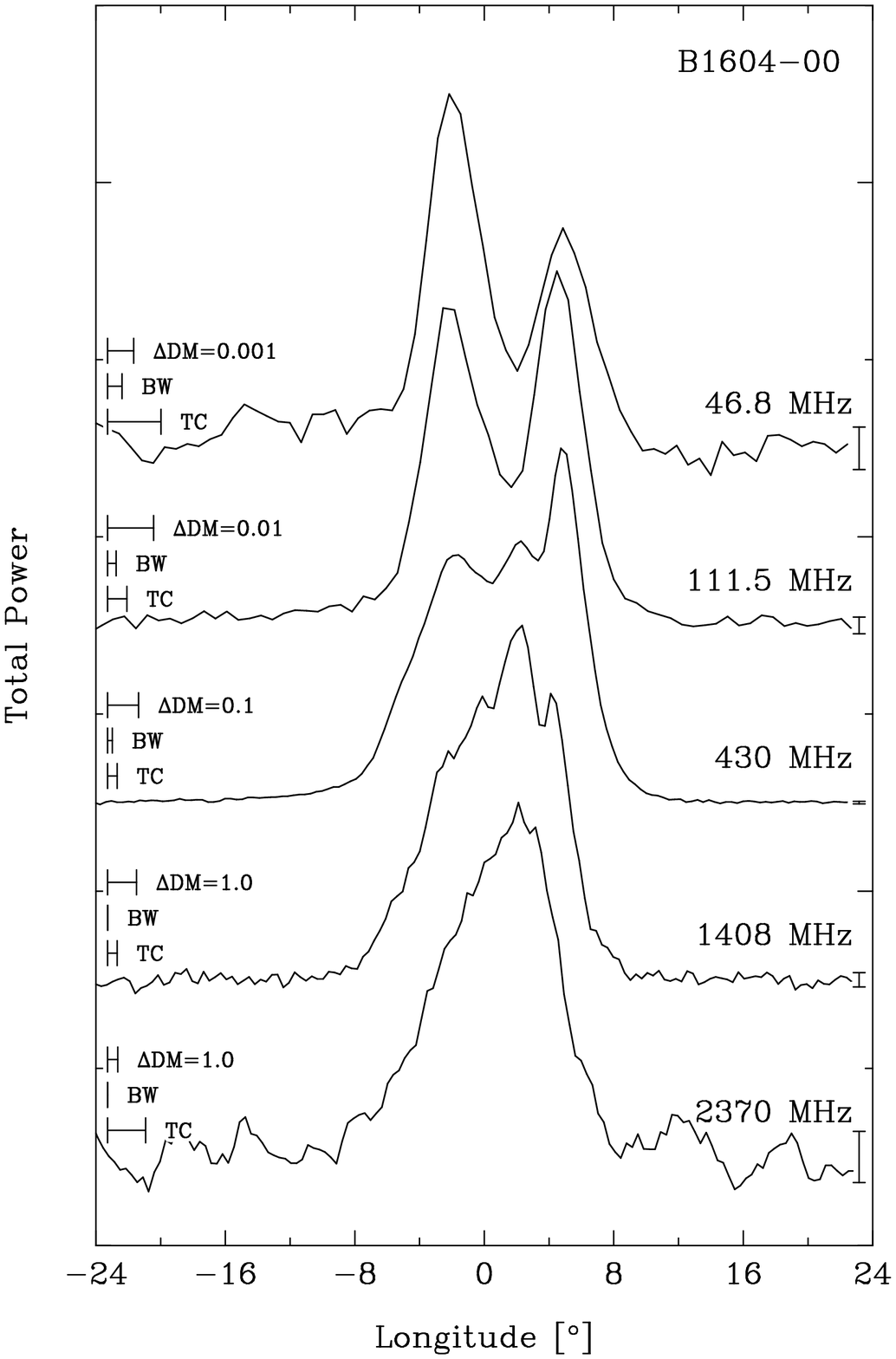}     
\pf{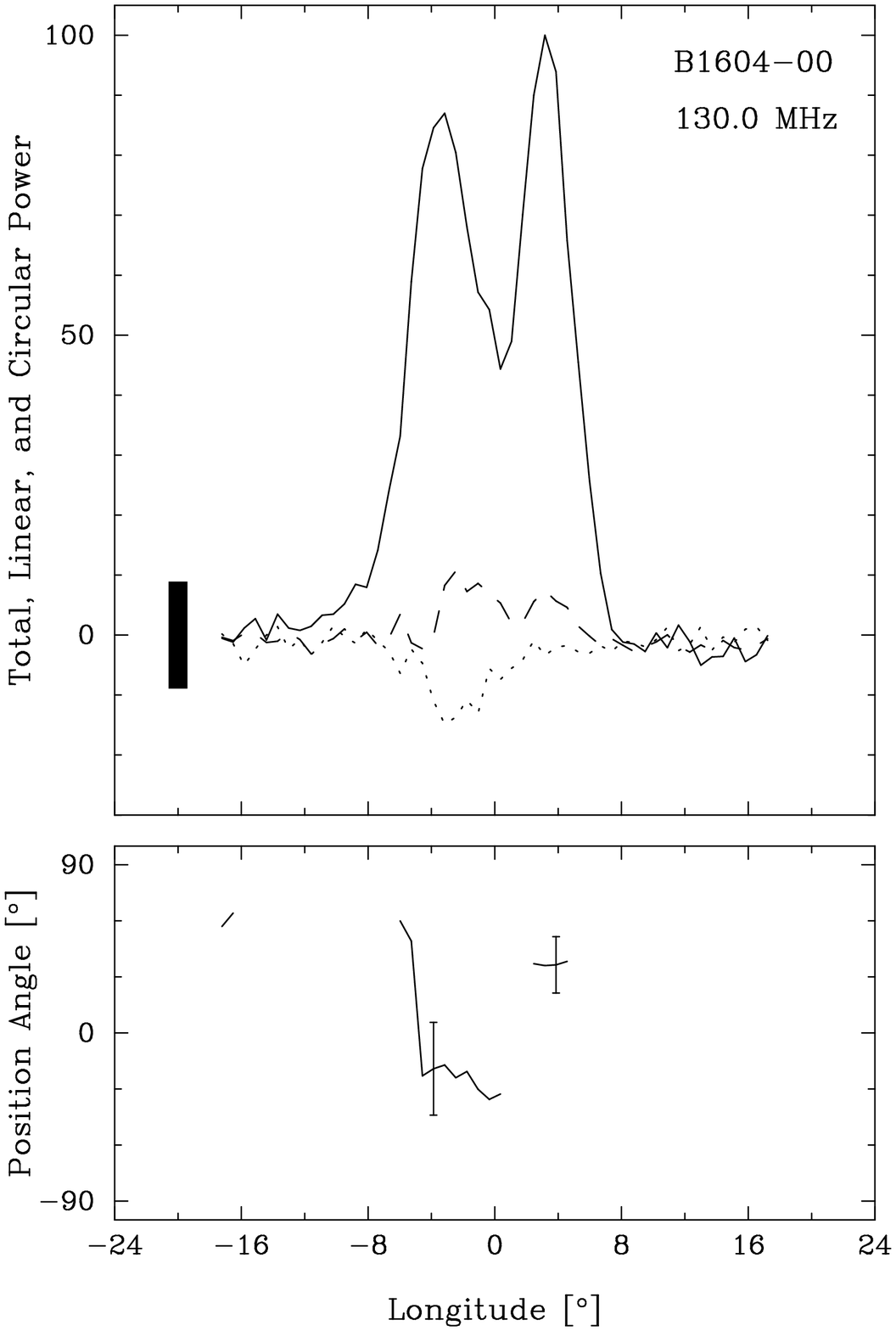} 
\pf{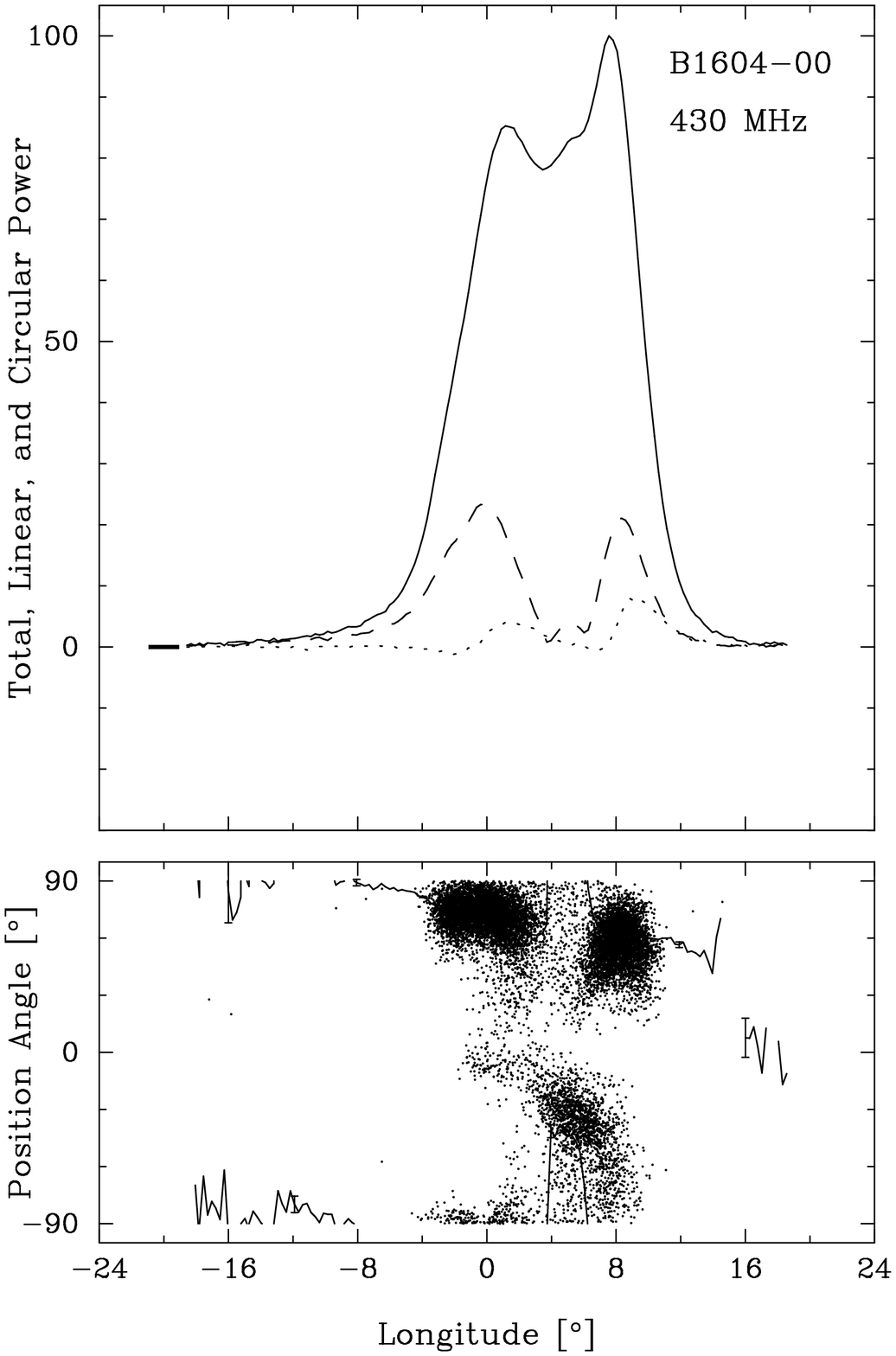} 
}
\quad \\
\centerline{
}
\caption{Multi-frequency profiles of B1530+27, B1612+07, B1633+24, B1541+09 and B1604$-$00, and polarization profiles of B1541+09 and B1604$-$00.}
\label{b8}
\end{figure}

\begin{figure}[htb] 
\centerline{ 
\pf{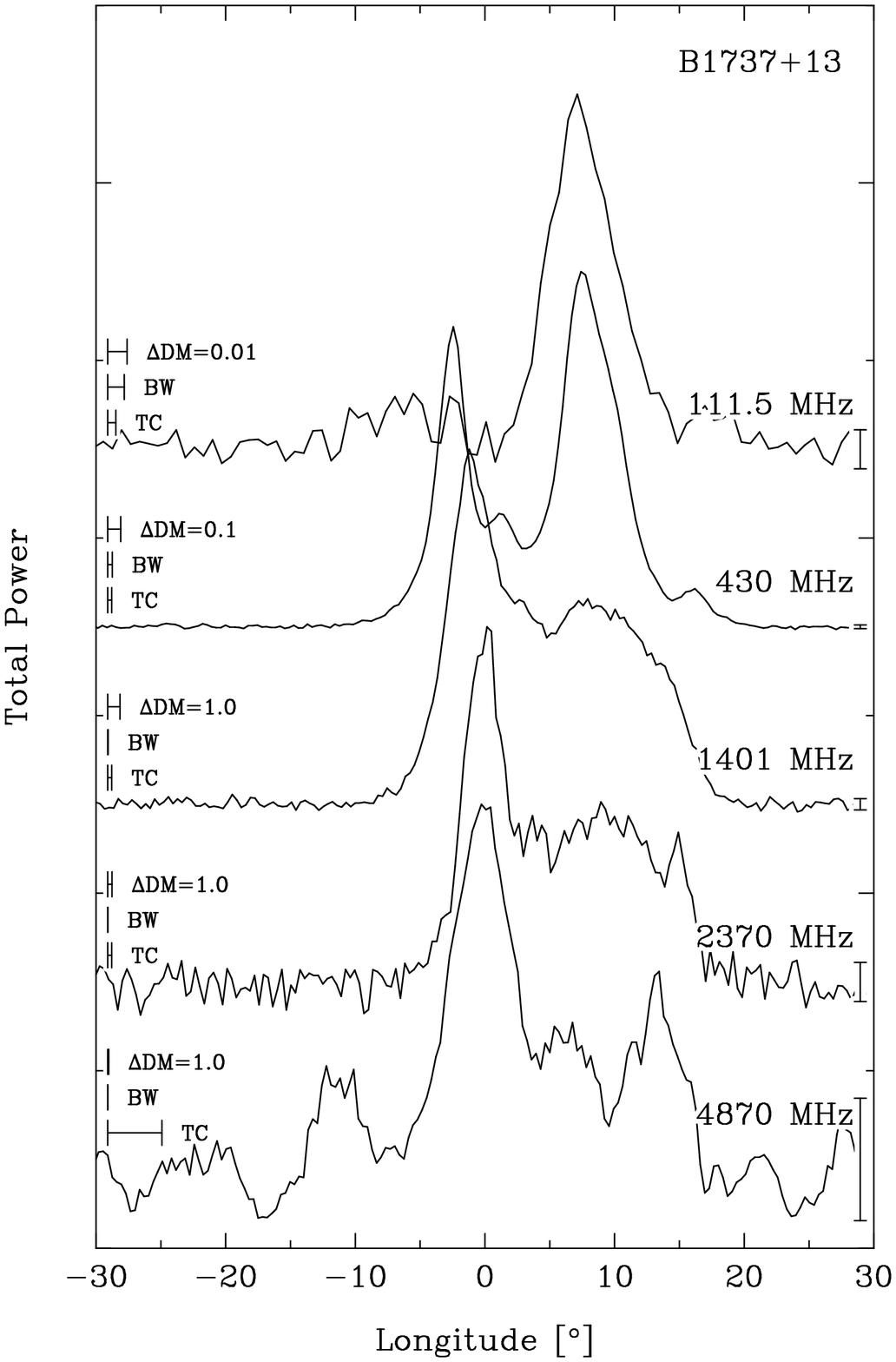}      
\pf{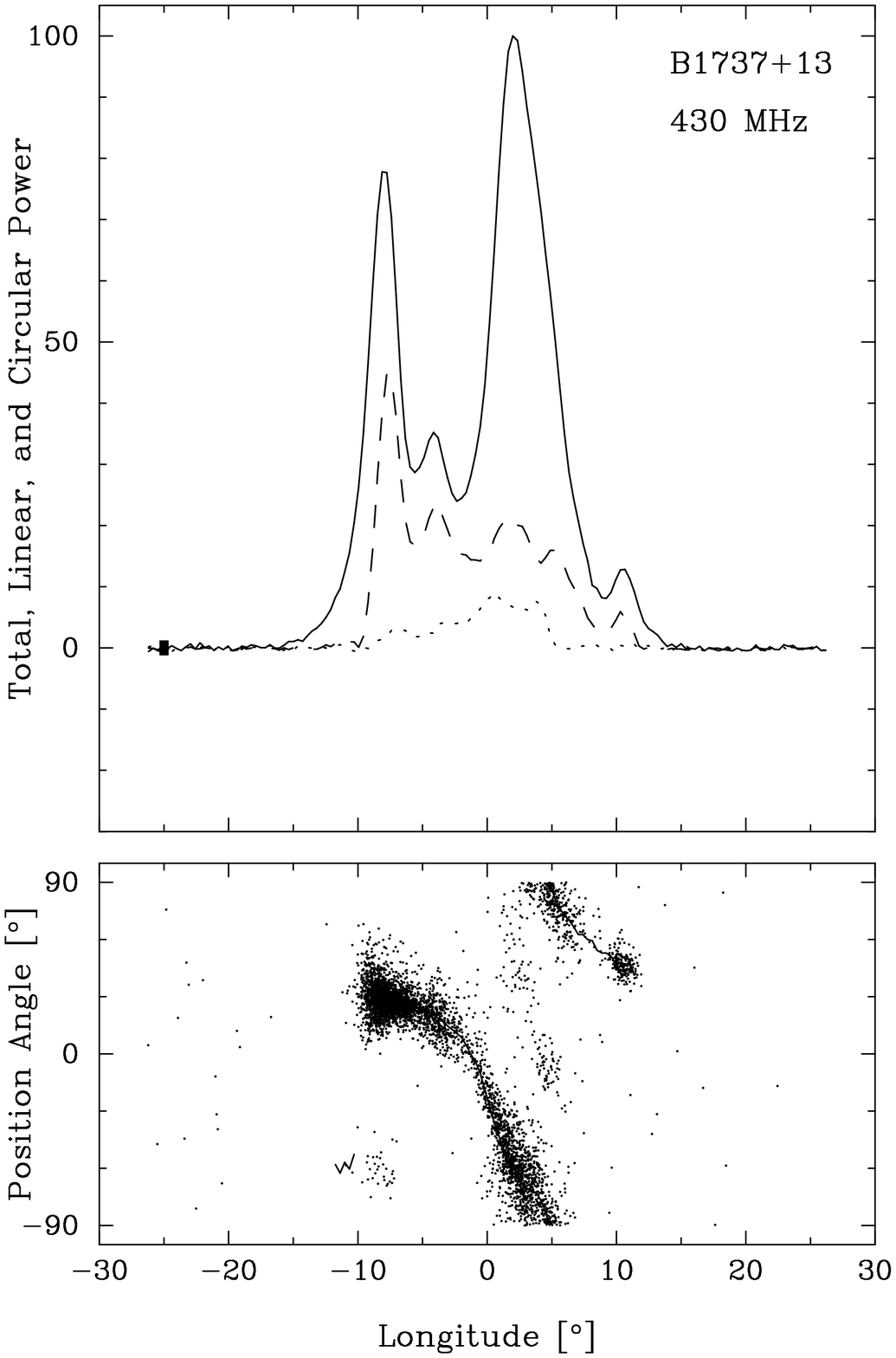}   
\pf{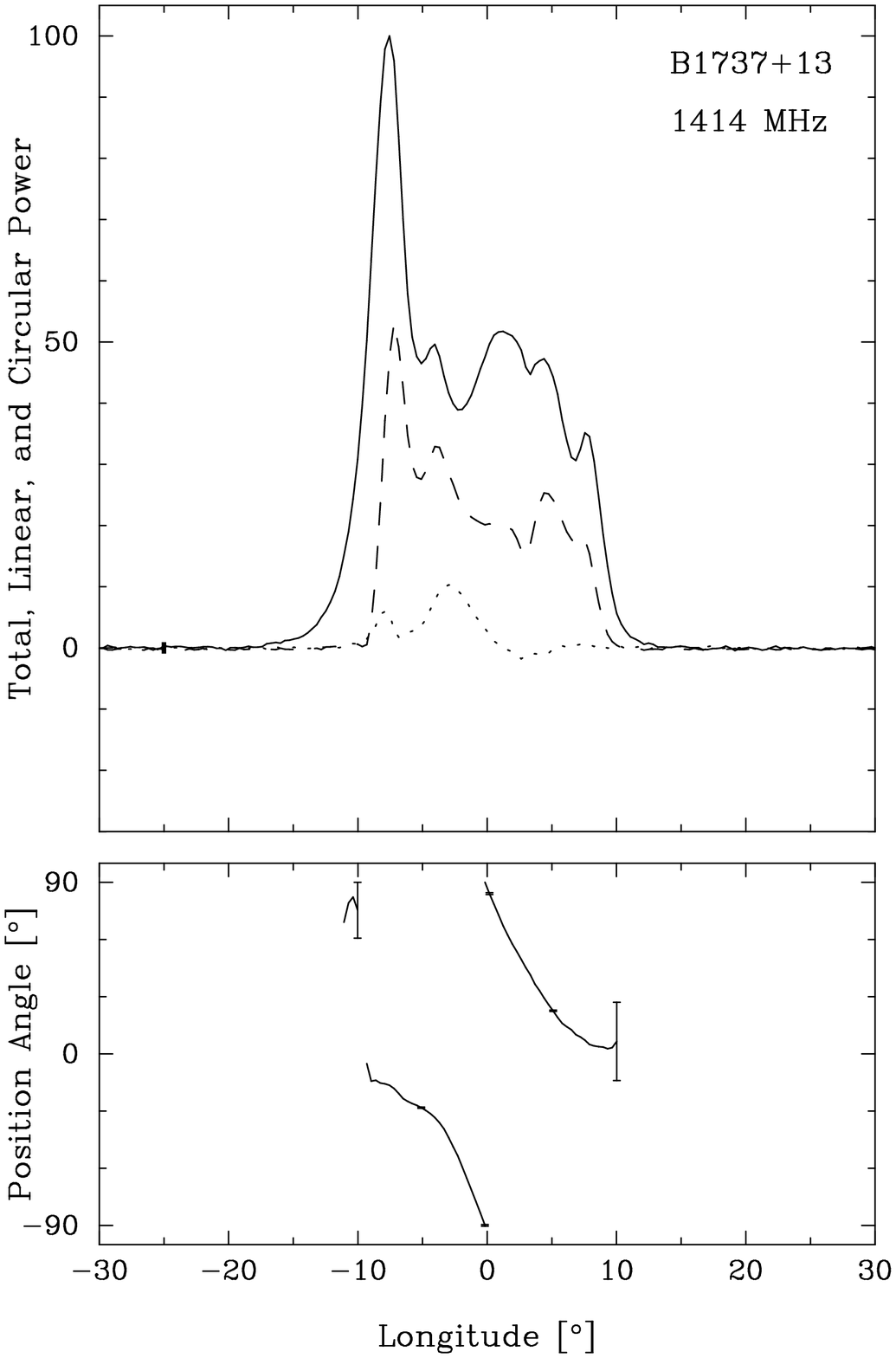}  
}
\quad \\
\centerline{
\pf{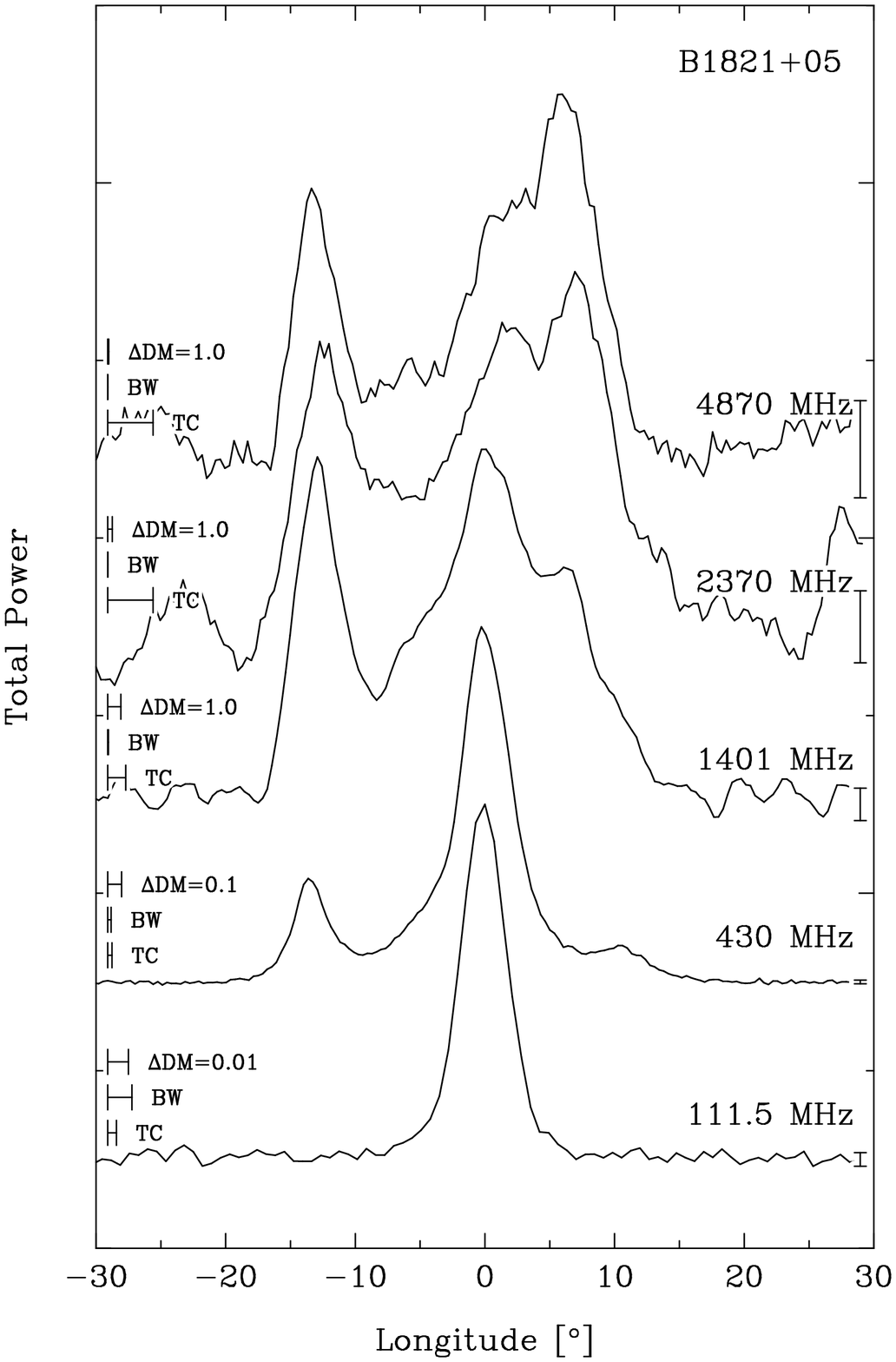}     
\pf{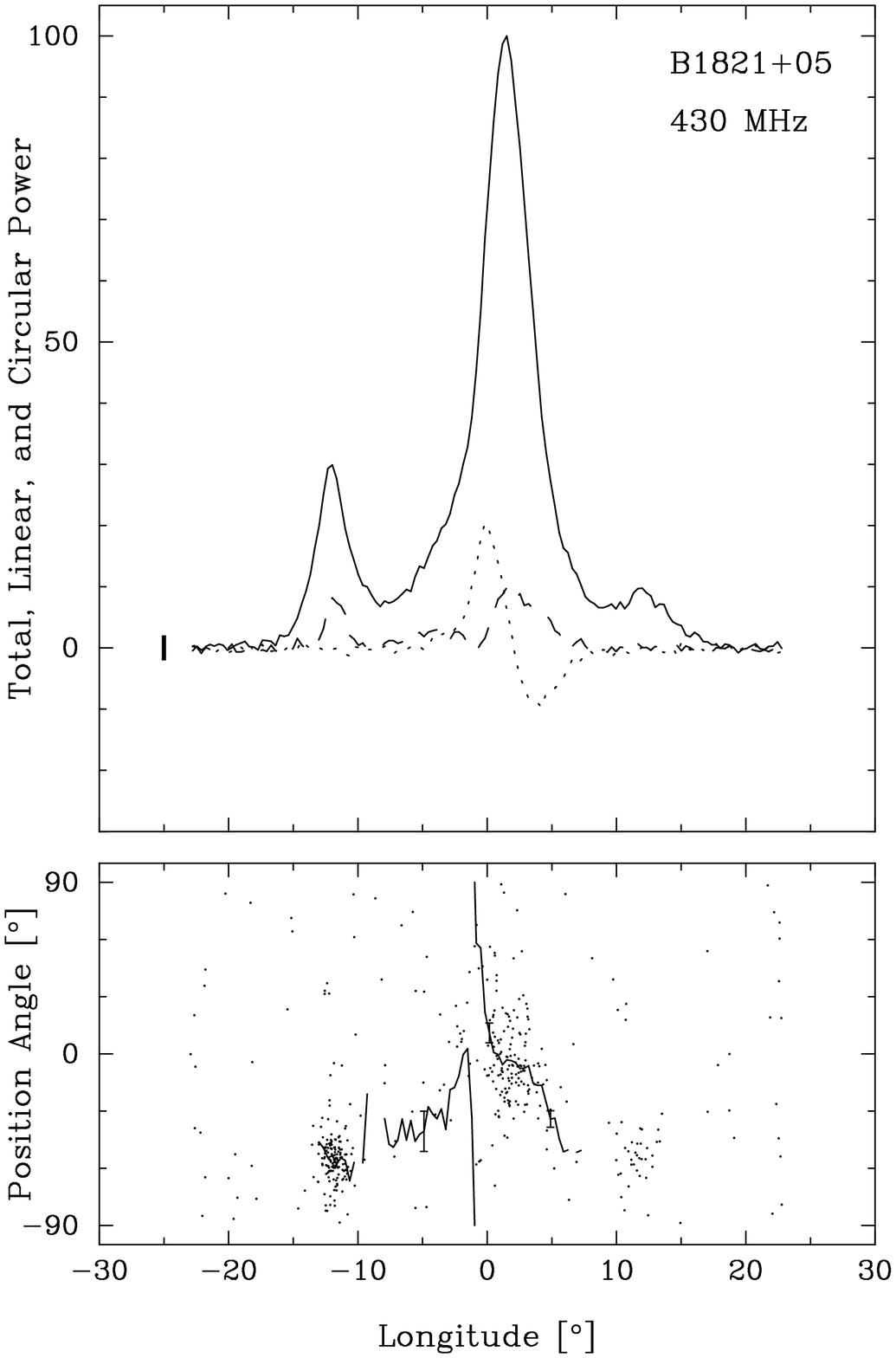} 
\pf{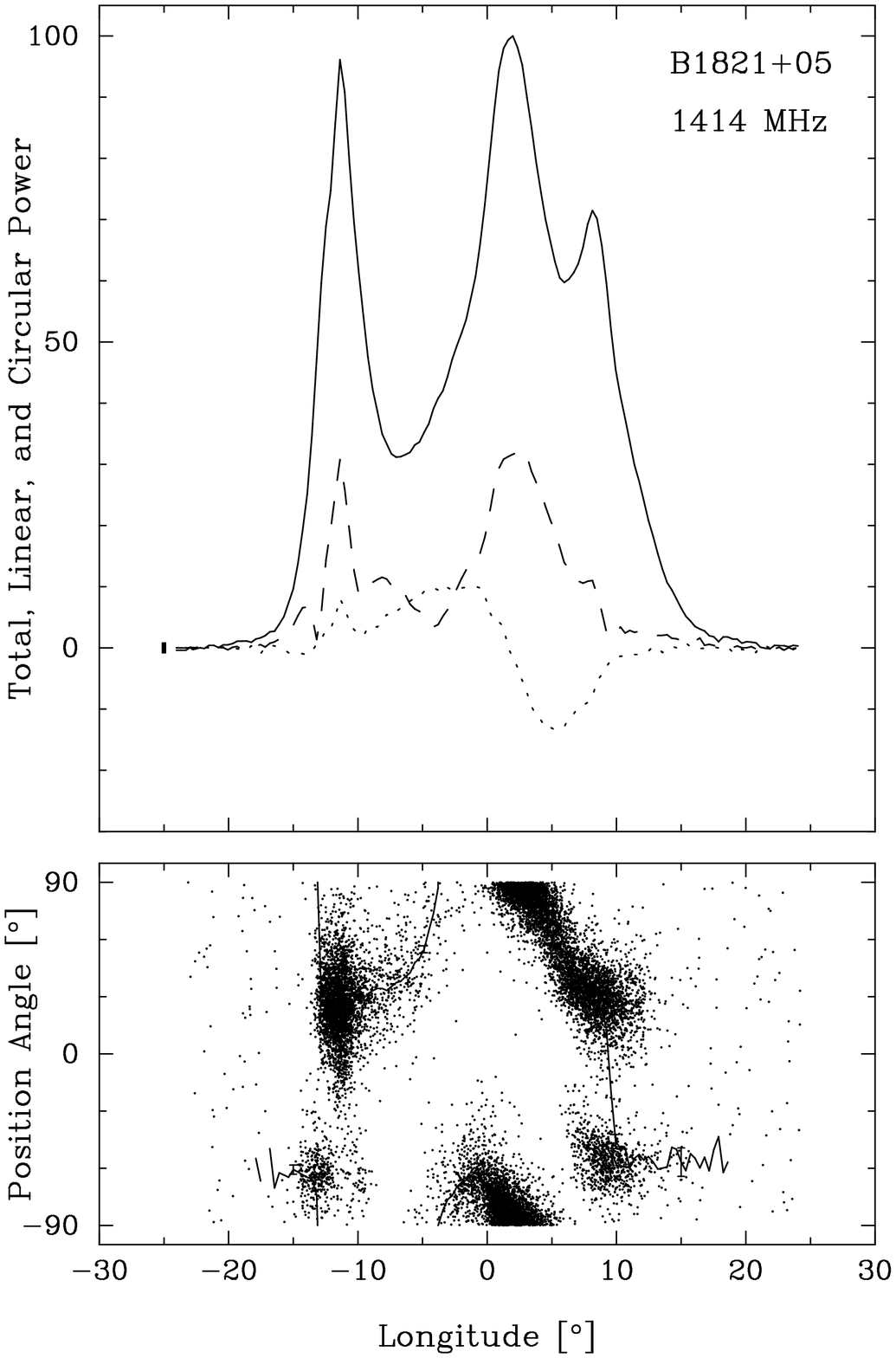} 
}
\quad \\
\centerline{
\pf{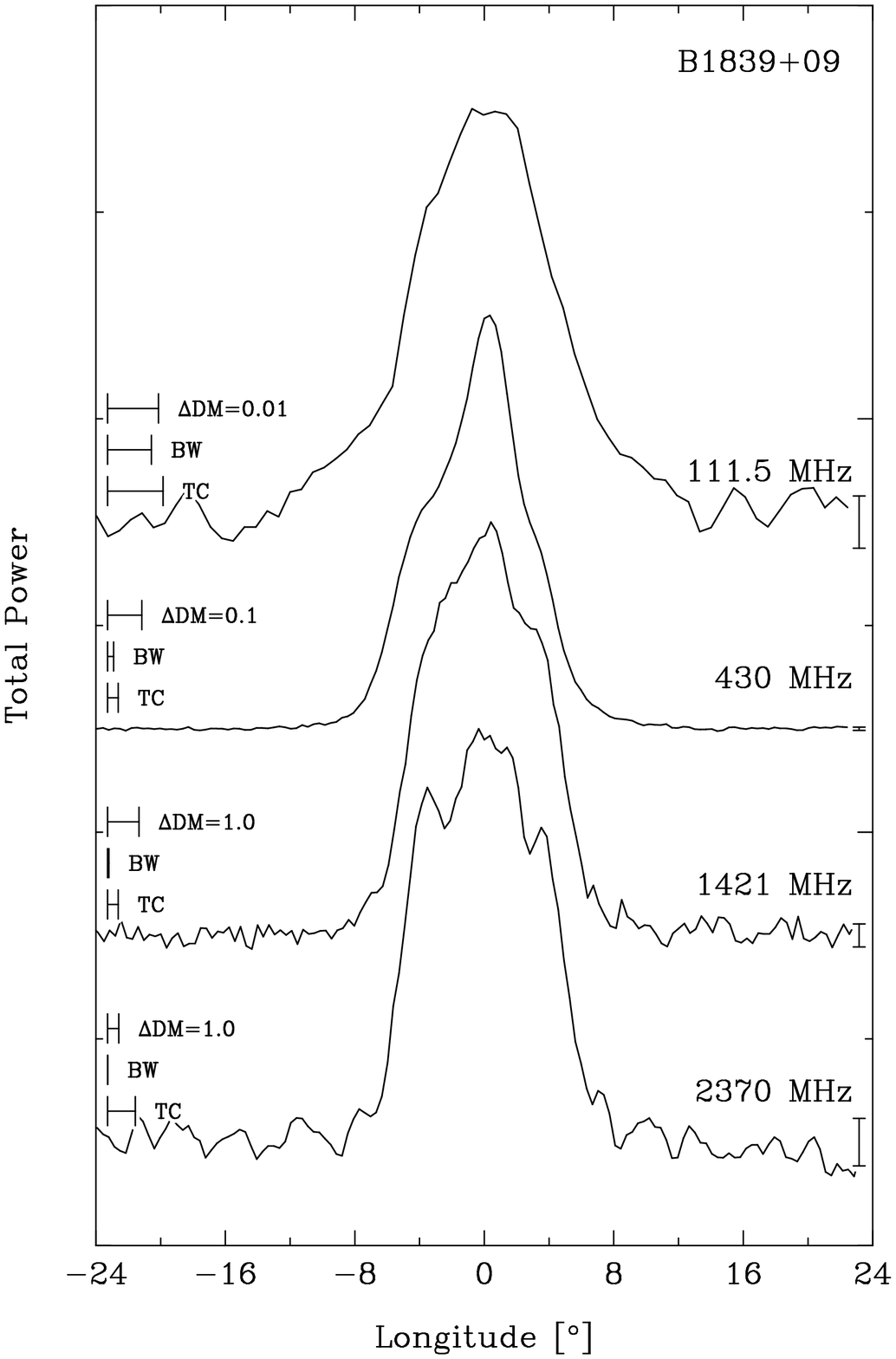}     
\pf{figs/dummy_fig.ps}   
\pf{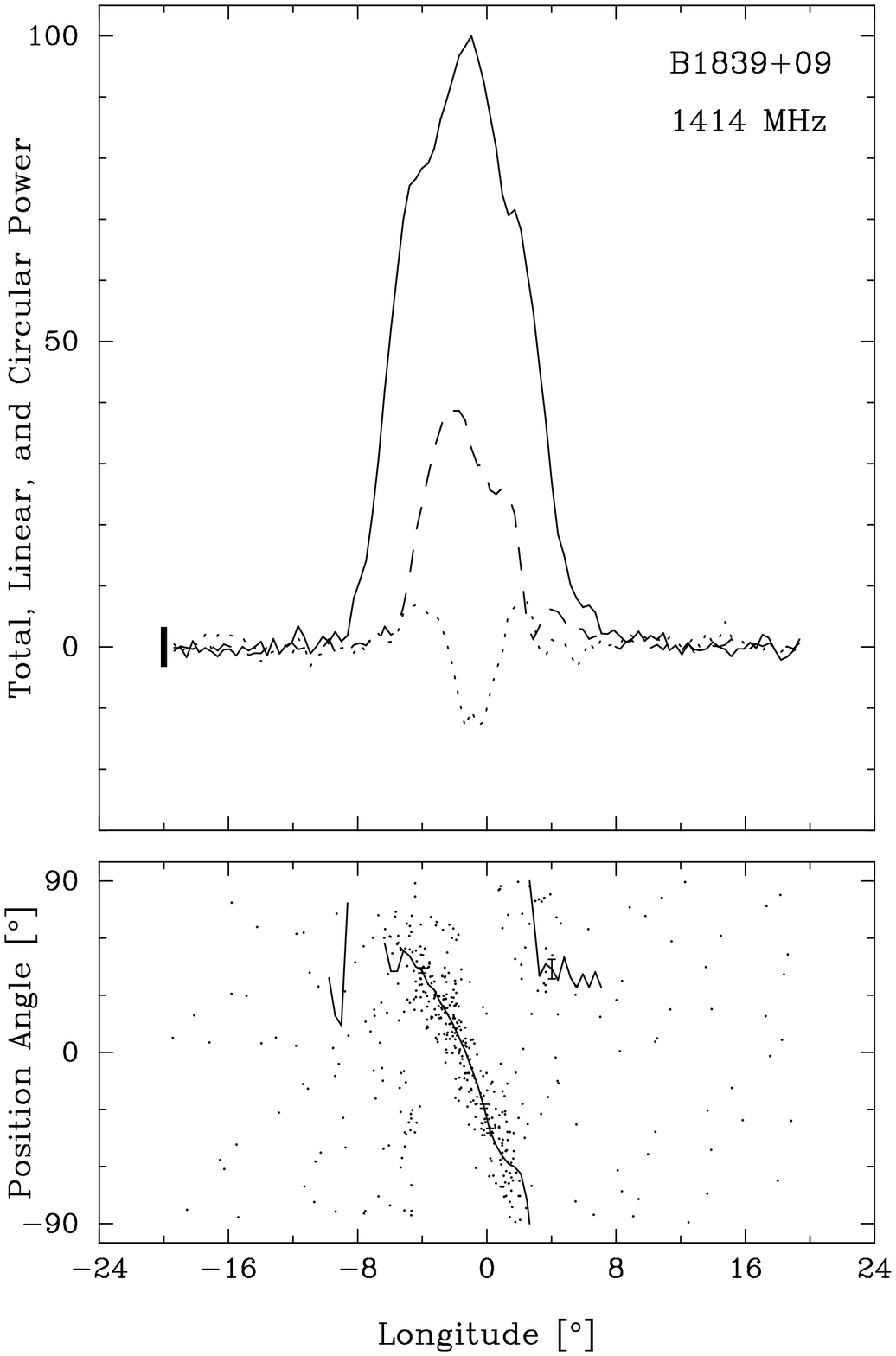} 
}
\caption{Multi-frequency and polarization profiles of B1737+13, B1821+05, and B1839+09.}
\label{b9}
\end{figure}

\begin{figure}[htb]
\centerline{
\pf{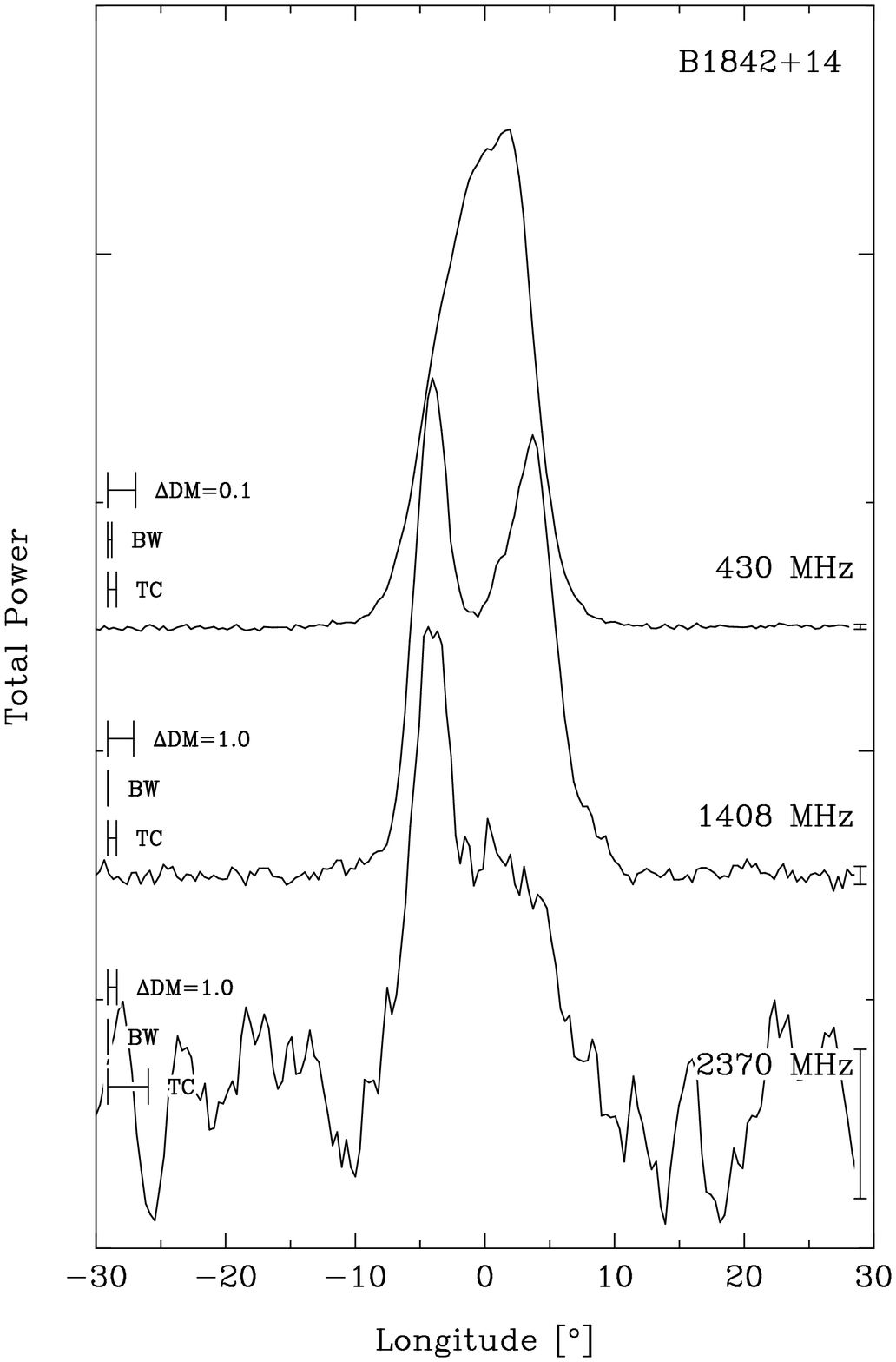}     
\pf{figs/dummy_fig.ps}   
\pf{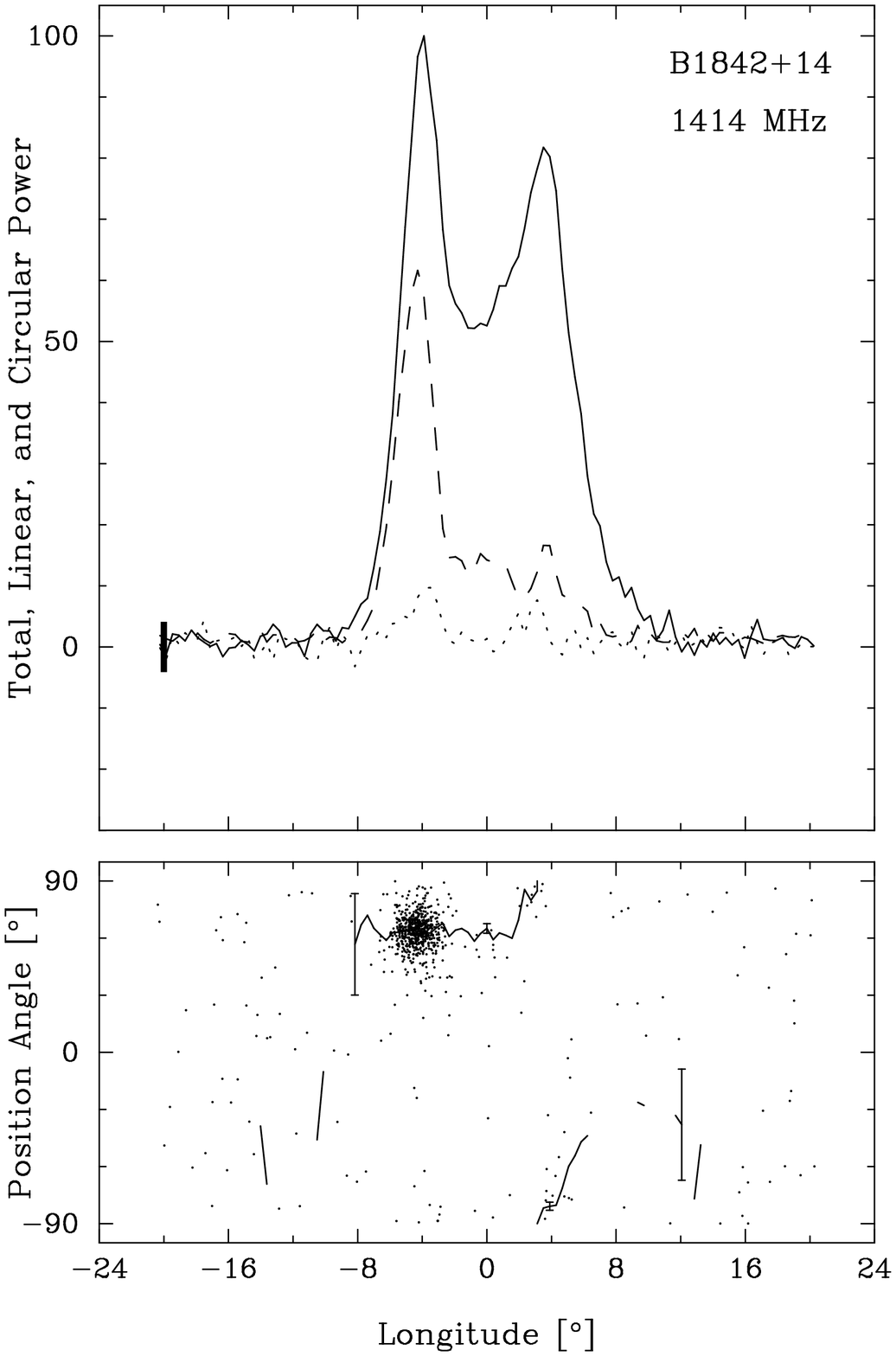} 
}
\quad \\
\centerline{
\pf{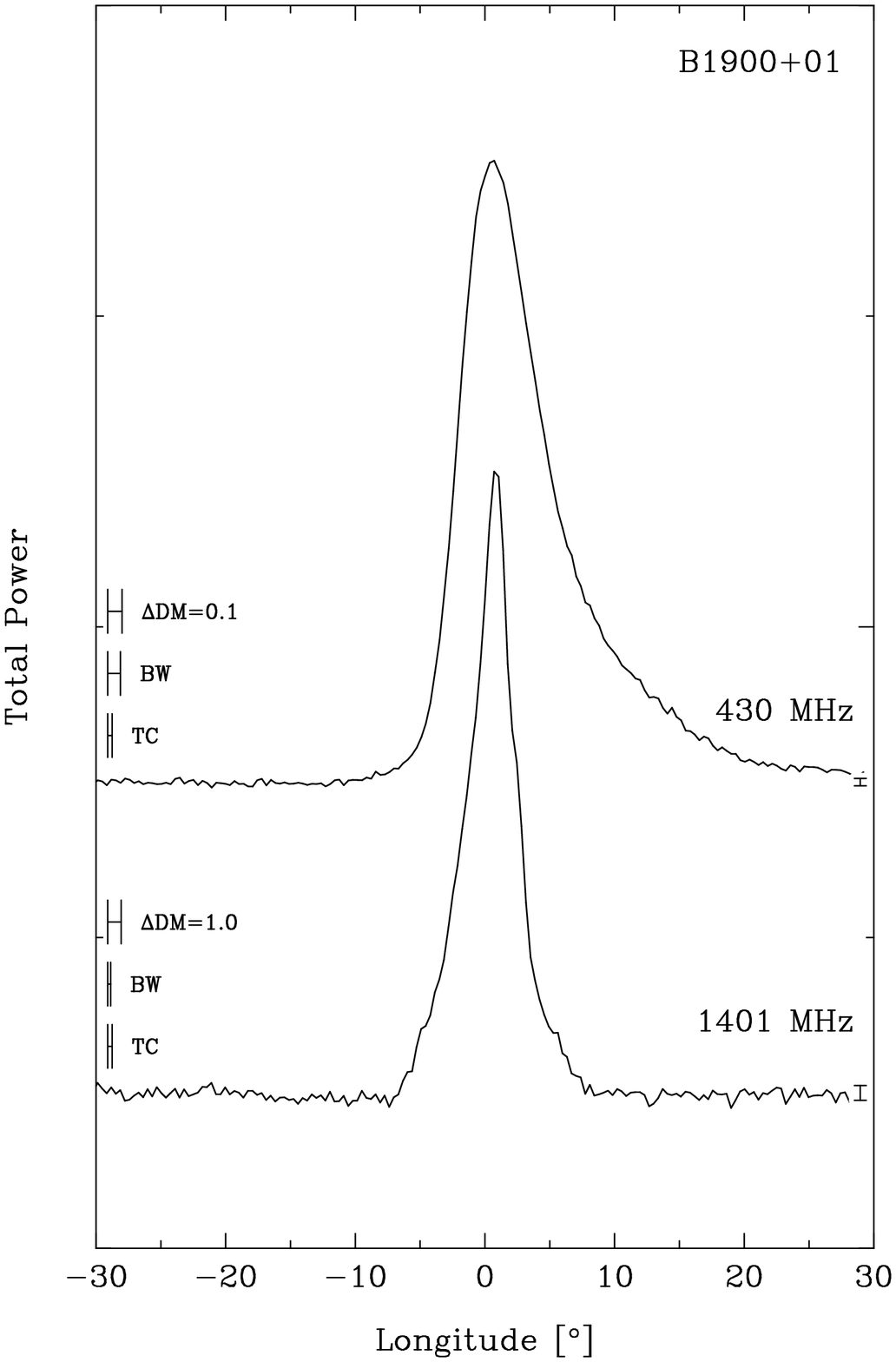}     
\pf{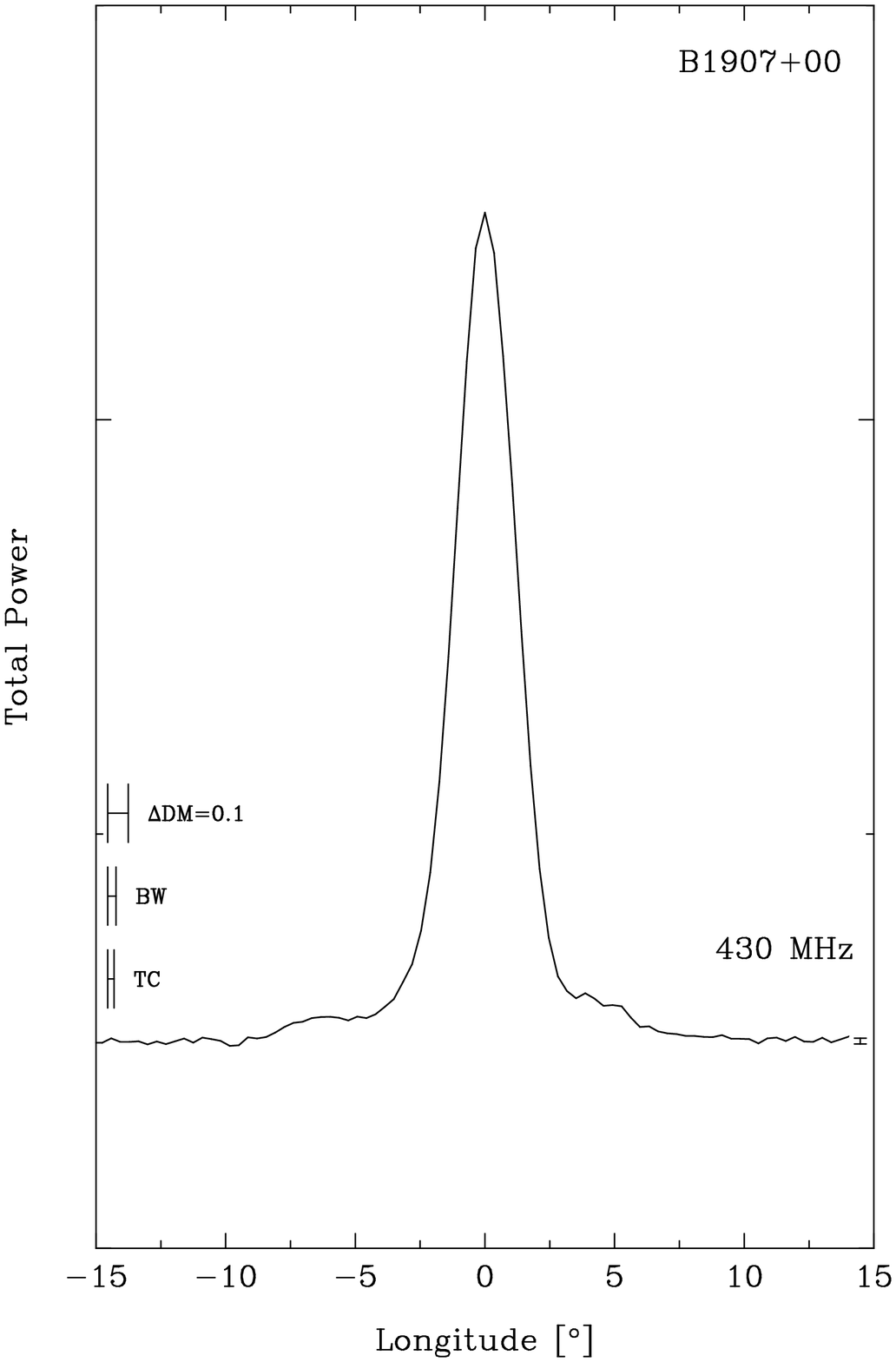}     
\pf{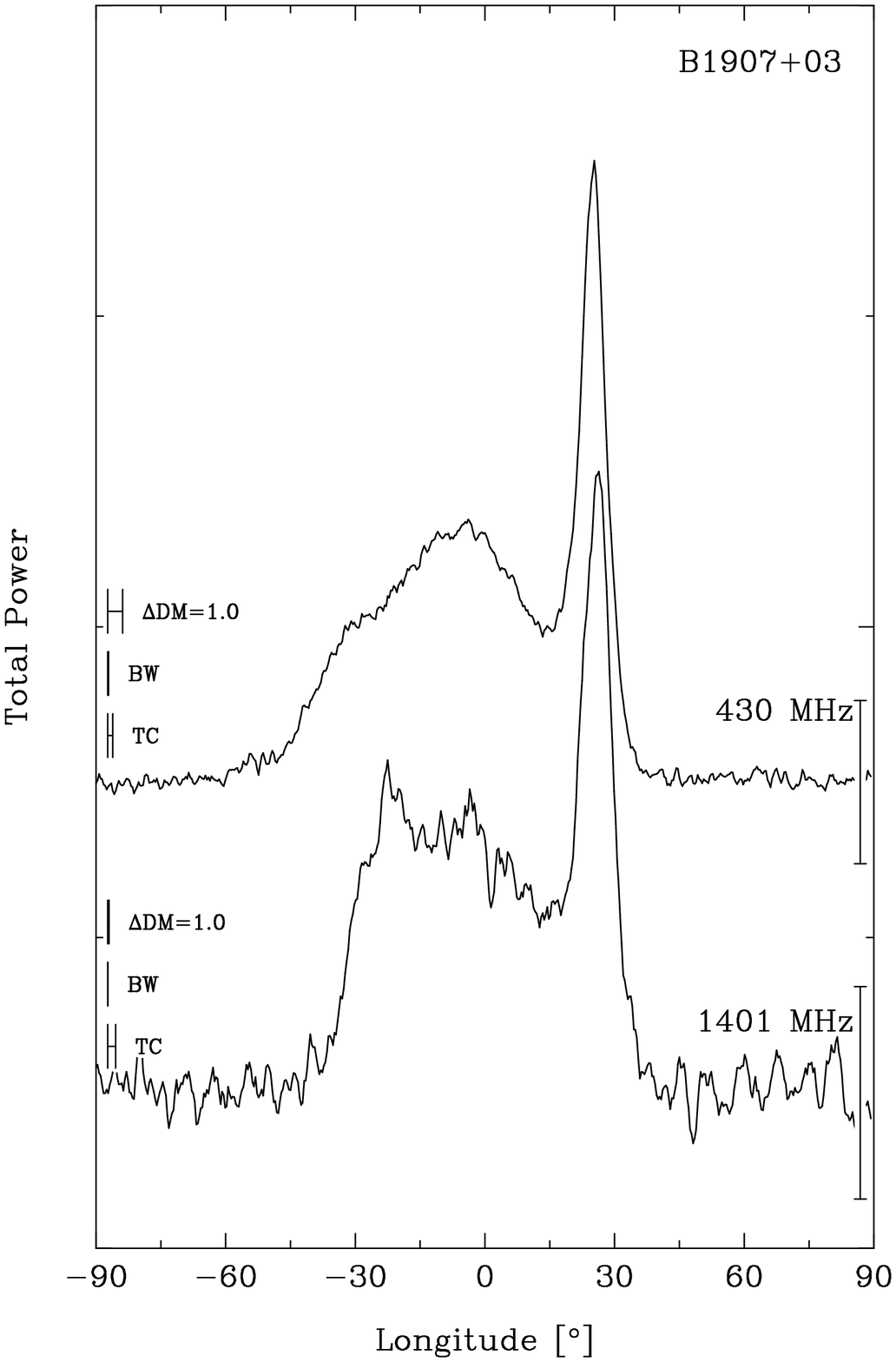}     
}
\quad \\
\centerline{
\pf{figs/dummy_fig.ps}     
\pf{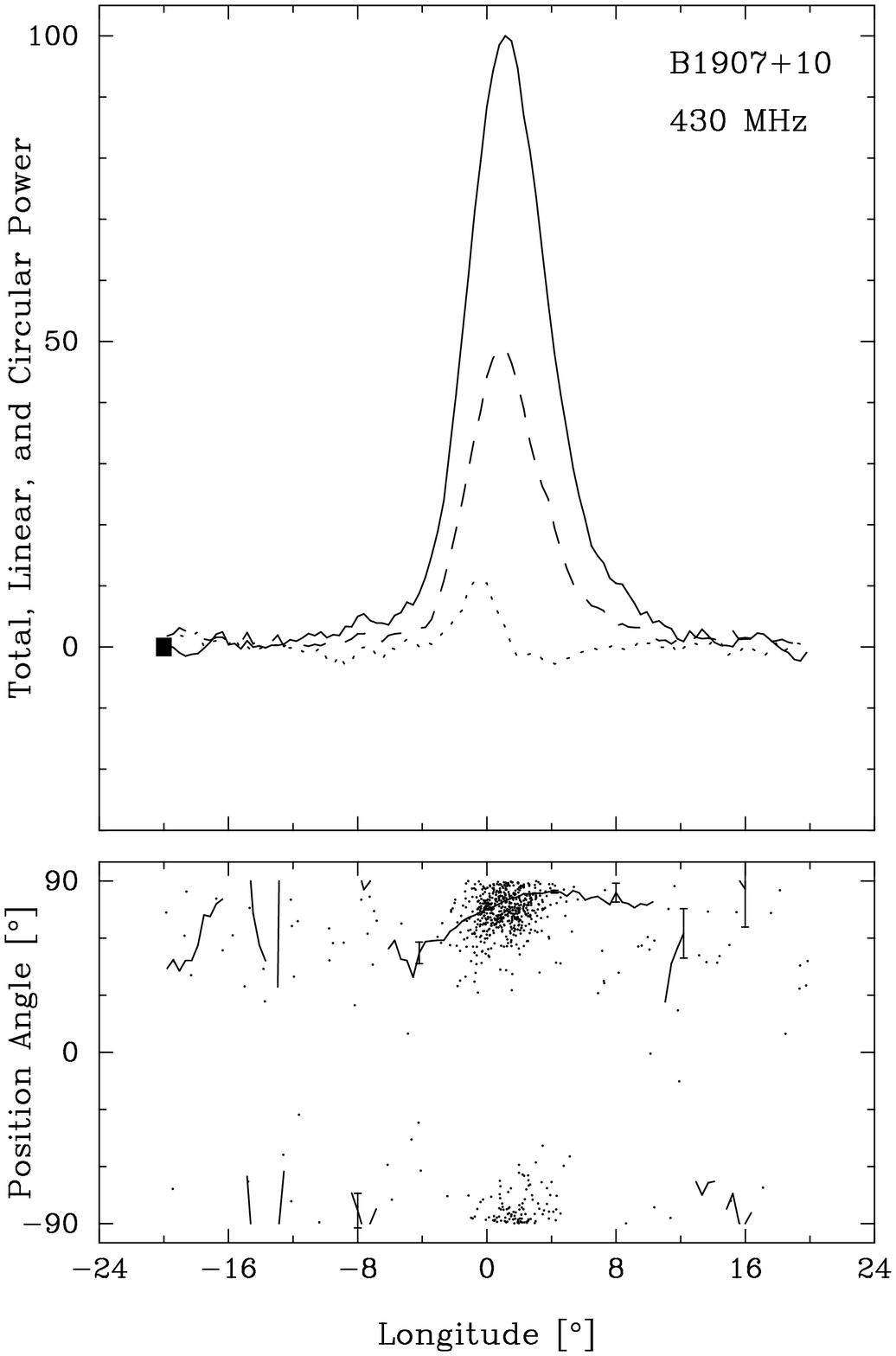} 
\pf{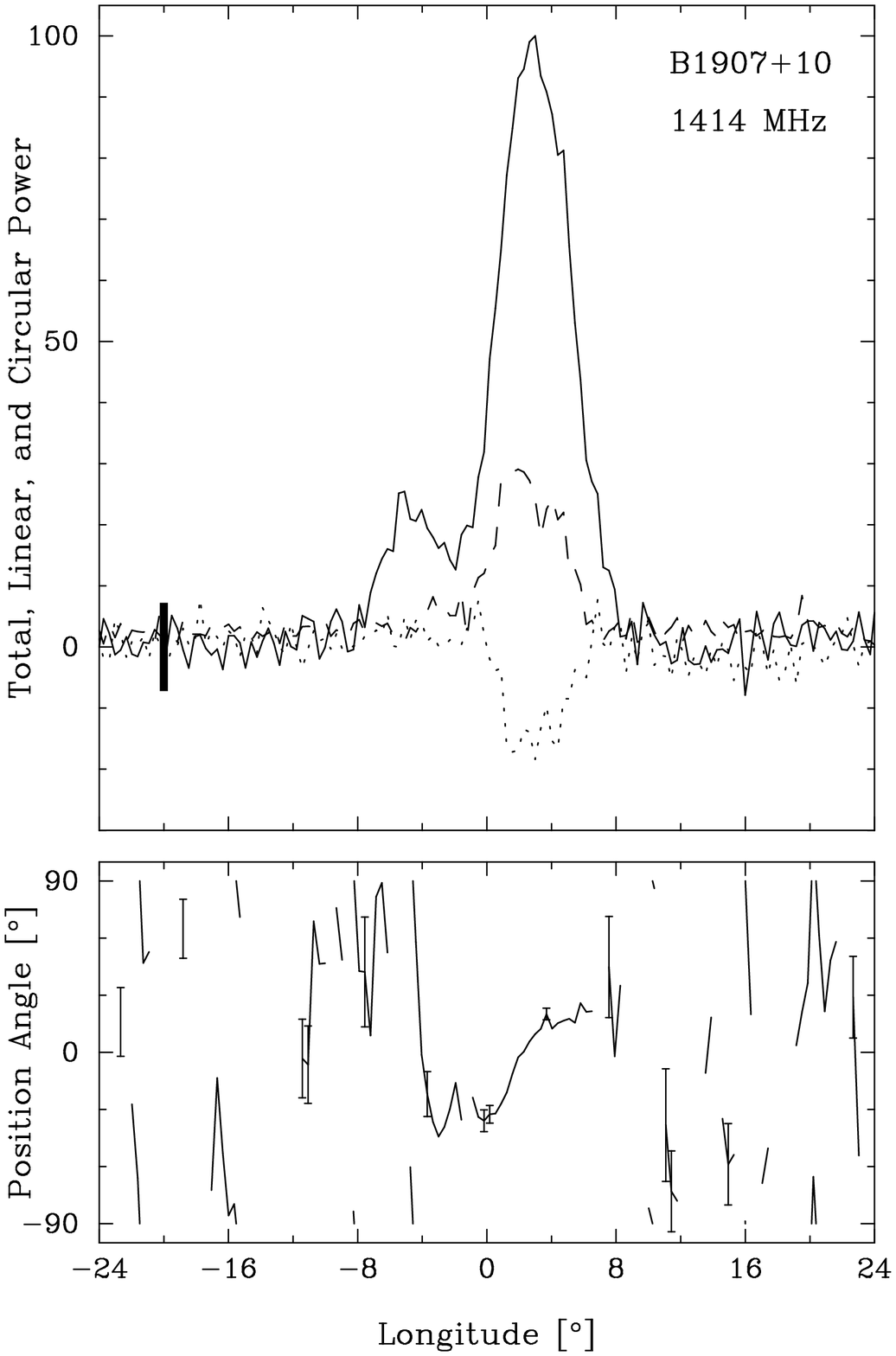} 
}
\caption{Multi-frequency profiles of B1842+14, B1900+01, B1907+00 and B1907+03, and polarization profiles of B1842+14 and B1907+10.}
\label{b10}
\end{figure}

\begin{figure}[htb]
\centerline{
\pf{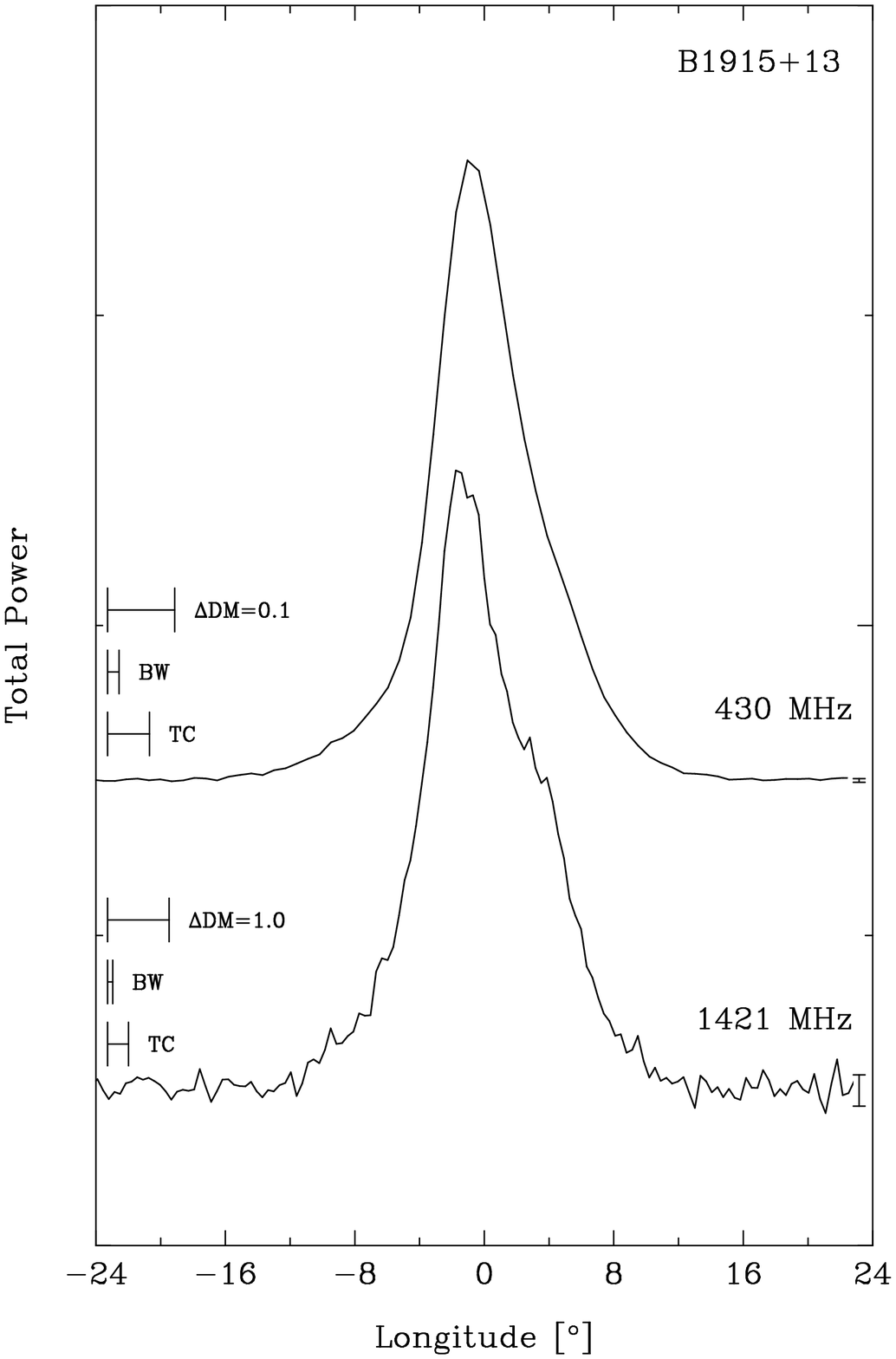}     
\pf{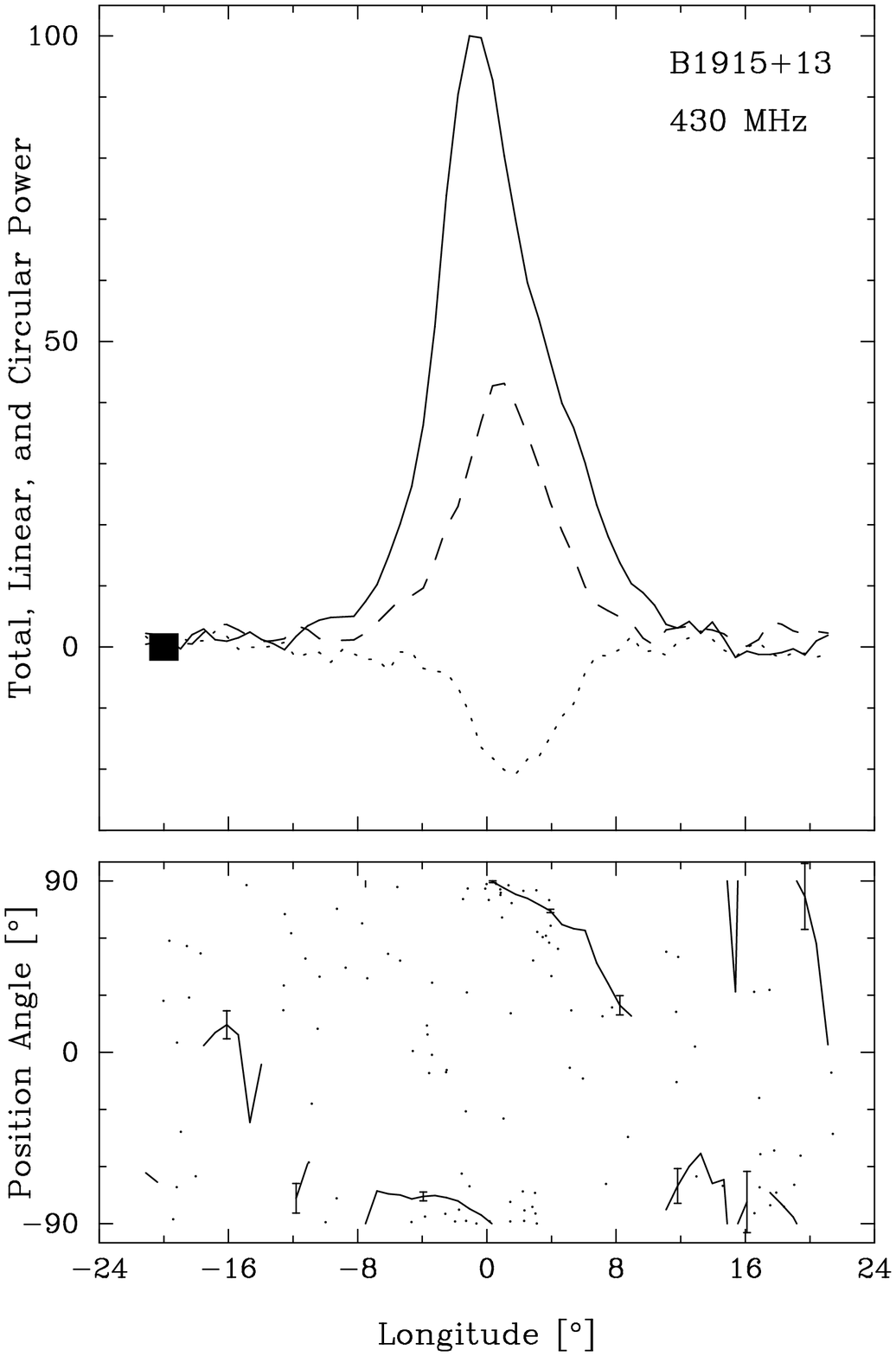} 
\pf{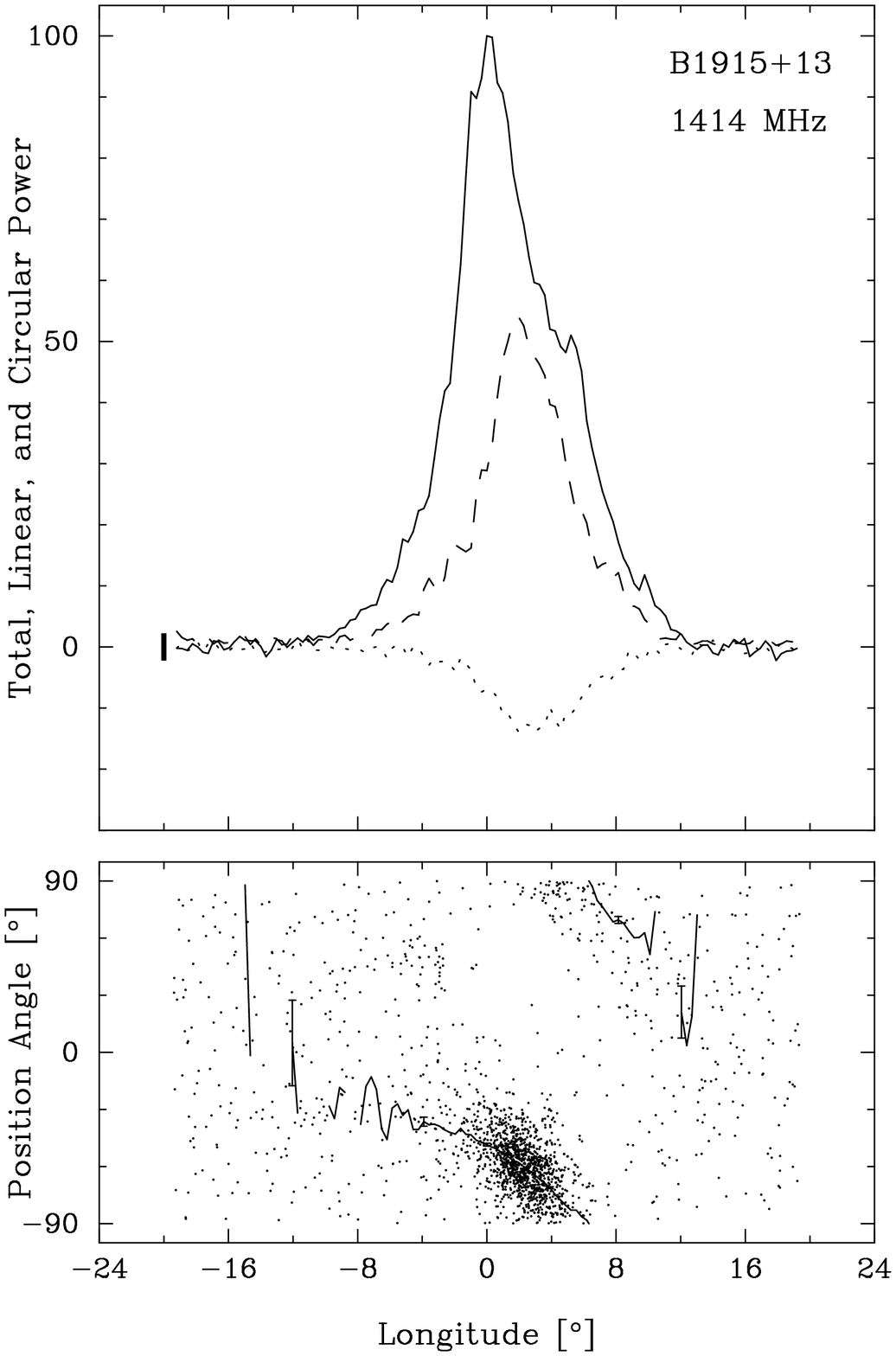} 
}
\quad \\
\centerline{
\pf{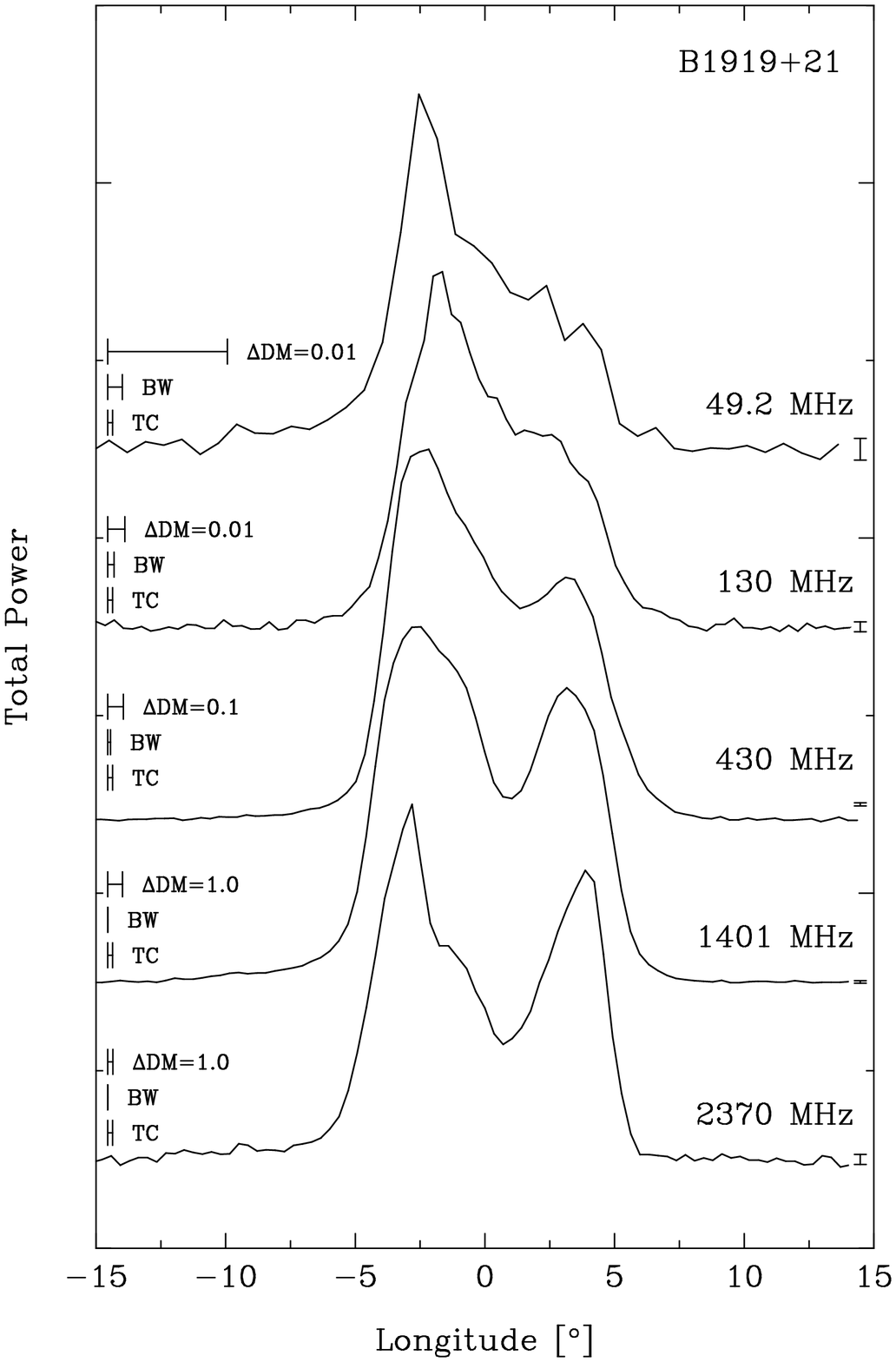}     
\pf{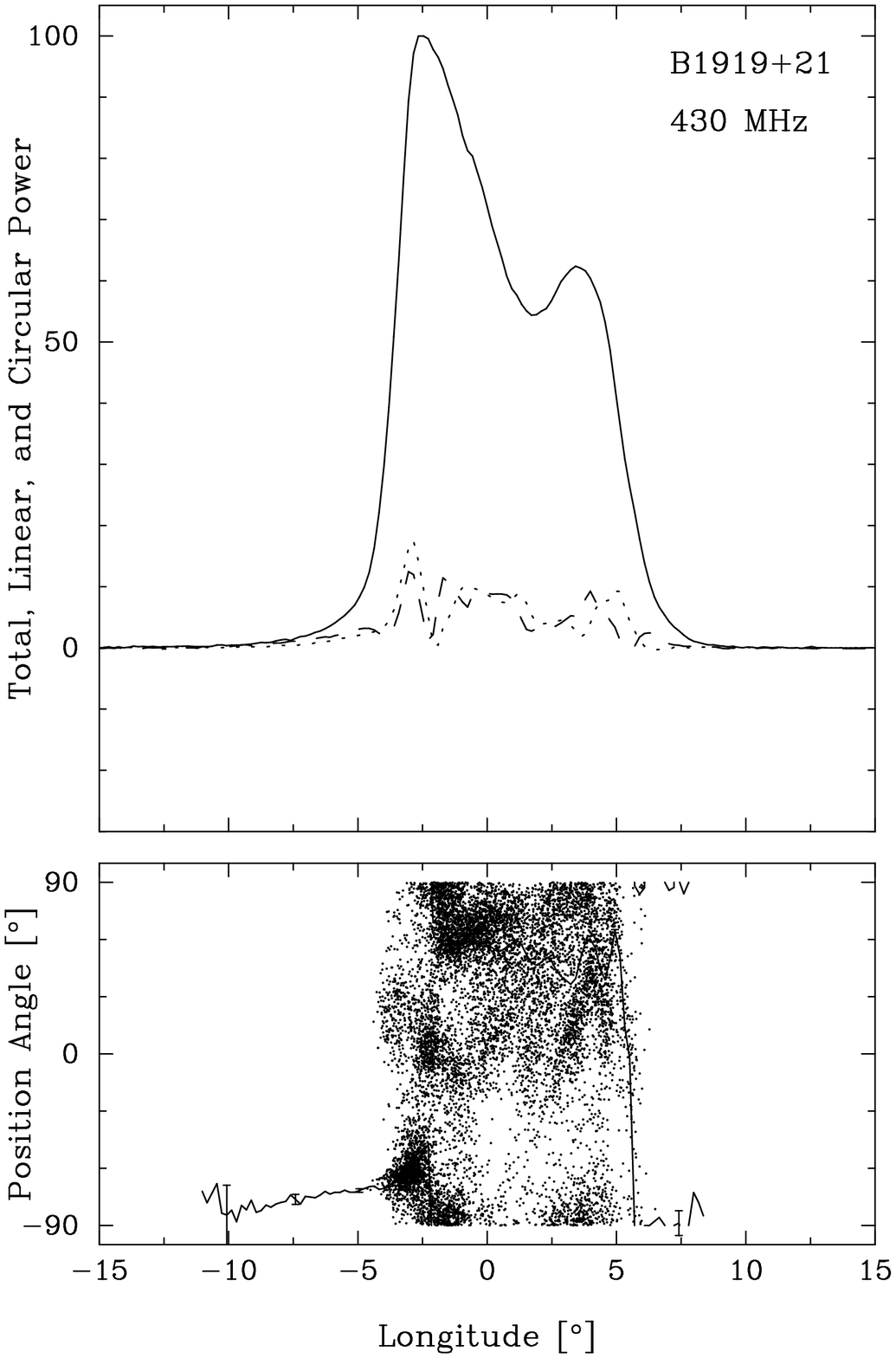} 
\pf{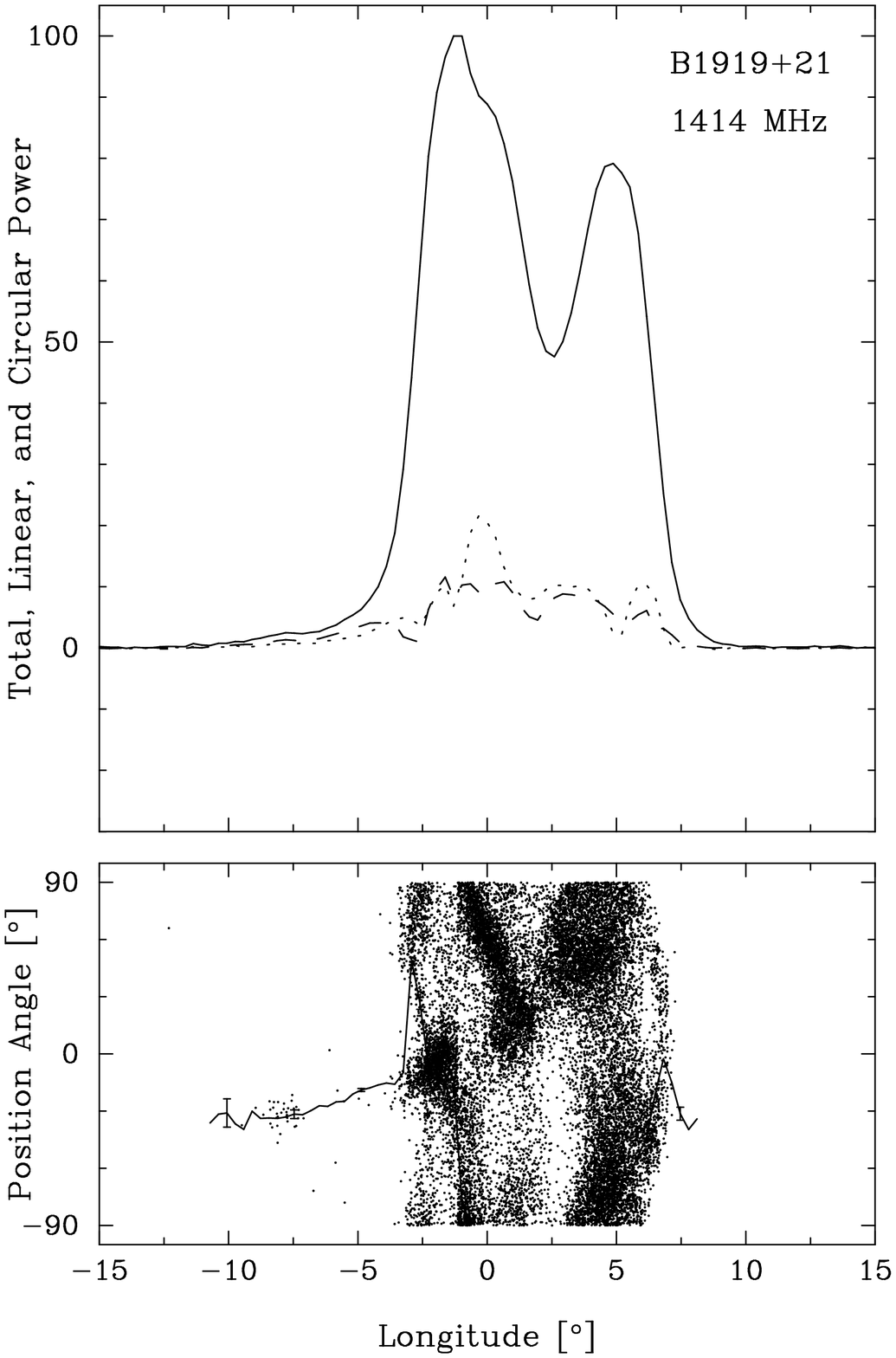} 
}
\quad \\
\centerline{
\pf{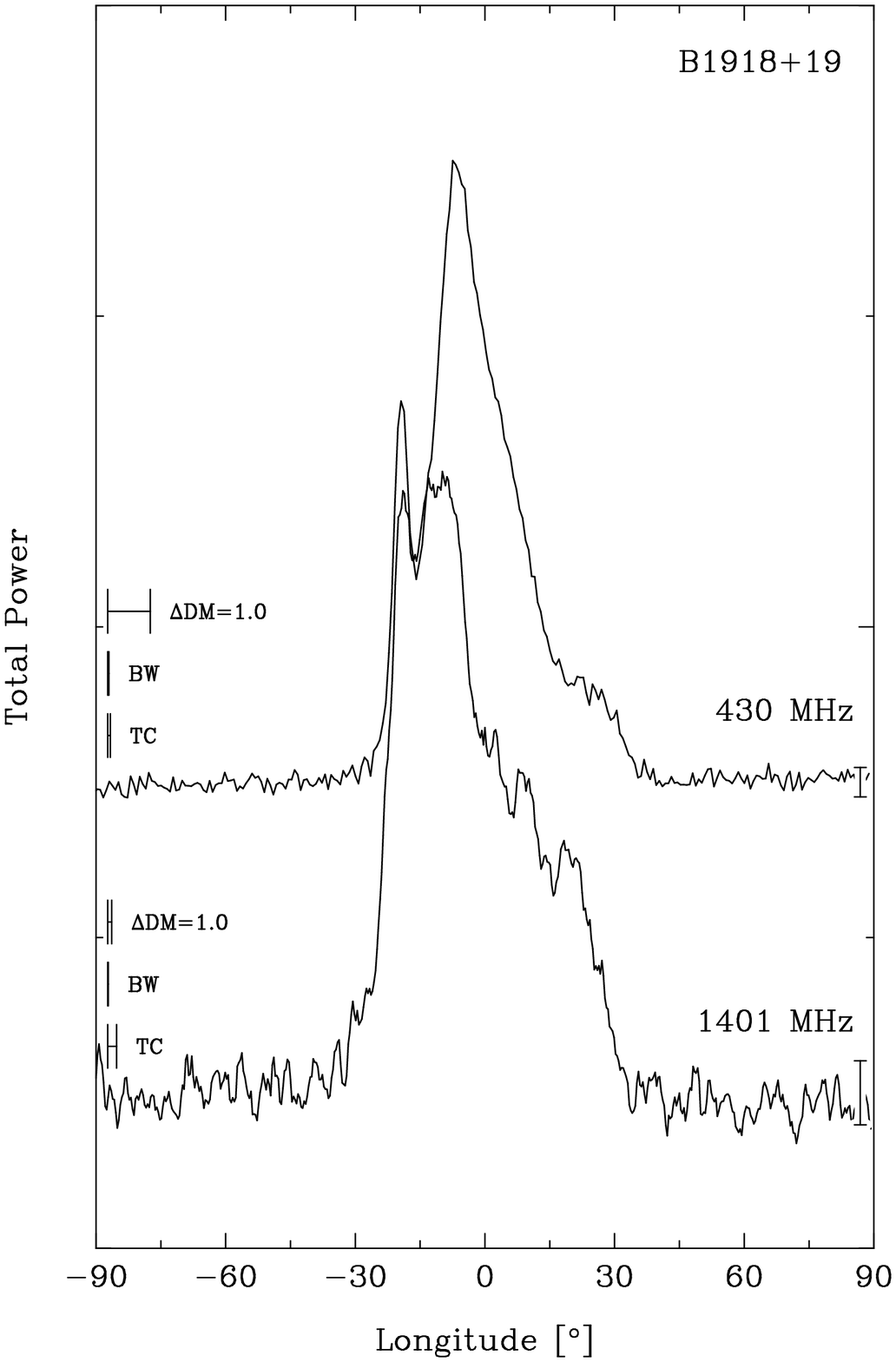}     
\pf{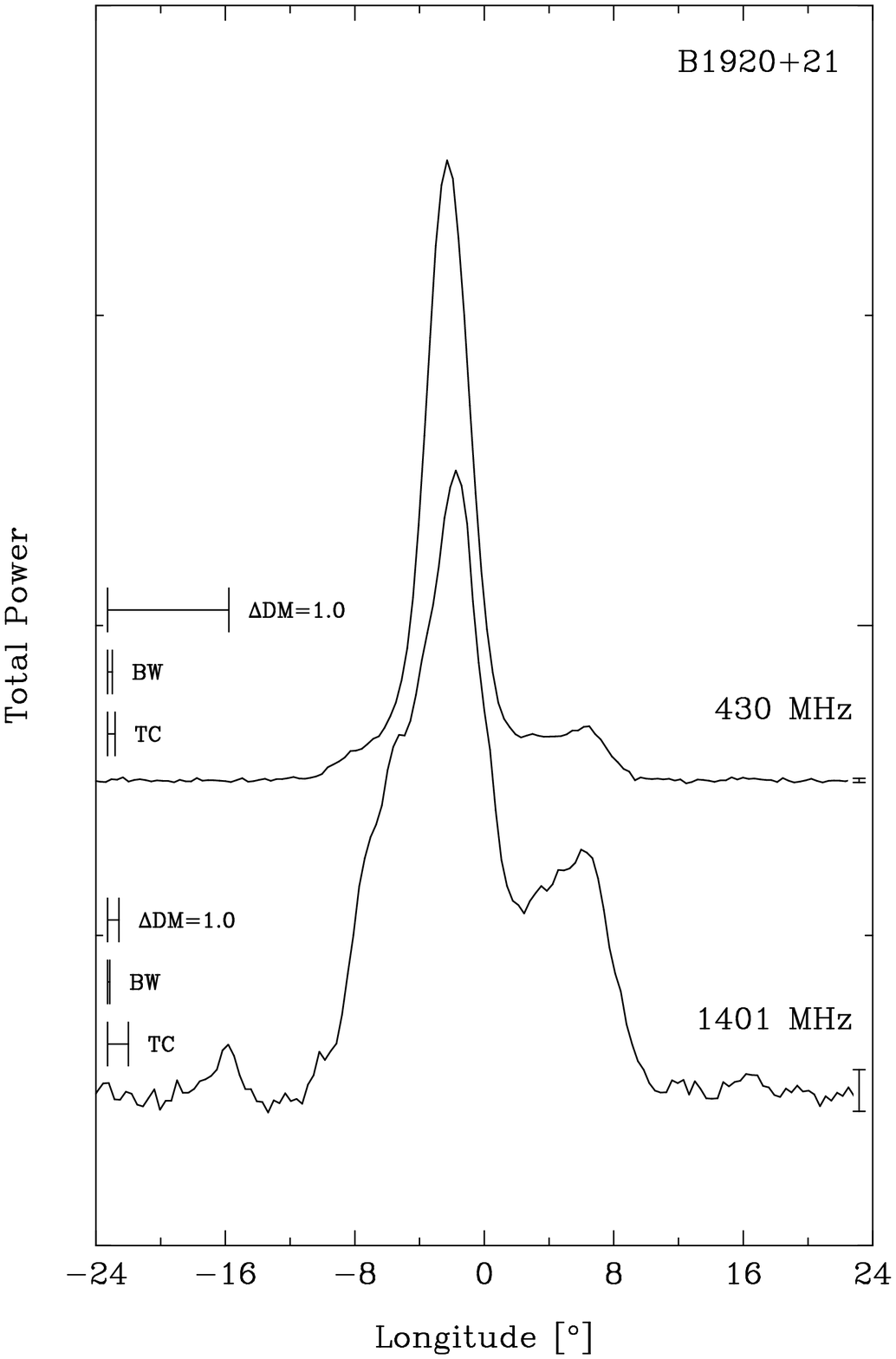}     
\pf{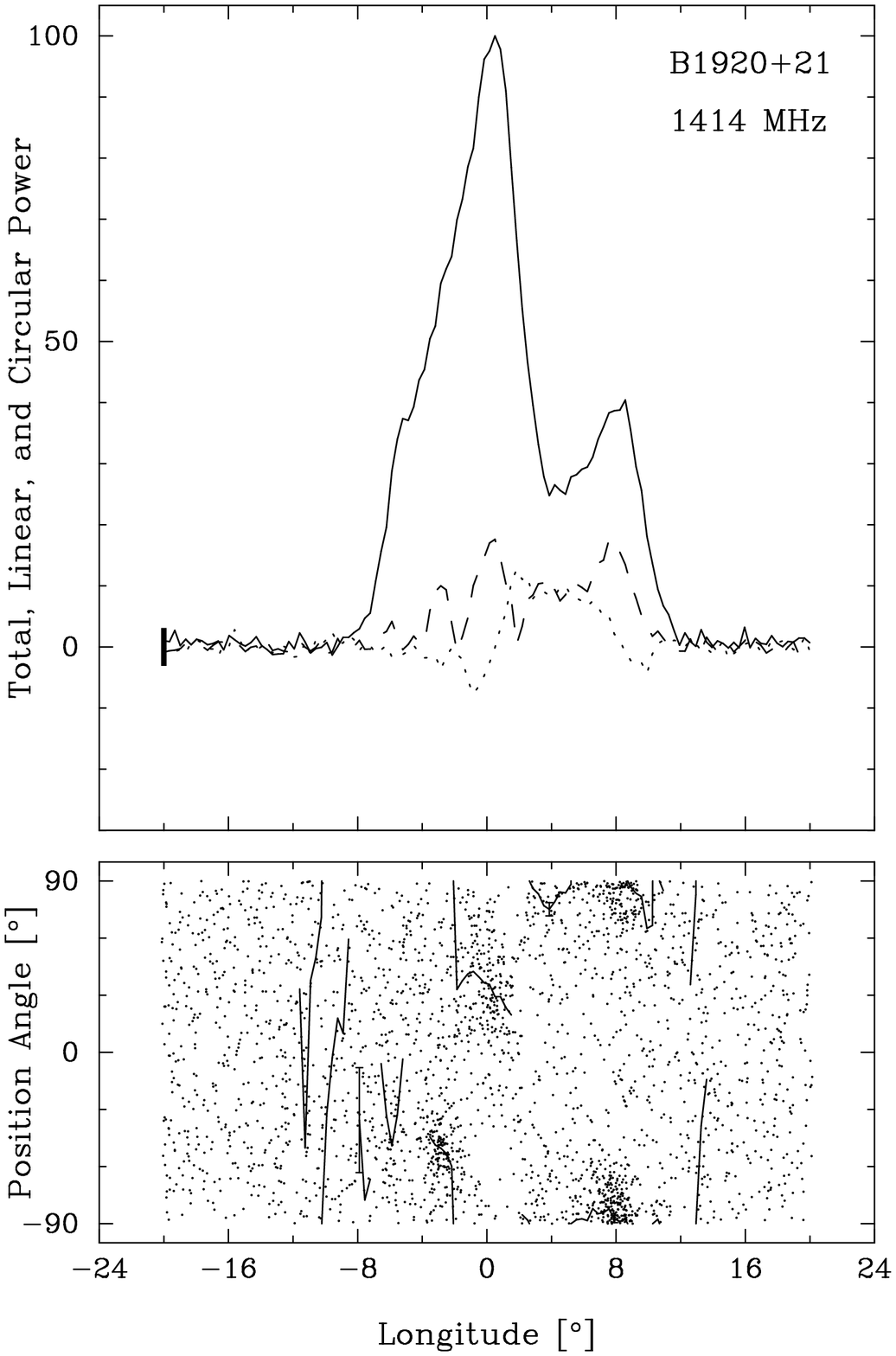} 
}
\caption{Multi-frequency profiles of B1915+13, B1919+21, B1918+19 and B1920+21, and polarization profiles of B1915+13, B1919+21 and B1920+21.}
\label{b11}
\end{figure}

\begin{figure}[htb]
\centerline{
\pf{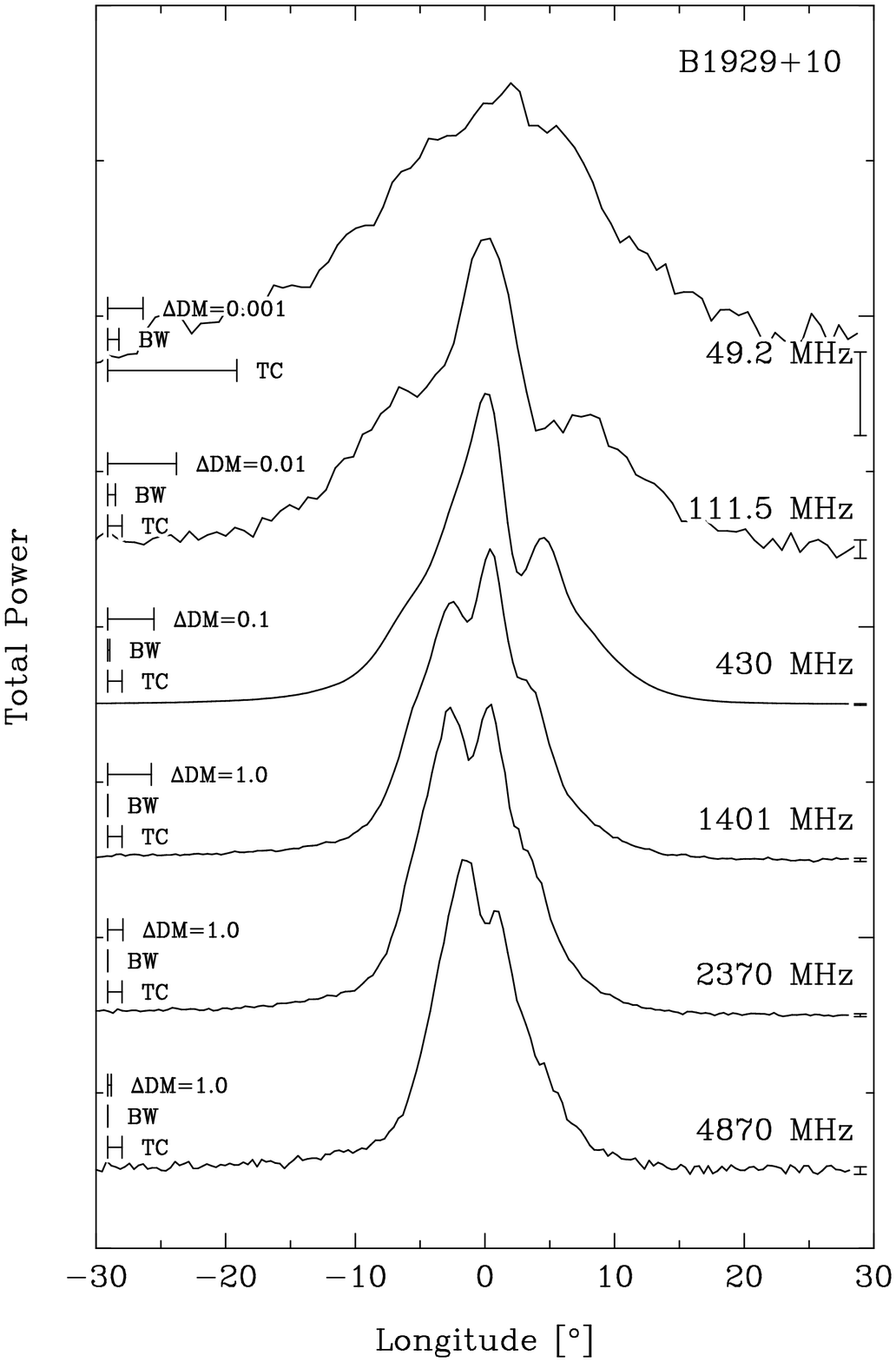}     
\pf{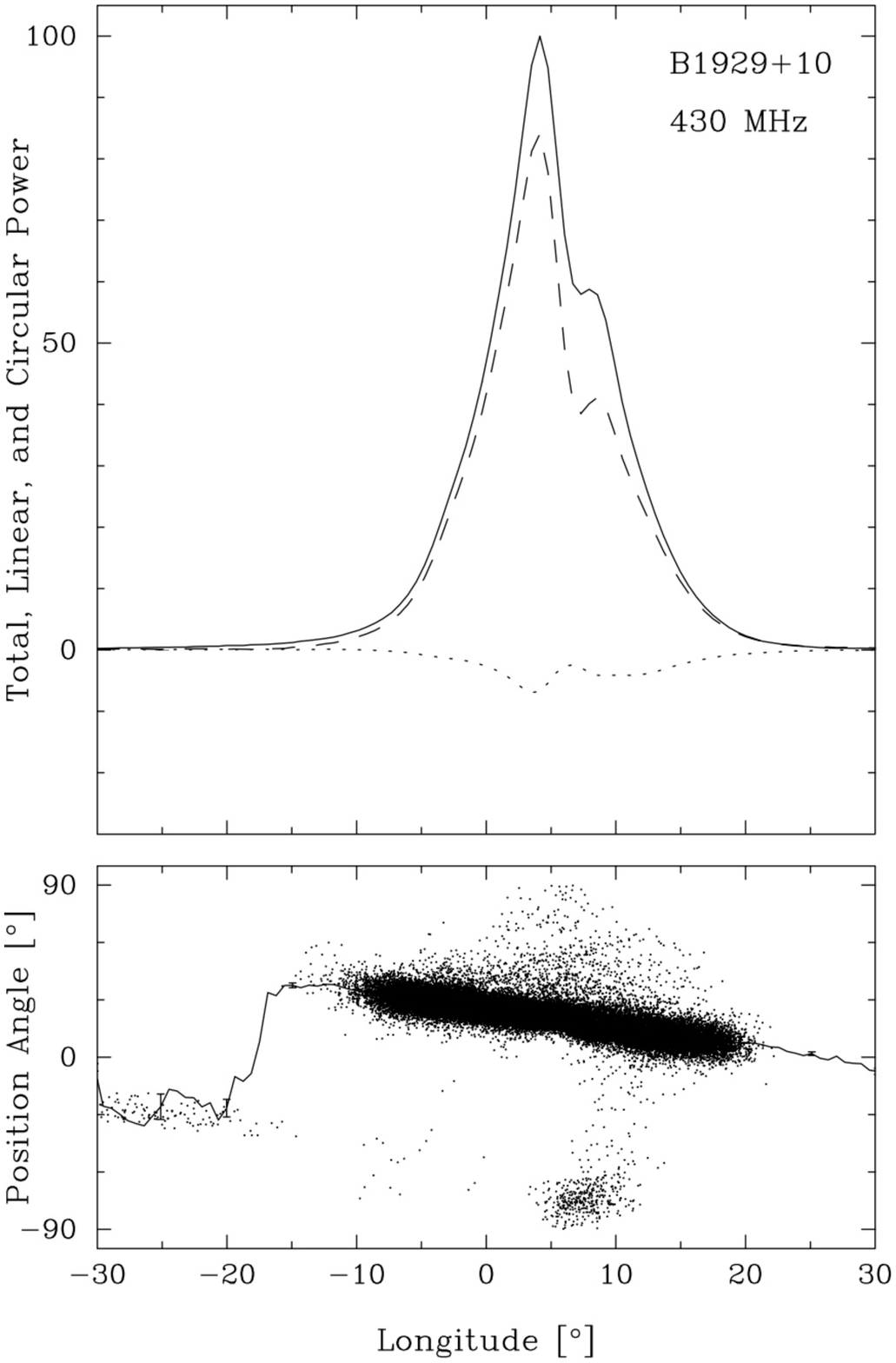} 
\pf{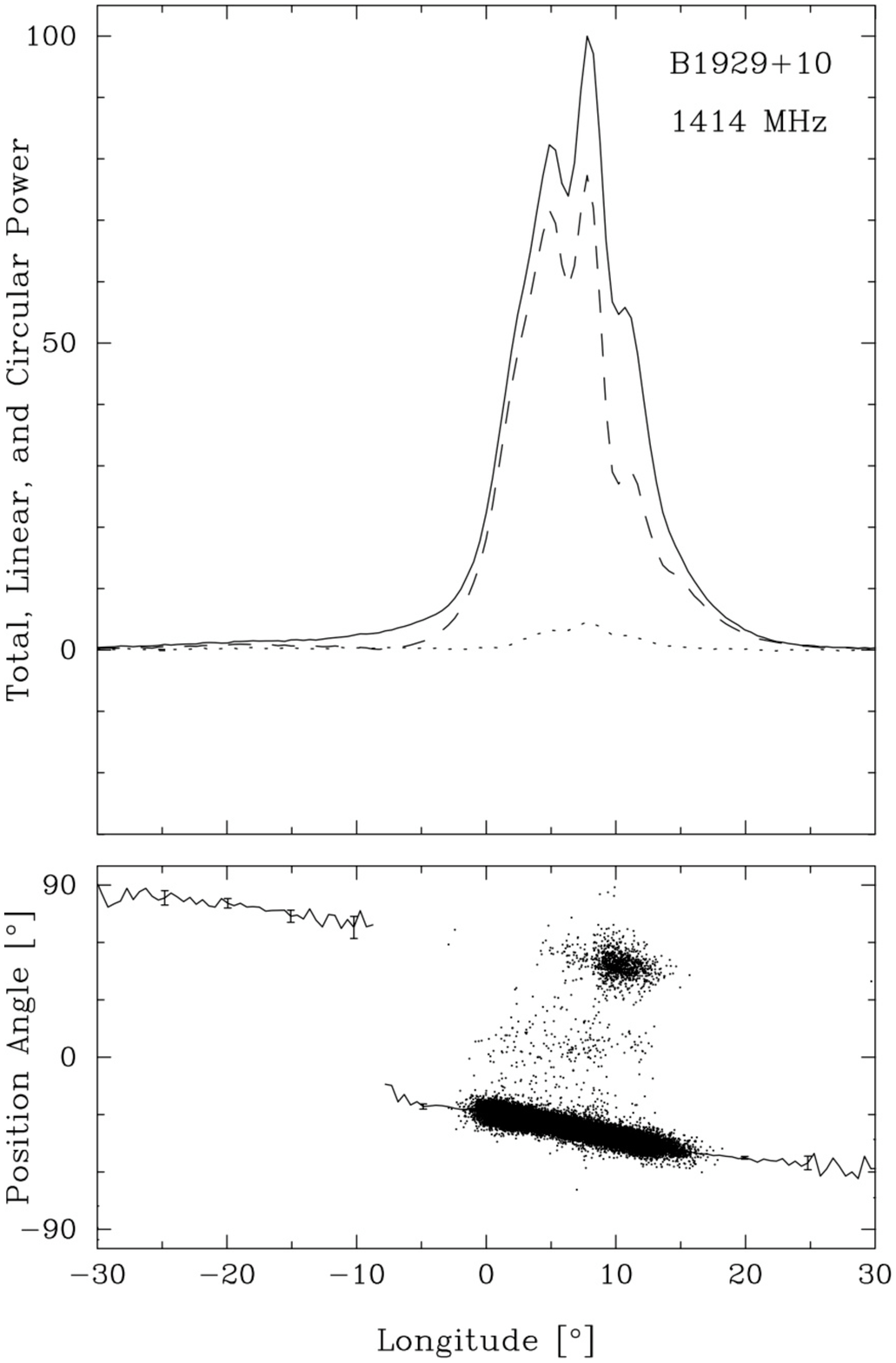} 
}
\quad \\
\centerline{
\pf{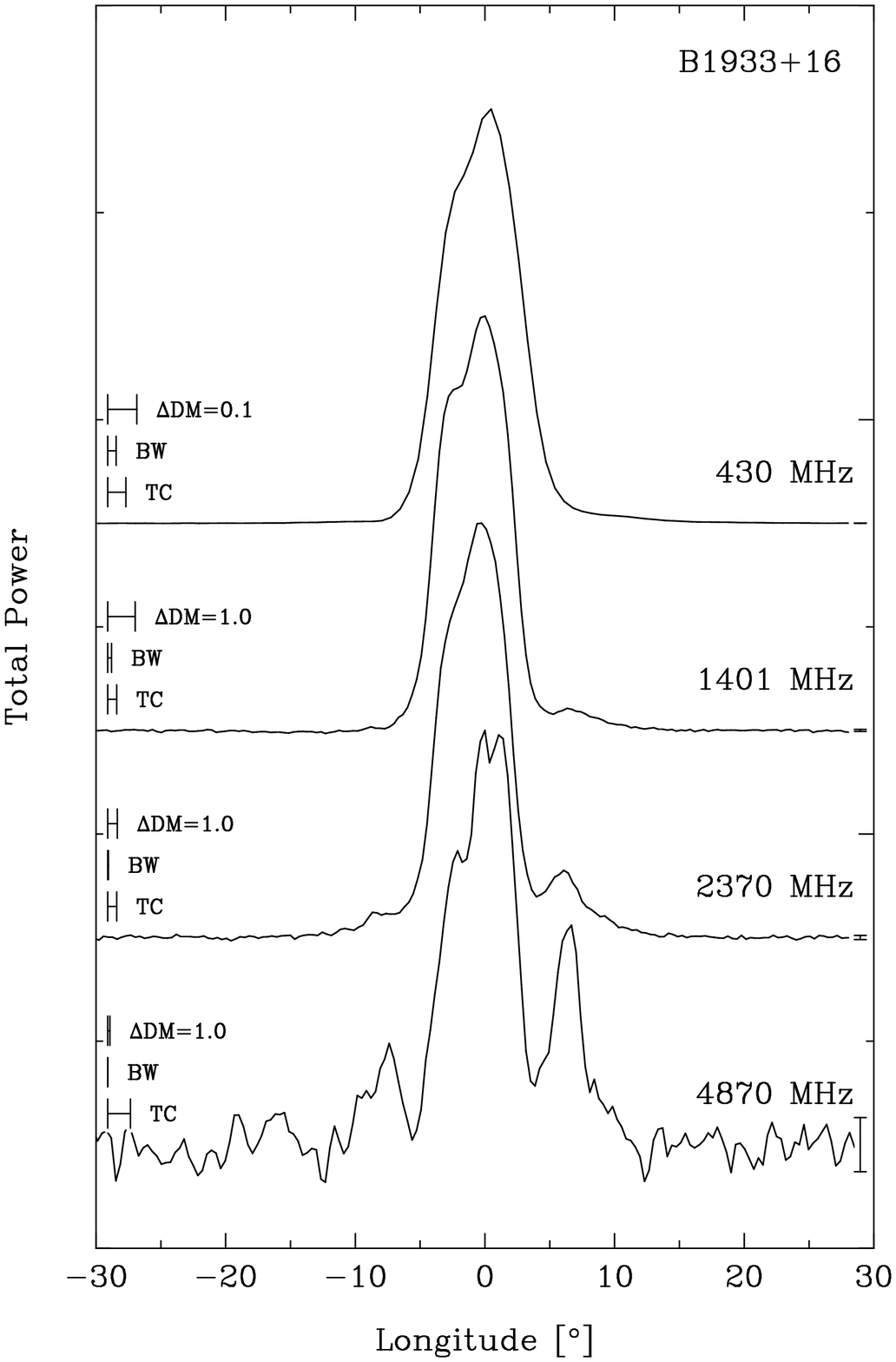}     
\pf{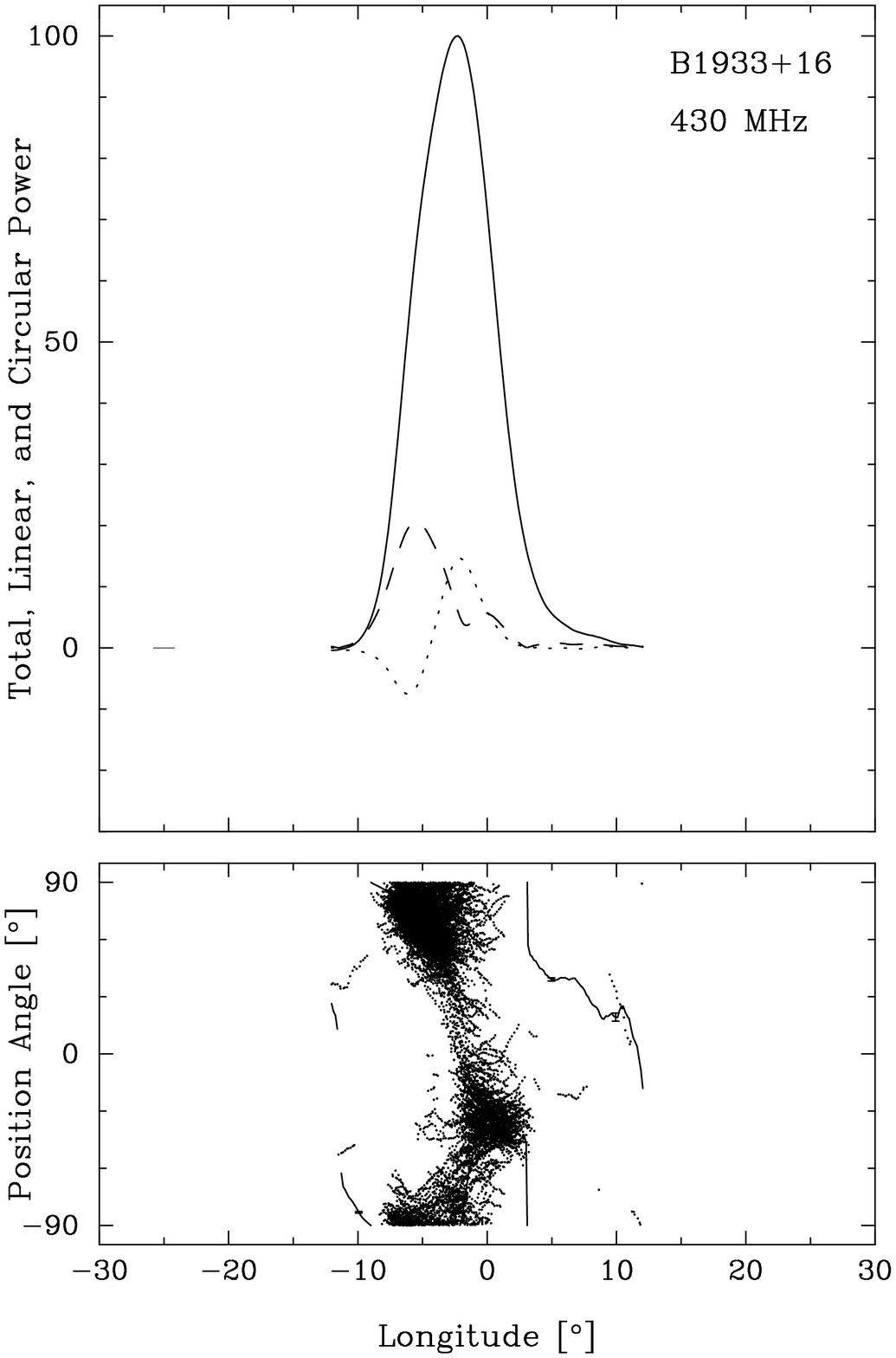} 
\pf{figs/dummy_fig.ps}     
}
\quad \\
\centerline{
\pf{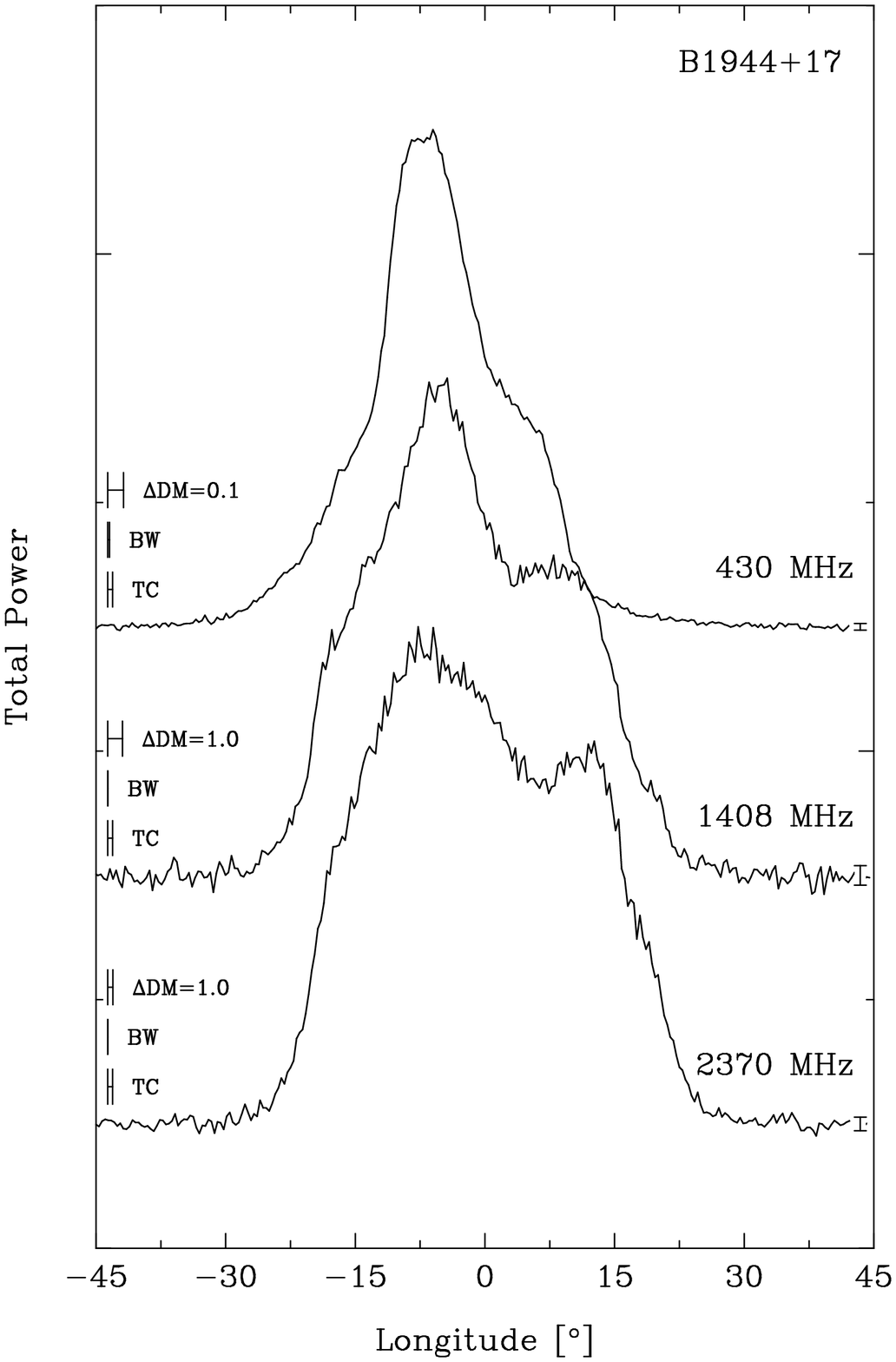}     
\pf{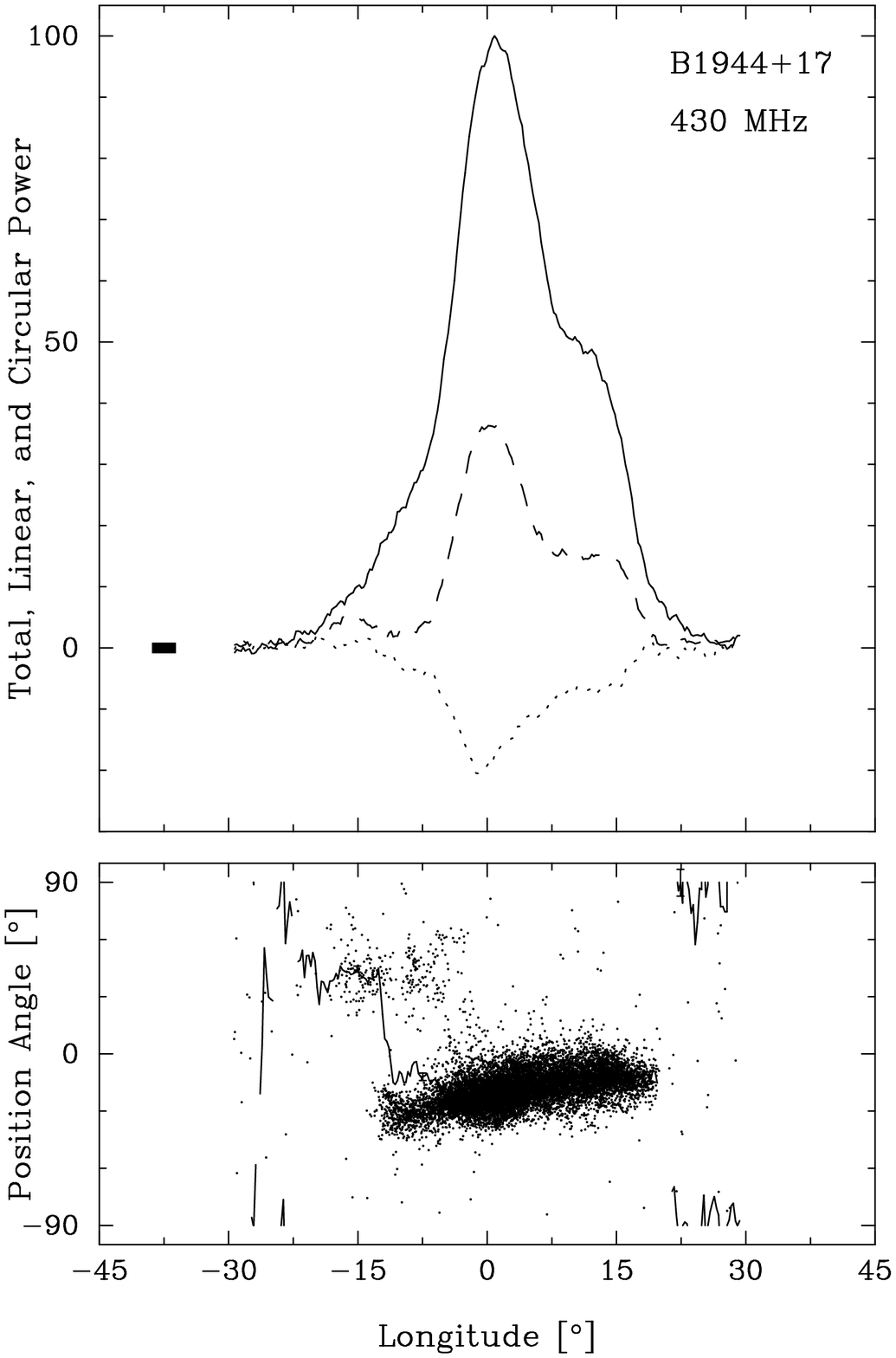} 
\pf{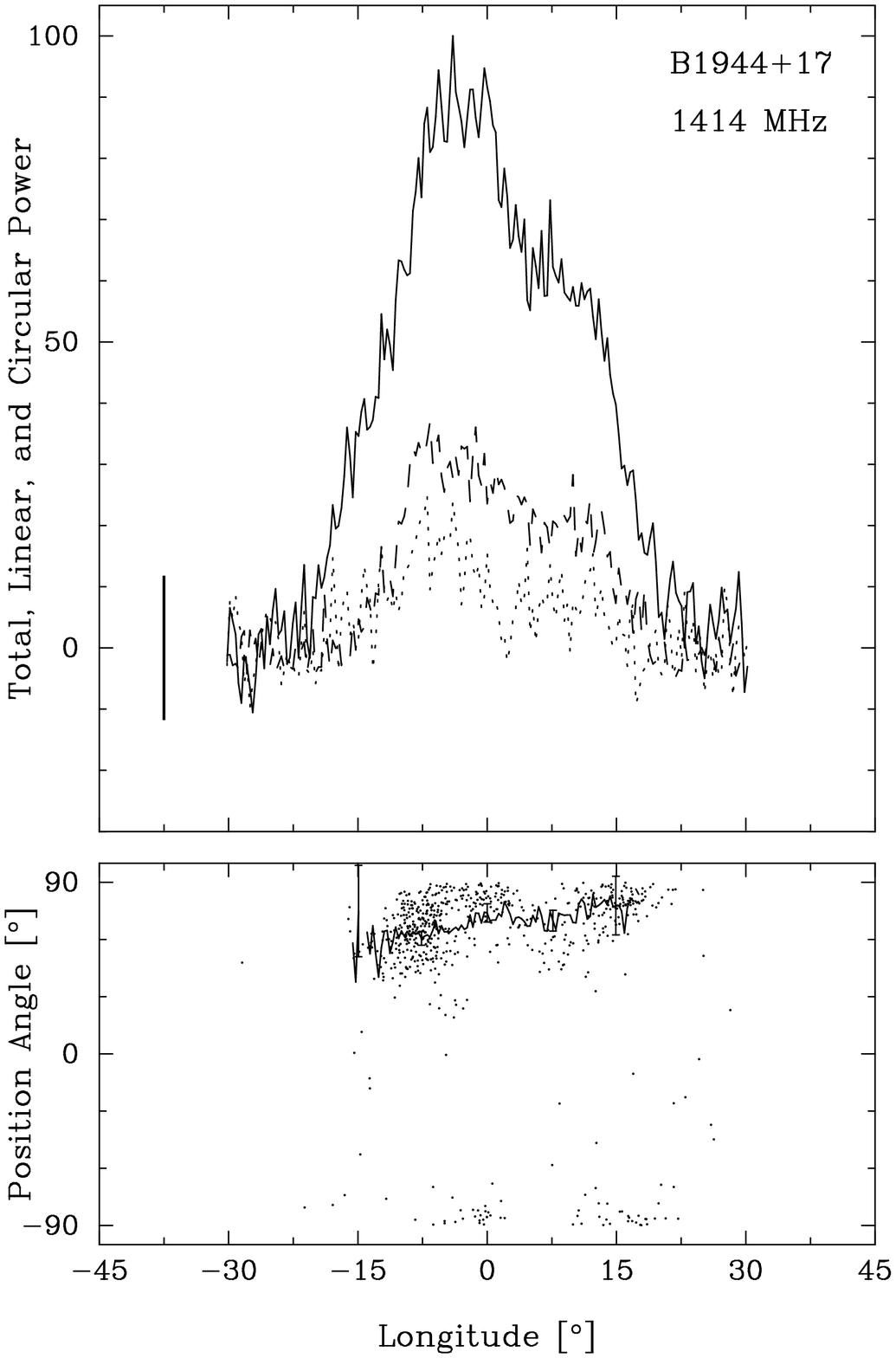} 
}
\caption{Multi-frequency and polarization  profiles of B1929+10, B1933+16 and B1944+17.}
\label{b12}
\end{figure}

\begin{figure}[htb]
\centerline{
\pf{figs/dummy_fig.ps}     
\pf{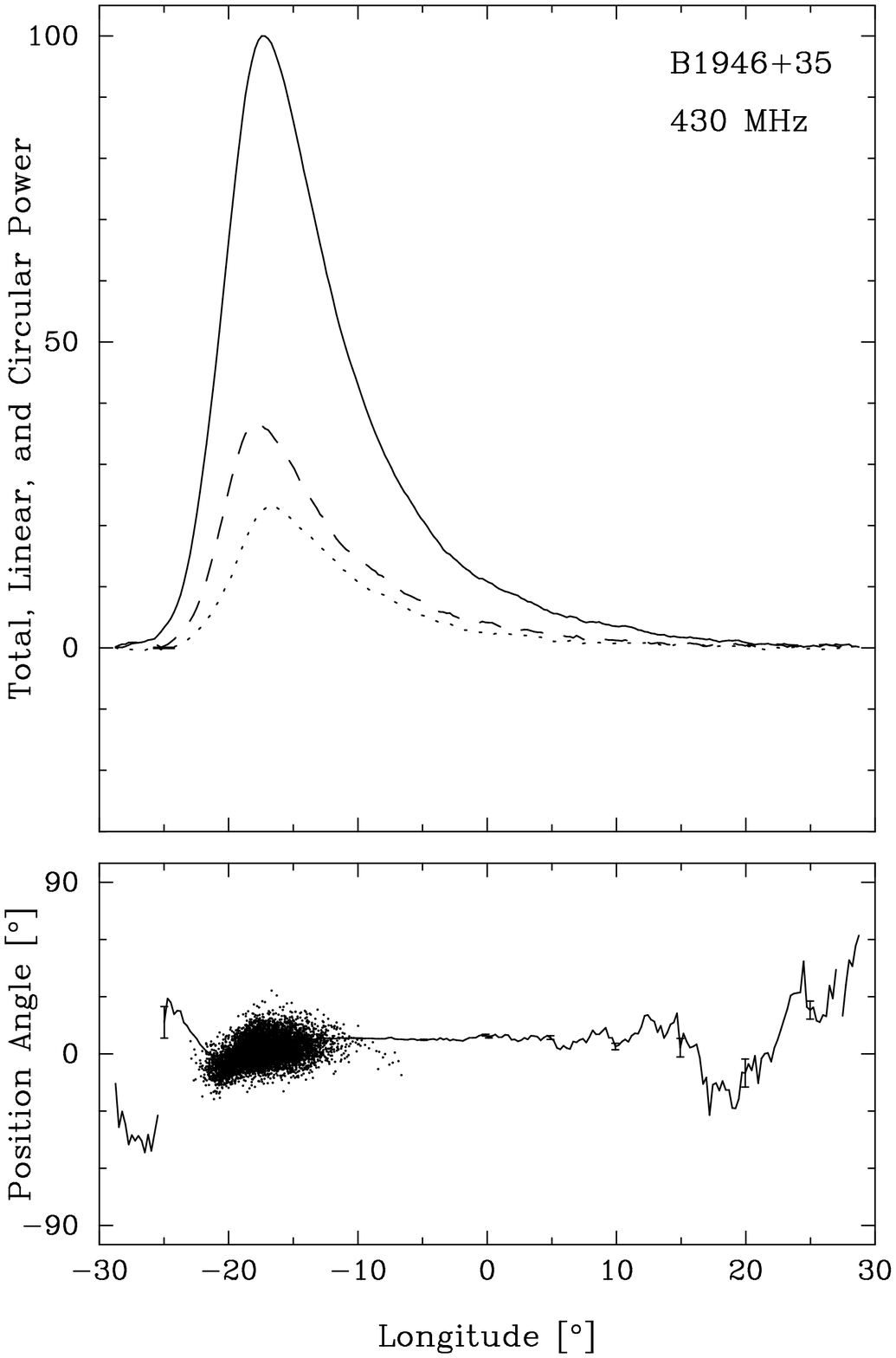} 
\pf{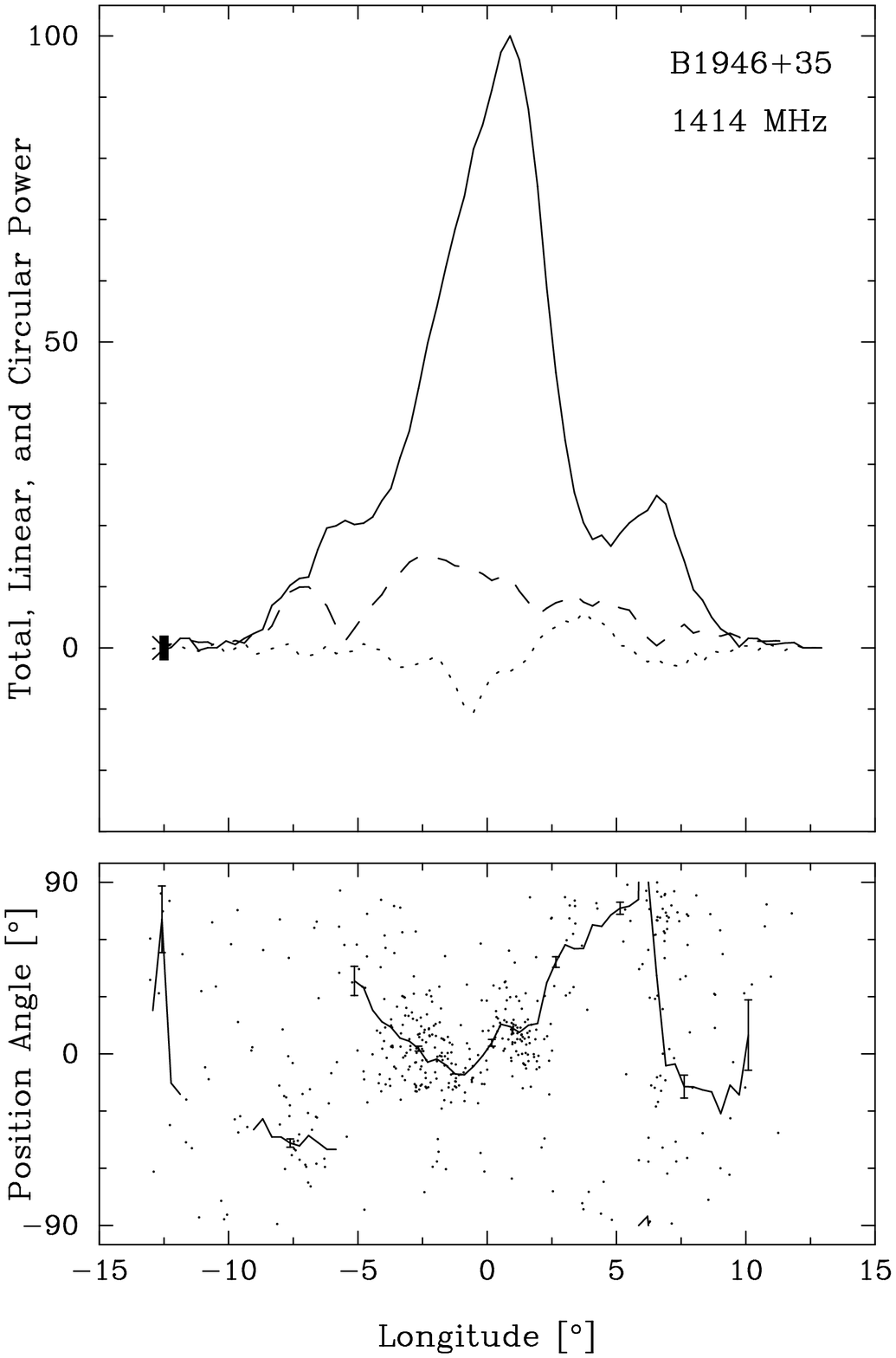} 
}
\quad \\
\centerline{
\pf{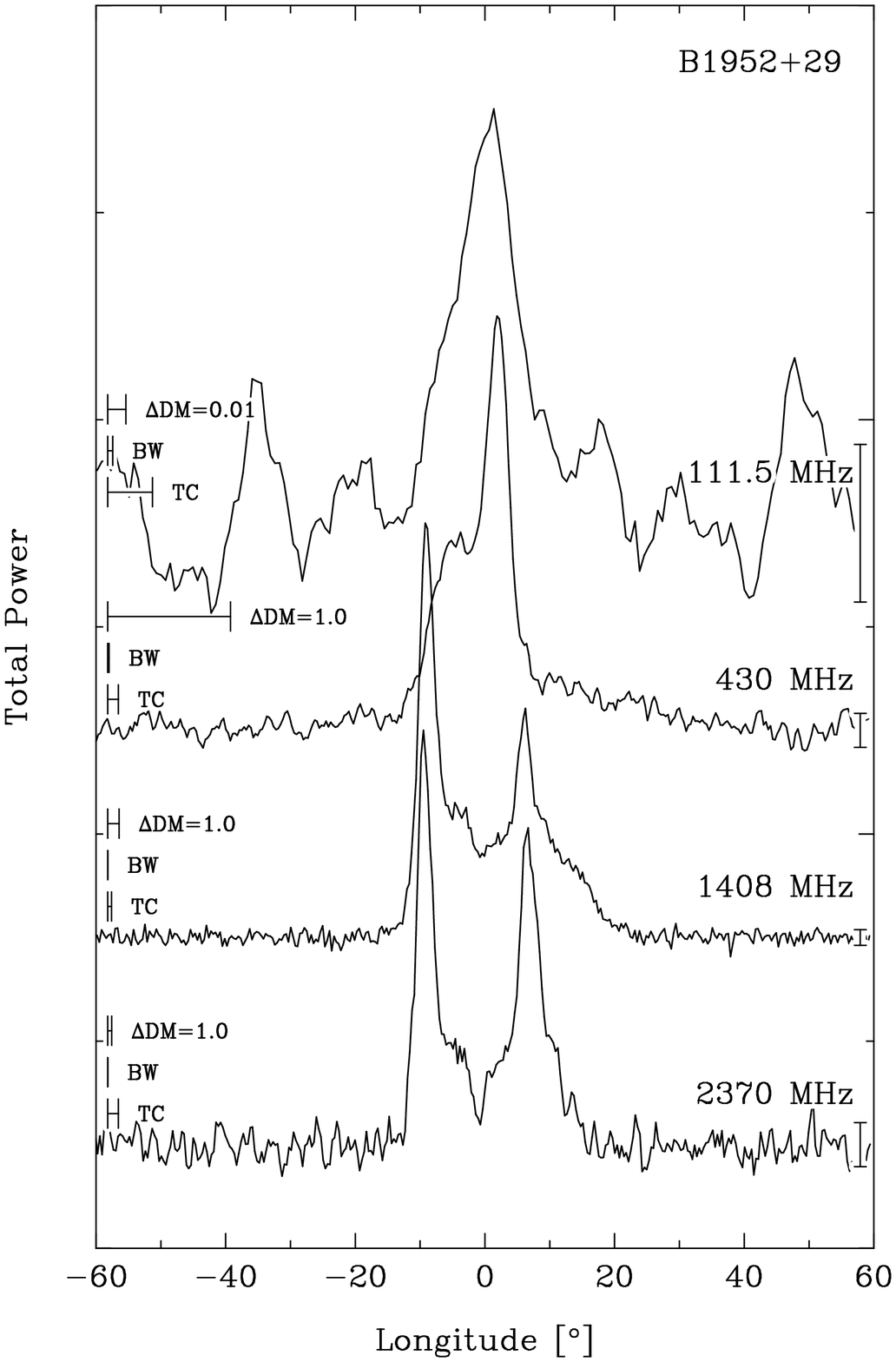}     
\pf{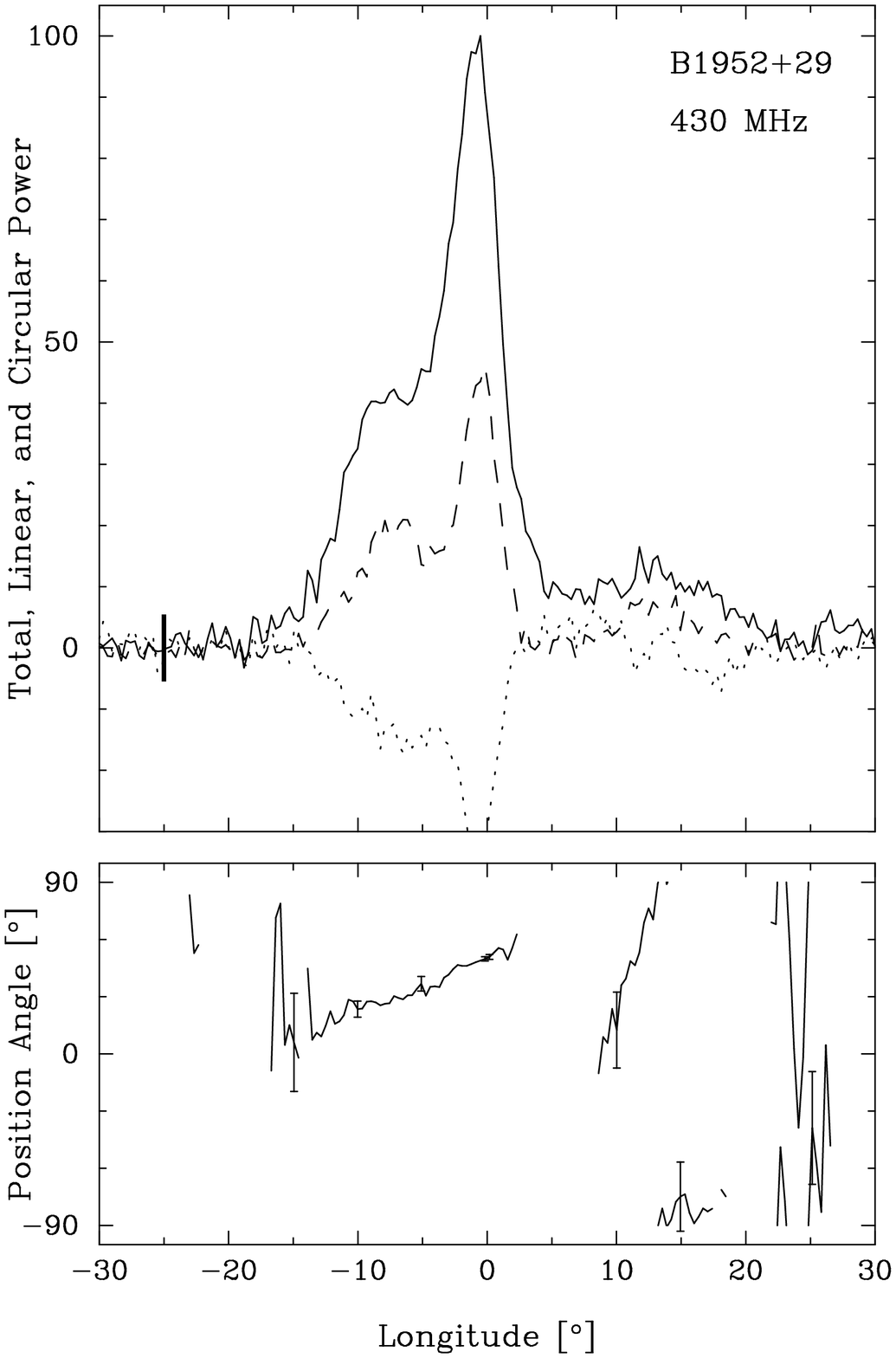}     
\pf{figs/dummy_fig.ps}     
}
\quad \\
\centerline{
\pf{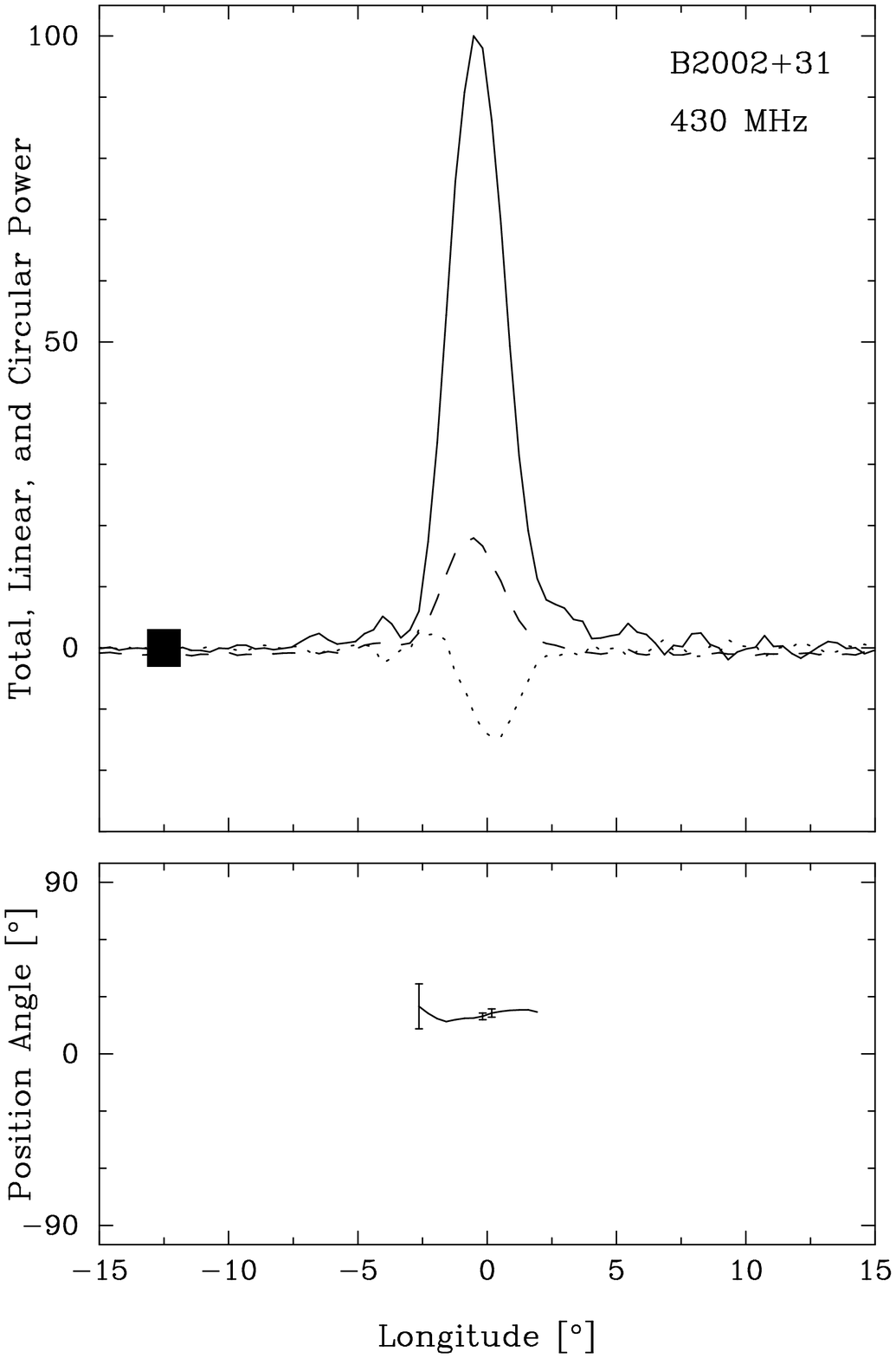}     
\pf{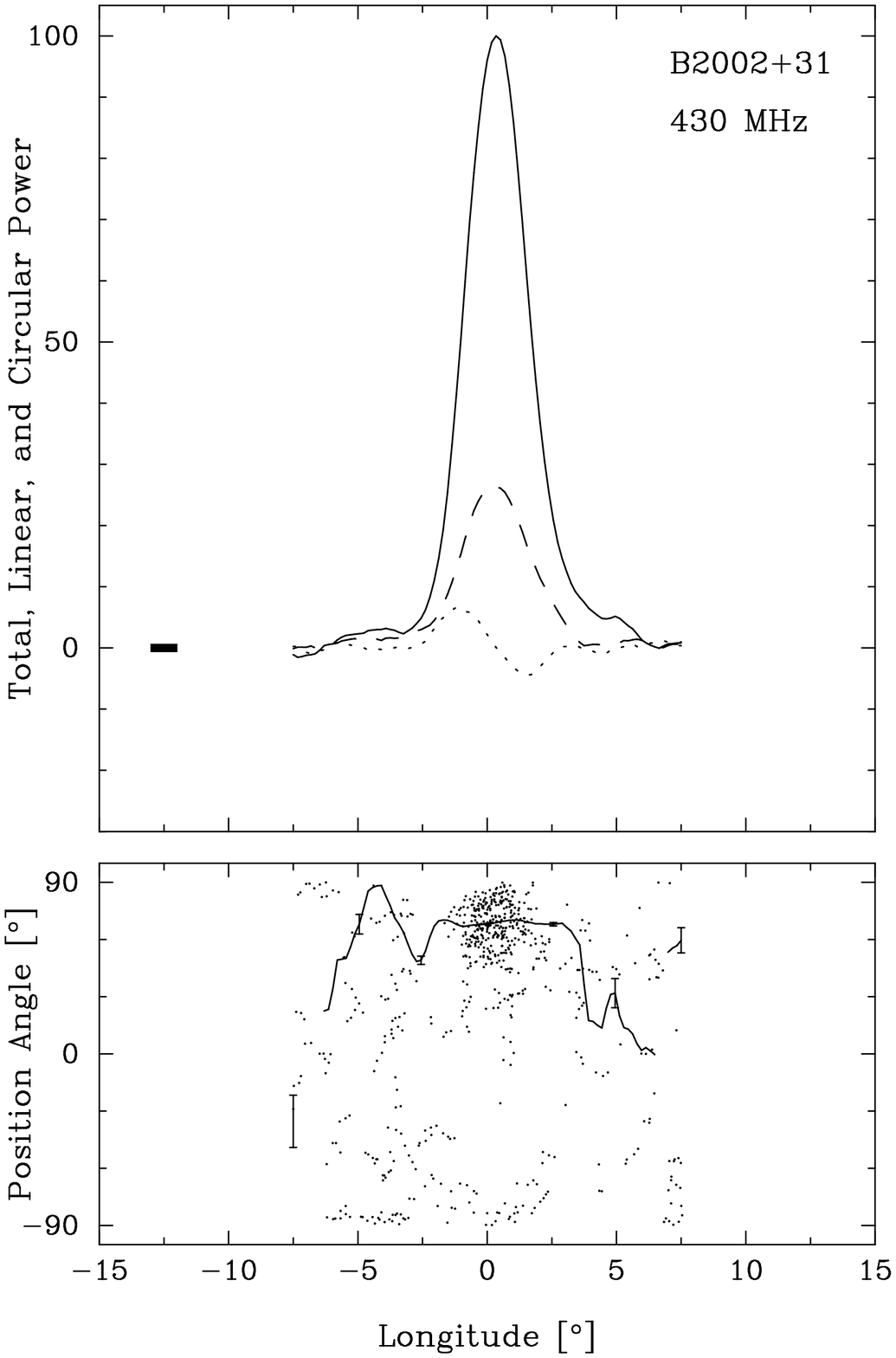} %
\pf{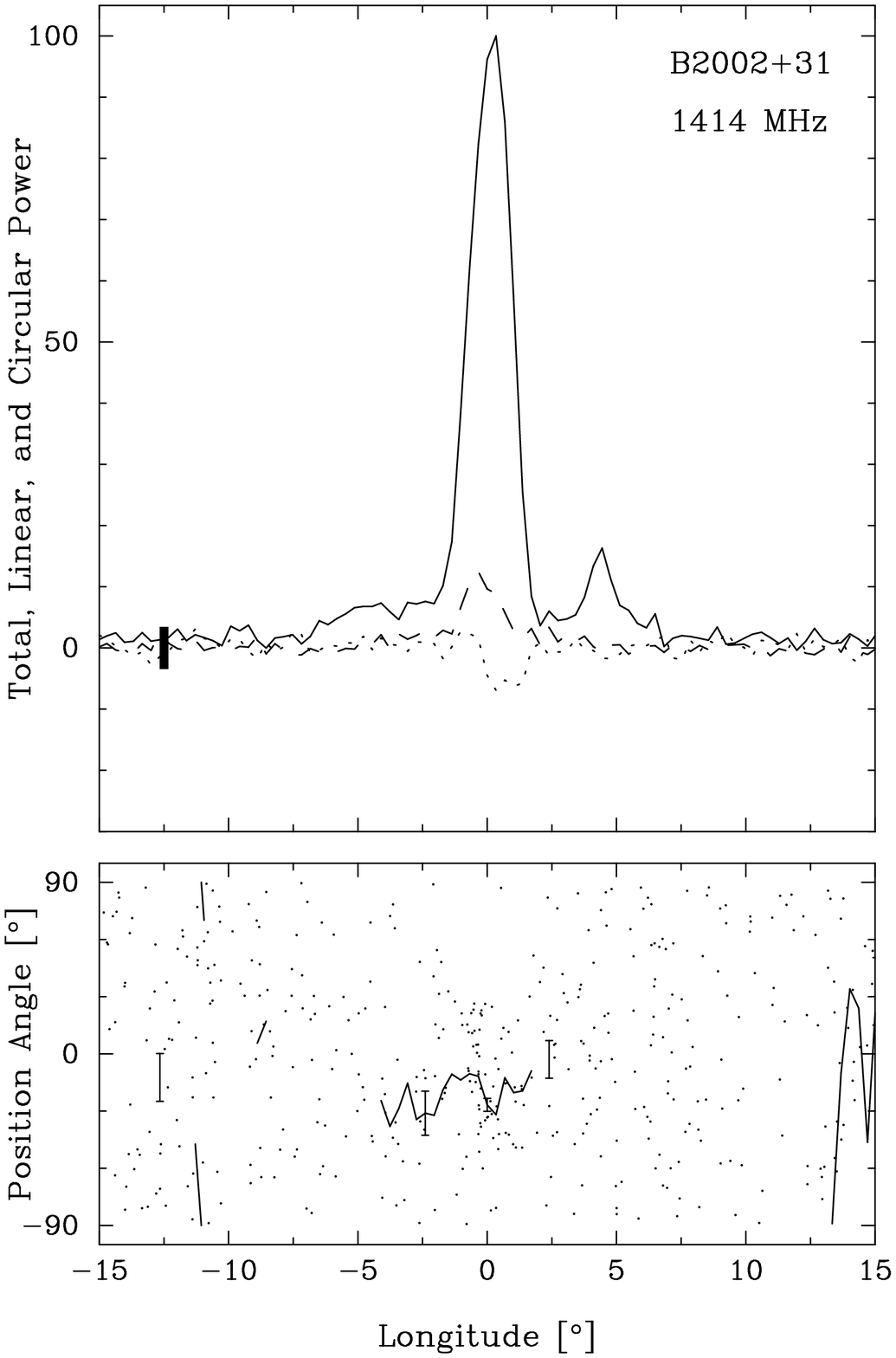} %
}
\caption{Multi-frequency and polarization profiles of B1946+35, B1952+29 and B2002+31.}
\label{b13}
\end{figure}

\begin{figure}[htb]
\centerline{
\pf{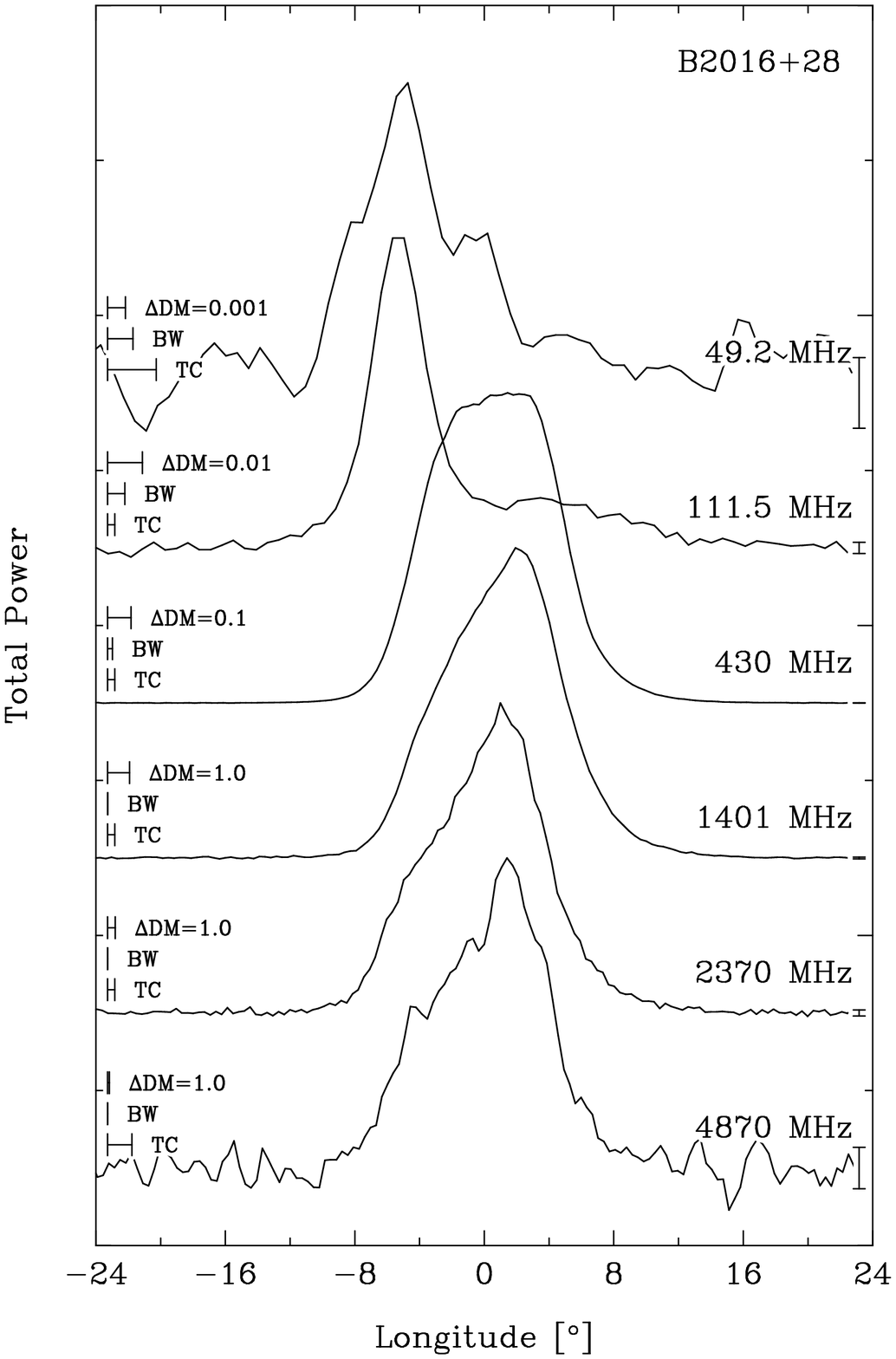}     
\pf{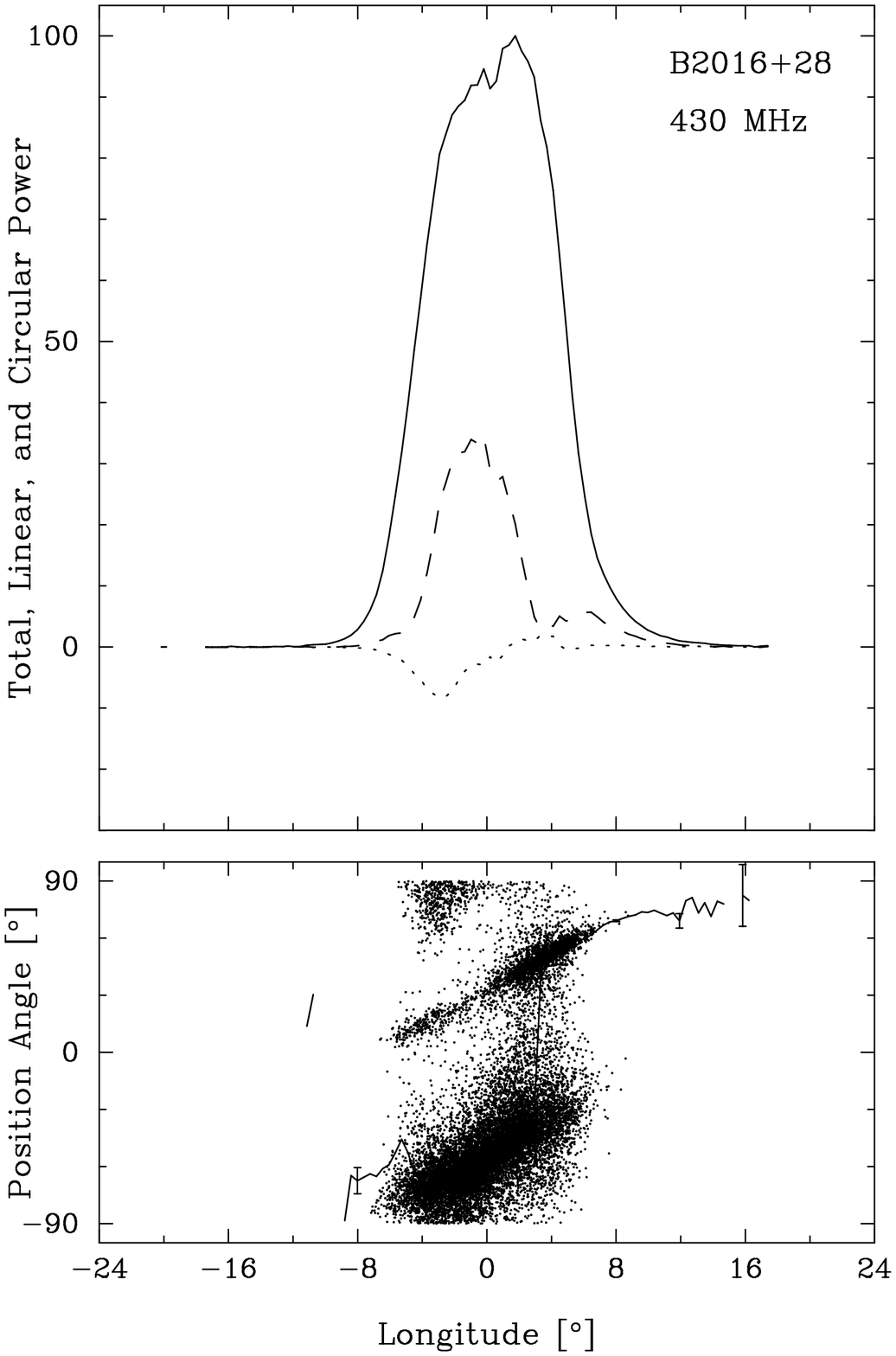} 
\pf{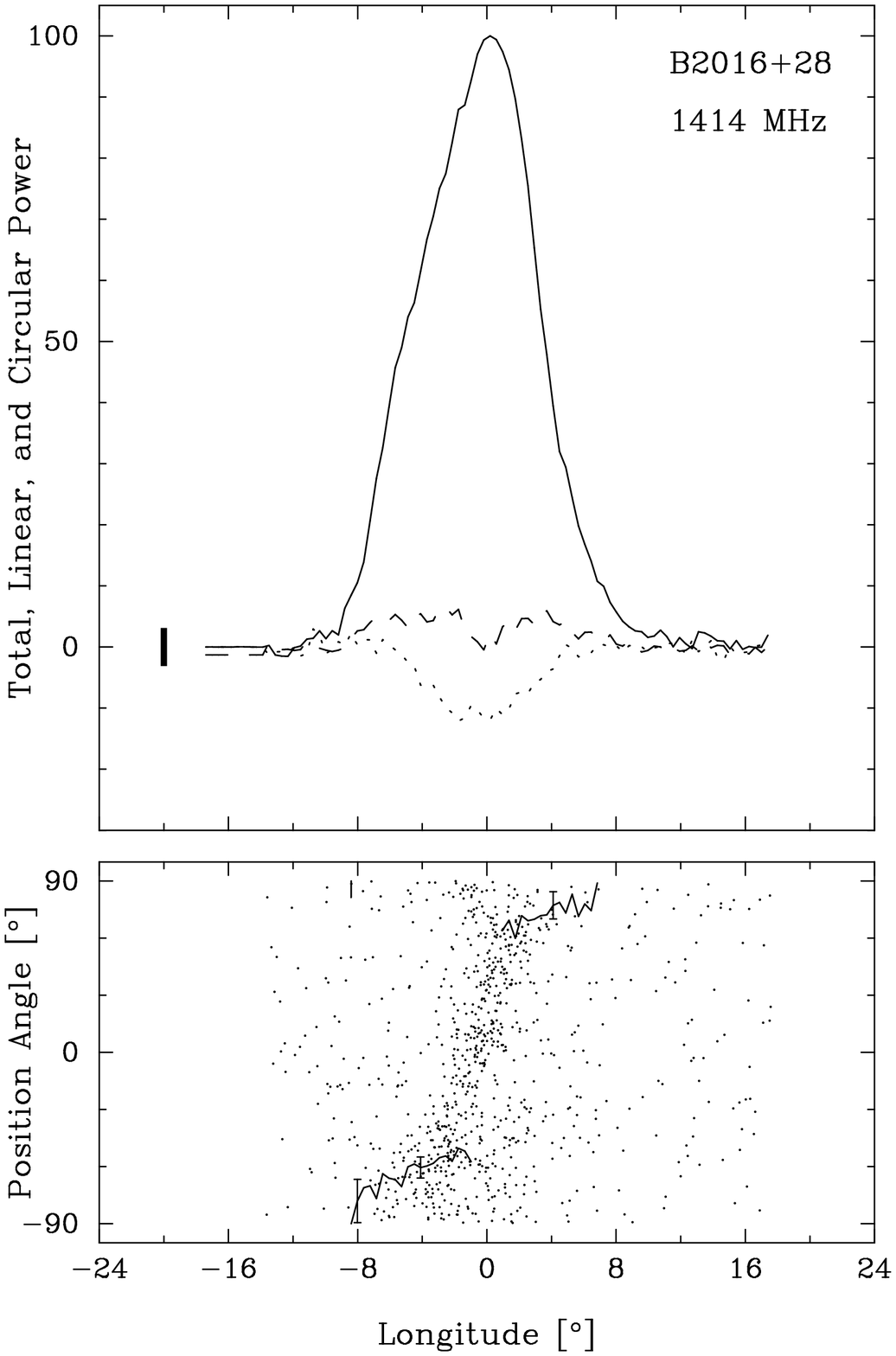} 
}
\quad \\
\centerline{
\pf{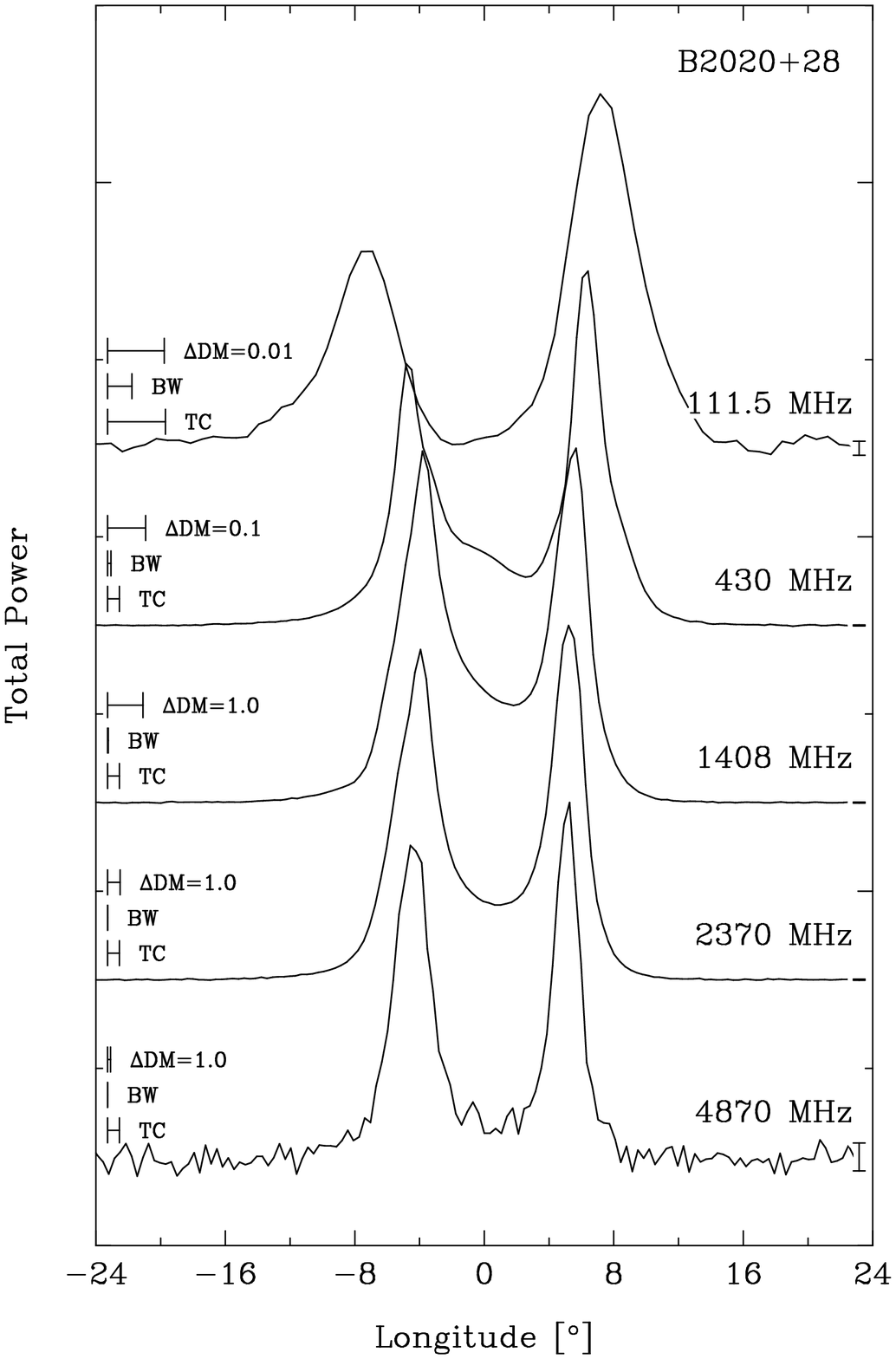}     
\pf{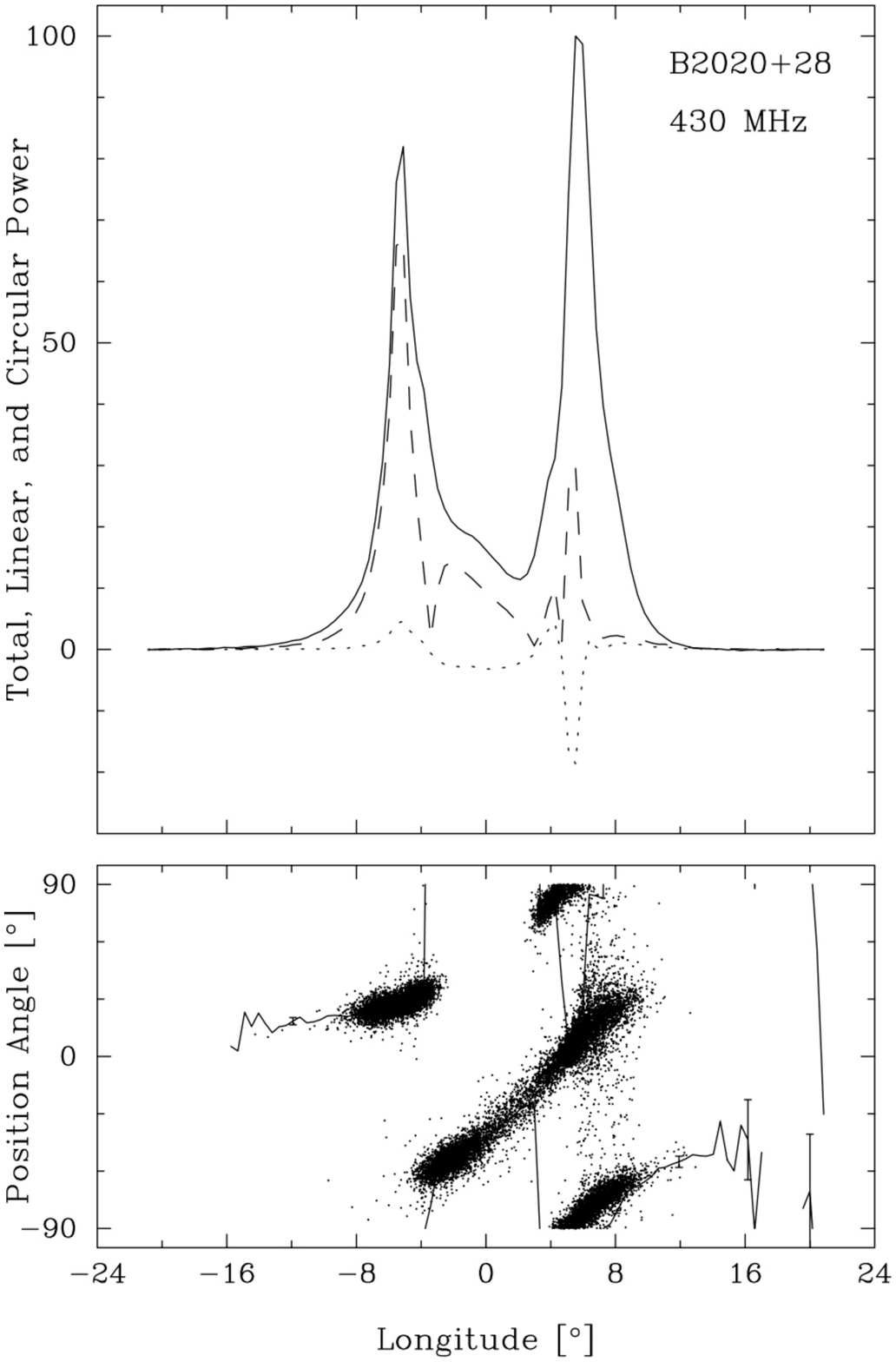} 
\pf{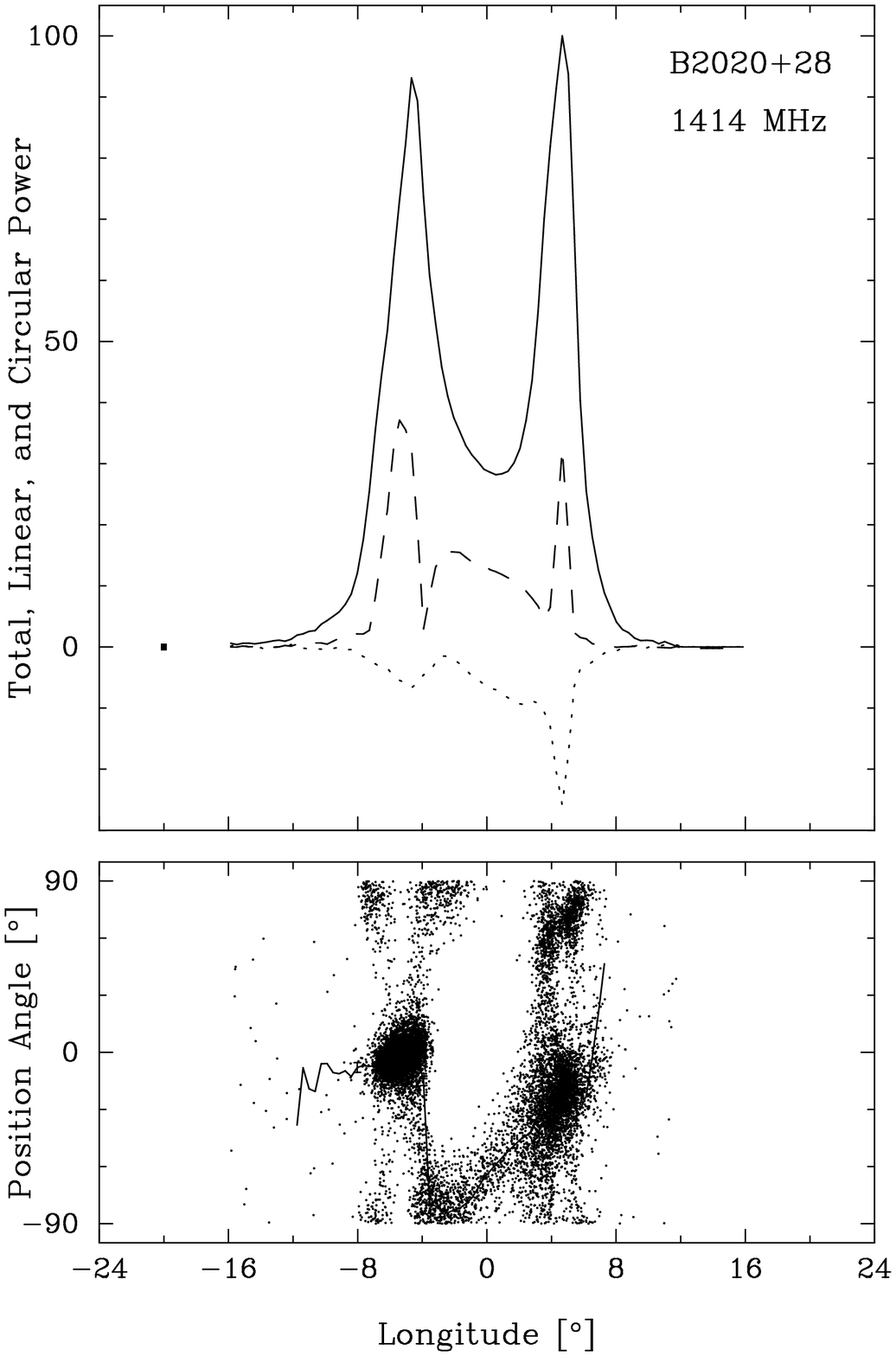} 
}
\quad \\
\centerline{
\pf{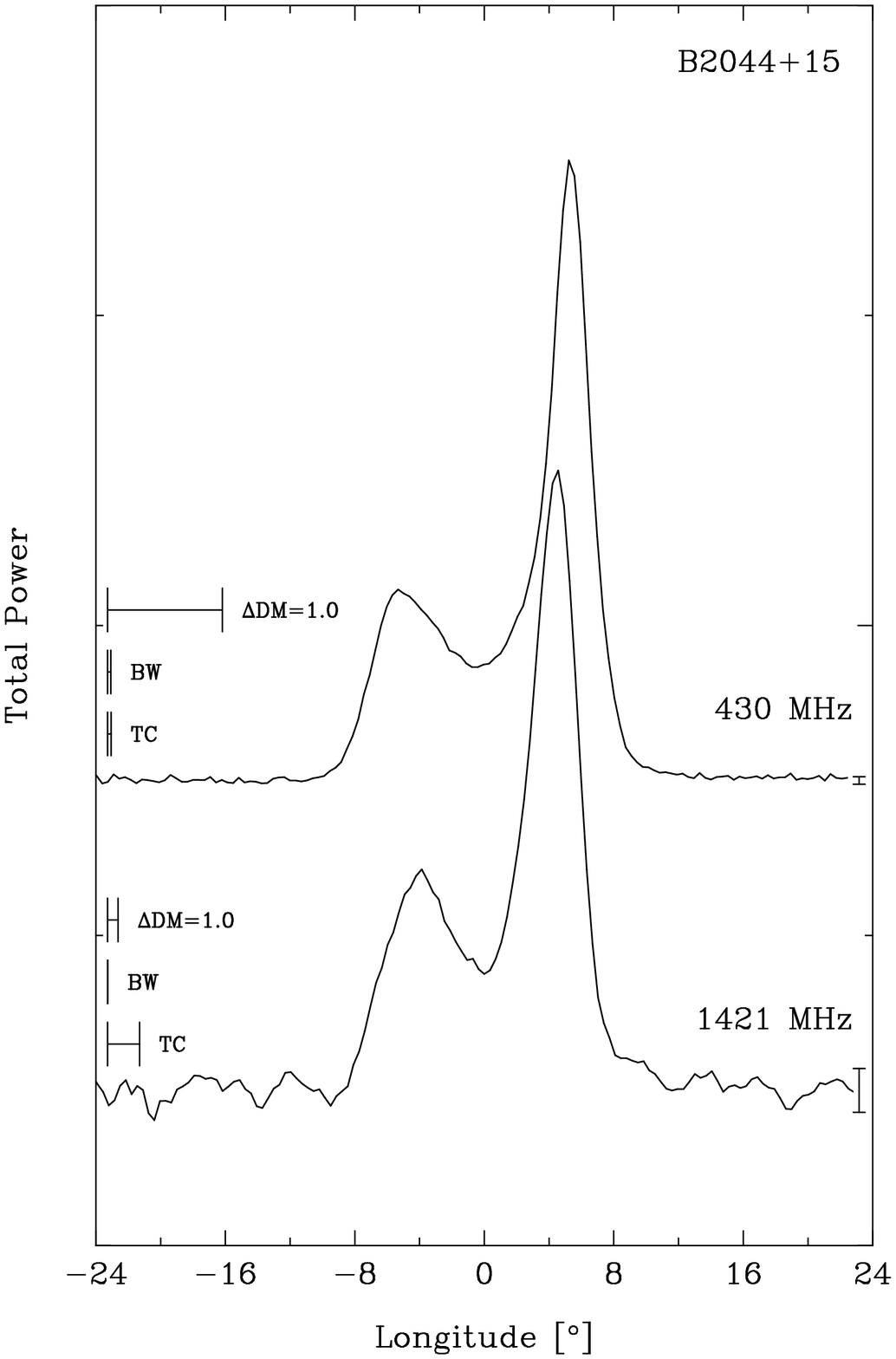}     
\pf{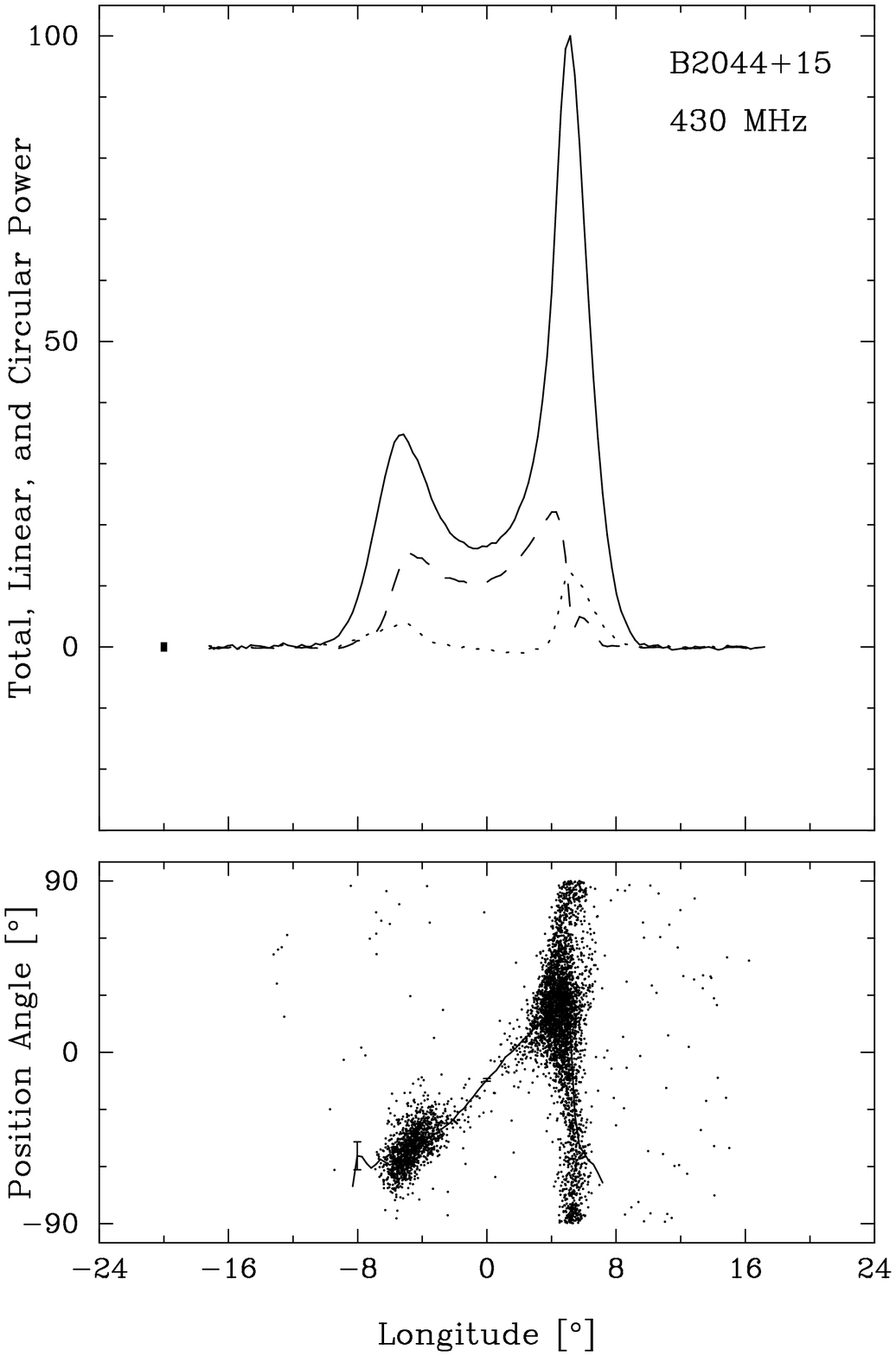} 
\pf{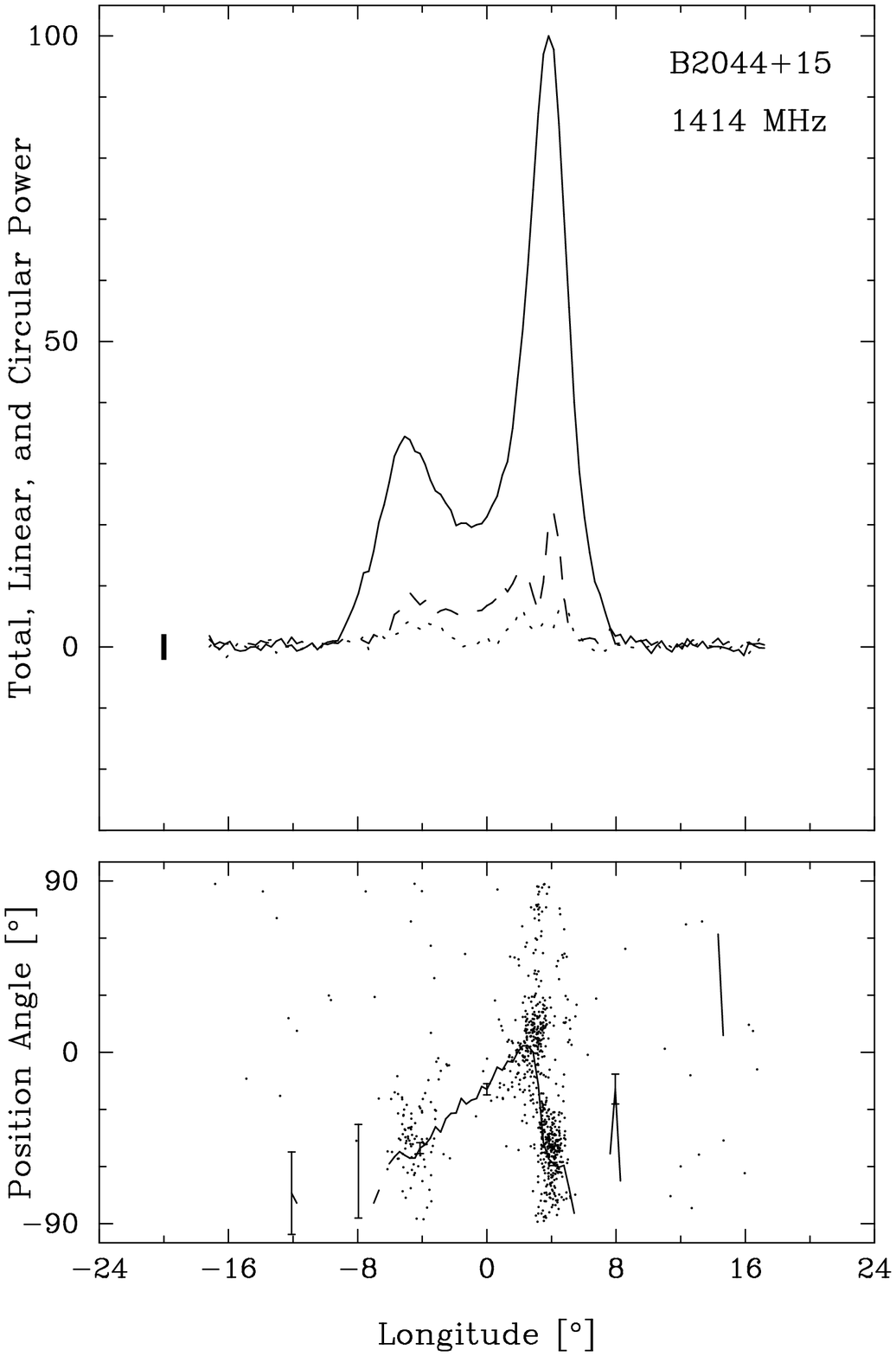} 
}
\caption{Multi-frequency and polarization profiles of B2016+28, B2020+28 and B2044+15.}
\label{b14}
\end{figure}

\begin{figure}[htb]
\centerline{
\pf{figs/dummy_fig.ps}     
\pf{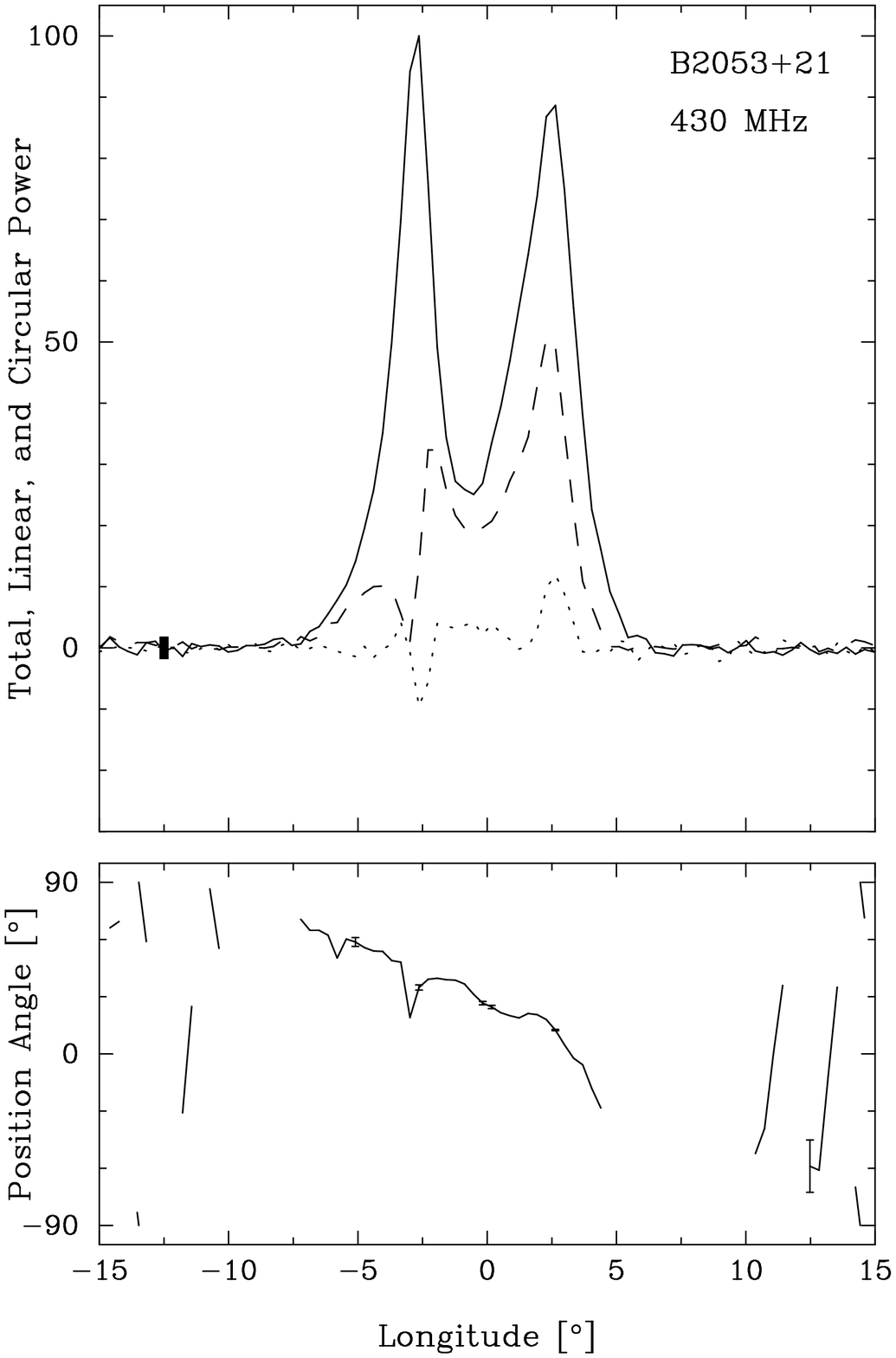} 
\pf{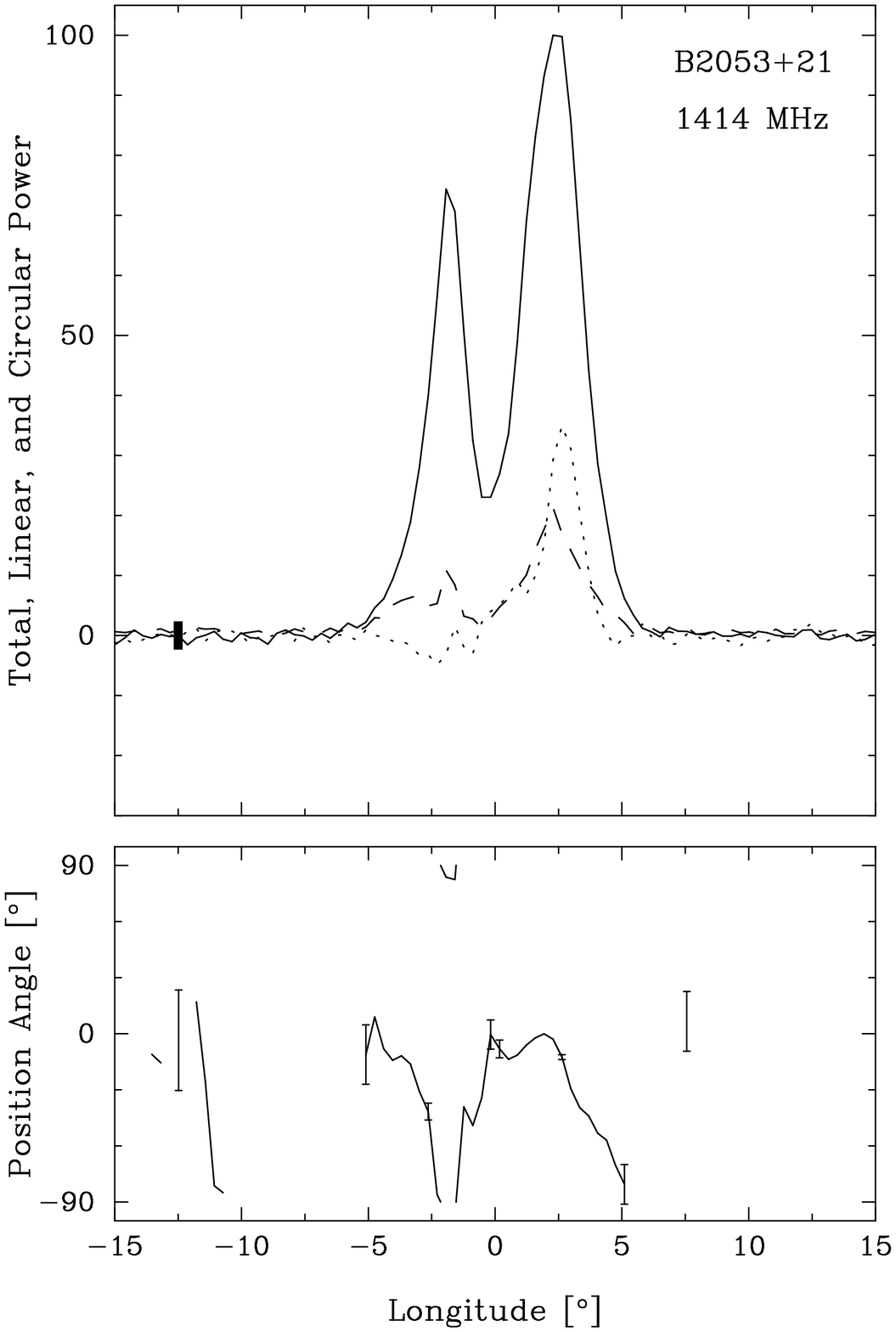} 
}
\quad \\
\centerline{
\pf{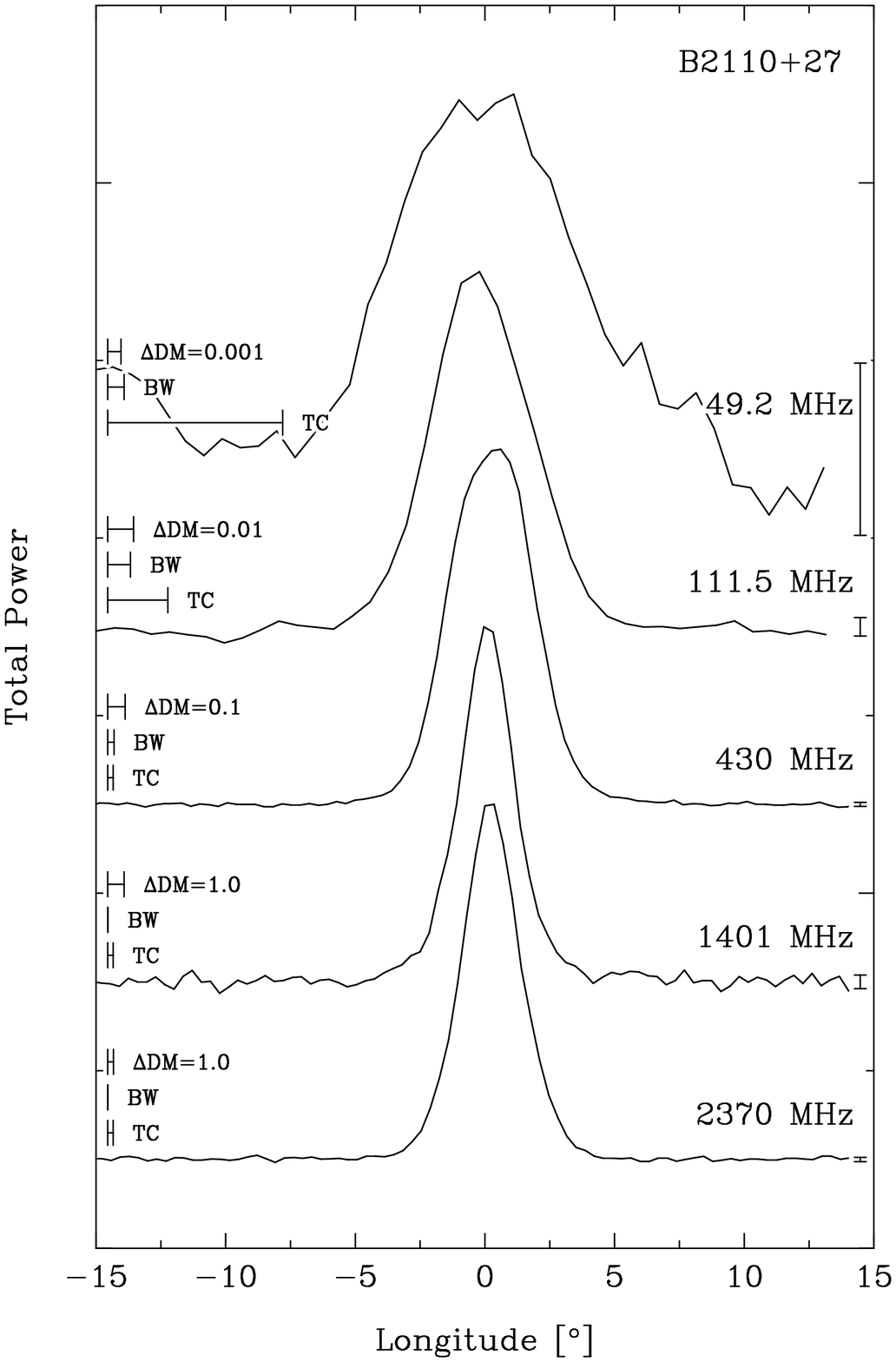}     
\pf{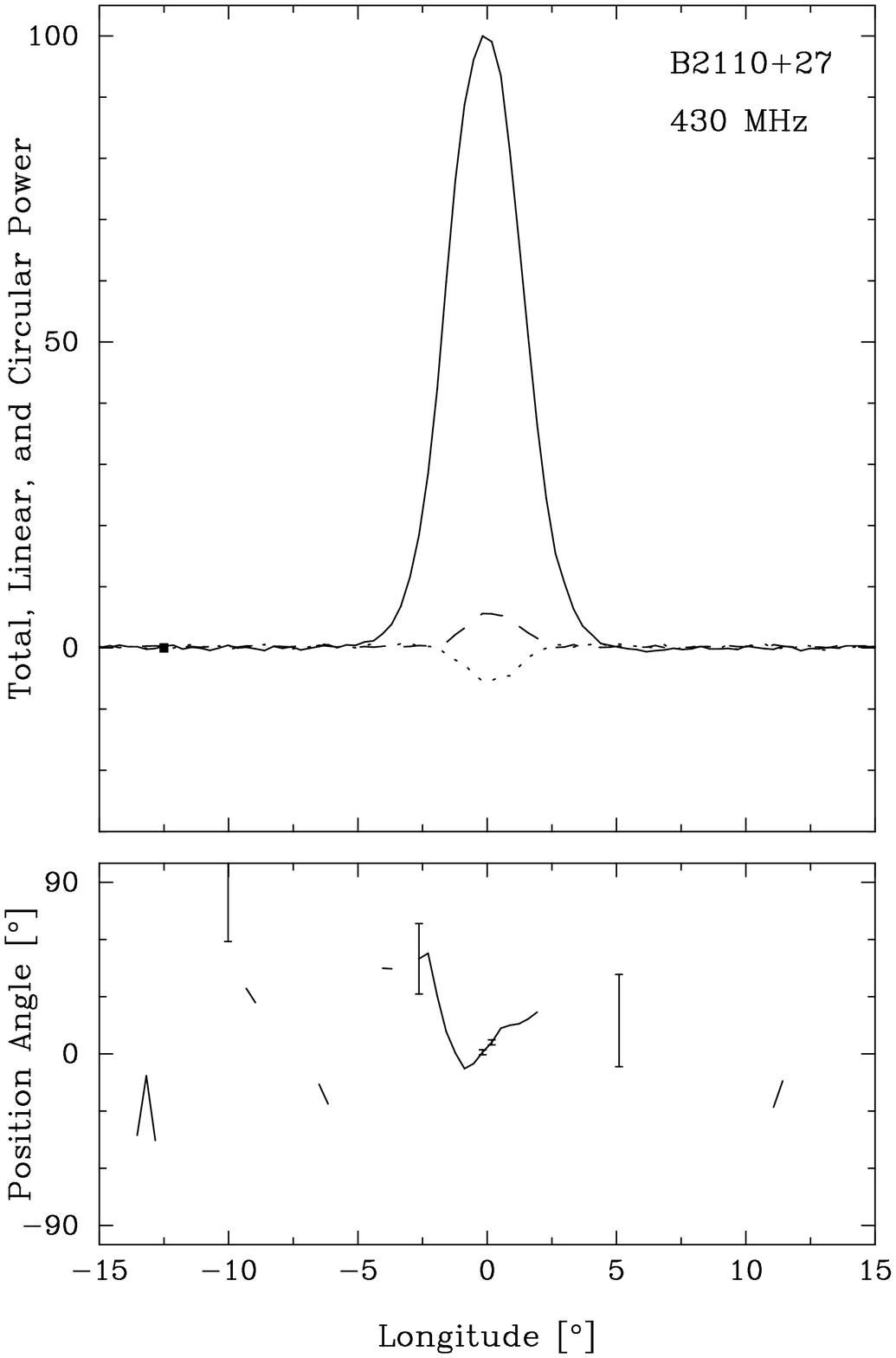} 
\pf{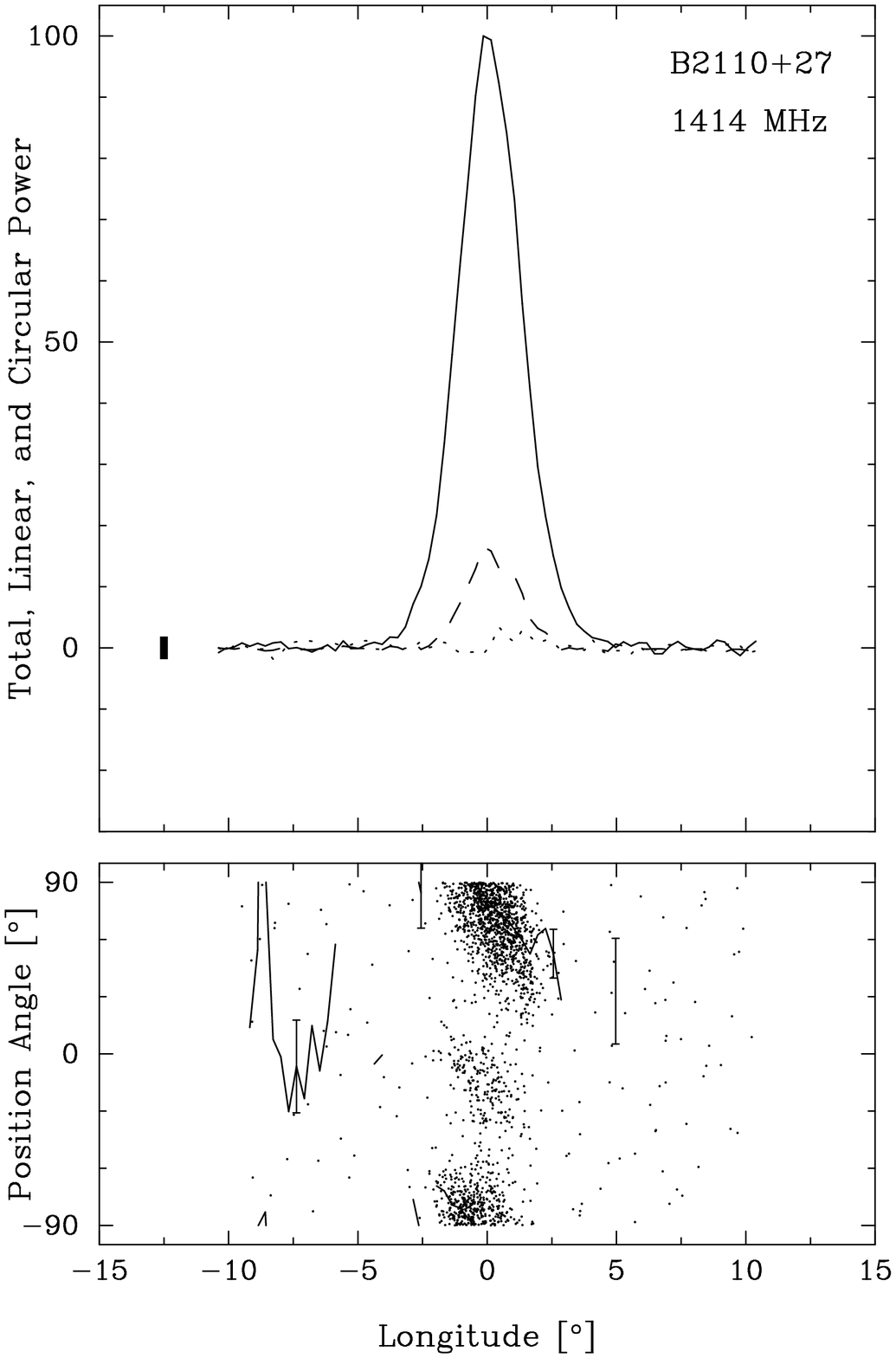} 
}
\quad \\
\centerline{
\pf{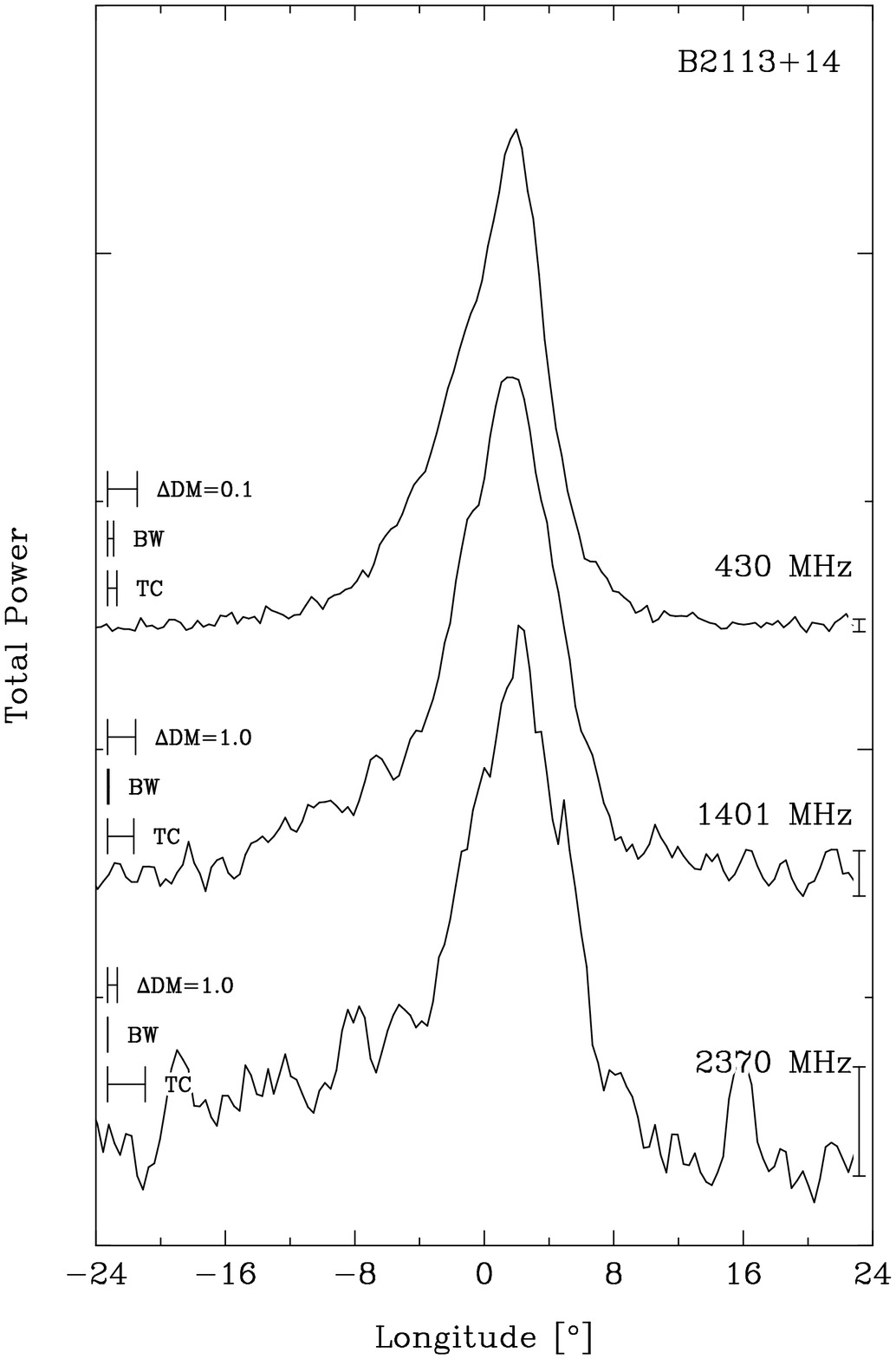}     
\pf{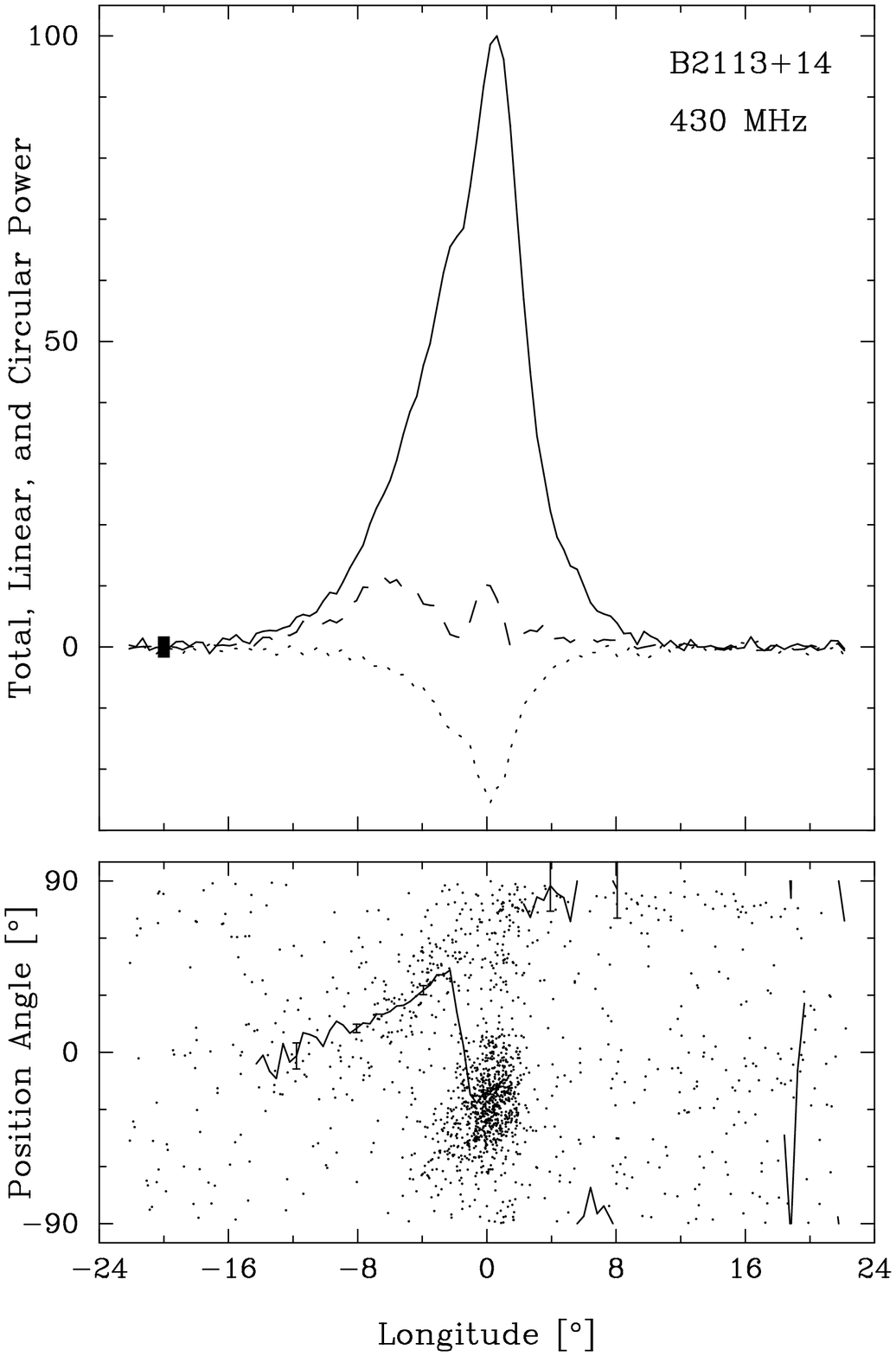} 
\pf{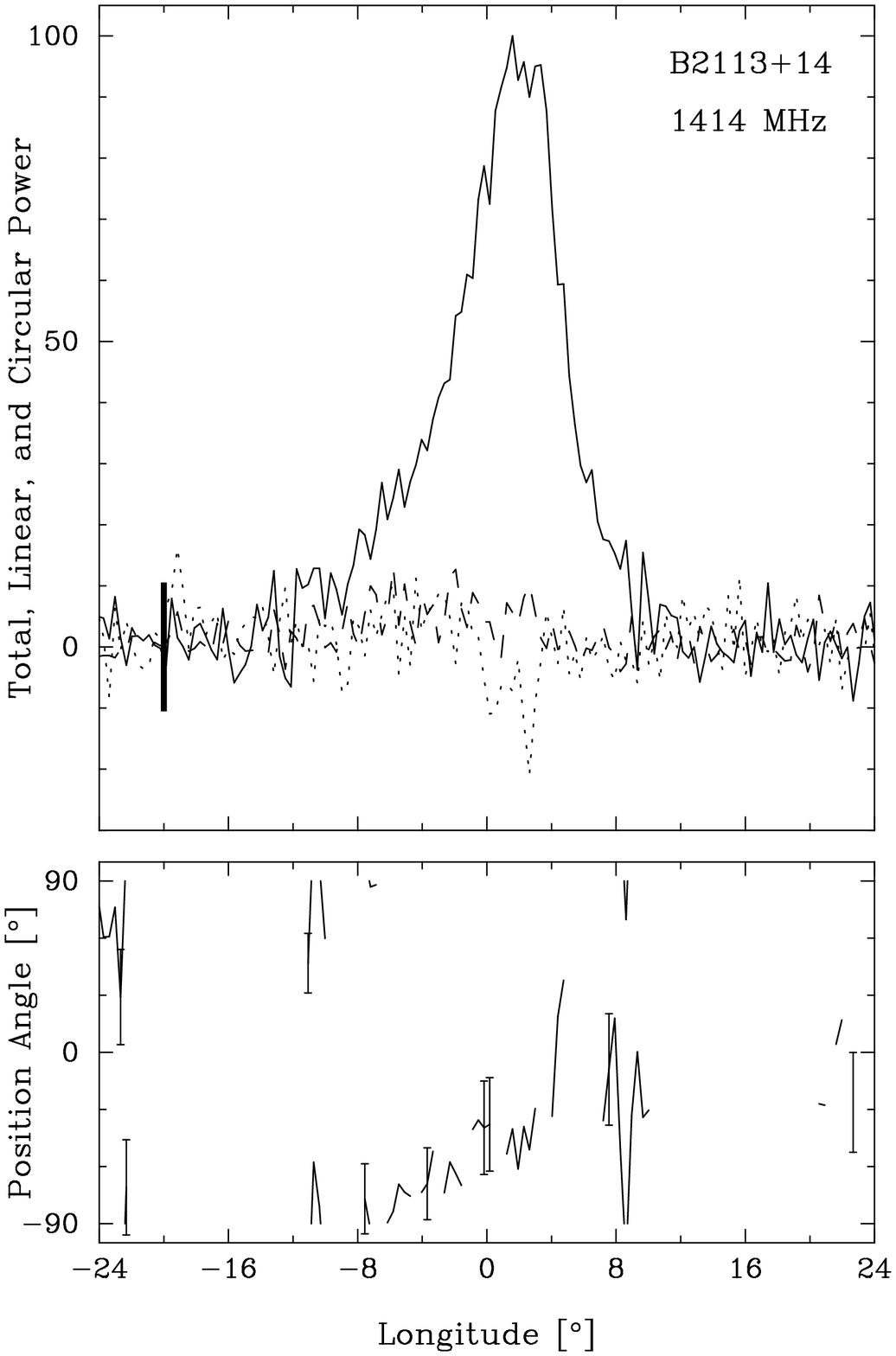} 
}
\caption{Multi-frequency and polarization profiles of B2053+21, B2110+27 and B2113+14.}
\label{b15}
\end{figure}

\begin{figure}[htb]
\centerline{
\pf{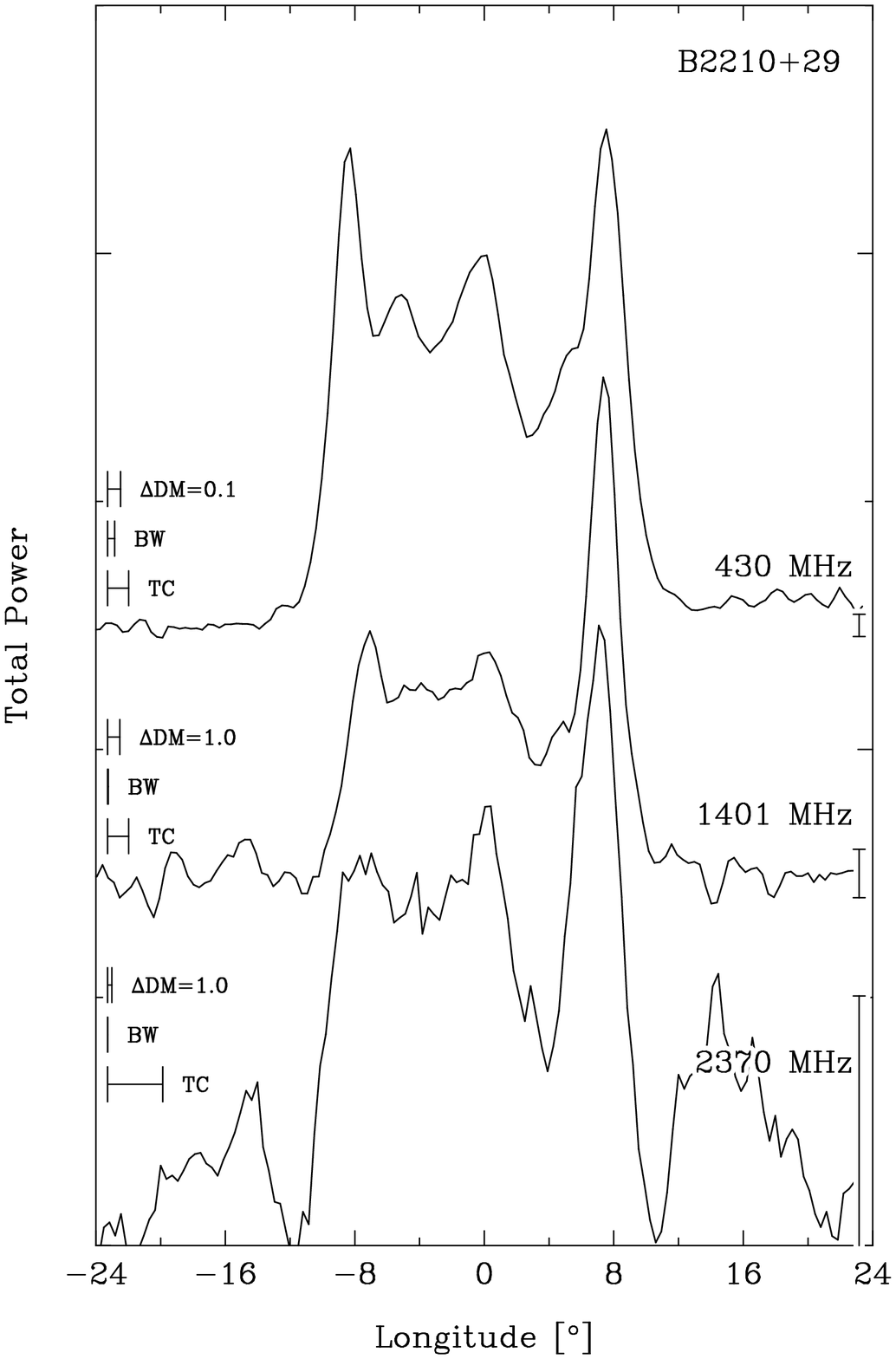}     
\pf{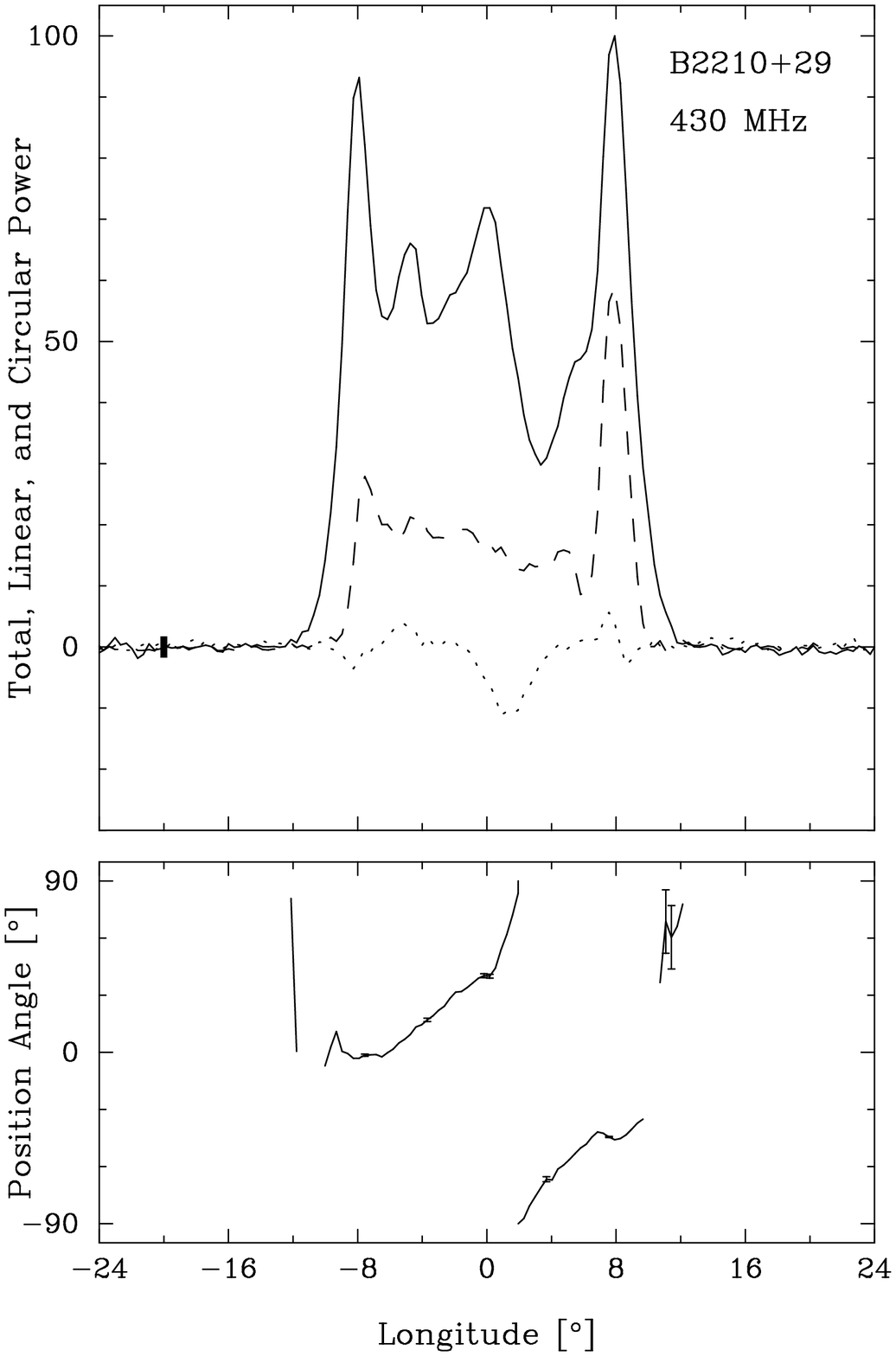} 
\pf{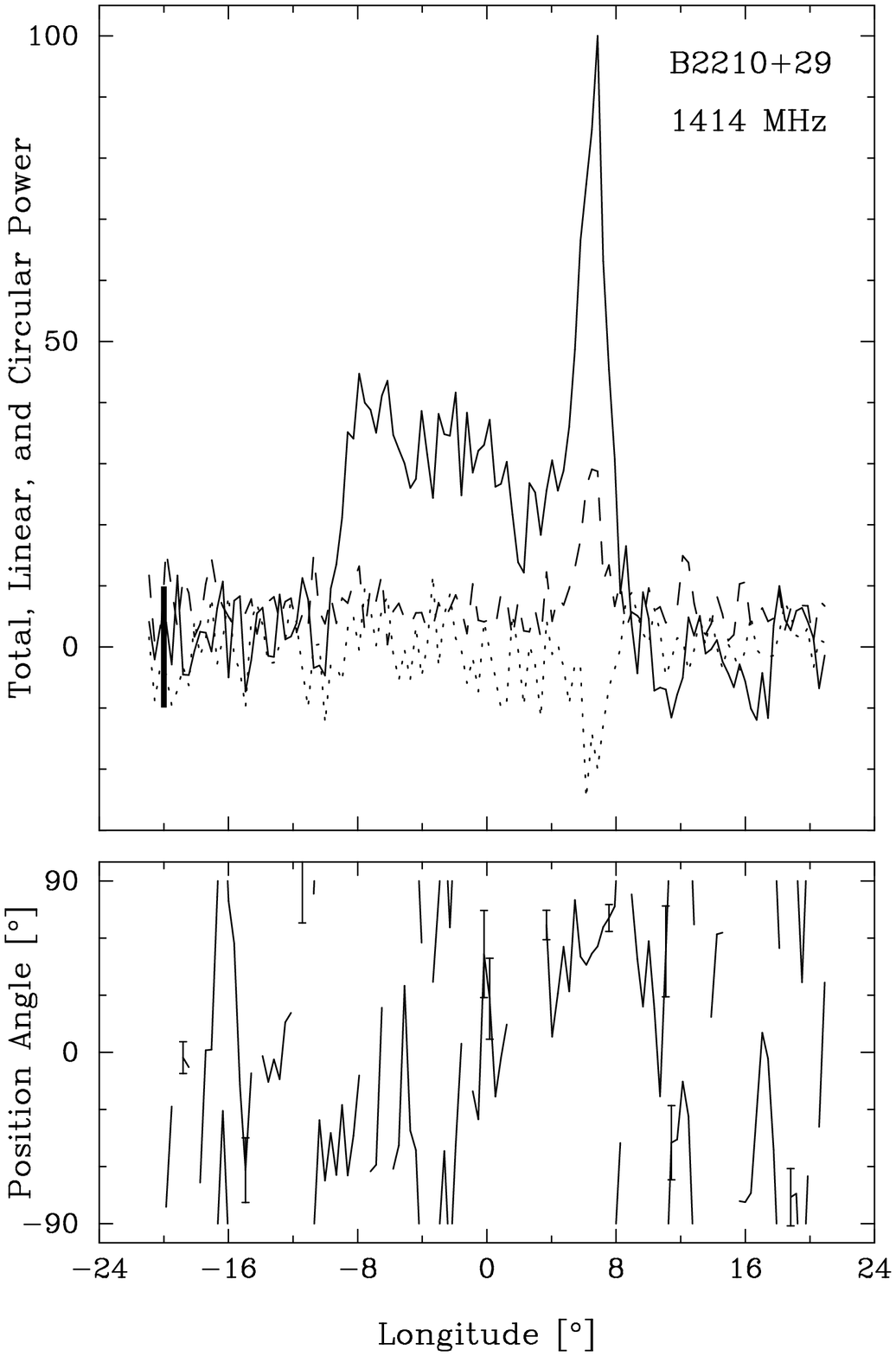} 
}
\quad \\
\centerline{
\pf{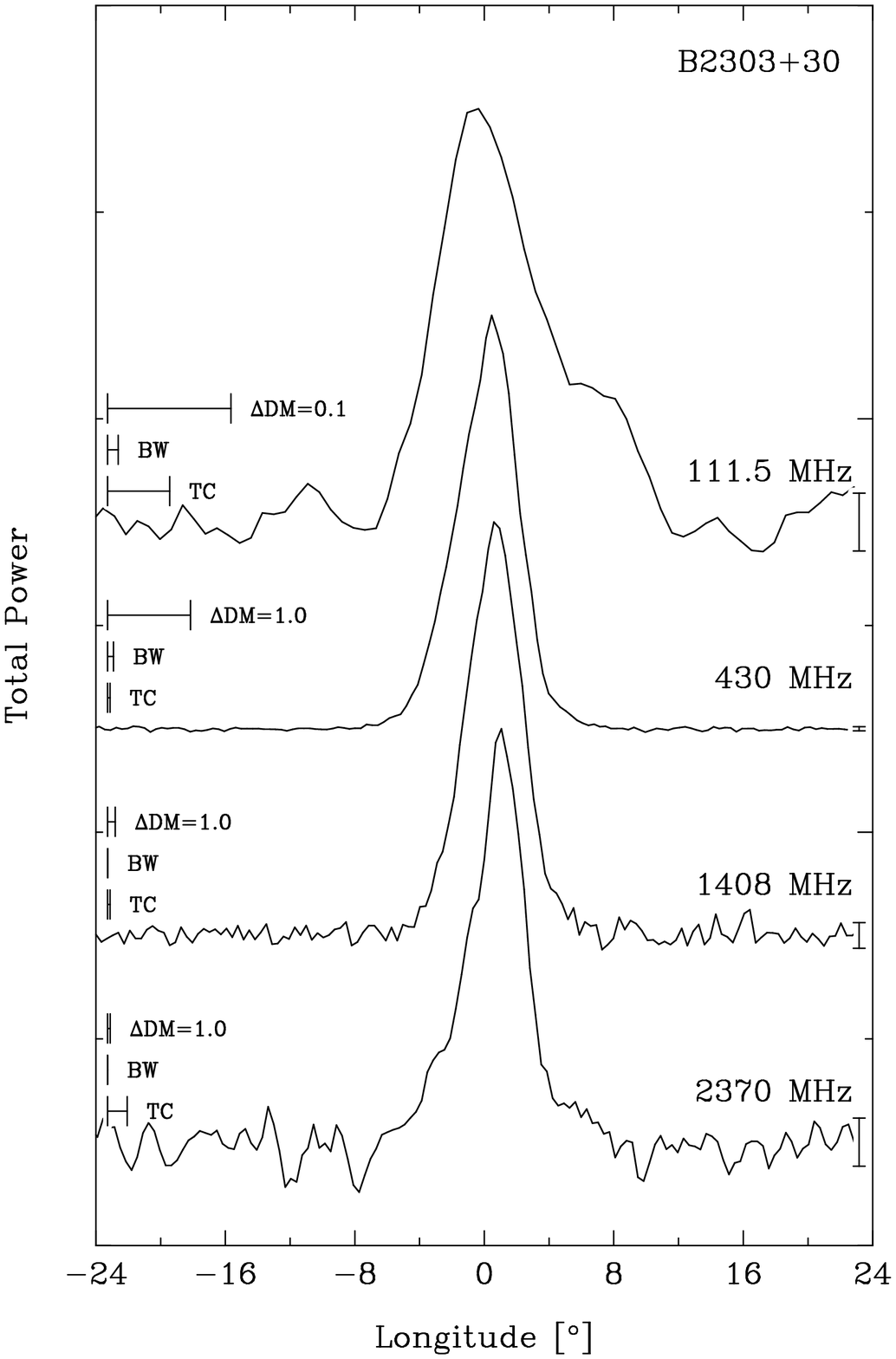}     
\pf{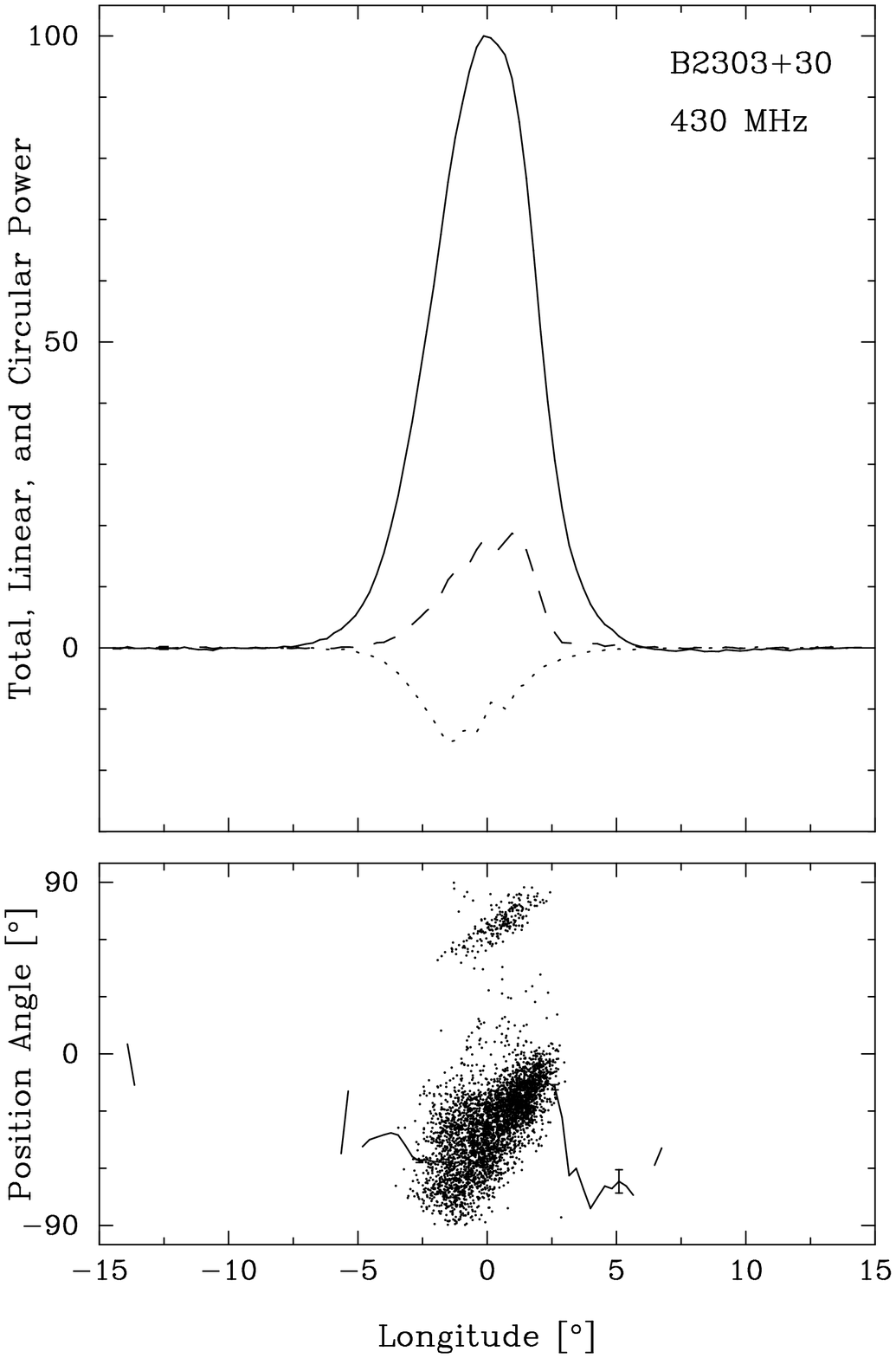} 
\pf{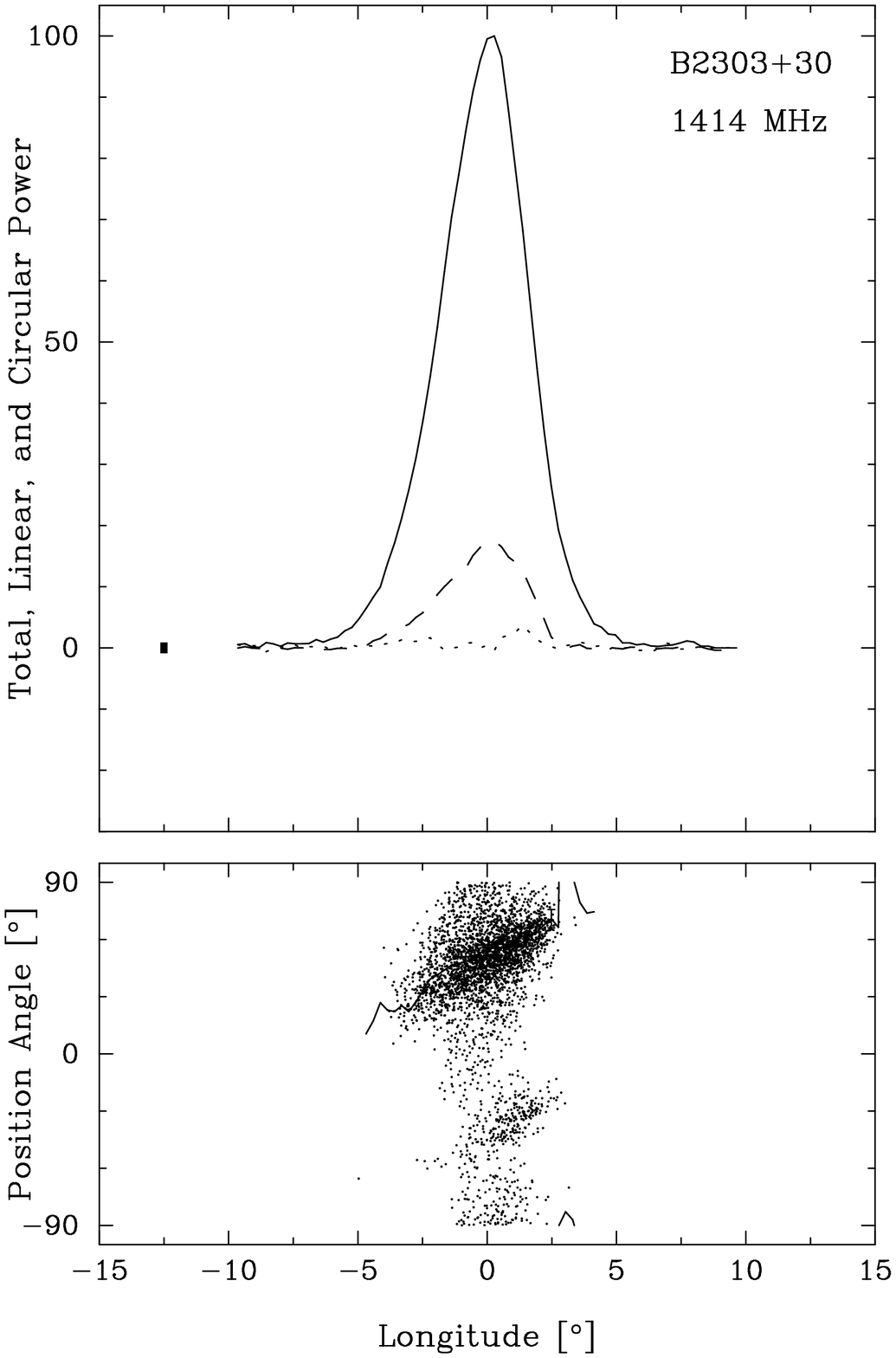} 
}
\quad \\
\centerline{
\pf{figs/dummy_fig.ps}     
\pf{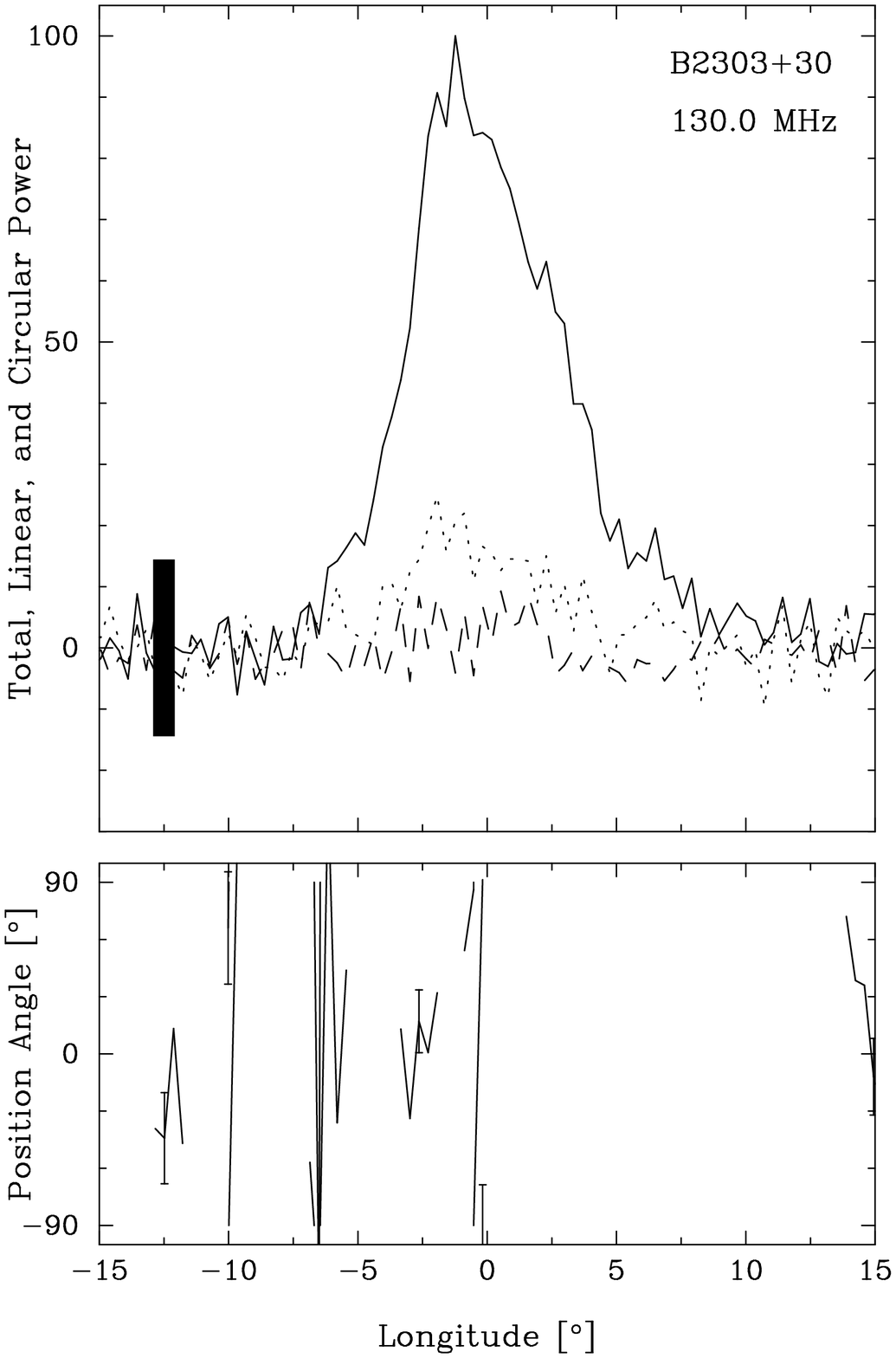} 
\pf{figs/dummy_fig.ps}     
}
\caption{Multi-frequency and polarization  profiles of B2210+29 and B2303+30.}
\label{b16}
\end{figure}

\begin{figure}[htb]
\centerline{
\pf{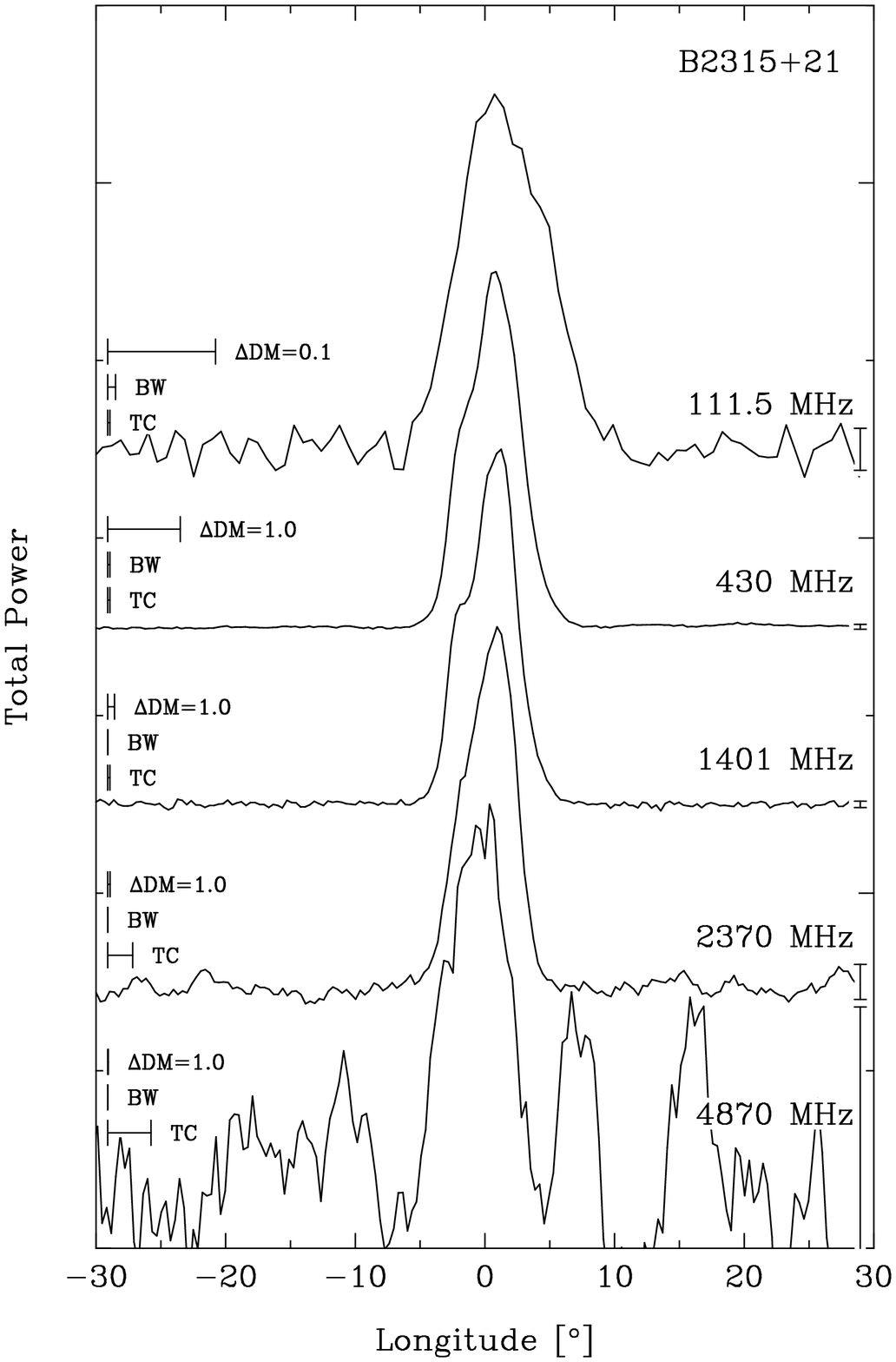}     
\pf{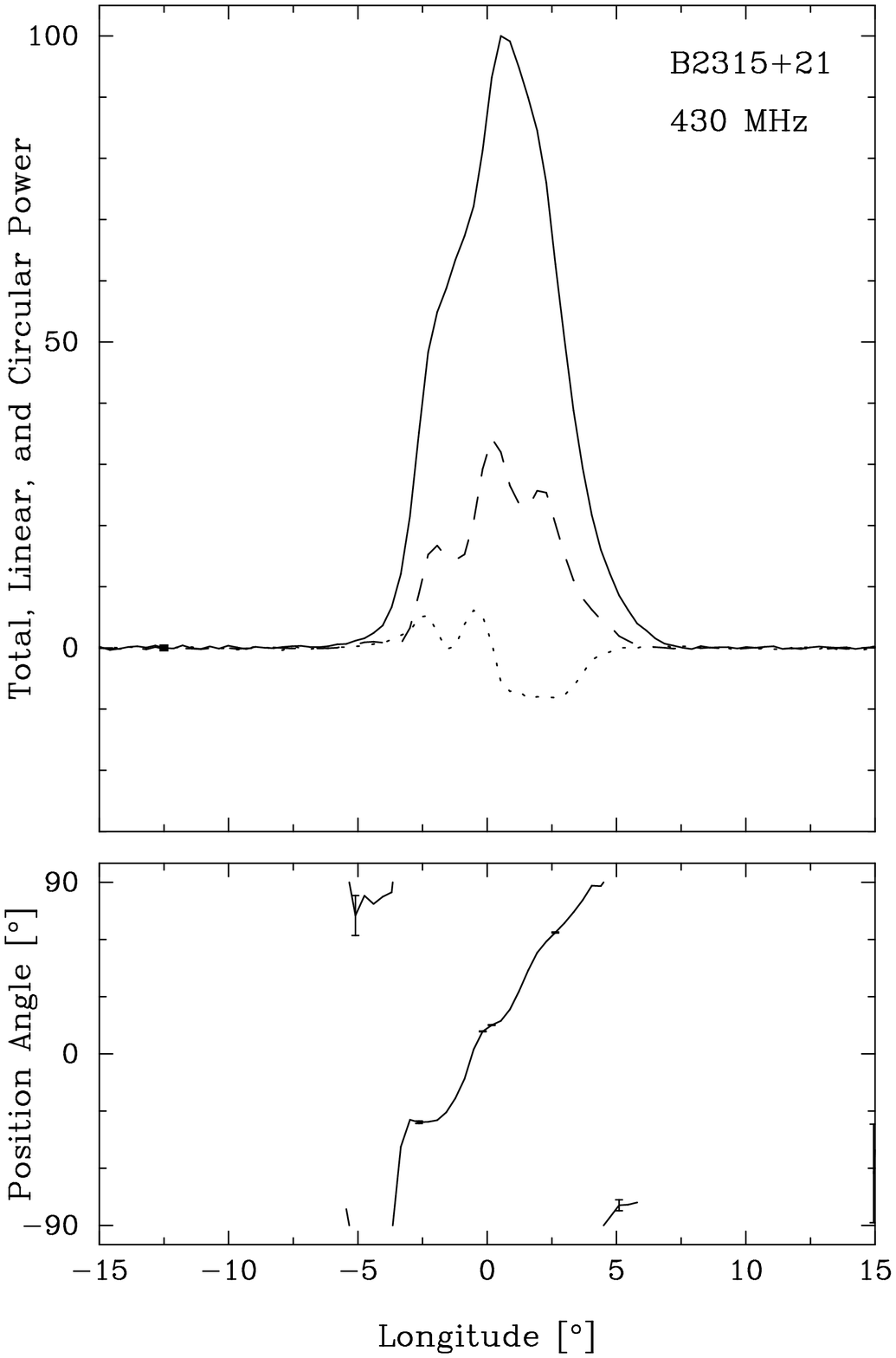} 
\pf{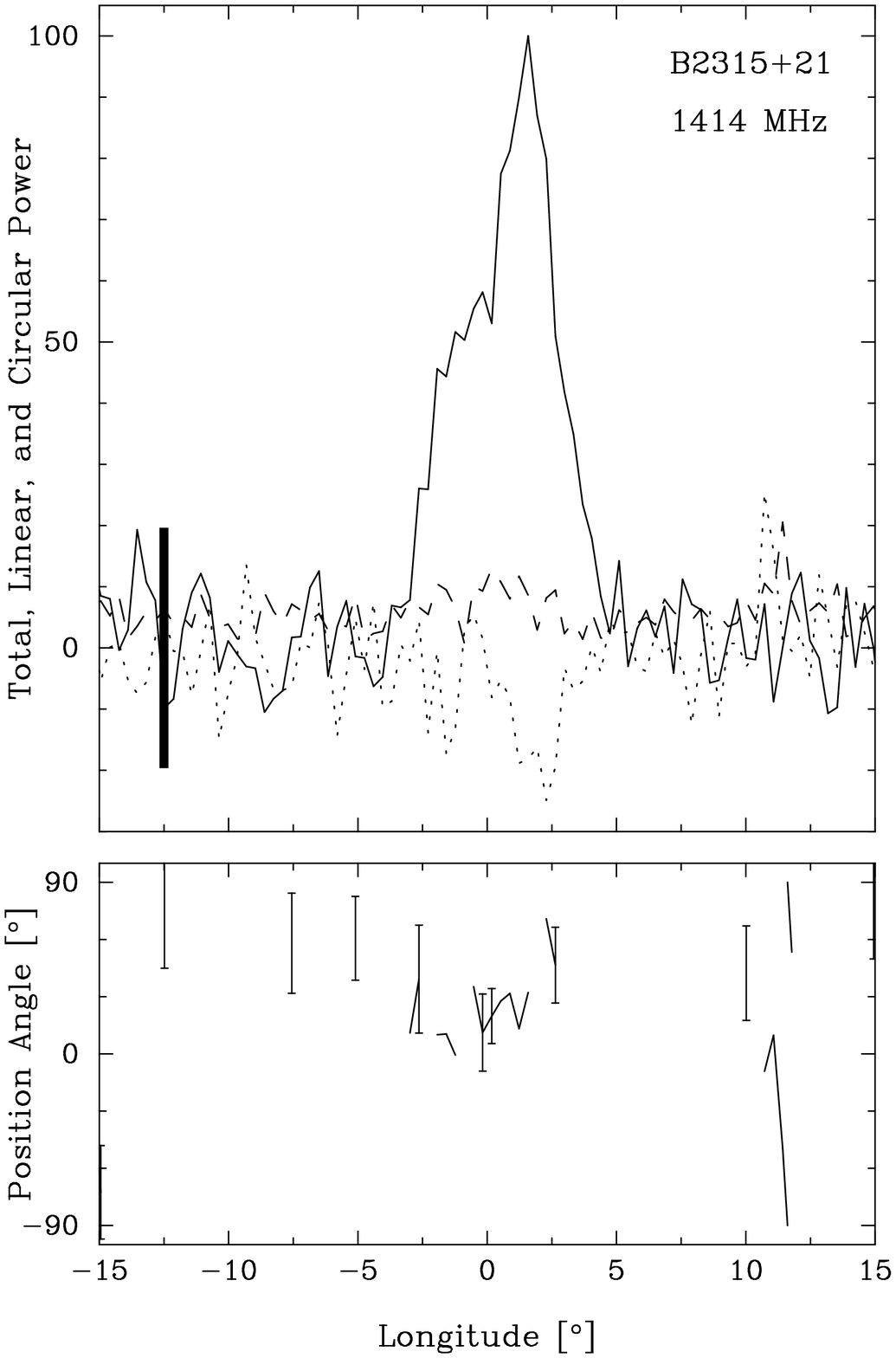} 
}
\caption{Multi-frequency and polarization profiles of B2315+21.}
\label{b17}
\end{figure}

\begin{figure}[htb]
\centerline{
\pf{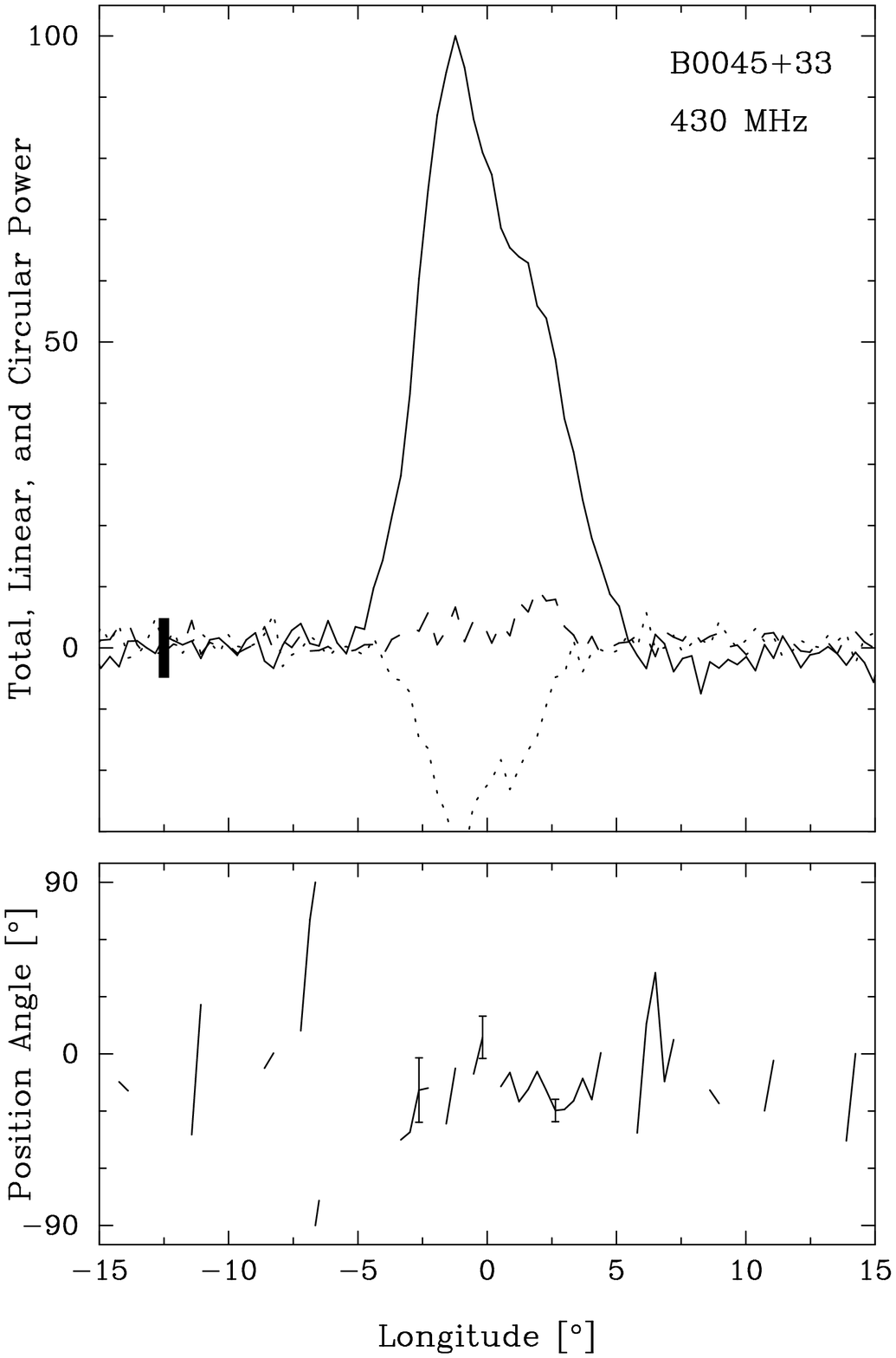} %
\pf{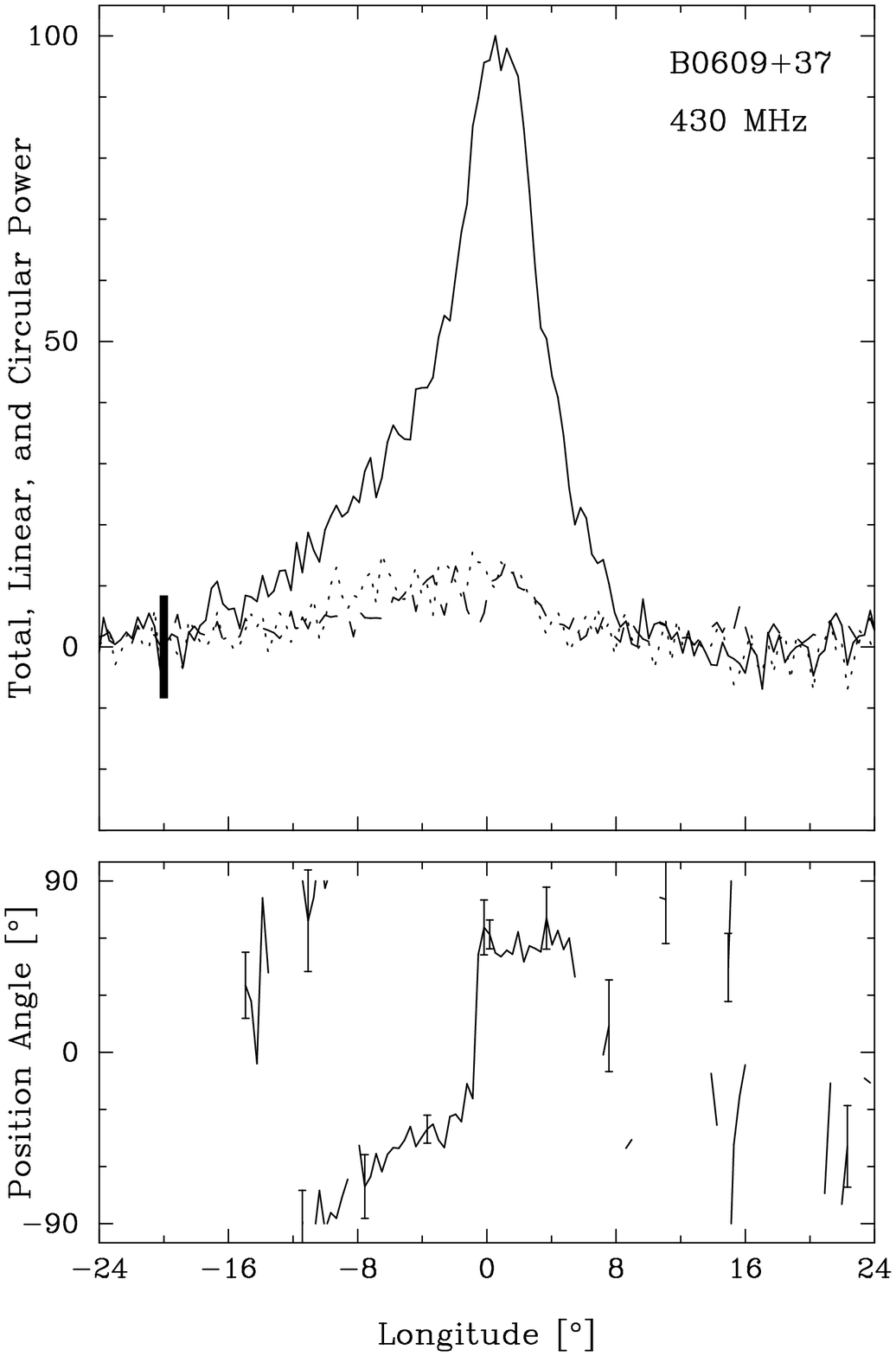} %
\pf{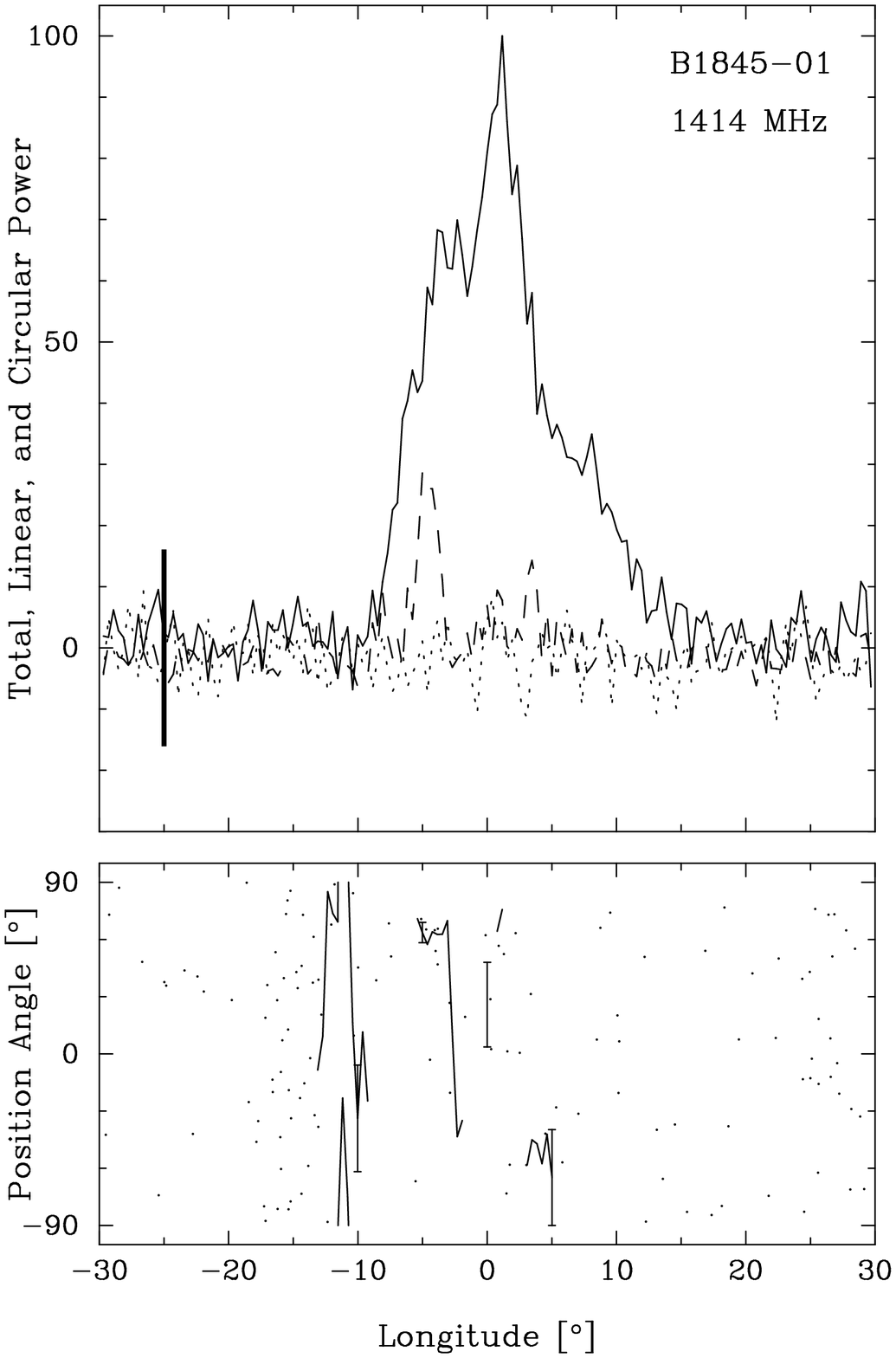} %
}
\quad \\
\centerline{
\pf{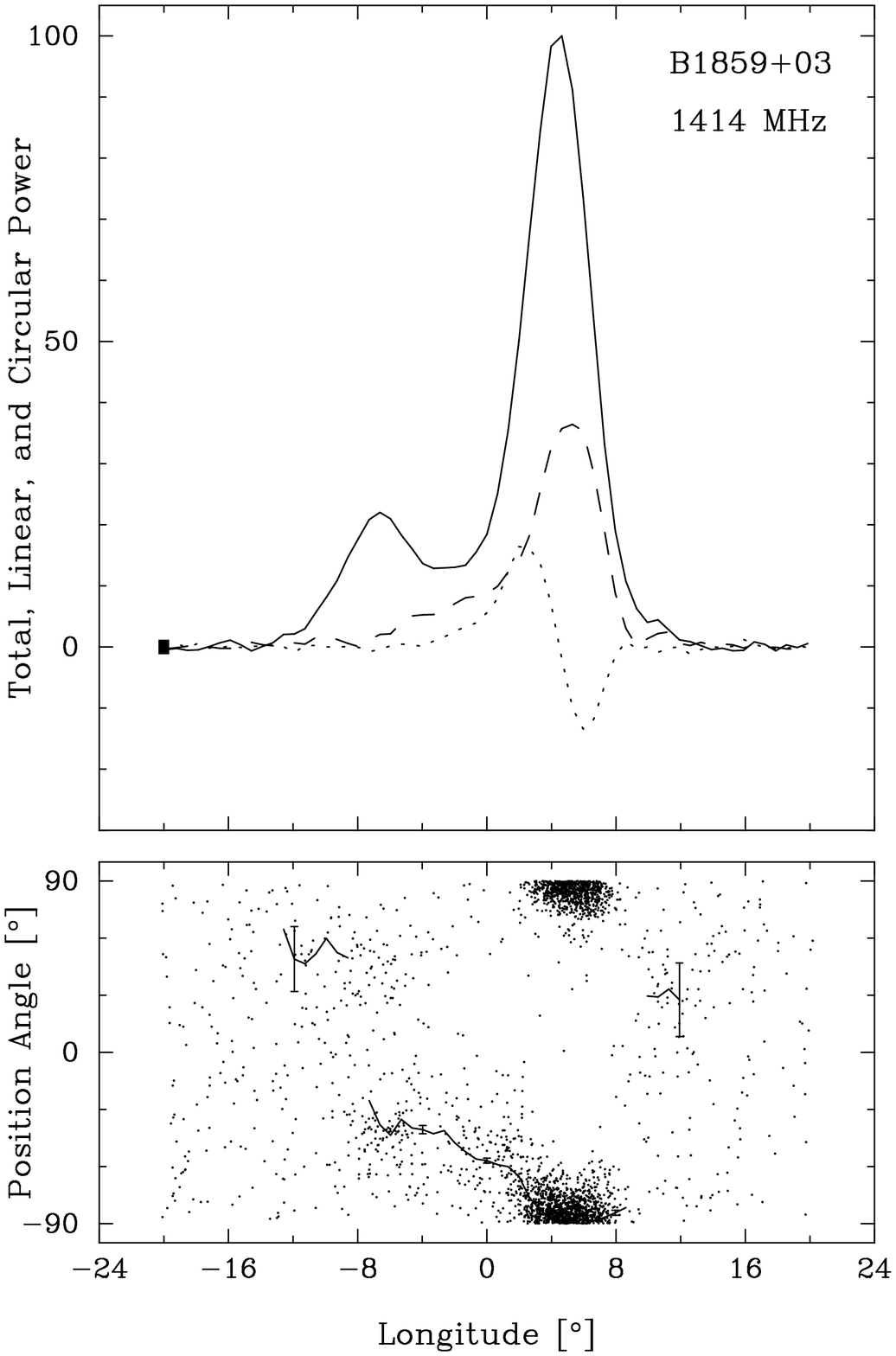} %
\pf{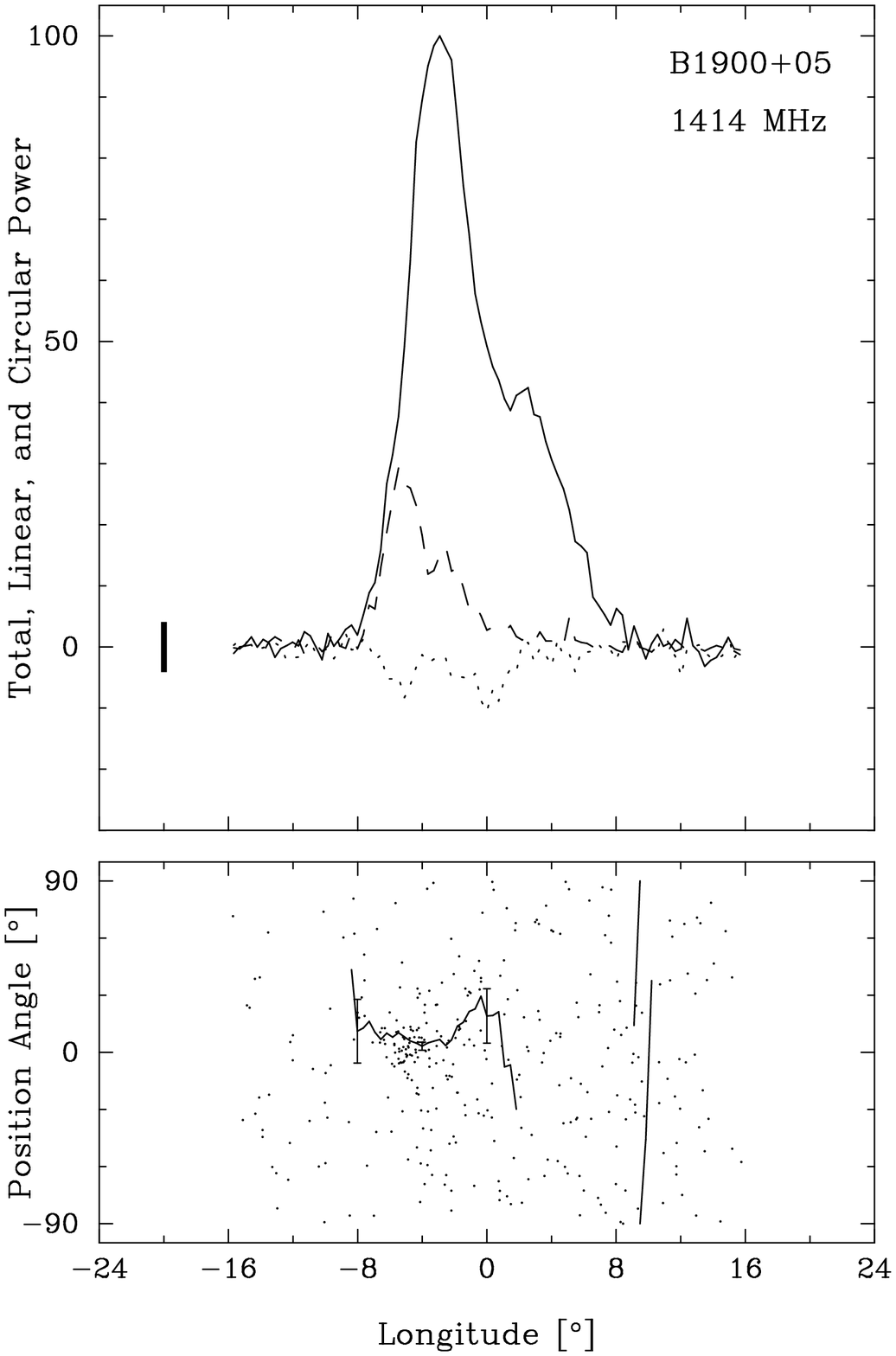} %
\pf{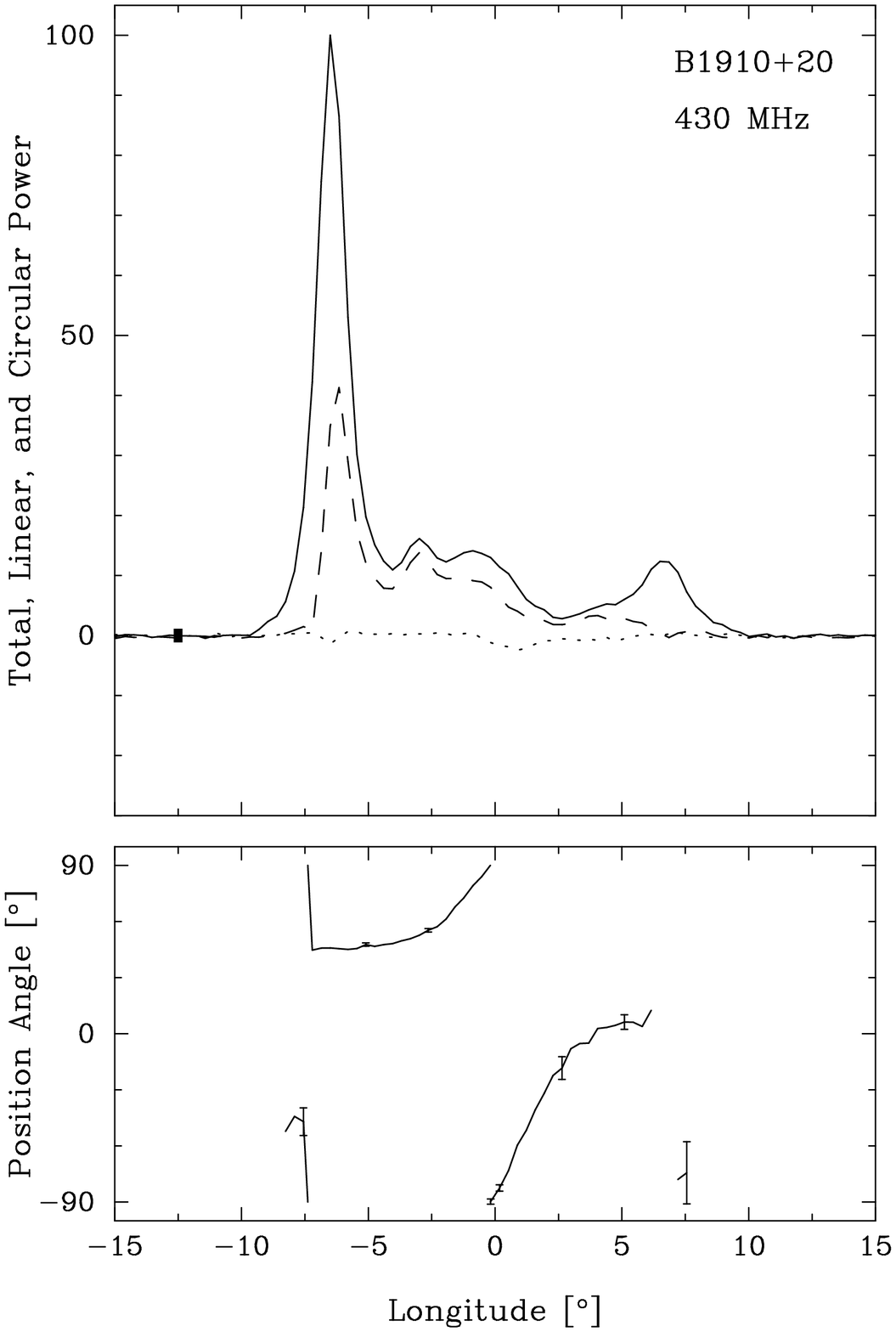} %
}
\quad \\
\centerline{
\pf{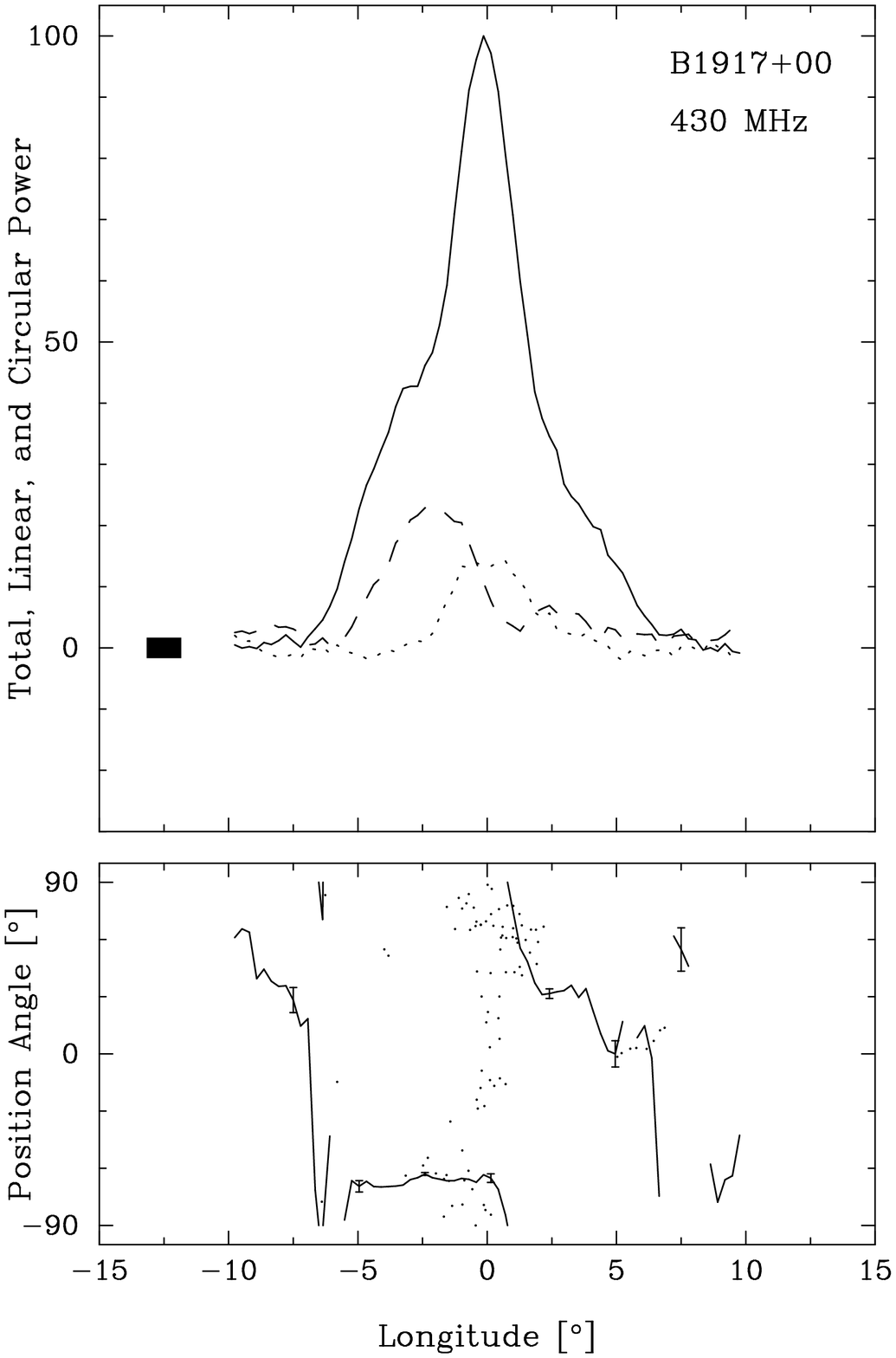} %
\pf{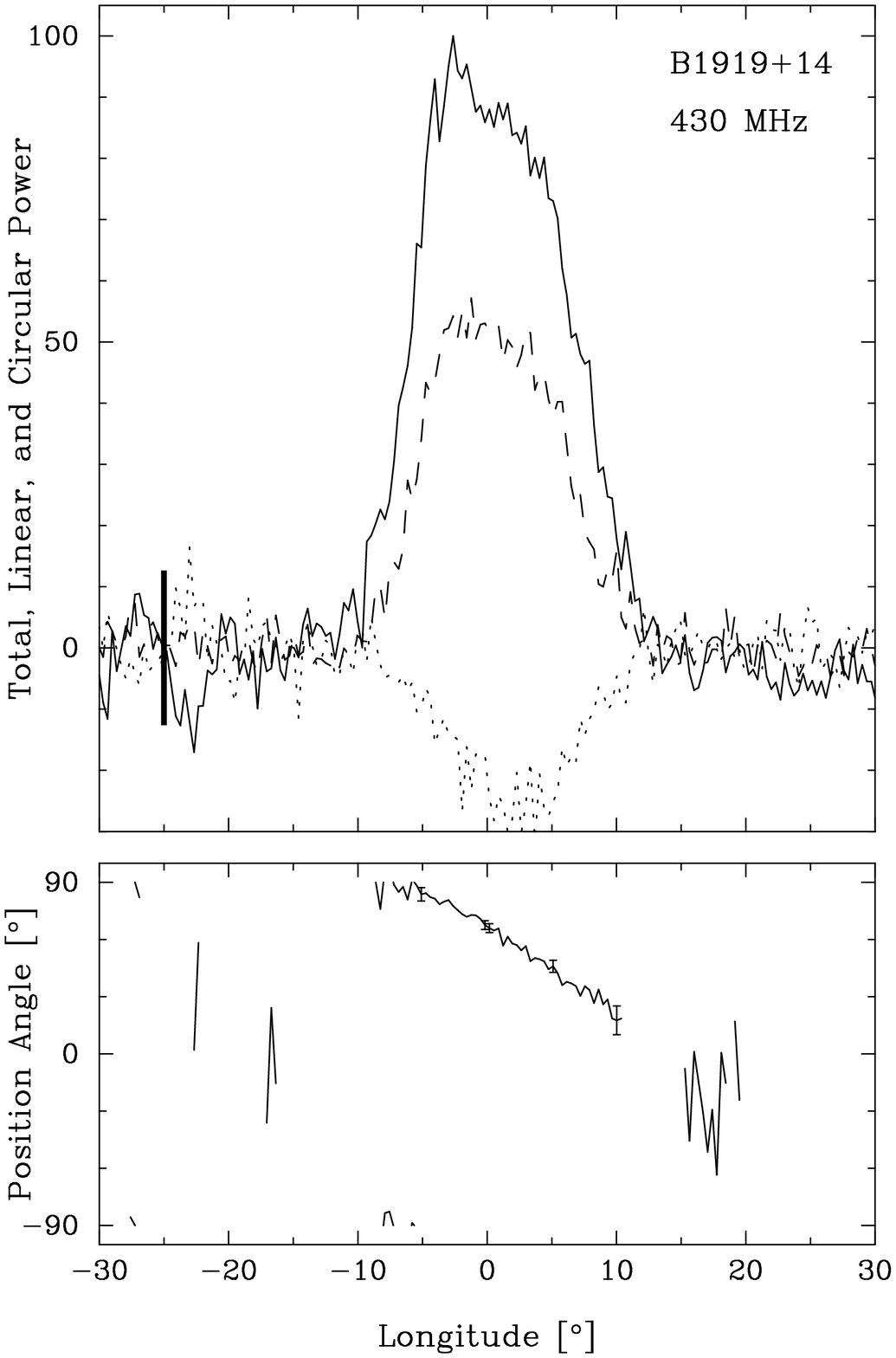} %
\pf{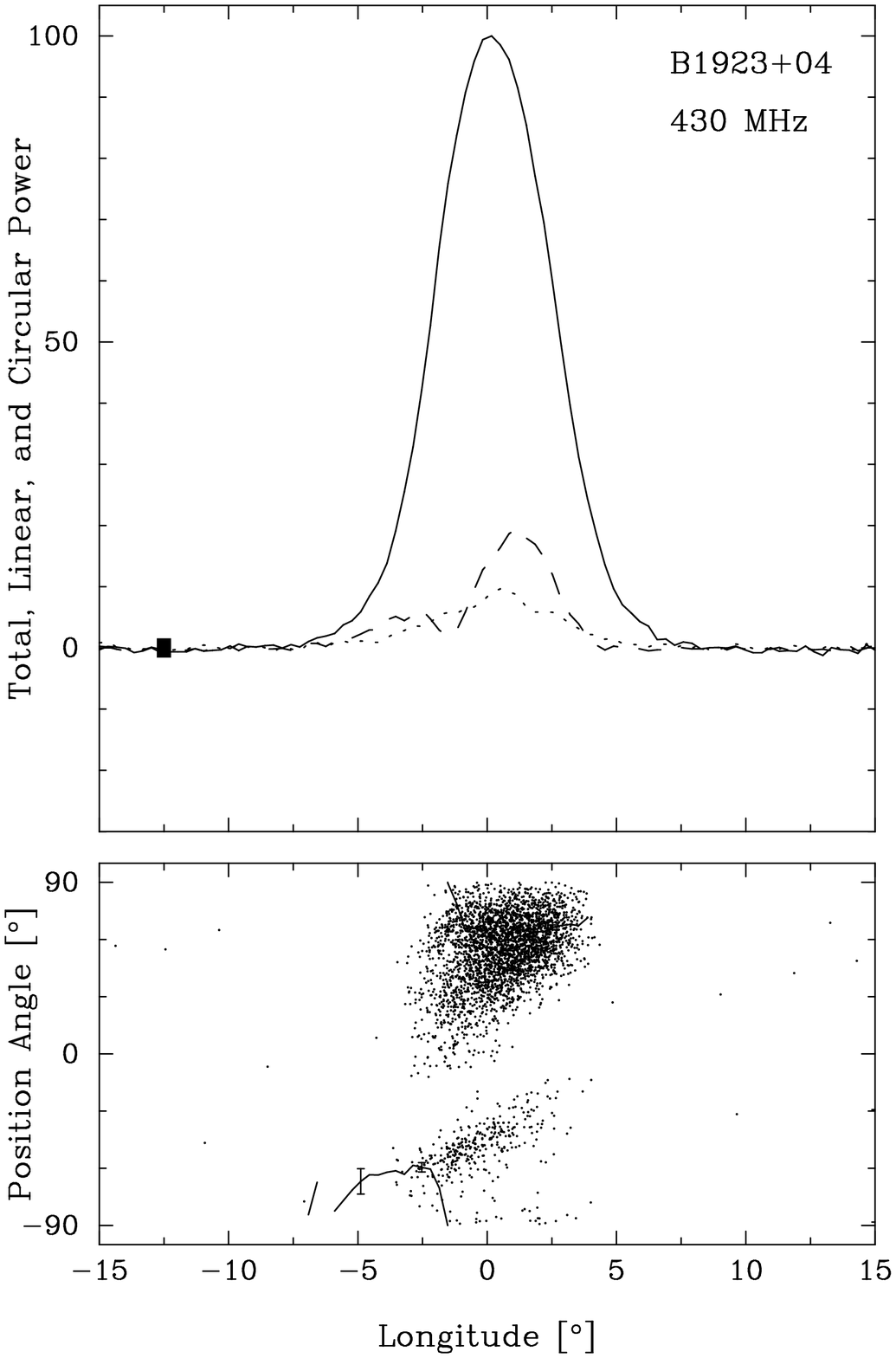} %
}
\caption{Polarization profiles of B0045+33, B0609+37, B1845$-$01, B1859+03, B1900+05, B1910+20, B1917+00, and B1919+14 and B1923+04.}
\label{p1}
\end{figure}

\begin{figure}[htb]
\centerline{
\pf{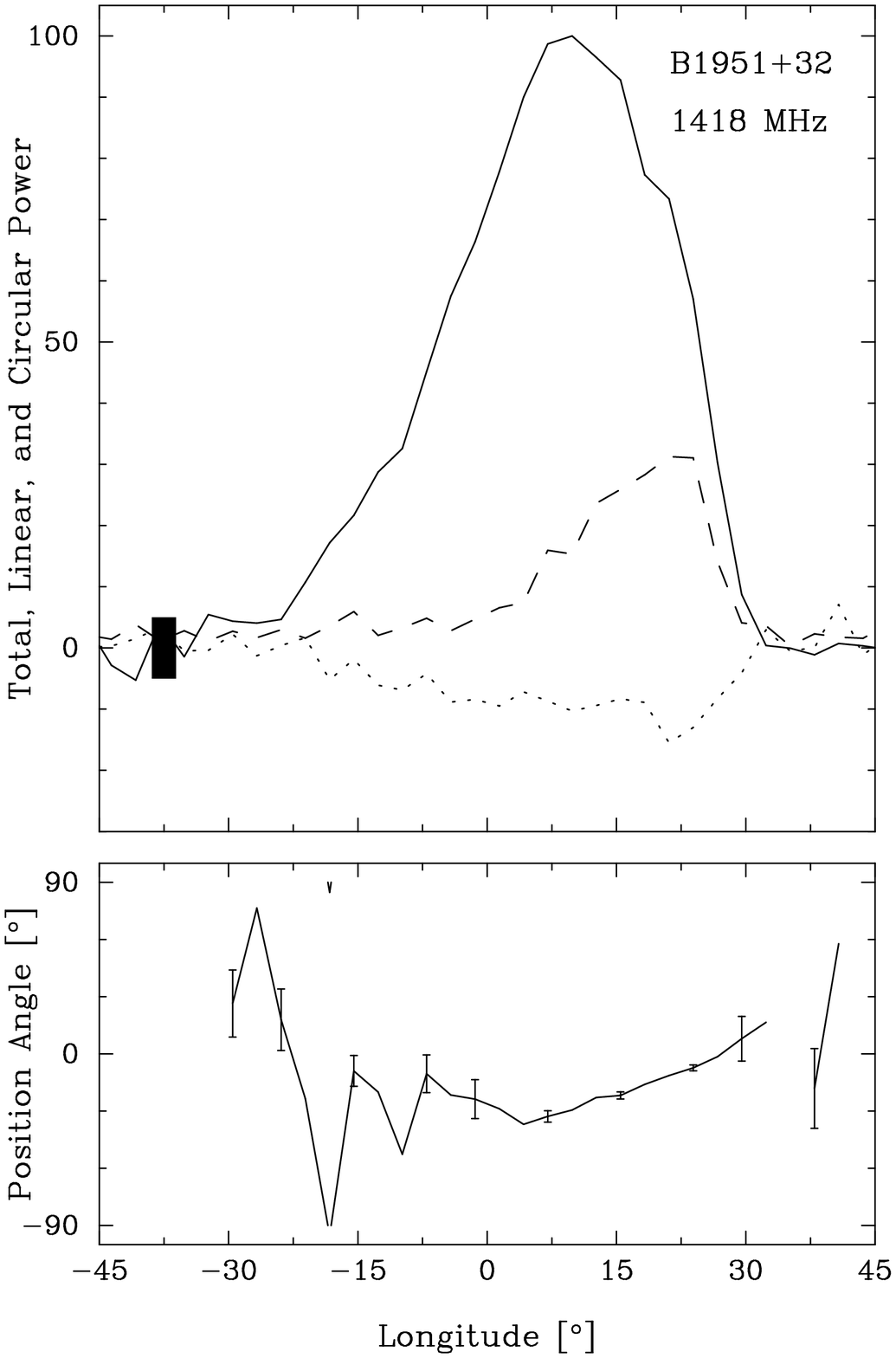} %
\pf{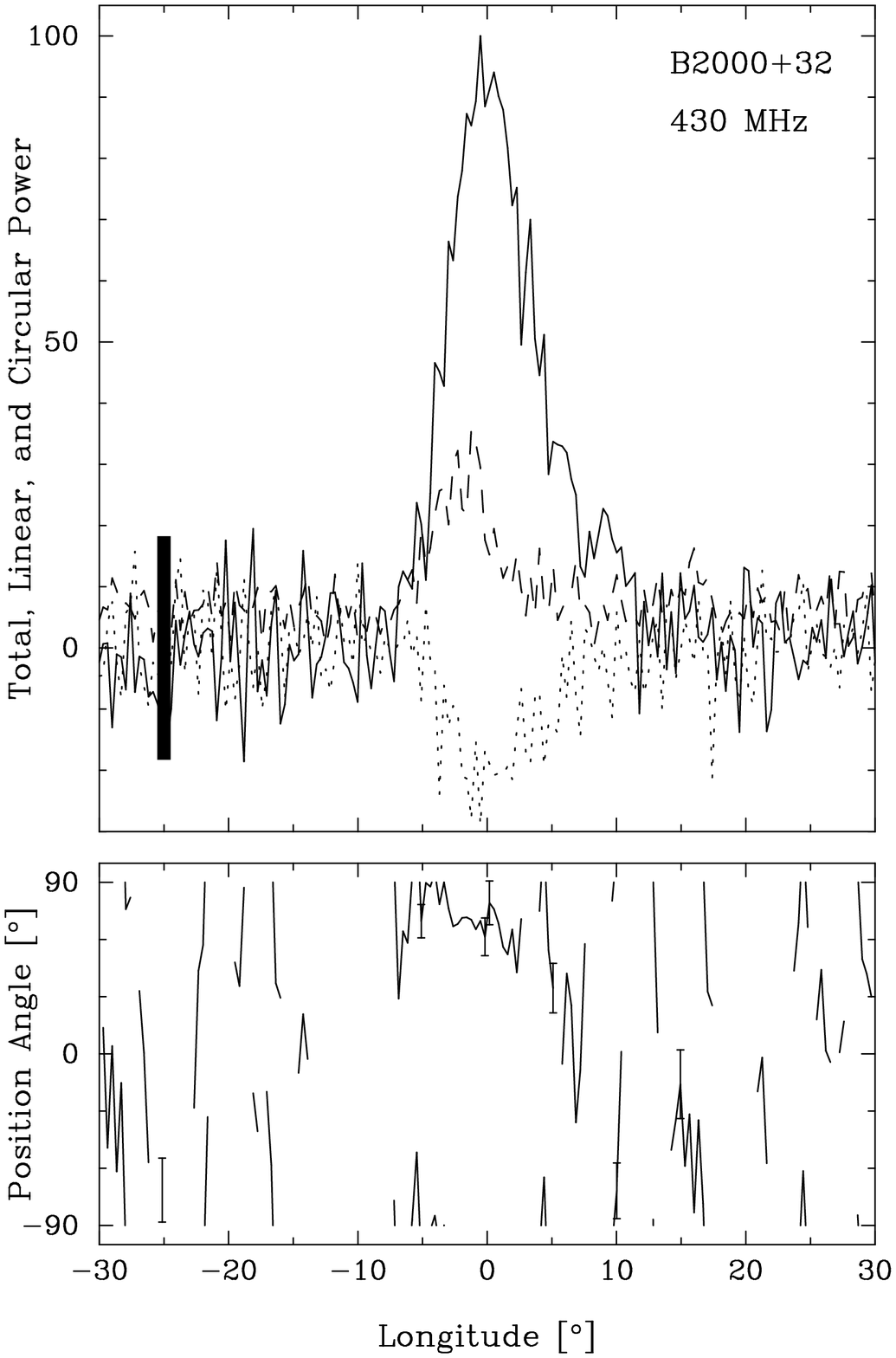} %
\pf{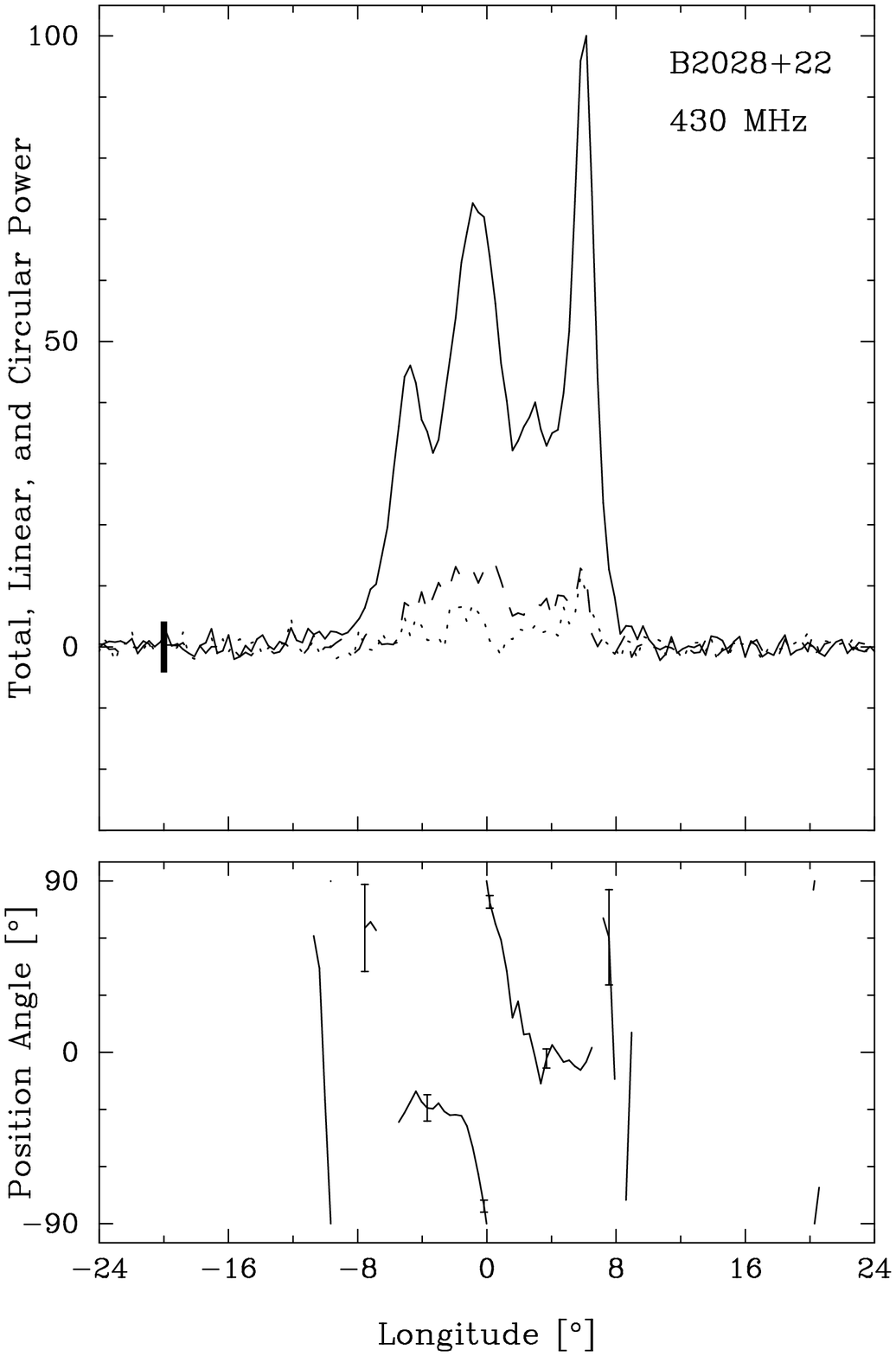} %
}
\quad \\
\centerline{
\pf{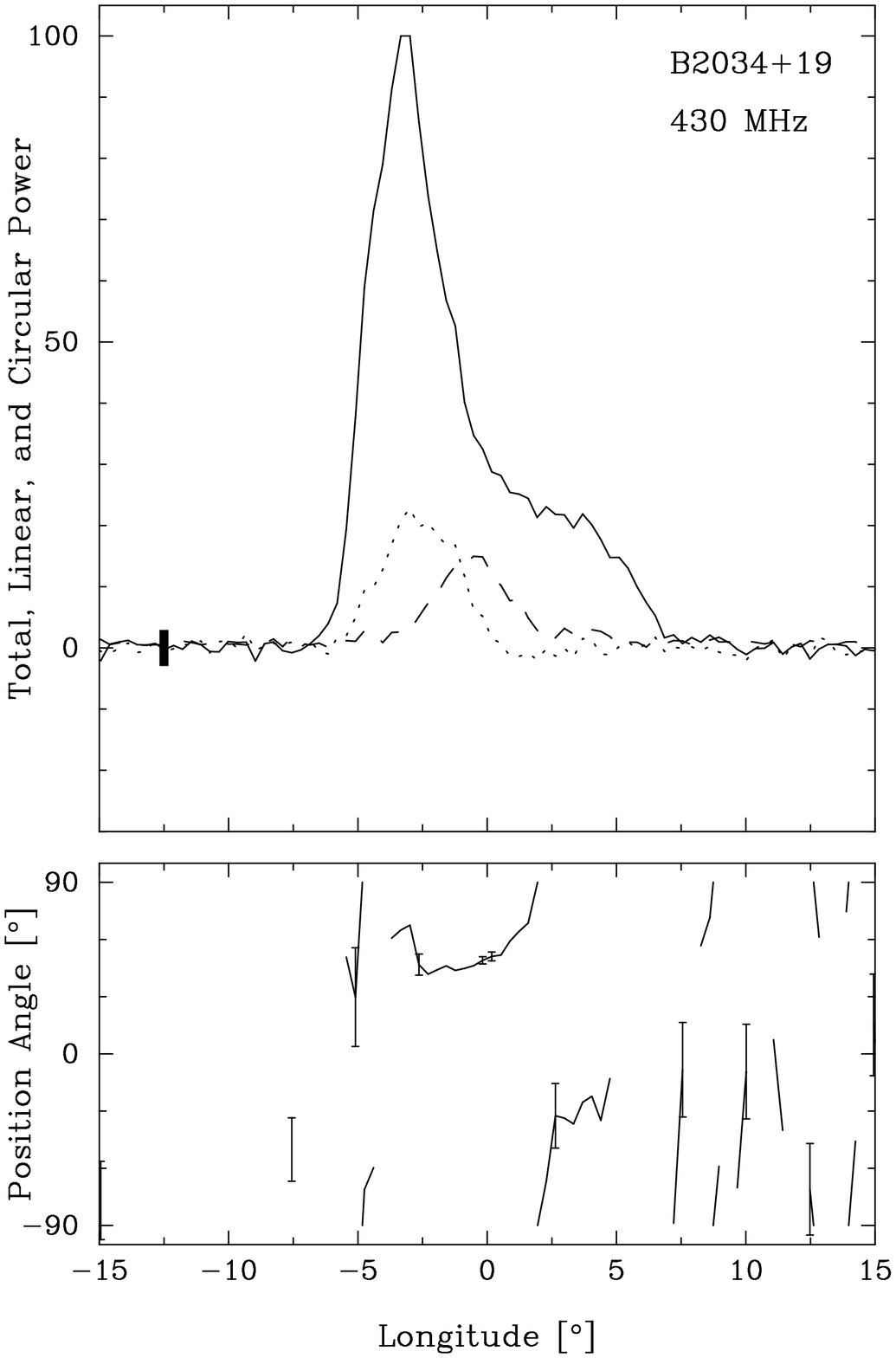} %
\pf{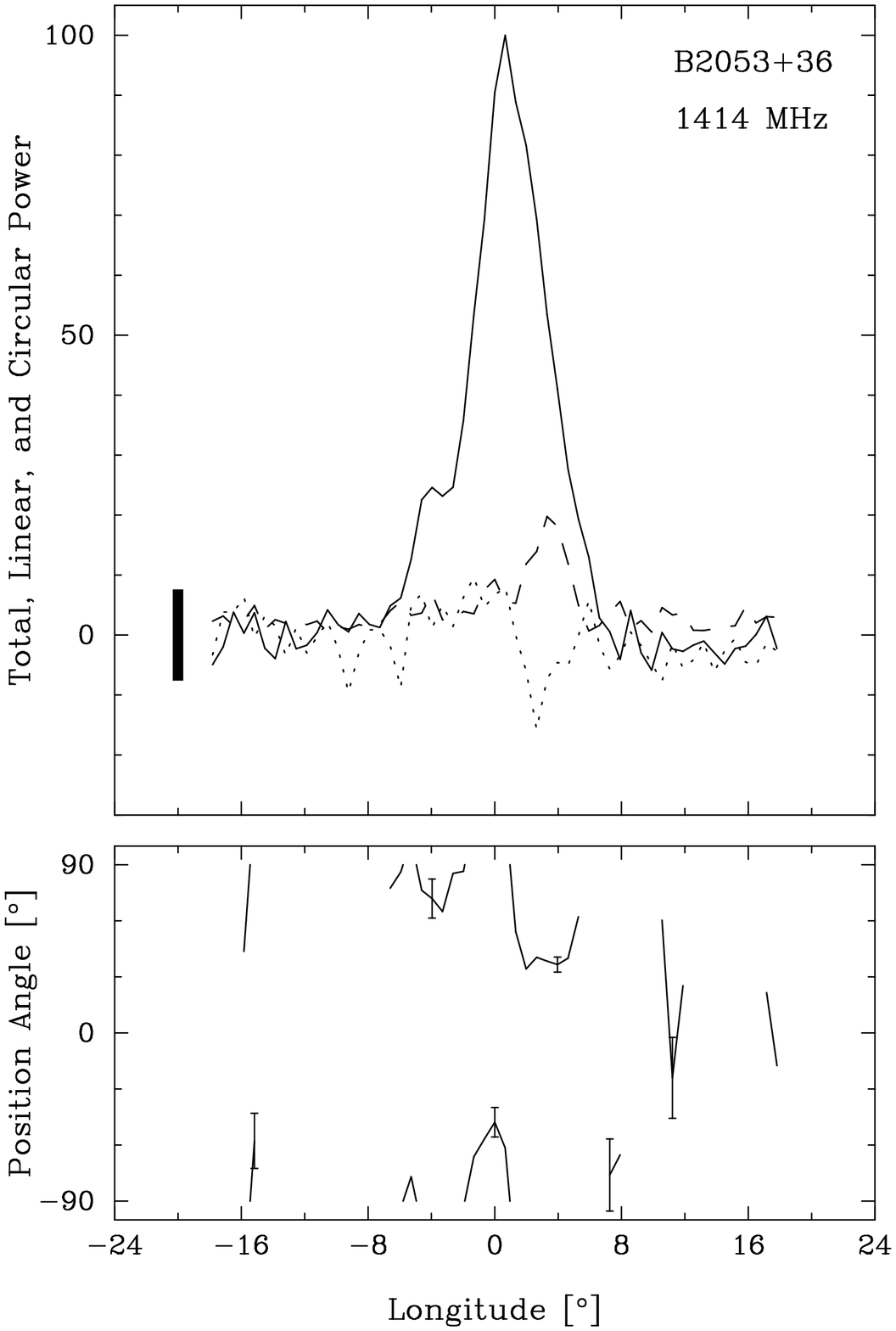} %
\pf{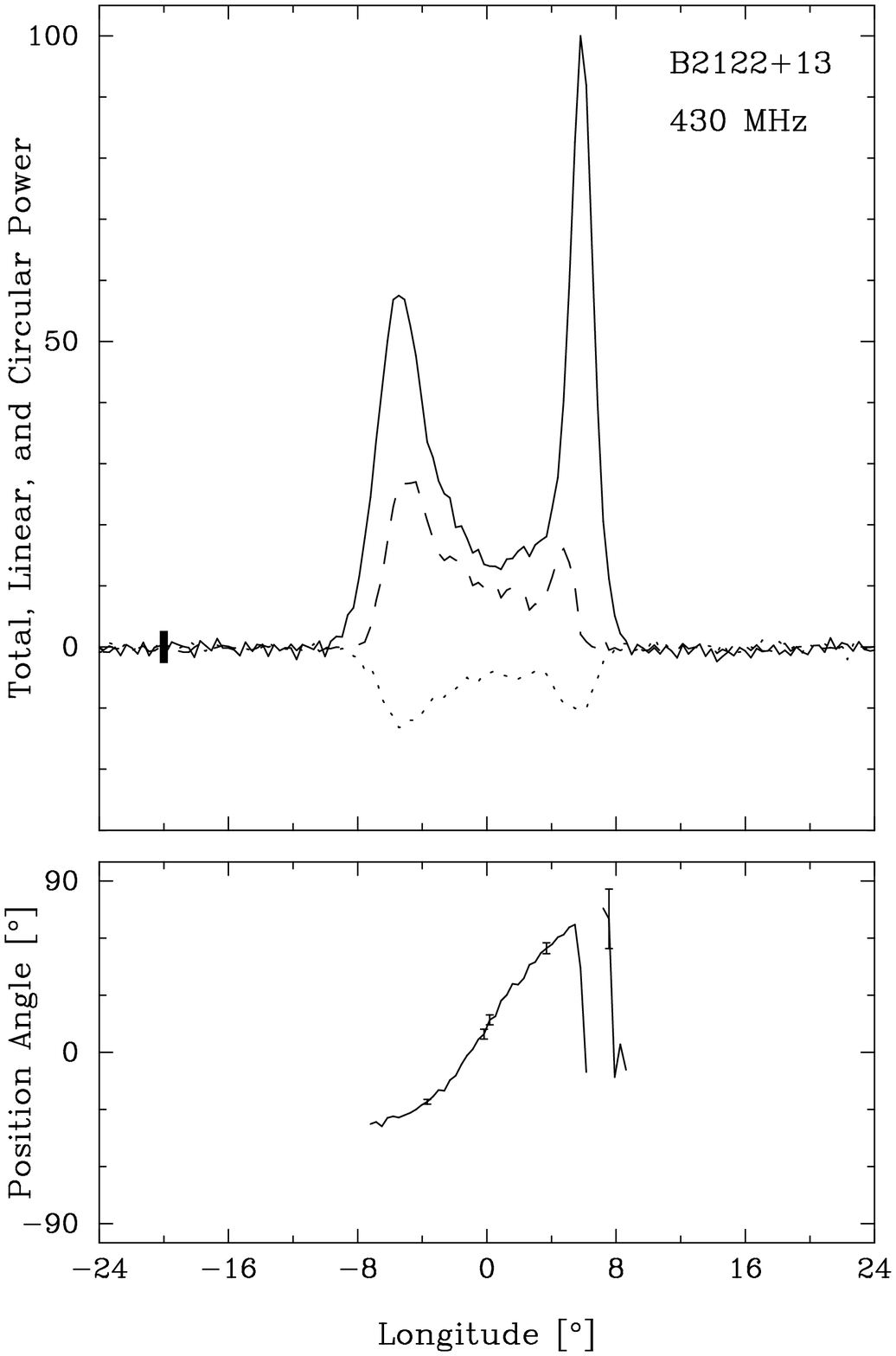} %
}
\caption{Polarization profiles of B1951+32, B2000+32, B2028+22, B2034+19, B2053+36, and B2122+13.}
\label{p2}
\end{figure}

The combined multi-frequency and polarimetry results are presented in 
approximate Right Ascension order in Figures \ref{b1} through \ref{b17} 
in order to facilitate intercomparison.  The multi-frequency profiles are 
normally positioned in the lefthhand columns and the 430- and 1400-MHz 
polarization displays in the central (C) and righthand (R) columns, where 
they are available---and in the half-dozen cases where both 111/130- and  
50-MHz polarimetry is also available, these are given in the center and 
righthand columns of the following row. Details of the multi-frequency 
alignments and polarimetry are given in Tables~\ref{table1} \& 
\ref{table_polarization}.  In several cases where no polarimetry was 
available, the multi-frequency results are interpolated into the sequence 
(Fig.~\ref{b8} (top row), Fig.~\ref{b10} center row, and Fig.~\ref{b11} bottom left).  
Finally, Figures \ref{p1}-\ref{p2} give polarimetry results for those stars 
where only single-frequency observations were carried out, and these 
in turn are described in Table~\ref{table_polarization_a}.  The time 
resolution of the multifrequency profiles is indicated by a set of three 
horizontal bars plotted on the left side of the profiles. The upper bar denotes 
the phase shift which would result from a change of dispersion measure of 
0.01, 0.1, or 1.0 pc cm$^{-3}$.  The middle bar, labeled ``BW'' shows the 
time resolution limited by the dispersion sweep 
time, $\tau_{\it DM}=\mbox{{\it DM}}\,\Delta\nu/(1.205\times 10^{-16}\nu^3)$, 
across the receiver bandwidth, $\Delta\nu$, tuned to frequency $\nu$. 
The bottom bar shows the integration time constant used for the plot. On 
the left side of the multifrequency profile plots an error bar is plotted to 
show the range of 2 standard deviations of the off-pulse noise level.

For the polarization plots the total intensity, linear and circular
polarizations are depicted by solid, dashed and dotted lines,
respectively. The solid vertical bars indicate the range of 3 standard
deviations of the off-pulse noise. Where present, the time resolution,
including the effects of dispersion across the receiver band and the
post integration time constant are shown by a horizontal bar. In the
lower part of each polarization plot the average linear polarization
position angle is shown wherever the linear polarization exceeds two
times the off-pulse noise level. 
Error bars for the position angle are shown for $\pm2\sigma$
uncertainties due to the estimation error of the linear polarization.
For many of the pulsars individual pulse records were available. For
these pulsars the position angle of each sample which exceeded 2 times
the off-pulse noise is plotted as a dot. This type of plot allows study
of polarization moding behavior (Manchester, Taylor \& Huguenin 1975, 
Backer \& Rankin 1980\nocite{mth1975,br1980}). 


\acknowledgements
We wish to acknowledge Dan Stinebring's assistance in making many of
the profile observations as well as making available some of our 1981
polarimetry and that of Amy Carlow, Vera Izvekova, N.\ Rathnasree,
Svetlana Suleymanova, and Kyriaki Xilouris in carrying out some of the
1990/1992 observations.  It is a pleasure to thank Phil Perillat without 
whose remarkable software for the Arecibo 40-MHz Correlator this work 
would have been impossible. We thank Mark McKinnon of NRAO for 
determining timing models for many of these pulsars, Dipanjan Mitra
for alignment of some of the profiles, the Pushchino pulsar group of
the Lebedev Physical Institute for 61 and 102-MHz profiles, and
J.\ A.\ Phillips for the 25-MHz profiles. The Arecibo Observatory is part
of the National Astronomy and Ionosphere Center, which is operated by
Cornell University for the National Science Foundation. This material
is based upon work supported by the National Science Foundation under
Grant  AST-8917722, AST-9618408, AST-9986754 and AST-0139641.


The following is a list of the pulsars mentioned to comply with AAS\TeX \S2.15.3. It doesn't print in the right place.
\objectname{B0045+33}
\objectname{B0301+19}
\objectname{B0523+11}
\objectname{B0525+21}
\objectname{B0540+23}
\objectname{B0609+37}
\objectname{B0611+22}
\objectname{B0626+24}
\objectname{B0656+14}
\objectname{B0751+32}
\objectname{B0820+02}
\objectname{B0823+26}
\objectname{B0834+06}
\objectname{B0919+06}
\objectname{B0940+16}
\objectname{B0943+10}
\objectname{B0950+08}
\objectname{B1133+16}
\objectname{B1237+25}
\objectname{B1530+27}
\objectname{B1541+09}
\objectname{B1604-00}
\objectname{B1612+07}
\objectname{B1633+24}
\objectname{B1737+13}
\objectname{B1821+05}
\objectname{B1822-09} !!!!!
\objectname{B1839+09}
\objectname{B1842+14}
\objectname{B1859+03}
\objectname{B1900+01}
\objectname{B1900+05}
\objectname{B1907+00}
\objectname{B1907+03}
\objectname{B1907+10}
\objectname{B1910+20}
\objectname{B1915+13}
\objectname{B1918+19}
\objectname{B1917+00}
\objectname{B1919+14}
\objectname{B1919+21}
\objectname{B1920+21}
\objectname{B1923+04}
\objectname{B1929+10}
\objectname{B1933+16}
\objectname{B1944+17}
\objectname{B1946+35}
\objectname{B1951+32}
\objectname{B1952+29}
\objectname{B2000+32}
\objectname{B2002+31}
\objectname{B2016+28}
\objectname{B2020+28}
\objectname{B2028+22}
\objectname{B2034+19}
\objectname{B2044+15}
\objectname{B2053+21}
\objectname{B2053+36}
\objectname{B2110+27}
\objectname{B2113+14}
\objectname{B2122+13}
\objectname{B2210+29}
\objectname{B2303+30}
\objectname{B2315+21}

\clearpage  

\clearpage  

\clearpage  

\clearpage  

\clearpage  

\clearpage  

\clearpage  

\clearpage  

\clearpage  

\clearpage  

\clearpage  

\clearpage  

\clearpage  

\clearpage  

\clearpage  

\clearpage  

\clearpage  

\clearpage  

\clearpage  

\clearpage  

\twocolumn
\section*{APPENDIX:  Polarimetric Calibration}
\begin{center}
{\rm Joanna M.\ Rankin, N.\ Rathnasree \& Kyriaki Xilouris}
\end{center}

While the Arecibo Observatory polarimetric feed systems were known to
be imperfect from the 1970s, only slowly were techniques developed to
measure these imperfections and ultimately to correct for them.  The
primary source of this imperfection results from the circumstance that
some power is mutually cross-coupled between the two radiometers.  At
Arecibo in 1992 all of the polarimeter systems involved a pair of
nominally left- and right-hand circularly polarized feeds, which were
connected to a pair of matched receivers with coherent local oscillators 
and in turn connected to the Arecibo 40-MHz Correlator, which served 
as a multiplying polarimeter.  

The basic effects of cross-coupling are well described in the Appendix of
Stinebring \etal\ (1984)\nocite{scrwb1984a}. Indeed, this work reports
not only the first technique for measuring the cross-coupling amplitude
and phase using pulsar observations, but also shows how these
measurements can be used to correct the measured Stokes parameters
for the distorting effects of the cross-coupling.  

Conceptually, we can understand that the cross-coupled power causes
the nominally circularly-polarized feeds to have an elliptically polarized
response.  In other words, while a purely circularly polarized feed
would have the same response to a linearly polarized signal at any
orientation (or polarization angle), the ellipticity results in angles
of maximum and minimum response 90$^\circ$ apart.  In order to
understand, then, the behavior of the polarimeter, we need to
understand the axial ratio and orientation of the elliptical response
of the left- and right-hand feeds.

Our analysis then extends that of the above paper in two significant
ways: First, we have made no assumptions about the behavior of the
feeds in terms of whether their response was ``orthogonal'' or
``coincident'' (see their eqs.\ A10 and A11), and second we have
determined the cross-coupling amplitude and phase as a function of
frequency over the entire usable band of the feed.  

In order to carry out our analyses, we computed raw Stokes parameters
for a pulsar with a region of nearly complete linear polarization;
polarized average profiles comprising a few hundred pulses were
computed over the full several hours that the pulsar could be tracked
in order to obtain the largest possible excursion of parallactic
angle.  Pulsar B1929+10 has an almost fully polarized leading edge,
which is excellent for this purpose as is virtually the entire profile
of B0656+14, though the latter is substantially weaker.  Then, we
recovered the left- and right-hand channel responses $\mathcal{L}$ 
and $\mathcal{R}$ through use of the Stokes parameter definitions
$\mathcal{L}= (I+V)/2$ and $\mathcal{R}=(I-V)/2$.  It was then possible
to fit sinusoidal curves to the observed dependences of $\mathcal{L}$
and $\mathcal{R}$ as functions of raw instrumental linear polarization 
angle (rotated with respect to the feed by the changing parallactic 
angle with hour angle) in order to determine both the axial ratio and 
the orientation of the cross-coupling-distorted linear response.  

This latter fitting procedure is nearly identical to that carried out
by Stinebring \etal\ (1984) in their Fig.\ 39, except that we fitted to
variations in $\mathcal{L}/L$ and $\mathcal{R}/L$ [where the total
linear power $L$ is $\sqrt{Q^2+U^2}$] rather than simply $V/L$.  It 
is also worth pointing out that these fits are to ratios of polarized
intensities, so that any variations in the amplitude of the pulsar
signal, due to scintillations or whatever, have no effect on the
values, only their uncertainties.  

The result then is that the fitted amplitudes determine the axial ratios 
of the elliptical response (complete circularity would produce no 
variation with raw feed position angle and imply an axial ratio of 
unity, whereas a completely linear response would imply an infinite 
axial ratio); whereas the phase of the variation determines the orientation
of the elliptical response relative to the polarization-angle ($PA$)
origin of the polarimeter.  In terms of Stinebring \etal\ (1984), this
simple procedure results in the determination of the crucial four
parameters, the respective cross-coupling amplitudes $\varepsilon_1$ 
and $\varepsilon_2$ and the cross-coupling phases $\psi_1$ and 
$\psi_2$ which first appear in their eq.\ (A2).  In particular, these 
quantities are the only variables upon which the transformation 
between the intrinsic Stokes parameters and those measured by the 
imperfect polarimeter depend.  

The results of our calibration efforts for the Arecibo 430-MHz line
feed and for the 21-cm line feed are given in Figures A1 and A2.  The
left- and right-hand cross-coupling amplitudes are plotted in the top
two panels and the left-hand phase in the bottom panel. In fact, we
determined both cross-coupling phases and found that they were 
always orthogonal within our measurement errors, so that only one 
need be specified.  The respective amplitudes, however, appear to 
exhibit subtle differences, though their overall behavior is very similar 
in the two cases.  

\begin{figure}
\centerline{\epsfig{figure=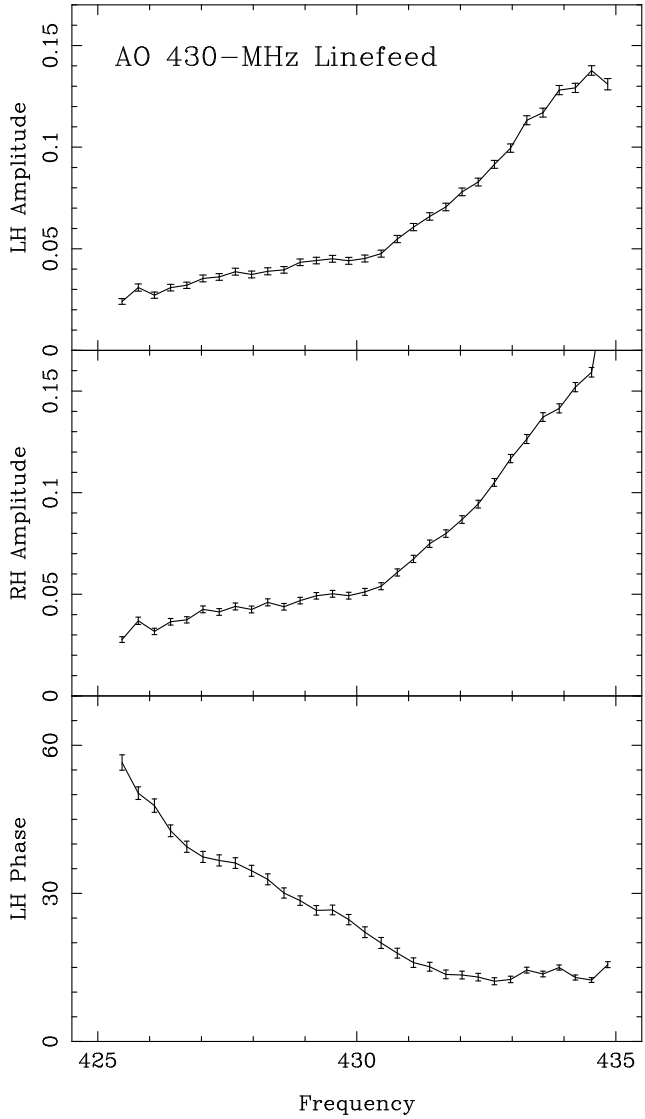,height=15cm}}
\caption{Left- (top panel) and right-hand (middle panel) cross-coupling
amplitudes and the left-hand cross-coupling phase (bottom panel) as a
function of frequency for the Arecibo Observatory 430-MHz line feed as
determined from a full-sky track of pulsar B1929+10 in October 1992.
The right-hand phase is orthogonal to the plotted left-hand phase within 
the observational errors, which represent one standard deviation.}
\end{figure}
\begin{figure}
\centerline{\epsfig{figure=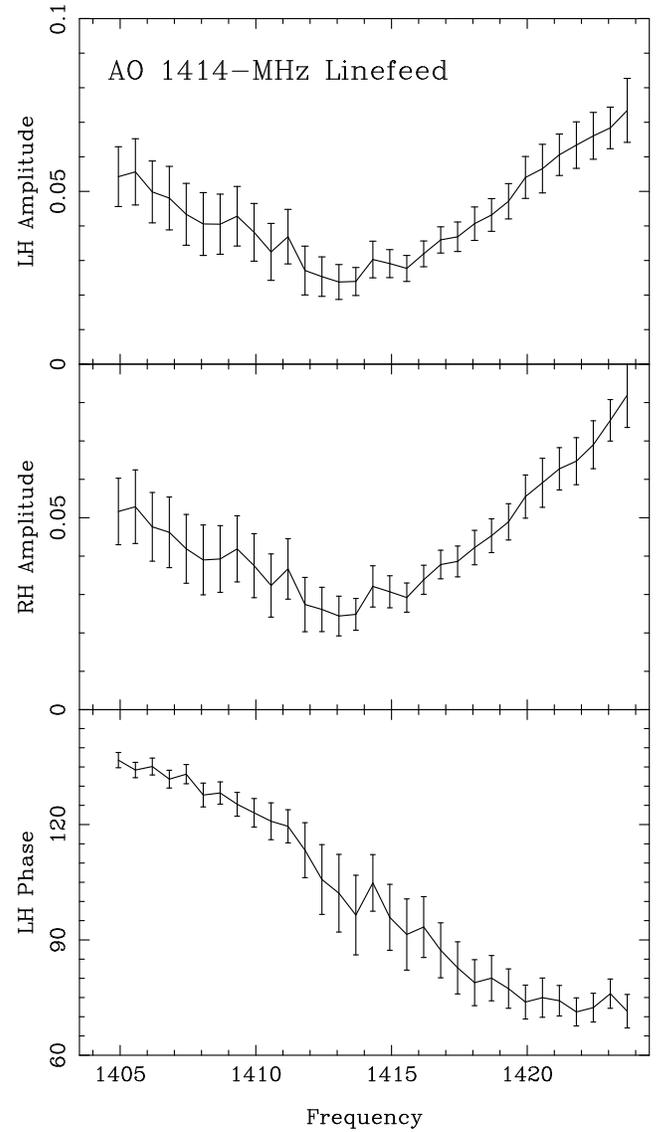,height=15cm}}
\caption{Cross-coupling amplitudes and phase for the Arecibo
Observatory 21-cm line feed, determined from a full-sky track of pulsar
B1929+10 in October 1992, as in Fig. A1.}
\end{figure}

During our October 1992 program observations we obtained several
different sets of full-sky tracks on pulsars B1929+10, B0656+14,
B1737+13 and B0525+21 as well as shorter tracks from which only the
cross-coupling phase could be estimated.  B1929+10 provided the 
most accurate results, and the two feeds appeared stable both from day 
to day in October 1992 as well as over the interval between several 
earlier preliminary observations in 1990 January and 1992 March, apart 
from the overall cross-coupling phase, which we believe results both 
from differences in our cabling setup for each observation as well 
(perhaps) as variations in the relative phase of the two receiver channels 
down from the platform to the control room.  Therefore, we were careful 
to determine the cross-coupling phase for each observation and depend 
on its stability only over a few hours time. 

Apparently, the 21-cm system exhibits a cross-coupling behavior common
to many feed systems in use at various observatories---that is, it exhibits 
minimum cross-coupling amplitudes near the design band center and a 
progress rise in cross-coupled power toward the band edges.  It was then 
surprising to see that the behavior of the 430-MHz system is very different, 
with minimum cross-coupled power near the lower band edge at 425 MHz 
and progressively rising cross-coupled power across the band.  Indeed, 
the level of cross-coupling at the upper band edge is fully 15\%, making it 
absolutely mandatory that observations at this frequency be corrected for 
the resulting distortion.  This also apparently explains a strange 
phenomenon in our older AO polarimetry:  we could not understand for 
many years why 430-MHz observations of B1929+10 in particular 
exhibited large variations in fractional circular polarization, between 
essentially zero and almost 40\%.  The reason we can now see is almost 
certainly that for such a low dispersion pulsar, the fractional circular would 
vary greatly depending upon how the scintillation-produced spectral 
variations excited the feed!  More highly dispersed stars tend to have 
more bright scintiles within our passband, and thus the cross-coupling 
distortion tended to average across them.  

Finally, we used the above cross-coupling information to correct the
measured Stokes parameters whenever possible.  If we write eq.\ (A5) 
of Stinebring \etal\ (1984) as ${\bf S^{\prime}}={\bf M}{\bf S}$, where ${\bf S}$
is the true Stokes vector ${\bf S^{\prime}}$ the measured one, and
${\bf M}$ is the 4$\times$4 M\"uller matrix relating the two, then our
correction of the Stokes parameters took the form 
\begin{equation}
{\bf S}(f)={\bf M^{-1}}[\varepsilon_1(f),\varepsilon_2(f),\psi_1(f),\psi_1(f)
\pm90^\circ]{\bf S^{\prime}}(f)  .
\end{equation}

Despite these careful efforts, our calibrations vary considerably in
quality.  The great bulk of the 430- and 1400-MHz observations were
fully calibrated both in terms of standard-source observations to
determine the relative left- and right-hand channel gains as well as
the cross-coupling distortion corrections outlined above.  However, in
some cases the standard-source pointings were corrupted by
interference, and in a few cases, our shorter trackings to determine
$\psi_1$ were unsuccessful.  We note these faults in 
Tables~\ref{table_polarization} \& \ref{table_polarization_a} above.

\end{document}